\documentclass[runningheads,envcountsame,envcountsect]{Springer_Nature_Lecture_Notes_in_Computer_Science/llncs}
\usepackage[T1]{fontenc}
\usepackage[svgnames]{xcolor}
\usepackage{amsmath}

\ifdefined\CFGKATNoAppendix
  \usepackage[appendix=strip,bibliography=common]{apxproof}
\else
  \usepackage[appendix=append,bibliography=common]{apxproof}
\fi

\usepackage{newtxtext}
\usepackage[varvw]{newtxmath}
\usepackage{graphicx}

\AtBeginDocument{%
  }
\allowdisplaybreaks%
\usepackage{mathpartir}
\usepackage{mathtools}
\usepackage[inline]{enumitem}
\usepackage{algorithm}
\usepackage{hyperref}
\hypersetup{pdfnewwindow=true}
\definecolor{LinkPurple}{cmyk}{0.55,1,0,0.15}
\definecolor{LinkDarkBlue}{cmyk}{1,0.58,0,0.21}
\hypersetup{colorlinks=true,linkcolor=LinkPurple,citecolor=LinkPurple,urlcolor=LinkDarkBlue,filecolor=LinkDarkBlue}
\usepackage{xspace}
\usepackage{cleveref}
\usepackage[noend]{algpseudocode}   
\usepackage[labelfont={bf},labelsep=period,font=small]{caption}
\newsavebox{\colR}

\usepackage{annotate-equations}
\usepackage{tabularx}
\usepackage{bussproofs}
\usepackage{wrapfig}
\usepackage[disable]{todonotes}

\usepackage{tikz}
\usetikzlibrary{positioning,arrows.meta,fit}

\tikzset{
	transOut/.style={
		-Stealth,
		double,
		double distance=1pt,
	},
	transVia/.style={
		-Stealth,
		thick,
	},
}

\newcolumntype{L}{>{\(}l<{\)}}   %
\newcolumntype{C}{>{\(}c<{\)}}   %
\newcolumntype{R}{>{\(}r<{\)}}   %
\newcolumntype{M}{L @{\quad} R @{\;} C @{\;} L} 
\newcolumntype{Y}{>{\centering\arraybackslash}X}

\newtheoremrep{theorem}{Theorem}[section]
\newtheoremrep{lemma}[theorem]{Lemma}
\newtheoremrep{corollary}[theorem]{Corollary}

\newcounter{ehfact}
\crefformat{ehfact}{#2#1#3}
\Crefformat{ehfact}{#2#1#3}
\crefrangeformat{ehfact}{#3#1#4--#5#2#6}
\Crefrangeformat{ehfact}{#3#1#4--#5#2#6}
\crefmultiformat{ehfact}{#2#1#3}{ and #2#1#3}{, #2#1#3}{, and #2#1#3}
\Crefmultiformat{ehfact}{#2#1#3}{ and #2#1#3}{, #2#1#3}{, and #2#1#3}
\newcommand{\lset}{\mathbb{L}\xspace}
\newcommand{\Lab}{\lset}
\newcommand{\acc}{\textbf{acc}\xspace}
\newcommand{\jmpc}{\textbf{jmp}\xspace}
\newcommand{\brkc}{\textbf{brk}\xspace}
\newcommand{\contc}{\textbf{cont}\xspace}
\newcommand{\retc}{\textbf{ret}\xspace}
\newcommand{\command}[1]{\ensuremath{\mathtt{#1}}\xspace}
\newcommand{\comBreak}{\command{break}}

\newcommand{\comCont}{\command{continue}}

\newcommand{\comGoto}{\command{goto}}
\newcommand{\comLabel}{\command{label}}
\newcommand{\comReturn}{\command{return}}

\newcommand{\At}{\ensuremath{\mathsf{At}}\xspace}
\newcommand{\dom}{\ensuremath{\mathrm{dom}}\xspace}
\newcommand{\Expm}{\ensuremath{\mathsf{Exp}}\xspace}
\newcommand{\Exp}{\ensuremath{\Expm_{+}}\xspace}
\newcommand{\Prog}{\ensuremath{\mathsf{Prog}}\xspace}

\newcommand{\BExp}{\ensuremath{\mathsf{BExp}}\xspace}

\newcommand{\tCref}[1]{\mbox{\textup{\Cref{#1}}}}
\newcommand{\CongRulesRef}{\hyperref[cong]{congruence-rules section of \Cref*{ax}}\xspace}
\newcommand{\langC}{\textsc{C}}
\newcommand{\BA}{\ensuremath{\mathsf{BA}}\xspace}
\newcommand{\eqBA}{=_{\BA}}
\newcommand{\neqBA}{\neq_{\BA}}
\newcommand{\leqBA}{\leq_{\BA}}
\newcommand{\seq}{\mathbin{;}}
\newcommand{\sys}{\textit{sys}\xspace}
\newcommand{\sysc}{\textit{sysc}\xspace}
\newcommand{\loopseq}{\vartriangleleft}
\newcommand{\CFGKAT}{\textsf{CF-GKAT}\xspace}
\newcommand{\final}{\ensuremath{\Omega}\xspace}
\newcommand{\finalmorph}{\ensuremath{\omega}\xspace}
\newcommand{\GKAT}{\textsf{GKAT}\xspace}
\newcommand{\KAT}{\ensuremath{\mathsf{KAT}}\xspace}
\newcommand{\NetKAT}{\ensuremath{\mathsf{NetKAT}}\xspace}
\newcommand{\KATb}{\textsf{KAT{+}B!}\xspace}
\newcommand{\resWith}[1]{\mathord{\mkern-2mu\downarrow_{#1}}}

\newcommand{\restrict}[2]{#1\!\upharpoonright_{#2}}
\newcommand{\Ax}[1]{\ensuremath{\mathsf{#1}}\xspace}
\newcommand{\coalg}[1]{%
	#1--\hskip0pt coalgebra%
}
\newcommand{\partialTo}{\rightharpoonup}
\newcommand{\assign}{\ensuremath{\leftarrow}\xspace}
\newcommand{\eqdef}{\ensuremath{\coloneq}\xspace}
\newcommand{\progeq}{\mathrel{\simeq_{\Prog}}}
\newcommand{\finalmap}{\mathord{!}}
\newcommand{\transSilent}{\rightsquigarrow}
\newcommand{\DefineAx}[2]{%
	\stepcounter{#2}%
	\expandafter\edef\csname #1\endcsname{\noexpand\Ax{#2\arabic{#2}}}%
}
\numberwithin{equation}{lemma}
\makeatletter
\newcommand{\RefAx}[1]{%
	\hyperref[\expandafter\@gobble\string#1]{#1}\xspace%
}
\makeatother
\providecommand{\RocqDocBase}{https://github.com/SpencerKoevering/cfgkat-rocq/blob/arxiv-v1}
\DeclareRobustCommand{\RocqIcon}{%
	\texorpdfstring{[\raisebox{-0.12ex}{\includegraphics[height=.75em]{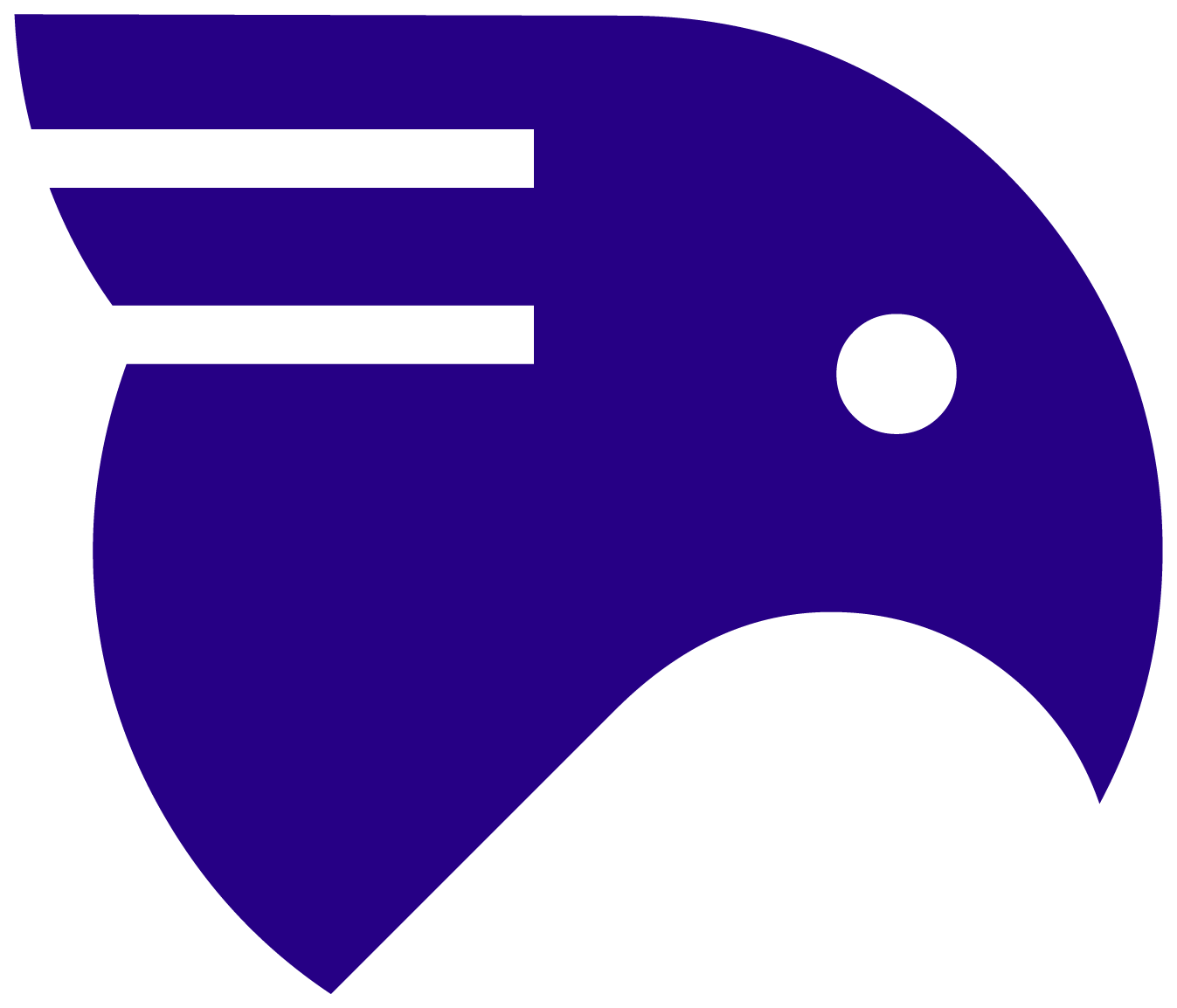}}]}{}%
}
\DeclareRobustCommand{\RocqRef}[1]{%
	\href{\RocqDocBase/#1}{\RocqIcon}%
}
\DeclareRobustCommand{\RocqRefMath}[1]{%
	\quad\text{\href{\RocqDocBase/#1}{\RocqIcon}}%
}
\makeatletter
\newcommand{\RocqDeclare}[3]{%
	\expandafter\def\csname RocqAnchor@#1\endcsname{#3}%
}
\newcommand{\RocqQed}[1]{%
	\ifcsname RocqAnchor@#1\endcsname
		\ifmmode
			\quad\text{\href{\RocqDocBase/\csname RocqAnchor@#1\endcsname}{\RocqIcon}}%
		\else
			\leavevmode\unskip\nobreak\hfill\href{\RocqDocBase/\csname RocqAnchor@#1\endcsname}{\RocqIcon}%
		\fi
	\else
		\PackageError{cfgkat-rocq-links}{Missing Rocq link declaration for #1}{Add \string\RocqDeclare{#1}{...} before using this Rocq link.}%
	\fi
}
\newcommand{\RocqQedRight}[1]{%
	\ifcsname RocqAnchor@#1\endcsname
		\tag*{\href{\RocqDocBase/\csname RocqAnchor@#1\endcsname}{\RocqIcon}}%
	\else
		\PackageError{cfgkat-rocq-links}{Missing Rocq link declaration for #1}{Add \string\RocqDeclare{#1}{...} before using this Rocq link.}%
	\fi
}
\DeclareRobustCommand{\RocqLabelRef}[1]{%
	\ifcsname RocqAnchor@#1\endcsname
		\RocqRef{\csname RocqAnchor@#1\endcsname}%
	\else
		\PackageError{cfgkat-rocq-links}{Missing Rocq link declaration for #1}{Add \string\RocqDeclare{#1}{...} before using this Rocq link.}%
	\fi
}
\DeclareRobustCommand{\RocqLabelRefMath}[1]{%
	\ifcsname RocqAnchor@#1\endcsname
		\RocqRefMath{\csname RocqAnchor@#1\endcsname}%
	\else
		\PackageError{cfgkat-rocq-links}{Missing Rocq link declaration for #1}{Add \string\RocqDeclare{#1}{...} before using this Rocq link.}%
	\fi
}
\DeclareRobustCommand{\RocqLinkedLabelText}[2]{%
	\ifcsname RocqAnchor@#1\endcsname
		\href{\RocqDocBase/\csname RocqAnchor@#1\endcsname}{#2}%
	\else
		\PackageError{cfgkat-rocq-links}{Missing Rocq link declaration for #1}{Add \string\RocqDeclare{#1}{...} before using this Rocq link.}%
	\fi
}
\makeatother
\RocqDeclare{Axindexclusivity}{BAIndicatorExclusivity}{soundness/BooleanAlgebraSoundness.v\#L83}
\RocqDeclare{Axindsum}{BAIndicatorSum}{soundness/BooleanAlgebraSoundness.v\#L87}
\RocqDeclare{ax}{CFGKATEquiv}{syntax/Axioms.v\#L15}

\RocqDeclare{Axidempotence}{EqG1}{syntax/Axioms.v\#L34}
\RocqDeclare{Axskewcomm}{EqG2}{syntax/Axioms.v\#L37}
\RocqDeclare{Axskewassoc}{EqG3}{syntax/Axioms.v\#L40}
\RocqDeclare{Axguardedness}{EqG4}{syntax/Axioms.v\#L44}
\RocqDeclare{Axrightdistr}{EqG5}{syntax/Axioms.v\#L47}

\RocqDeclare{Axassoc}{EqS1}{syntax/Axioms.v\#L51}
\RocqDeclare{Axabsleft}{EqS2}{syntax/Axioms.v\#L54}
\RocqDeclare{Axneutleft}{EqS3}{syntax/Axioms.v\#L57}
\RocqDeclare{Axneutright}{EqS4}{syntax/Axioms.v\#L60}
\RocqDeclare{Axtestcomb}{EqS5}{syntax/Axioms.v\#L63}
\RocqDeclare{Axrightseqannret}{EqS6}{syntax/Axioms.v\#L66}
\RocqDeclare{Axrightseqanncont}{EqS7}{syntax/Axioms.v\#L69}
\RocqDeclare{Axrightseqannbrk}{EqS8}{syntax/Axioms.v\#L72}
\RocqDeclare{Axrightseqanngoto}{EqS9}{syntax/Axioms.v\#L75}

\RocqDeclare{Axneutleftloop}{EqL1}{syntax/Axioms.v\#L79}
\RocqDeclare{Axrightloopannabort}{EqL2}{syntax/Axioms.v\#L82}
\RocqDeclare{Axrightloopannret}{EqL3}{syntax/Axioms.v\#L85}
\RocqDeclare{Axrightloopanngoto}{EqL4}{syntax/Axioms.v\#L88}
\RocqDeclare{Axbreak}{EqL5}{syntax/Axioms.v\#L91}
\RocqDeclare{Axcontinue}{EqL6}{syntax/Axioms.v\#L94}
\RocqDeclare{Axloopassocact}{EqL7}{syntax/Axioms.v\#L97}
\RocqDeclare{Axloopassocassign}{EqL8}{syntax/Axioms.v\#L100}
\RocqDeclare{Axloopdistributivity}{EqL9}{syntax/Axioms.v\#L103}
\RocqDeclare{Axloopunfold}{EqL10}{syntax/Axioms.v\#L106}
\RocqDeclare{Axlooptighten}{EqL11}{syntax/Axioms.v\#L109}

\RocqDeclare{Axindcommact}{EqA1}{syntax/Axioms.v\#L114}
\RocqDeclare{Axindval}{EqA2}{syntax/Axioms.v\#L117}
\RocqDeclare{Axindred}{EqA3}{syntax/Axioms.v\#L120}
\RocqDeclare{Axindannih}{EqA4}{syntax/Axioms.v\#L123}
\RocqDeclare{Axindret}{EqA5}{syntax/Axioms.v\#L126}
\RocqDeclare{Axinddistr}{EqA6}{syntax/Axioms.v\#L129}
\RocqDeclare{Axindabort}{EqA7}{syntax/Axioms.v\#L133}
\RocqDeclare{Axindindcomm}{EqA8}{syntax/Axioms.v\#L136}
\RocqDeclare{Axindinddistr}{EqA9}{syntax/Axioms.v\#L140}

\RocqDeclare{Axweaken}{EqC1}{syntax/Axioms.v\#L146}
\RocqDeclare{Axreplacement}{EqC2}{syntax/Axioms.v\#L150}
\RocqDeclare{Axredundancy}{EqC3}{syntax/Axioms.v\#L154}
\RocqDeclare{Axrenaming}{EqC4}{syntax/Axioms.v\#L161}
\RocqDeclare{Axsubstitution}{EqC5}{syntax/Axioms.v\#L166}
\RocqDeclare{Axunprbra}{EqC6}{syntax/Axioms.v\#L170}
\RocqDeclare{Axuniqueness}{EqC7}{syntax/Axioms.v\#L175}

\RocqDeclare{Axcongguardedchoice}{EqChoiceCong}{syntax/Axioms.v\#L187}
\RocqDeclare{Axcongboolguardedchoice}{EqBoolChoiceCong}{syntax/Axioms.v\#L192}
\RocqDeclare{Axcongboolassertion}{EqBoolAssertCong}{syntax/Axioms.v\#L196}
\RocqDeclare{Axcongboolloop}{EqBoolLoopCong}{syntax/Axioms.v\#L200}
\RocqDeclare{Axcongrightseq}{EqSeqRightCong}{syntax/Axioms.v\#L204}
\RocqDeclare{Axcongrightloopseq}{EqLoopSeqRightCong}{syntax/Axioms.v\#L209}
\RocqDeclare{Axcongleftseqempty}{EqSeqLeftEmptyCong}{syntax/Axioms.v\#L214}
\RocqDeclare{Axcongleftloopseqempty}{EqLoopSeqLeftEmptyCong}{syntax/Axioms.v\#L218}
\RocqDeclare{Axcongloopbodyempty}{EqLoopBodyEmptyCong}{syntax/Axioms.v\#L222}
\RocqDeclare{Axcongloopbody}{EqLoopBodyCong}{syntax/Axioms.v\#L226}

\RocqDeclare{Fcoalg}{CFGKATCoalgebra}{completeness/CFGKATCoalgebra.v\#L76}
\RocqDeclare{hom}{CFGKATCoalgebraHomomorphism}{completeness/CFGKATCoalgebra.v\#L79}
\RocqDeclare{bisimulation}{CFGKATAutomatonBisimulation}{semantics/Bisimulation.v\#L39}
\RocqDeclare{syntax}{CFGKATProgramSyntax}{semantics/LabelExtraction.v\#L7}
\RocqDeclare{labelextraction}{label_extraction}{semantics/LabelExtraction.v\#L198}
\RocqDeclare{Booleansemantics}{eval_test}{syntax/Preliminaries.v\#L331}
\RocqDeclare{operationalsemantics}{CFGKATExpressionDynamics}{semantics/ExpressionSemantics.v\#L361}
\RocqDeclare{resolutiondef}{ResolveCFGKATAutomatonDynamics}{semantics/AutomataResolution.v\#L107}
\RocqDeclare{silentTrans}{CFGKATSilentTransition}{semantics/Bisimulation.v\#L745}
\RocqDeclare{rejectingsetdef}{cfgkat_rejecting_set}{semantics/Bisimulation.v\#L778}
\RocqDeclare{smallerderivation}{CFGKATSyntacticSilentTransition_finite_smaller}{semantics/Bisimulation.v\#L1458}
\RocqDeclare{rejectingsetaborts}{rejecting_set_resolves_reject}{semantics/Bisimulation.v\#L1533}
\RocqDeclare{strongbisim}{CFGKATStrongBisimulation}{semantics/Bisimulation.v\#L818}
\RocqDeclare{strongbisimtobisim}{strong_bisim_to_bisim}{semantics/Bisimulation.v\#L1834}
\RocqDeclare{supsyssolsubsyssynt}{sup_sys_sol_sub_sys_synt_bisim}{soundness/ContextualAxioms.v\#L472}
\RocqDeclare{Booleansoundness}{BooleanAlgebra_sound}{soundness/BooleanAlgebraSoundness.v\#L152}
\RocqDeclare{synsub}{subst_goto}{soundness/ContextualDefinitions.v\#L114}
\RocqDeclare{rightmostdef}{RHS}{soundness/ContextualDefinitions.v\#L158}
\RocqDeclare{guardedsemdef}{guarded}{soundness/ContextualDefinitions.v\#L298}
\RocqDeclare{notransitionnogoto}{no_goto_expr_next}{soundness/ContextualAxioms.v\#L337}
\RocqDeclare{def:relation-lifting}{lifted_relation}{semantics/Bisimulation.v\#L17}
\RocqDeclare{thm:substitute-preserve-non-jump}{C5_simulation}{soundness/ContextualAxioms.v\#L1208}
\RocqDeclare{indnotrefbool}{test_indicator_free_update}{soundness/AssignmentLaws.v\#L140}
\RocqDeclare{renamingsemlem}{rename_goto_dynamics}{soundness/ContextualAxioms.v\#L1585}
\RocqDeclare{common-suffix-lifting}{C1_bisim}{soundness/ContextualAxioms.v\#L584}
\RocqDeclare{thm:sound}{CFGKATEquiv_sound}{soundness/Soundness.v\#L73}
\RocqDeclare{sctransform-derivation}{short_circuit_transform}{examples/ShortCircuitTransform.v\#L291}
\RocqDeclare{complassign/test}{complete_assignment}{completeness/FundamentalTheorem.v\#L56}
\RocqDeclare{atomsum}{atom_partition}{completeness/BooleanAlgebraCompleteness.v\#L675}
\RocqDeclare{atomsdisagree}{atom_to_tests_disagree}{completeness/BooleanAlgebraCompleteness.v\#L1159}
\RocqDeclare{indboollem}{complete_test_partition_aux}{completeness/BooleanAlgebraCompleteness.v\#L715}
\RocqDeclare{inddisagree}{complete_tests_disagree}{completeness/BooleanAlgebraCompleteness.v\#L1056}
\RocqDeclare{boolnormleqlem}{minterm_test_reduce}{completeness/BooleanAlgebraCompleteness.v\#L1355}
\RocqDeclare{Booleannormalform}{boolean_dnf_BA}{completeness/BooleanAlgebraCompleteness.v\#L1460}
\RocqDeclare{atomscorr}{BA_atom_iff_AtIY_test}{completeness/BooleanAlgebraCompleteness.v\#L1477}
\RocqDeclare{Booleancompleteness}{Boolean_denotation_completeness}{completeness/BooleanAlgebraCompleteness.v\#L1534}
\RocqDeclare{ggdedsum}{guarded_sum_all}{completeness/FundamentalTheorem.v\#L213}
\RocqDeclare{gensumwd}{guarded_sum_well_defined}{completeness/FundamentalTheorem.v\#L1098}
\RocqDeclare{biplusboollem}{biplusboollem}{completeness/FundamentalTheorem.v\#L1432}
\RocqDeclare{branchsel}{branchsel}{completeness/FundamentalTheorem.v\#L395}
\RocqDeclare{testdistr}{testdistr}{completeness/FundamentalTheorem.v\#L462}
\RocqDeclare{antiskewcom}{antiskewcom}{completeness/FundamentalTheorem.v\#L545}
\RocqDeclare{neutralright}{neutralright}{completeness/FundamentalTheorem.v\#L331}
\RocqDeclare{loopassoctest}{loopassoctest}{completeness/FundamentalTheorem.v\#L576}
\RocqDeclare{loopseqisseqtest}{loopseqisseqtest}{completeness/FundamentalTheorem.v\#L589}
\RocqDeclare{loopseqisseqasasign}{loopseqisseqassign_on}{completeness/FundamentalTheorem.v\#L284}
\RocqDeclare{loopseqisseqact}{loopseqisseqact}{completeness/FundamentalTheorem.v\#L273}
\RocqDeclare{inddistr}{inddistr}{completeness/FundamentalTheorem.v\#L601}
\RocqDeclare{indcommtest}{indcommtest}{completeness/FundamentalTheorem.v\#L647}
\RocqDeclare{idem+}{idem_plus}{completeness/FundamentalTheorem.v\#L1453}
\RocqDeclare{indval+}{indval_plus_on}{completeness/FundamentalTheorem.v\#L957}
\RocqDeclare{indred+}{indred_plus_on}{completeness/FundamentalTheorem.v\#L1016}
\RocqDeclare{indannih+}{indannih_plus_on}{completeness/FundamentalTheorem.v\#L802}
\RocqDeclare{guardedness+}{guardedness_plus_on}{completeness/FundamentalTheorem.v\#L1074}
\RocqDeclare{loopassocassign+}{loopassocassign_plus_on}{completeness/FundamentalTheorem.v\#L698}
\RocqDeclare{testdistr+}{testdistr_plus_on}{completeness/FundamentalTheorem.v\#L1321}
\RocqDeclare{branchsel+}{branchsel_plus_on}{completeness/FundamentalTheorem.v\#L1361}
\RocqDeclare{sys}{sys}{completeness/FundamentalTheorem.v\#L1466}
\RocqDeclare{sysannih}{sysannih}{completeness/FundamentalTheorem.v\#L1490}
\RocqDeclare{fundthm}{fundamental_theorem}{completeness/FundamentalTheorem.v\#L2094}
\RocqDeclare{acc-cont-loop-invariant}{acc_cont_loop_invariant}{semantics/TransitionFunction.v\#L520}
\RocqDeclare{L11lemmaFTloop}{L11lemmaFTloop_largest}{completeness/FundamentalTheorem.v\#L1987}
\RocqDeclare{ggdedsumprodinj}{ggdedsumprodinj}{completeness/FundamentalTheorem.v\#L1262}
\RocqDeclare{iteratedloweringlem}{ResolveCFGKATAutomatonDynamicsBoundJump}{semantics/AutomataResolution.v\#L279}
\RocqDeclare{exprsubcofinite}{expr_raw_nexts}{completeness/gtnflem2.v\#L174}
\RocqDeclare{gtnflem1}{gtnflem1}{completeness/gtnflem1.v\#L3602}
\RocqDeclare{syscYdef}{sysc_Y}{completeness/gtnfthm.v\#L40}
\RocqDeclare{syscdef}{sysc}{completeness/gtnflem2.v\#L528}
\RocqDeclare{gtnflem2}{gtnflem2}{completeness/gtnflem2.v\#L2013}
\RocqDeclare{gtnfthm}{gtnfthm}{completeness/gtnfthm.v\#L2036}
\RocqDeclare{gtnfcor}{gtnfcor}{completeness/gtnfthm.v\#L2202}
\RocqDeclare{canonicalenv}{canonical_env}{completeness/CFGKATCoalgebra.v\#L112}
\RocqDeclare{hompressol}{hompressol}{completeness/Completeness.v\#L1827}
\RocqDeclare{ebotsemenv}{ebotsemenv}{completeness/Completeness.v\#L70}
\RocqDeclare{completeness}{completeness}{completeness/Completeness.v\#L2274}
\RocqDeclare{semanticlemmaindrem}{semanticlemmaindrem}{examples/GhostVariableElimination.v\#L156}
\RocqDeclare{wellnesteddef}{well_nested}{examples/GhostVariableElimination.v\#L441}
\RocqDeclare{nestinglem}{nestinglem}{examples/GhostVariableElimination.v\#L673}
\RocqDeclare{indremoval}{ghost_variable_elimination}{examples/GhostVariableElimination.v\#L951}

\RocqDeclare{looptighten_invariant}{looptighten_invariant}{examples/UsefulFacts.v\#L102}
\RocqDeclare{loopuniqueafterlargest}{loop_unique_closed_after_largest_invariant_exit_pair}{examples/UsefulFacts.v\#L797}
\RocqDeclare{loopseqdnr}{indpushloopseq}{examples/UsefulFacts.v\#L1322}
\RocqDeclare{assignmentduplication}{assignment_duplication}{examples/UsefulFacts.v\#L2451}
\RocqDeclare{partialselection}{partialselection}{examples/UsefulFacts.v\#L2579}
\RocqDeclare{assignmentbranchselectiontrue}{assign_selects_implied_branch}{examples/UsefulFacts.v\#L2612}
\RocqDeclare{assignmentbranchselectionfalse}{assign_skips_disjoint_branch}{examples/UsefulFacts.v\#L2630}
\RocqDeclare{assignmentprefixenvironment}{assignment_prefix_environment}{examples/UsefulFacts.v\#L3441}
\RocqDeclare{iteratedenvironmentreplacement}{iterated_environment_replacement}{examples/UsefulFacts.v\#L3617}

\RocqDeclare{ehzeroentryloopequiv}{ehzeroentryloopequiv}{examples/ErosaHendrenLoopCase.v\#L1351}
\RocqDeclare{ehmeetinthemiddle}{ehmeetinthemiddle}{examples/ErosaHendrenLoopCase.v\#L2196}
\RocqDeclare{erosa-hendren-loop-case}{erosa_hendren_loop_case}{examples/ErosaHendrenLoopCase.v\#L3135}

\DefineAx{Axindexclusivity}{B}
\DefineAx{Axindsum}{B}

\DefineAx{Axidempotence}{G}
\DefineAx{Axskewcomm}{G}
\DefineAx{Axskewassoc}{G}
\DefineAx{Axguardedness}{G}
\DefineAx{Axrightdistr}{G}

\DefineAx{Axassoc}{S}
\DefineAx{Axabsleft}{S}
\DefineAx{Axneutleft}{S}
\DefineAx{Axneutright}{S}
\DefineAx{Axtestcomb}{S}
\DefineAx{Axrightseqannret}{S}
\DefineAx{Axrightseqanncont}{S}
\DefineAx{Axrightseqannbrk}{S}
\DefineAx{Axrightseqanngoto}{S}

\DefineAx{Axneutleftloop}{L}
\DefineAx{Axrightloopannabort}{L}
\DefineAx{Axrightloopannret}{L}
\DefineAx{Axrightloopanngoto}{L}
\DefineAx{Axbreak}{L}
\DefineAx{Axcontinue}{L}
\DefineAx{Axloopassocact}{L}
\DefineAx{Axloopassocassign}{L}
\DefineAx{Axloopdistributivity}{L}
\DefineAx{Axloopunfold}{L}
\DefineAx{Axlooptighten}{L}

\DefineAx{Axindcommact}{A}
\DefineAx{Axindval}{A}
\DefineAx{Axindred}{A}
\DefineAx{Axindannih}{A}
\DefineAx{Axindret}{A}
\DefineAx{Axinddistr}{A}
\DefineAx{Axindabort}{A}
\DefineAx{Axindindcomm}{A}
\DefineAx{Axindinddistr}{A}

\DefineAx{Axweaken}{C}
\DefineAx{Axreplacement}{C}
\DefineAx{Axredundancy}{C}
\DefineAx{Axrenaming}{C}
\DefineAx{Axsubstitution}{C}
\DefineAx{Axunprbra}{C}
\DefineAx{Axuniqueness}{C}

\DefineAx{Axcongrightseq}{CG}
\DefineAx{Axcongrightloopseq}{CG}
\DefineAx{Axcongloopbody}{CG}
\DefineAx{Axcongguardedchoice}{CG}
\DefineAx{Axcongboolguardedchoice}{CG}
\DefineAx{Axcongboolassertion}{CG}
\DefineAx{Axcongboolloop}{CG}
\DefineAx{Axcongleftseqempty}{CG}
\DefineAx{Axcongleftloopseqempty}{CG}
\DefineAx{Axcongloopbodyempty}{CG}

\providecommand{\Axindexclusivity}{}
\providecommand{\Axindsum}{}

\providecommand{\Axidempotence}{}
\providecommand{\Axskewcomm}{}
\providecommand{\Axskewassoc}{}
\providecommand{\Axguardedness}{}
\providecommand{\Axrightdistr}{}

\providecommand{\Axassoc}{}
\providecommand{\Axabsleft}{}
\providecommand{\Axneutleft}{}
\providecommand{\Axneutright}{}
\providecommand{\Axneutleftloop}{}
\providecommand{\Axrightloopannabort}{}
\providecommand{\Axtestcomb}{}
\providecommand{\Axrightseqannret}{}
\providecommand{\Axrightseqanncont}{}
\providecommand{\Axrightseqannbrk}{}
\providecommand{\Axrightseqanngoto}{}
\providecommand{\Axrightloopannret}{}
\providecommand{\Axrightloopanngoto}{}
\providecommand{\Axbreak}{}
\providecommand{\Axcontinue}{}

\providecommand{\Axloopassocact}{}
\providecommand{\Axloopassocassign}{}
\providecommand{\Axloopdistributivity}{}
\providecommand{\Axloopunfold}{}
\providecommand{\Axlooptighten}{}

\providecommand{\Axinddistr}{}
\providecommand{\Axindindcomm}{}
\providecommand{\Axindcommact}{}
\providecommand{\Axindval}{}
\providecommand{\Axindred}{}
\providecommand{\Axindannih}{}
\providecommand{\Axindret}{}
\providecommand{\Axindabort}{}
\providecommand{\Axindinddistr}{}

\providecommand{\Axweaken}{}
\providecommand{\Axredundancy}{}
\providecommand{\Axreplacement}{}
\providecommand{\Axsubstitution}{}
\providecommand{\Axrenaming}{}
\providecommand{\Axunprbra}{}
\providecommand{\Axuniqueness}{}

\providecommand{\Axcongrightseq}{}
\providecommand{\Axcongrightloopseq}{}
\providecommand{\Axcongloopbody}{}
\providecommand{\Axcongguardedchoice}{}
\providecommand{\Axcongboolguardedchoice}{}
\providecommand{\Axcongboolassertion}{}
\providecommand{\Axcongboolloop}{}
\providecommand{\Axcongleftseqempty}{}
\providecommand{\Axcongleftloopseqempty}{}
\providecommand{\Axcongloopbodyempty}{}

\makeatletter
\renewcommand\tagform@[1]{\maketag@@@{\ignorespaces#1\unskip\@@italiccorr}}
\makeatother

\creflabelformat{equation}{#2\textup{#1}#3}
\newcommand{\tr}[3][]{%
	\allowbreak
	\mathrel{%
		\ifthenelse{\equal{#2}{}}%
		{\xrightarrow{#2}_{#3}^{#1}}%
		{\xrightarrow{#2}_{#3}^{\raisebox{-1ex}{\scriptsize\(#1\)}}}%
	}%
}

\makeatletter
\newcommand{\bigplus}{%
	\DOTSB\mathop{\mathpalette\mattos@bigplus\relax}\slimits@%
}
\newcommand\mattos@bigplus[2]{%
	\vcenter{\hbox{%
			\sbox\z@{\(#1\sum\)}%
			\resizebox{!}{0.9\dimexpr\ht\z@+\dp\z@}{\raisebox{\depth}{\(\m@th#1+\)}}%
	}}%
	\vphantom{\sum}%
}
\makeatother

\makeatletter
\newcommand{\figmathpartight}{%
	\let\MathparLineskip\mpr@lesslineskip
	\setlength{\abovedisplayskip}{0.6ex}%
	\setlength{\belowdisplayskip}{0.6ex}%
	\setlength{\abovedisplayshortskip}{0.6ex}%
	\setlength{\belowdisplayshortskip}{0.6ex}}
\makeatother

\newif\ifanonymous
\anonymousfalse

\begin{document}
\title{\CFGKAT: A Complete Calculus \texorpdfstring{\\}{} for Non-Local Control Flow}
\titlerunning{\CFGKAT}

\ifanonymous
\author{Anonymous Author(s)}
\authorrunning{Anonymous Author(s)}
\institute{Anonymous Institution(s)}
\else
\author{Spencer Van Koevering\inst{1}\orcidID{0009-0008-5026-8060} \and
Alexandra Silva\inst{1}\orcidID{0000-0001-5014-9784} \and
Cheng Zhang\inst{2}\orcidID{0000-0002-8197-6181} \and
Tobias Kapp\'{e}\inst{3}\orcidID{0000-0002-6068-880X}}
\authorrunning{S. Van Koevering et al.}
\institute{Cornell University, Ithaca, NY, USA\\
\email{\{sv493,alexandra.silva\}@cornell.edu} \and
Worcester Polytechnic Institute, Worcester, MA, USA\\
\email{czhang13@wpi.edu} \and
Leiden University, Leiden, The Netherlands\\
\email{t.w.j.kappe@liacs.leidenuniv.nl}}
\fi

\maketitle

\begin{abstract}

\CFGKAT is an extension of guarded Kleene Algebra with Tests (\GKAT) that adds non-local control flow (e.g., \comBreak, \comGoto) while maintaining an efficient decision procedure. An axiomatization of equivalence for \CFGKAT presents unique barriers created by the presence of non-local control flow and contextual semantics of \comGoto which compromises compositional reasoning. 
In this paper, we overcome these barriers and present a \emph{sound and complete} axiomatization of program equivalence in \CFGKAT.
We verify the loop case of the Erosa--Hendren \comGoto\ elimination algorithm, a difficult syntactic metatheorem which requires reasoning about fixpoints, control-flow contexts, and mutable state.
We mechanize the metatheory and case study in Rocq.
\end{abstract}

\begin{toappendix}
	\section{Coalgebra}\label{app:coalgebra}
	We review some relevant definitions and facts from universal coalgebra \cite{universalcoalg}. We assume that the reader has some familiarity with category theory.
	\begin{definition}[{\cite{universalcoalg}} \RocqLabelRef{Fcoalg}]\label{Fcoalg}
		Given a functor \(F\), an \coalg{\ensuremath{F}} is a pair \((X, \chi)\) of a set \(X\) and a function \(\chi: X \to FX\). The set \(X\) is called the carrier, or set of states, and the function \(\chi\) is called the transition function.
	\end{definition}
	\begin{definition}[{\cite{universalcoalg}} \RocqLabelRef{hom}]\label{hom}
		Let \((X,\gamma)\) and \((Y,\delta)\) be \coalg{F}s. 
		A \emph{coalgebra homomorphism} from \((X,\gamma)\) to \((Y,\delta)\) is a morphism
		\(f : X \to Y\) such that
		\[
		\delta \circ f \;=\; Ff \circ \gamma.
		\]
	\end{definition}
	\begin{definition}[{\cite{universalcoalg}}]\label{bisimulation}
		Let \((X, \gamma)\) and \((Y, \delta)\) be \coalg{F}s. 
		A \emph{bisimulation} between \(X\) and \(Y\) is a relation
		\[
		R \subseteq X \times Y
		\]
		such that whenever \((x,y) \in R\), the transitions of \(x\) and \(y\) are related. That is, \(\exists m \in F(R)\) such that \(F(\pi_1)(m) = \gamma(x)\) and \(F(\pi_2)(m) = \delta(y)\).
	\end{definition}
	\begin{remark}
		A bisimulation can be equivalently defined as a coalgebra \((X \times Y, \zeta)\) such that \(\pi_1:X\times Y \to X\) and \(\pi_2:X\times Y \to Y\) are coalgebra homomorphisms \cite{universalcoalg}.
	\end{remark}
	\begin{lemma}\label{Fpreserveswp}
		The functor \(F = (K + \Sigma \times \text{Id})^\At\) preserves weak pullbacks.
	\end{lemma}
	\begin{proof}
		Follows immediately from \cite[Example~4.10; Lemma~4.11]{gummcoalgweaklimit}.
	\end{proof}
	\begin{lemma}
		There is a coalgebra \((\final, \finalmorph)\) which is final in \(\mathit{Coalg}_F\) for \(F = (K + \Sigma \times \text{Id})^\At\).
	\end{lemma}
	\begin{proof} 
		Follows immediately from \cite[Theorem 10.1]{universalcoalg}.
	\end{proof}
	\begin{lemma}[Coinduction {\cite[Theorem 3.4.1]{jacobs2017introduction}}]\label{coinduction}
		\(s,\tau \in \final\) are bisimilar if and only if \(s=\tau\).
	\end{lemma}
\end{toappendix}

\newcommand{\algSCTarget}{\begin{algorithm}[H]\caption{Source}
			\begin{algorithmic}
				\If{\((a \land b) \lor c \)}
				 \(\hat e\)
				\textbf{else}
				\(\hat f\)
				\EndIf
			\end{algorithmic}
\end{algorithm}}
\newcommand{\algSCTransformed}{\begin{algorithm}[H]\caption{Target}
	\begin{algorithmic}
				\If{\(a\)}
				\If{\(b\)}
				\State \comLabel \(\ell\)
				\State \(\hat e\)
				\Else
				\State \comLabel k
				\If{\(c\)}
				 \comGoto~\(\ell\)
				\textbf{else } \(\hat f\)
				\EndIf
				\EndIf
				\Else
				\State \comGoto~k
				\EndIf
			\end{algorithmic}
\end{algorithm}}
\newcommand{\algSource}{\begin{algorithm}[H]\caption{Source}
  \begin{algorithmic}
        \If{\(c\)}
          \State \(\comGoto~\ell\)
        \EndIf
        \State \(\hat e\)
        \While{\(b\)}
          \State \(\hat f\)
          \State \(\comLabel~\ell\)
          \State \(\hat g\)
        \EndWhile
      \end{algorithmic}
\end{algorithm}}
\newcommand{\algTransformed}{\begin{algorithm}[H]\caption{Target}
      \begin{algorithmic}
		\If{\(c\)}
		  \State \(x \gets 1\)
		\Else
		  \State \(x \gets 0\)
		\EndIf
        \If{\(x=0\)}
          \State \(\hat e\)
        \EndIf
        \While{\(b \lor {x = 1}\)}
          \If{\({x = 1}\)}
            \State \(\comGoto~\ell\)
          \EndIf
          \State \(\hat f\)
          \State \(\comLabel~\ell\)
          \State \(x \gets {0}\)
          \State \(\hat g\)
        \EndWhile
      \end{algorithmic}
\end{algorithm}}

\section{Introduction}
Control flow transformations are a common tool to improve code efficiency~\cite{jumpthreading}, locality~\cite{10.1145/197405.197406}, or brevity~\cite{nomoregoto}.
As these transformations are often applied in complex contexts, it is important to have verification methods, and different reasoning frameworks have been proposed, ranging from translation validation~\cite{DBLP:conf/tacas/PnueliSS98,10.1145/358438.349314,647482.728262} to verified compilation~\cite{10.1145/1538788.1538814}.
One prominent such framework is based on Kleene Algebra with Tests (\KAT)~\cite{kozenkat97}, in which many transformations of control flow, such as loop unrolling and hoisting, can be verified algebraically and automated using automata-based methods~\cite{kozenkat97,kozenbohmjacopini,kozen_coalgebraic_2017}.

\begin{wrapfigure}{r}{0.35\textwidth}  %
    \vspace{-18pt}
    \begin{minipage}{.35\textwidth}
        \setlength{\intextsep}{0pt}%
        \algSCTarget
        \vspace{2pt}
        \algSCTransformed
    \end{minipage}
    \caption{A Short-Circuit Transform}\label{sctransform}
    \vspace{-18pt}
\end{wrapfigure}
A key feature of \KAT is that by abstracting away the concrete meaning of program actions and tests, it enables lightweight compositional reasoning about equivalences that depend purely on a program's control flow, i.e., the actions that take place, their order, and the conditions that enable them.
More recently,~\cite{gkat} studied Guarded \KAT (\GKAT), the fragment of \KAT that focuses on structured control flow, i.e., traditional programs with traditional \textbf{while} loops, and showed that --- in contrast with \KAT --- equivalence can be verified rather efficiently.
Moreover, \GKAT comes with a set of laws that encode facts about structured control flow that are intuitive to many programmers (e.g., $\text{\textbf{if }} b \text{\textbf{ then }} p \text{\textbf{ else }} q \equiv \text{\textbf{if not }} b \text{\textbf{ then }} q \text{\textbf{ else }} p$), which makes it amenable to axiomatic reasoning.

There are however important transformations that involve \emph{non-local control flow}, caused by for instance \comGoto or \comBreak, that cannot be captured by \GKAT.
As a first example of such a transformation, consider \Cref{sctransform}, where the \textbf{\textit{short-circuit transform}} is applied to a branch with a nested condition, a common transformation applied by a compiler to avoid having to fully evaluate the condition.
This is the standard behavior of conditionals in \langC~\cite[\S5.1.2.4.4]{clang}.
Although this transformation might look innocent, it can still be tricky to verify.%
\footnote{%
    In fact, the paper from which this example is adapted contains a bug:~\cite[Fig.~6c]{transformationspaper} is wired incorrectly, with all true branches wired into the next conditional and the false branches wired into the overall false branch, rather than the false branch of either conjunct wired to the disjunct and the true branch of the conjunction wired to acceptance.
}
For a more complex transformation, consider \Cref{fig:eh-loop-transform}, where a \comGoto statement is pushed inside a loop using an auxiliary variable $x$.
This transformation is part of an algorithm to eliminate \comGoto statements~\cite{taming}.

To enable reasoning about non-local control flow, Zhang et al.~\cite{cfgkat} introduced \CFGKAT, an extension of \GKAT with \comGoto, \comBreak, \comReturn, and a limited form of variables.
\CFGKAT admits an efficient decision procedure for equivalence, which allows it to \emph{validate} concrete instances of the short-circuit transformation.
But it is far from obvious how to extend the axiomatization of \GKAT to the richer language of \CFGKAT.
Such an axiomatization would allow one to \emph{verify} the short-circuit transformation in general, or even equivalences predicated on other equivalences.

To better understand the challenges that underpin axiomatic reasoning in the presence of non-local control flow, consider the programs \((\comLabel~\ell)\seq p\) and \((\comLabel~k)\seq p\).
One might expect these to be equivalent, since labels have no observable behavior.
However, if we proceed on that basis, then swapping out constituent parts of a program for equivalent programs (a property known as \emph{congruence}) would be unsound --- as it would imply
\[
\text{\textbf{if }} b \text{\textbf{ then }} (\comLabel~\ell) \seq p \text{\textbf{ else }} \comGoto~\ell
\;\progeq\;
\text{\textbf{if }} b \text{\textbf{ then }} (\comLabel~k) \seq p \text{\textbf{ else }} \comGoto~\ell.
\]%
The loss of congruence is a serious problem, as it prevents ``zooming in'' on relevant parts of the program, and reasoning locally.

 We address this by ornamenting equivalences between programs with \emph{environments}, which contain information about where labels point to.
The axiomatic system that arises as a result largely preserves the algebraic reasoning style of \GKAT.
A fortunate side-effect of this design choice is that we gain a compact way to represent recursive systems of equations.
This allows us to phrase a powerful rule stating uniqueness of solutions, which comes in useful when proving properties of loops, \comGoto, or even programs that involve both.
Indeed, we establish that our axioms are \emph{complete}, i.e., they are powerful enough to prove all valid equivalences, while avoiding the infinitary axiom scheme that is traditionally required for completeness results in \GKAT~\cite{gkat,gkatcoequations,rozowski2023probabilistic,van2025weighted}.

Our work also fills a gap in existing literature, by joining together different lines of work surrounding \KAT, and answering questions raised there.
Almost two decades ago, Kozen and Tseng~\cite{kozenbohmjacopini} showed that traditional (while-based) deterministic control flow could not capture all deterministic behavior.
Concurrently, Kozen~\cite{kozennonlocal} showed that \comGoto~and multi-level \comBreak can be simulated in \KAT by leveraging its primitives for non-deterministic composition.
However, several questions were left open for further work, including
\begin{enumerate*}[label={(Q\arabic*)}]
    \item\label{qderivative}whether non-local control-flow constructs can be given a derivative-based operational semantics;
    \item\label{qhypotheses}whether it is possible to reason about non-local control flow in the presence of other equational hypotheses;
    \item\label{qprimitives}whether non-local control-flow constructs can be treated as primitives; and
    \item\label{qaxiomatization}whether the axiomatization of \GKAT could be extended to include non-local control flow, while retaining the efficient decision procedure.
\end{enumerate*}
After a brief overview in \Cref{sec:overview}, we addresses these questions by providing:
\begin{wrapfigure}{r}{0.25\textwidth}  %
    \vspace{-1em}
    \centering
    \begin{minipage}{.275\textwidth}
        \centering
        \setlength{\intextsep}{0pt}%
        \algSource
        \vspace{2pt}
        \algTransformed
    \end{minipage}
    \caption{Program transformation which moves a \comGoto\ into a loop \cite{taming}}
    \label{fig:eh-loop-transform}
    \vspace{-2.1em}
\end{wrapfigure}
\begin{enumerate}[leftmargin=*,nosep]\vspace{-1em}
	\item
    An operational (derivative-based) semantics of \CFGKAT~\ref{qderivative}, a language that natively includes non-local control flow~\ref{qprimitives} in \Cref{sec:semantics};
	\item
    A sound and complete axiomatization~\ref{qaxiomatization} of bisimulation equivalence of programs in \CFGKAT, that supports reasoning under hypotheses~\ref{qhypotheses}, in \Cref{sec:axioms,sec:completeness}.
\end{enumerate}
To further support these results, we additionally:
\begin{enumerate}[resume,leftmargin=*,nosep]
	\item
    Leverage the completeness result to verify ghost variable elimination, a folklore result notoriously difficult to prove axiomatically, in \Cref{sec:examples};
	\item  
	Provide a fully certified proof for the loop case in the Erosa--Hendren goto elimination algorithm (\Cref{fig:eh-loop-transform}) as a metatheorem derived by our syntactic calculus, in \Cref{sec:examples}.
\end{enumerate}
We conclude with a discussion of related work and future directions in \Cref{sec:related,sec:conclusion}. 

\smallskip
\noindent\textit{\textbf {Mechanization}}
All results in this paper --- including soundness (\Cref{thm:sound}),
completeness (\Cref{completeness}), the fundamental theorem (\Cref{fundthm}), 
and the case studies of \Cref{sec:examples} --- are formalized in Rocq. 
The checked results rely exclusively on standard Rocq library principles.
Statements in the paper are annotated with \RocqIcon{} links into our Rocq development, available at {\urlstyle{same}\href{https://github.com/SpencerKoevering/cfgkat-rocq}{\nolinkurl{github.com/SpencerKoevering/cfgkat-rocq}}} and archived on Zenodo~\cite{cfgkatrocq}.
The formalization was produced with LLM assistance, based on handwritten proofs and a Rocq skeleton of the key definitions as a starting point.
We then manually refactored it to produce a more readable and reusable artifact. 
Beyond certifying the paper's results, the development is structured to support future proofs in \CFGKAT: the axioms of \Cref{ax} are exposed as a reusable proof interface, and the case studies are built on these axioms.
\ifanonymous
We also provide the original pen and paper proof in the appendix, but because of their length, we do not expect the reviewers to read them in full
\fi

\section{Overview}\label{sec:overview}
We now give a brief and informal overview of our main results and technical setup, and in particular of \emph{contextual} reasoning captures the behavior of non-local control flow.
\paragraph{\bf Syntax}
The syntax of \CFGKAT closely resembles imperative programs; it is parameterized over a finite set of tests \(T\), primitive actions \(\Sigma\), labels \(\lset\), variables \(Y\), and values \(I\), combined using standard imperative constructs (cf. \Cref{syntax}).
Unary operators take precedence over binary operators, and \(;\) as well as \(\loopseq\) take (equal) precedence over \(+_b\).
\begin{figure}[t]
	\captionsetup{justification=raggedright,singlelinecheck=false}
	\newcommand{\gramlhs}[1]{\multicolumn{3}{@{}l@{}}{#1 \eqdef}}%
	\noindent
	\begin{minipage}[t]{0.46\linewidth}
		\(\begin{array}[t]{@{\hspace{0.8em}}c@{\hspace{0.35em}}l@{\quad}l@{}}
			\gramlhs{b, c \in \BExp} \\
			&\textbf{0}& \textit{false} \\
			\mid& \textbf{1} & \textit{true} \\
			\mid &y = i&  \textit{indicator test, \(y \in Y, i \in I\)} \\
			\mid &t \in T & \textit{basic test} \\
			\mid &b \land c & \textit{conjunction} \\
			\mid &b \lor c & \textit{disjunction} \\
			\mid &\lnot b & \textit{negation}
		\end{array}\)
		\par\vspace{.4ex}
		\caption{Syntax of Boolean expressions \(\BExp\) (above) and \CFGKAT program expressions \(\Exp\) (on the right) parametric on finite sets of tests \(T\), program actions \(\Sigma\), indicator variables \(Y\), and labels \(\lset\).~\RocqLabelRef{syntax}
		}%
		\label{syntax}
	\end{minipage}\hfill
	\begin{minipage}[t]{0.5\linewidth}
		\(\begin{array}[t]{@{\hspace{0.8em}}c@{\hspace{0.35em}}l@{\quad}l@{}}
			\gramlhs{e, f \in \Exp} \\
			&b \in \BExp & \textit{assert b}\\
			\mid &p \in \Sigma& \textit{do p}\\
			\mid &\comReturn & \textit{return}\\
			\mid &\comBreak & \textit{break}\\
			\mid &\comCont & \textit{continue}\\
			\mid &\comGoto~\ell& \textit{goto, \(\ell \in \lset\)}\\
			\mid &\comLabel~\ell& \textit{label target for goto, \(\ell \in \lset\)}\\
			\mid &e+_b f & \textit{if}\ b\  \textit{then}\  e\ \textit{else}\ f\\
			\mid &e\seq f & \textit{do}\ e\ \textit{then do}\ f\\
			\mid &e^{(b)} & \textit{while}\ b\ \textit{do}\ e\\
			\mid &y\assign i & \textit{set}\ y\ \textit{to}\ i,\text{ \(y \in Y, i \in I\)} \\
			\mid &e\loopseq f & \textit{do body}\ e\ \textit{then resume}\ f%
		\end{array}\)
	\end{minipage}\vspace{-1em}
\end{figure}
The only non-standard operators are indicator variables and the loop sequencing operator \(\loopseq\). 
We sketch these briefly; a more precise semantics follows in~\Cref{sec:op-sem}.

First, indicator variables are special in that they are separate from any variables that may be used by the uninterpreted programs in \(\Sigma\).
Analogous to the \emph{mutable tests} of~\cite{katb!}, indicator variables play the role of a ``flag'' in the program.
They may be set or compared to a constant value, but they do not interact with primitive actions \(p \in \Sigma\) or tests \(t \in T\).
In the past, such variables have been used to eliminate \comGoto statements~\cite{ammarguellat-1992,gussoni-etal-2020,bohmjacopini,ashcroft-manna-1972}.

Second, to define the semantics of \comBreak, which terminates the innermost loop, and \comCont, which restarts the current loop, we introduce \(\loopseq\).
This operator identifies the parent loop of the unfolded loop bodies to handle the end-of-loop composition: intuitively, the expression \(e\) in \(e \loopseq f\) behaves like the loop body unfolded from the parent \(f\), that is, \(\comBreak\) within \(e\) will terminate \(f\), and \(\comCont\) will restart \(f\).

Of course, not all expressions represent well-formed programs. In particular, a program cannot contain two identical \(\comLabel~\ell\) statements, as it would not be clear where \(\comGoto~\ell\) would point.
We focus on \emph{well-formed} programs that obey this restriction.
It should be noted that we \emph{do} allow \comGoto~statements without a corresponding \comLabel.
We call such a \comGoto~``free'' as it is otherwise ``bound'' by a unique \comLabel statement.
\paragraph{\bf Semantics}		
To build a compositional equational theory, we separate a \emph{program} \(e\) into two parts: a \emph{reduced expression} \(e_-\) which is simply the program with label statements removed,
and an environment \(\sigma(e)\) which gives meanings to labels.
Formally, a reduced expression \(e_- \in \Expm\) is given by \(e_{-} \eqdef e[1/\comLabel~\ell\mid \ell \in \lset]\) where the bracket denotes substitutions. 
An environment \(\sigma(e)\) is a partial map from labels \(\lset\) into reduced expressions \(\Expm\); we often use \(E\) to denote a generic environment.
For instance, if \(e\) is the second program in \Cref{sctransform}, then \(\sigma(e)(k)\) amounts to \(\text{\textbf{if }} c \text{\textbf{ then }} \comGoto~\ell \text{\textbf{ else }} f\). %
Not all labels need to be bound, hence environments are often partial functions.

This separation of expressions and environments not only allows us to develop an elegant axiomatic system, as outlined in the next subsection, but also streamlines the definition of semantics into several steps: 
first, label extraction in~\Cref{sec:label-extraction} constructs the environment \(\sigma(e)\); 
then, the semantics for the reduced expression \(e_{-}\) are constructed using derivatives in~\Cref{operationalsemantics}, like in standard \GKAT~\cite{gkatcoequations}; 
finally, the meaning of the environment is added back into the semantics through jump resolution in~\Cref{resolutiondef}.

One crucial distinction with existing work~\cite{cfgkat,symb-cfgkat} is that we focus on bisimulation equivalence~\cite{gkatcoequations}, a stricter notion of equivalence that will distinguish actions on non-terminating branches. 
This notion of equivalence is more realistic for our purpose (see~\Cref{sec:op-sem} for a detailed comparison), but also introduces significant challenges in our proofs, as we now need to reason about potentially infinite execution traces, a challenge we overcome by the use of several techniques from universal coalgebra~\cite{universalcoalg}.
\paragraph{\bf Axiomatization}
Because of the separation of reduced expressions and environments, our axiomatization adapts most of the non-contextual axioms directly from~\cite{gkat}.
These include axioms that govern branching, such as \((e +_b f) +_c g \equiv e +_{b\land c} (f +_c g)\).
However, our loop unfolding axiom \(e^{(b)} \equiv e \loopseq e^{(b)} +_b 1\) uses loop sequencing, since the \GKAT unfolding rule \(e^{(b)} \equiv e \seq e^{(b)} +_b 1\) is unsound when \(e\) contains \(\comBreak\) or \(\comCont\).

We also add axioms for indicator variables.
Although these axioms are simple to state, e.g., \(y \assign i; y \assign j \equiv y \assign j\), they capture a delicate semantic phenomenon: while indicator values are observable at termination, the action of assigning to an indicator is hidden.
This leads to intricate semantics.
For example, a loop may iterate repeatedly by updating an indicator variable, before performing an observable action.

To involve the programs corresponding to the labels, we instrument equivalences with environments by writing \(E \vdash e \equiv f \dashv F\), read as ``\(e\) in environment \(E\) is equivalent to \(f\) in environment \(F\)''.
The non-contextual laws discussed above are valid in \emph{any} environment, 
\begin{wrapfigure}{r}{5.8cm}
	\vspace{-16pt}%
	\centering
	{\small\(\displaystyle
		\inferrule[Substitution]
		{\comGoto~\ell \text{ is rightmost in } e}
		{E,(\ell = e_1) \vdash e[e_1/\comGoto~\ell] \equiv e \dashv E,(\ell = e_1)}
	\)}
	\caption{The substitution axiom for \comGoto}\label{substitutionexax}%
	\vspace*{-18pt}
\end{wrapfigure}
and hence they were omitted.
But our system also includes contextual axioms, like the substitution axiom on the right, which states that, in some circumstances, a program from the environment may be substituted for a corresponding \comGoto~statement.

 Contextual axioms take some care; for instance, unlike function calls, execution does not return to the site of a \comGoto~statement once its target has successfully terminated.
Hence, on an intuitive level, replacing \(\comGoto~\ell\) with a program representing what happens starting from \(\ell\) is sound only when no actions appear syntactically ``after'' \(\comGoto~\ell\); this explains the side condition on the substitution axiom in \Cref{substitutionexax}.
Additional difficulties arise from self-reference, when the code starting from a given label \(\ell\) contains a \comGoto~statement that points to \(\ell\) itself, possibly even before an action is executed.

Finally, our axiomatization includes a number of congruence rules, which allow for limited compositional reasoning.
To ensure soundness in the presence of \comGoto, these rules are restricted; still, they suffice for completeness and practical use cases.
\paragraph{\bf Completeness}
The gold standard for any axiomatization is completeness, which states that all valid equivalences can be proved. 
We achieve this by gradually transforming expressions and their environments into a \emph{contextual normal form}, which closely mirrors the bisimulation that defines our equivalence.
In contrast with prior arguments for completeness of \GKAT variants, which involve meta-reasoning about systems of guarded equations and their solutions, our proof internalizes these systems into the control-flow environment, which makes for a more compact argument.
Interestingly, the normal form bears a striking resemblance to the legacy language SNOBOL \cite{snobol}, where each statement is a triple consisting of a \comLabel, a rule, and a series of conditional \comGoto statements.
\paragraph{\bf Case Studies}
To show syntactic reasoning in our system and how to apply the completeness result, we conclude with some case studies, ranging from a simple 
short-circuit transformation, to ghost variable elimination, and finally 
 the case of the transformation in \Cref{fig:eh-loop-transform}~\cite{taming}.
This transformation uses many implicit facts, including the aforementioned ghost variable elimination, which are difficult to derive in spite of their conceptual simplicity.
The overall proof of correctness is nontrivial because it must carefully manage loops and changes to the bindings of labels.
It is worth noting that the techniques to address these are general, and can be applied elsewhere.

\section{Semantics}\label{sec:semantics}
In this section we develop the operational semantics, starting with some needed preliminary material on label extraction and \CFGKAT coalgebras.
\subsection{Label Extraction and Boolean Semantics}\label{sec:label-extraction}
Label extraction generates an environment \(\sigma(e): \lset \partialTo \Expm\) given a program \(e\), similar to how this was done for \KAT~\cite{kozennonlocal}.
Leveraging the fact that no label appears more than once, we can take a well-formed program \(e \in \Exp\) and a label \(\ell \in \lset\), and produce the unique reduced expression that would be executed upon encountering \(\comGoto~\ell\).

\begin{definition}[\RocqLabelRef{labelextraction}]\label{labelextraction}
    With every program $e \in \Exp$, we associate a partial function $\sigma(e)$ from $\lset$ to $\Expm$, writing $\sigma(e)_\ell$ for the value of $\sigma(e)$ on label $\ell$, as follows.
	Let \(\mathsf{Lab}(e)\) be the set of labels occurring in \(e\).
    We set $\sigma(\comLabel\,\ell)_\ell := 1$, and
	\begingroup
\renewcommand{\arraystretch}{0.9}
\newcommand{\sigcases}[4]{%
  \left\{
  \begin{array}{@{}l@{\hspace{0.5em}}l@{}}%
    #1 & #2,\\[-0.25ex]
    #3 & #4
  \end{array}
  \right.%
}
\[
\begin{array}{@{}r@{\;}c@{\;}l@{\hspace{1em}}r@{\;}c@{\;}l@{}}%
\sigma(e+_b f)_\ell
  &\eqdef&
  \sigcases
    {\sigma(e)_\ell}
    {\ell\in\Lab(e)}
    {\sigma(f)_\ell}
    {\ell\in\Lab(f)}
&
\sigma(e\seq f)_\ell
  &\eqdef&
  \sigcases
    {\sigma(e)_\ell\seq f_{-}}
    {\ell\in\Lab(e)}
    {\sigma(f)_\ell}
    {\ell\in\Lab(f)}
\\[0.35em]
\sigma(e^{(b)})_\ell
  &\eqdef& \sigma(e)_\ell \loopseq (e^{(b)})_{-}\, \text{(if $\ell\in\Lab(e)$)}
&
\sigma(e\loopseq f)_\ell
  &\eqdef&
  \sigcases
    {\sigma(e)_\ell\loopseq f_{-}}
    {\ell\in\Lab(e)}
    {\sigma(f)_\ell}
    {\ell\in\Lab(f)}
\end{array}
\]%
\endgroup
In all other cases, $\sigma(e)_\ell$ is not defined.
\end{definition}

We can now elaborate on our earlier intuition: programs \(e\) and \(f\), possibly including labels, are equivalent (written \(e \progeq f\)) when \(\sigma(e) \vdash e_- \equiv f_- \dashv \sigma(f)\) --- that is, when their reduced forms are equivalent in their respective environments.

\medskip
To define an operational semantics on reduced \CFGKAT expressions, we must first give a semantics to Boolean expressions, assigning \(b \in \BExp\) to a set \(\langle b \rangle  \subseteq \At \times I^Y\), where \(\At = 2^T\) is the set of \emph{atoms}, and \(I^Y\) is the set of functions from variables in \(Y\) to values in \(I\) (\emph{valuations}).
Intuitively, \((\alpha, \pi) \in \langle b \rangle\) means \(b\) holds on any state where (1)~tests in $\alpha$ are true, (2)~tests outside $\alpha$ are false, and (3)~each  \(y\) has value \(\pi(y)\).

\begin{definition}[Boolean Semantics \RocqLabelRef{Booleansemantics}]\label{Booleansemantics}
    For \(b \in \BExp\), we define \(\langle b \rangle \subseteq \At \times I^Y\) by:
	\begin{mathpar}
        \langle 1 \rangle \eqdef \At \times I^Y
        \and
        \langle 0 \rangle \eqdef \emptyset
        \and
        \langle t \rangle \eqdef \{\alpha \mid t \in \alpha\}\times I^Y
        \and
        \langle b \land c \rangle \eqdef \langle b \rangle \cap \langle  c \rangle
        \and
		\langle b \lor c \rangle \eqdef \langle b \rangle \cup \langle  c \rangle
        \and
        \langle  \lnot b \rangle \eqdef (\At \times I^Y) \setminus \langle b \rangle
        \and
		\langle x=i \rangle \eqdef \At \times \{\pi \in I^Y \mid \pi(x) = i\}
    \end{mathpar}
\end{definition}
Later, we may treat an atom \(\alpha \in \At\) as the Boolean expression \(\bigwedge_{t \in T} t_\alpha\), where \(t_\alpha = t\) when \(t \in \alpha\), and \(t_\alpha = \neg t\) otherwise; this is well-defined up to Boolean equivalence.

\subsection{\coalg{\CFGKAT}s}\label[subsection]{sec:cfgkat-coalg}
Our operational semantics is defined as a transition system, formalized as a coalgebra (see the work of Rutten~\cite{universalcoalg} for a thorough introduction).
The development in the main text is phrased in elementary terms, with more details in \ifdefined\CFGKATNoAppendix{} the appendix of the full version\else\Cref{app:coalgebra}\fi.

\begin{definition}
	A \coalg{\CFGKAT} is a transition system that consists of a (potentially infinite) set \(S\) of states, and a transition function \(\delta: S \to (K + \Sigma\times S)^\At\), where \(K\) is the set of \emph{continuations}
    \(
        K \eqdef \{\bot, \retc\} + (\{\acc, \brkc, \contc\}+ \{\jmpc~\ell \mid\ell\in \lset\})\times I^Y
    \).
\end{definition}

For a \coalg{\CFGKAT} \((S, \delta)\), \(s \in S\), and \(\alpha \in \At\), we use these shorthands:
\begin{mathpar}
\delta(s)_\alpha = \delta(s)(\alpha)
\and
\delta(s)_\alpha = (p, s') \Leftrightarrow s \tr[\delta]{p}{\alpha} s'
\and
\delta(s)_\alpha=k \in K \Leftrightarrow s \tr[\delta]{}{\alpha} k
\end{mathpar}
In the sequel, we write \(s \hookrightarrow_\alpha^\delta r\) for \(r \in K + \Sigma \times S\) to capture both types of transitions.
Formally, \(s \hookrightarrow_\alpha^\delta k \in K\) denotes \(s \tr[\delta]{}{\alpha} k \in K\), and \(s \hookrightarrow_\alpha^\delta (p, s')\) denotes \(s \tr[\delta]{p}{\alpha} s'\).

Intuitively, \(s \tr[\delta]{p}{\alpha} s'\) indicates that when \(\alpha\) holds, the state \(s\) transitions into the state \(s'\) while executing \(p\).  
Moreover, \(s \tr[\delta]{}{\alpha} k\) with \(k \in K\) means that when \(\alpha\) holds, \(s\) immediately encounters a continuation \(k\), whose meaning is as follows:
\(k = \bot\) signals execution failure/divergence;
\(k = \retc\) indicates termination with \comReturn;
\(k = (\brkc, \pi')\) or \(k = (\contc, \pi')\) means we break out of or restart the parent loop respectively;
and \(k = (\jmpc~\ell, \pi')\) signals a jump to label \(\ell\) --- where in the latter three cases, $\pi'$ holds the values of the indicator variables to be used when resuming the program.

Most whole programs end in \(\bot\) or \(\retc\) with no indicator state attached (hiding the indicator values), while the other continuations occur only in partial programs.
Yet, partial programs are essential for compositional reasoning, as we often want to use equivalences between partial programs to infer equivalences of whole programs.

We can now define our notion of equivalence between \coalg{\CFGKAT}s, which is exactly the standard notion of bisimulation derived from the relevant functor~\cite{universalcoalg}.
\begin{definition}[Bisimulation \RocqLabelRef{bisimulation}]\label{bisimulationwat}
	A \emph{bisimulation} between \coalg{\CFGKAT}s \((S, \delta)\) and \((U, \eta)\) is a relation \(R \subseteq S \times U\) such that if \(sRu\) and \(\alpha \in \At\), either \(s \tr[\delta]{}{\alpha} k\) and \(u \tr[\eta]{}{\alpha} k\) for some \(k \in K\); or \(s \tr[\delta]{p}{\alpha} s'\) and \(u \tr[\eta]{p}{\alpha} u'\) such that \(s'Ru'\).
	If there exists a bisimulation $R$ such that $sRu$, we call them \emph{bisimilar} and write \(s \sim u\).

\end{definition}

\subsection{Operational Semantics}\label{sec:op-sem}
Our operational semantics takes the form of a \coalg{\CFGKAT} \((\Expm \times I^Y, \partial)\), where the states are reduced \CFGKAT expressions paired with an indicator state, and where \(\partial\) is the function generated by the rules in \Cref{operationalsemantics}, where we drop the superscript \(\partial\) to reduce notational clutter.
We call \((\Expm \times I^Y, \partial)\) the \emph{syntactic \coalg{\CFGKAT}}.

Transitions not derivable with the rules in~\Cref{operationalsemantics} are defined to output failure/divergence \(\bot\).
For instance, when \((\alpha, \pi) \not\in \langle b \rangle\), no rule applies to \((b, \pi)\), so \((b, \pi) \tr{}{\alpha} \bot\).
A less trivial consequence of this convention comes from the final rule.
Consider the transition of \(1^{(1)}\):
to derive \((1^{(1)}, \pi) \hookrightarrow_{\alpha} r\), we are required to show
\(
  (1 \loopseq 1^{(1)}, \pi) \hookrightarrow_{\alpha} r,
\)
which in turn requires
\(
  (1^{(1)}, \pi) \hookrightarrow_{\alpha} r
\)
as a premise.
Thus, no finite derivation of \((1^{(1)}, \pi) \hookrightarrow_{\alpha} r\) exists, which leads us to conclude
\(
  (1^{(1)}, \pi) \hookrightarrow_{\alpha} \bot
\).

The identification of unproductive infinite loops with immediate rejection matches both trace-based~\cite{gkat} and bisimulation-based semantics~\cite{gkatcoequations} of \GKAT.
However, unproductive loops in \CFGKAT can also occur if toggling between values for an indicator variable without performing any primitive action, e.g., \(e = {(x \leftarrow 1 +_{x = 0} x \leftarrow 0)}^{(\lnot x = 2)}\).

\begin{figure}[t]
	\centering
	\begingroup
	\setlength{\tabcolsep}{0.6em}
	\setlength{\fboxsep}{0.45em}
	\setlength{\fboxrule}{0.4pt}
	\figmathpartight
	\newcommand{\opsemheader}[1]{%
		\centering
		\textsc{#1}\par
		\vspace{-2.2ex}
	}%
	\begin{tabularx}{\linewidth}{|Y|Y|}
		\hline
		\multicolumn{2}{|p{\dimexpr\linewidth-2\tabcolsep-2\arrayrulewidth\relax}|}{%
			\begin{minipage}[t]{\linewidth}
				\opsemheader{Primitives}
				\begin{mathpar}
					\inferrule{(\alpha, \pi) \in \langle b \rangle}
						{(b, \pi)\tr{}{\alpha}(\acc, \pi)}
					\and
					\inferrule{ }
						{(x\assign i, \pi)\tr{}{\alpha}(\acc, \pi[x\leftarrow i])}
					\and
					\inferrule{p \in \Sigma}
						{(p, \pi)\tr{p}{\alpha}(1, \pi)}
					\and
					\inferrule{ }
						{(\comBreak, \pi)\tr{}{\alpha}(\brkc, \pi)}
					\and
					\inferrule{ }
						{(\comCont, \pi)\tr{}{\alpha}(\contc, \pi)}
					\and
					\inferrule{ }
						{(\comReturn, \pi)\tr{}{\alpha}\retc}
					\and
					\inferrule{\ell \in \lset}
						{(\comGoto~\ell, \pi)\tr{}{\alpha}(\jmpc~\ell, \pi)}
				\end{mathpar}
			\end{minipage}
		}\\
		\hline
		\begin{minipage}[t]{\hsize}
			\opsemheader{Guarded Choice}
			\begin{mathpar}
				\inferrule{
					(\alpha, \pi) \in \langle b \rangle \\
					(e, \pi)\hookrightarrow_\alpha r
				}{(e+_b f, \pi)\hookrightarrow_\alpha r}
				\and
				\inferrule{
					(\alpha, \pi) \in \langle \lnot b \rangle \\
					(f, \pi)\hookrightarrow_\alpha r
				}{(e+_b f, \pi)\hookrightarrow_\alpha r}
			\end{mathpar}
		\end{minipage}
		&
		\begin{minipage}[t]{\hsize}
			\opsemheader{While Loop}
			\begin{mathpar}
				\inferrule{(\alpha, \pi) \in \langle \lnot b \rangle}
					{(e^{(b)}, \pi)\tr{}{\alpha}(\acc, \pi)}
				\and
				\inferrule{
					(\alpha, \pi) \in \langle b \rangle \\
					(e\loopseq e^{(b)}, \pi)\hookrightarrow_\alpha r
				}{(e^{(b)}, \pi)\hookrightarrow_\alpha r}
			\end{mathpar}
		\end{minipage}
		\\
		\hline
		\begin{minipage}[t]{\hsize}
			\opsemheader{Sequential Composition}
			\begin{mathpar}
				\inferrule{
					(e, \pi)\tr{}{\alpha}(\acc, \pi^\prime) \\
					(f, \pi^\prime)\hookrightarrow_\alpha r
				}{(e\seq f, \pi)\hookrightarrow_\alpha r}
				\and
				\inferrule{(e, \pi)\tr{p}{\alpha}(e', \pi^\prime)}
					{(e\seq f, \pi)\tr{p}{\alpha}(e'\seq f, \pi^\prime)}
				\and
				\inferrule{
					(e, \pi)\tr{}{\alpha}k \\
					k \in K_{\mathrm{seq}}
				}{(e\seq f, \pi)\tr{}{\alpha}k}
			\end{mathpar}
			\begin{center}
			\fbox{%
			\begin{minipage}{\dimexpr\linewidth-2\fboxsep-2\fboxrule\relax}
				\[
				\begin{aligned}
					K_{\mathrm{seq}}
						&\eqdef \{\retc, \bot\}+(\{\brkc, \contc\} \\
						&\quad+\{\jmpc~\ell \mid \ell \in \lset\})\times I^Y,\\
					K_{\mathrm{loop}}
						&\eqdef \{\retc, \bot\}+\{\jmpc~\ell \mid \ell \in \lset\}\times I^Y.
				\end{aligned}
				\]
			\end{minipage}}
			\end{center}
		\end{minipage}
		&
		\begin{minipage}[t]{\hsize}
			\opsemheader{Loop Composition}
			\begin{mathpar}
				\inferrule{
					(e, \pi)\tr{}{\alpha}k \\
					k \in K_{\mathrm{loop}}
				}{(e\loopseq f, \pi)\tr{}{\alpha}k}
				\and
				\inferrule{(e, \pi)\tr{}{\alpha}(\brkc, \pi^\prime)}
					{(e\loopseq f, \pi)\tr{}{\alpha}(\acc, \pi^\prime)}
				\and
				\inferrule{(e, \pi)\tr{p}{\alpha}(e', \pi^\prime)}
					{(e\loopseq f, \pi)\tr{p}{\alpha}(e'\loopseq f, \pi^\prime)}
				\and
				\inferrule{
					(e, \pi)\tr{}{\alpha}(c, \pi^\prime) \\
					c \in \{\contc, \acc\} \\
					(f, \pi^\prime)\hookrightarrow_\alpha r
				}{(e\loopseq f, \pi)\hookrightarrow_\alpha r}
			\end{mathpar}
		\end{minipage}
		\\
		\hline
	\end{tabularx}
	\endgroup
		\caption{%
	        The operational semantics of \CFGKAT.
	        We use \(\pi[x \leftarrow i]\) to denote the function which is equal to \(\pi\) everywhere except at \(x\), where it is \(i\).
	        Any inputs that do not yield an outcome by a finite application of these rules are defined to yield failure/divergence \(\bot\).\RocqLabelRef{operationalsemantics}
	    }%
		\label{operationalsemantics}
	\end{figure}

\paragraph{Jump Resolution}
Jump resolution takes a \coalg{\CFGKAT} and a \emph{jump map} to recursively resolve \comGoto statements, thus imbuing them with their intended semantics.

	\begin{definition}[Jump Resolution \RocqLabelRef{resolutiondef}]\label{resolutiondef}
		Given a \coalg{\CFGKAT} \((S, \delta)\) and a jump map \(\gamma: L \times I^Y \to S\) with \(L \subseteq \lset\), we define \((S, \delta)\resWith{\gamma} = (S, \delta\resWith{\gamma})\), where \(\delta\resWith{\gamma}\) is defined by the following rules:
		\begin{mathpar}
		\inferrule{%
            \delta(s)_\alpha \notin \{\jmpc~\ell \mid \ell \in L\} \times I^Y
        }{%
            \delta\resWith{\gamma}(s)_\alpha = \delta(s)_\alpha
        }
		\and  
		\inferrule{%
            \delta(s)_\alpha = (\jmpc~\ell, \sigma) \\
			\delta\resWith{\gamma}(\gamma(\ell, \sigma))_\alpha = r
        }{%
            \delta\resWith{\gamma}(s)_\alpha = r
        }
	\end{mathpar}
	When \(s \in S\) and \(\alpha \in \At\) are not covered by these rules, \(\delta\resWith{\gamma}(s)_\alpha = \bot\).
    In particular, in case of infinite jumps, e.g., \(\gamma(\ell, \pi) = s\) and \(\delta(s)_\alpha = (\jmpc~\ell, \pi)\), the output is \(\bot\).
\end{definition}
When discussing bisimilarity in resolved \coalg{\CFGKAT}s, we write $(s)\resWith{\gamma} \sim (u)\resWith{\zeta}$ to denote that $s \sim u$ in the coalgebras resolved using $\gamma$ and $\zeta$, respectively.

\paragraph{Derivative and Loop Sequencing}
Our current semantics is defined via derivatives, similar to the approach of~Schmid et al.~\cite{gkatcoequations}, rather than the Thompson construction given by~Zhang et al.~\cite{cfgkat}.
This change not only resolves an open question left by~Kozen~\cite{kozennonlocal}, but also streamlines a number of constructions as predicted in the same work.
Specifically, we no longer require notions like \emph{start dynamics}, \emph{uniform continuation}, 
or \emph{iteration lifting}: all of these are stored in the 
derivation trees of transitions.
However, 
\begin{wrapfigure}{r}{4.5cm}
	\vspace{-3em}
	\begin{minipage}{4.5cm}\centering
	\begin{equation*}
	\eqnmarkbox[blue]{body}{\comBreak} \loopseq \eqnmarkbox[red]{loop}{(p; \comBreak)^{(1)}}
	\end{equation*}
	\tikzset{annotate equations/arrow/.style={->}}
	\annotatetwo[yshift=-5pt]{below, label below}{body}{loop}{$\comBreak$ belongs to $(p; \comBreak)^{(1)}$}
	\end{minipage}
	\vspace{-2em}
\end{wrapfigure}
designing the derivative also has challenges: the loop $(p; \comBreak)^{(1)}$ will first execute $p$, reaching $\comBreak \loopseq (p; \comBreak)^{(1)}$, then it exits the loop and accepts.
Notice the behavior of loop sequencing is different from sequencing: $\comBreak; (p; \comBreak)^{(1)}$ outputs a $\brkc$ continuation instead of accepting.
This difference is because loop sequencing is also a \emph{scoping operator}, connecting the remaining loop body with the parent loop, which allows $\comBreak$/$\comCont$ in the loop body to correctly exit/restart the parent loop.
Intuitively, $\comBreak; e$ means “break, then run ($e$),” whereas $\comBreak \loopseq e$ means “break out of ($e$).” 
This scoping also elegantly enables the encoding of $\command{do}~e~\command{while}~b$ loops as $e \loopseq e^{(b)}$, even in the presence of $\comBreak$ and $\comCont$, and enables the compact invariant specification used in \RefAx{\Axlooptighten}.

Thanks to these simplifications, we can extend the previous semantics with \contc continuations (produced by \comCont) and multiple indicator variables with ease.

\paragraph{Internal States}
Each state in the syntactic coalgebra includes an indicator valuation, whereas Zhang et al.~\cite{cfgkat} push indicator valuations into the state when lowering \CFGKAT automata into \GKAT automata.
Although the resulting signatures are essentially the same, our presentation eliminates this step in the semantic construction.

\begin{wrapfigure}{r}{3.1cm}
	\vspace{-2.2em}
	\centering
	\fbox{\(\begin{array}{@{}l@{}}
		(\command{sendEmail()})^{(1)}\\[0.3ex]
		(\command{sleep()})^{(1)}
	\end{array}\)}
	\vspace{-1.2em}
\end{wrapfigure}
\paragraph{Bisimulation Equivalence}
Previous works~\cite{cfgkat,symb-cfgkat} prioritize \emph{finite-trace} semantics, which do not distinguish infinite executions, such as the two programs on the right.
Intuitively, finite trace semantics requires the effect of the program's actions to be hidden until it terminates.
This assumption holds in various domains, like numerical computing, but not in general.
To provide a generic framework without relying on purity assumptions, we opt to use the more strict bisimulation equivalence.
In fact, whether or not programs that fail (crash or diverge) now are equivalent to programs that inevitably fail later is the only difference between trace and bisimulation equivalence in \GKAT~\cite{gkatcoequations}. 

\begin{toappendix}
	
\subsection{Intermediate Bisimulation}

As we have noted in~\tCref{sec:cfgkat-coalg}, continuations of a sub-program are often not observable in a large program.
This property provides us with a desirable notion of equality: internal states are not reflected in the overall control flow.
Yet, it also makes the proof of bisimulation challenging, since crucial boundaries like internal jumps and loop restarts are now silent.
We introduce \emph{intermediate bisimulation}, as a proof technique to recover this fine-grained reasoning.
Specifically, it provides an alternative proof obligation to show silent transitions progress into related states.
With this technique, we can explicitly reason about \emph{silent transitions}, even when they resolve to \(\bot\).

First, we generalize jump resolution on syntactic \coalg{\CFGKAT}: if \(\vec{\gamma}\) is a sequence of jump maps, then \((S, \delta)\resWith{\vec{\gamma}} = (S, \delta)\) when \(\vec{\gamma}\) is empty, and \((S, \delta)\resWith{\vec{\gamma}\zeta} = (S, \delta)\resWith{\vec{\gamma}}\resWith{\zeta}\).
We abbreviate \(\tr[\partial\resWith{\vec{\gamma}}]{}{\alpha}\) by \(\tr[\vec{\gamma}]{}{\alpha}\).

\begin{definition}[Silent Transitions \RocqLabelRef{silentTrans}]\label{silentTrans}
    Let \(\vec{\gamma}\) be a sequence of jump maps and \(\alpha \in \At\).
    Let the silent \(\alpha\)-transition for \(\vec{\gamma}\), denoted \(\transSilent_{\alpha}^{\vec{\gamma}}\), be the smallest relation on \(\Expm \times I^Y\) satisfying the following rules, where \(c \in \{ \acc, \contc \}\) and \(\zeta\) is a jump map: we omit the empty sequence \(\vec{\gamma}\) in the first two rules.
    \begin{center}
    \setlength{\tabcolsep}{0pt}
    \begin{tabularx}{\linewidth}{@{}>{\centering\arraybackslash}X@{\quad}>{\centering\arraybackslash}X@{}}
        \(\displaystyle
        \inferrule{%
            (\alpha, \pi) \in \langle b \rangle \\
            (e, \pi) \tr{}{\alpha} (c, \sigma)
        }{%
            (e^{(b)}, \pi)
            \transSilent_{\alpha}
            (e^{(b)}, \sigma)
        }\)
        &
        \(\displaystyle
        \inferrule{%
            (e', \pi) \tr{}{\alpha} (c, \sigma)
        }{%
            (e'\loopseq e^{(b)}, \pi)
            \transSilent_{\alpha}
            (e^{(b)}, \sigma)
        }\)
        \\[2ex]
        \(\displaystyle
        \inferrule{%
            (e, \pi) \tr[\vec{\gamma}]{}{\alpha} (\jmpc~\ell, \pi') \\
            \zeta(\ell, \pi') = (f, \sigma)
        }{%
            (e, \pi)
            \transSilent_\alpha^{\vec{\gamma}\zeta}
            (f, \sigma)
        }\)
        &
        \(\displaystyle
        \inferrule{%
            (e, \pi)
            \transSilent_\alpha^{\vec{\gamma}}
            (f, \sigma)
        }{%
            (e, \pi)
            \transSilent_\alpha^{\vec{\gamma}\zeta}
            (f, \sigma)
        }\)
    \end{tabularx}
    \end{center}
\end{definition}

The first two rules correspond to recursion as a result of loops, and the third comes from jump resolution; they are meant to materialize the internal states that appear in the recursive rules.

There are two important properties about silent transitions that will make manipulating the semantics significantly easier.
First, any state where silent transitions get ``stuck'' will always reject.%
\begin{definition}[Rejecting Set \RocqLabelRef{rejectingsetdef}]\label{rejectingsetdef}
A set of states \(N \subseteq \Expm \times I^Y\) is \emph{rejecting} for \(\alpha\) in \((\Expm, \partial)\resWith{\vec{\gamma}}\) if for all \(s \in N\), either \(s \tr[\vec{\gamma}]{}{\alpha}
 \bot\) or there exists some \(s' \in N\) such that \(s \transSilent_{\alpha}^{\vec{\gamma}} s'\).

\end{definition}	
We note that when we say one derivation is smaller than another we mean that it contains fewer nodes (rule applications).
    
\begin{lemma}[\RocqLabelRef{smallerderivation}]\label{smallerderivation}
    Let \(s, s' \in (\Expm \times I^Y, \partial)\resWith{\vec{\gamma}}\) be states of a resolved syntactic \coalg{\CFGKAT} and \(s \transSilent_{\alpha}^{\vec{\gamma}} s'\), then \(s \hookrightarrow_\alpha^{\vec{\gamma}} r\) is derivable if and only if \(s' \hookrightarrow_\alpha^{\vec{\gamma}} r\) is derivable. Furthermore if both are derivable, then there exists a derivation of \(s' \hookrightarrow_\alpha^{\vec{\gamma}} r\) which is smaller than any derivation of \(s \hookrightarrow_\alpha^{\vec{\gamma}} r\).

\end{lemma}
\begin{proof}
	
	We prove this by induction on the rules.
	
	The first three are base cases, and the last is the only inductive case. First we note that strict subderivations have strictly fewer nodes than the derivation which they are a part of.
    
   If the silent transition is from the loop case of \Cref{silentTrans}, then \(s = (e^{(b)}, \pi), s' = (e^{(b)}, \sigma)\) for some \(e, b, \pi, \sigma\) such that \((\alpha, \pi) \in \langle b \rangle\) and \((e, \pi) \tr{}{\alpha} (c, \sigma)\), with \(c \in \{\acc, \contc\}\). Then for any derivation of \(s \hookrightarrow_{\alpha}^{\vec{\gamma}} r\) it must be of the following shape for some \(r'\)
    \begin{prooftree}\def\ScoreOverhang{1pt}
        \AxiomC{\((\alpha, \pi) \in \langle b \rangle\)}
        \AxiomC{\dots}
        \UnaryInfC{\((e, \pi) \tr{}{\alpha} (c, \sigma)\)}
        \AxiomC{\dots}
        \UnaryInfC{\((e^{(b)}, \sigma) \hookrightarrow_\alpha r'\)}
        \BinaryInfC{\((e\loopseq e^{(b)}, \pi) \hookrightarrow_\alpha r'\)}
        \BinaryInfC{\((e^{(b)}, \pi) \hookrightarrow_\alpha r'\)}
    \end{prooftree}
 	Hence, there is a derivation of the fact that \(s' \hookrightarrow_\alpha r\) for each derivation of \(s \hookrightarrow_\alpha r\).
    
    If the silent transition is from the loop sequencing case of \Cref{silentTrans}, then \(s = (e'\loopseq e^{(b)}, \pi), s' = (e^{(b)}, \sigma)\) for some \(e, e', b, \pi, \sigma\) such that \((\alpha, \pi) \in \langle b \rangle\) and \((e', \pi) \tr{}{\alpha} (c, \sigma)\), with \(c \in \{\acc, \contc\}\). Then for any derivation of \(s \hookrightarrow_{\alpha} r\) it must be of the following shape for some \(r'\)
    \begin{prooftree}\def\ScoreOverhang{1pt}
        \AxiomC{\dots}
        \UnaryInfC{\((e', \pi) \tr{}{\alpha} (c, \sigma)\)}
        \AxiomC{\dots}
        \UnaryInfC{\((e^{(b)}, \sigma) \hookrightarrow_\alpha r'\)}
        \BinaryInfC{\((e'\loopseq e^{(b)}, \pi) \hookrightarrow_\alpha r'\)}
    \end{prooftree}
    Hence, there is a derivation of the fact that \(s' \hookrightarrow_\alpha r\) for each derivation of \(s \hookrightarrow_\alpha r\).
    
    If the silent transition is from the jump resolution case of \Cref{silentTrans}, then \(s \tr[\vec{\gamma}]{}{\alpha} (\jmpc~\ell, \sigma)\) for some \(\sigma\) and \(\zeta(\ell, \sigma) = s'\). For any derivation of \(s \hookrightarrow_{\alpha}^{\vec{\gamma}} r\) in \(S\resWith{\gamma}\) it must be of the shape
    \begin{prooftree}\def\ScoreOverhang{1pt}
        \AxiomC{\(\partial\resWith{\vec{\gamma}}(s)_\alpha = (\jmpc~\ell, \sigma)\)}
        \AxiomC{\(\dots\)}
        \UnaryInfC{\(\partial\resWith{\vec{\gamma}\zeta}(\zeta(\ell, \sigma))_\alpha = r\)}
        \BinaryInfC{\(\partial\resWith{\vec{\gamma}\zeta}(s)_\alpha = r\)}
    \end{prooftree}
    Hence, there is a smaller derivation of the fact that \(\zeta(\ell, \sigma) \hookrightarrow_\alpha^{\vec{\gamma}\zeta} r\) and \(\zeta(\ell, \sigma) = s'\).
    
    For the backwards direction for the base cases we construct the same derivation as the forwards direction with the given derivation of \(s' \hookrightarrow_\alpha r\).
    
    By the above proof trees, for each derivation of \(s \hookrightarrow_{\alpha} r\), there exists a smaller derivation of \(s' \hookrightarrow_{\alpha} r\) and the sub-derivation relation is well-founded. Hence, if there exists a derivation of each, there exists a derivation of \(s' \hookrightarrow_{\alpha} r\) which is smaller than any derivation of \(s \hookrightarrow_{\alpha} r\).

    Finally, for the weakening case. For the forwards direction suppose \(s=(e, \pi) \transSilent_\alpha^{\vec{\gamma}} (f, \sigma)=s'\) and \(s \hookrightarrow_\alpha^{\vec{\gamma}} r\). By the inductive hypothesis there exists a derivation of \(s' \hookrightarrow_\alpha^{\vec{\gamma}} r\) which is smaller than any derivation of \(s\) \(\hookrightarrow_\alpha^{\vec{\gamma}} r\). First suppose \(r \neq (\jmpc~\ell, \nu) \in \dom(\zeta)\). Then by \Cref{resolutiondef} 
    \begin{center}
    \def\ScoreOverhang{1pt}
    	\AxiomC{\(\dots\)}
    	\UnaryInfC{\(\partial\resWith{\vec{\gamma}}(s)_\alpha = r\)}
    	\UnaryInfC{\(\partial\resWith{\vec{\gamma}\zeta}(s)_\alpha = r\)}
    \DisplayProof{}
    \quad and \quad
    \def\ScoreOverhang{1pt}
    	\AxiomC{\(\dots\)}
    	\UnaryInfC{\(\partial\resWith{\vec{\gamma}}(s')_\alpha = r\)}
    	\UnaryInfC{\(\partial\resWith{\vec{\gamma}\zeta}(s')_\alpha = r\)}
    \DisplayProof
    \end{center}
    The backwards direction can be seen by the same argument using the fact that there exists a derivation of \(s' \hookrightarrow_\alpha^{\vec{\gamma}} r\). Since each derivation has been extended by a single step, the tree for \(s'\) remains smaller than any derivation for \(s\).
    
    If \(r = (\jmpc~\ell, \nu) \in \text{dom}(\zeta)\), then by \Cref{resolutiondef} 
    \begin{prooftree}\def\ScoreOverhang{1pt}
    	\AxiomC{\(\dots\)}
    	\UnaryInfC{\(\partial\resWith{\vec{\gamma}}(s)_\alpha = (\jmpc~\ell, \nu)\)}
    	\AxiomC{\(\dots\)}
    	\UnaryInfC{\(\partial\resWith{\vec{\gamma}\zeta}(\jmpc~\ell, \nu)_\alpha = r\)}
    	\BinaryInfC{\(\partial\resWith{\vec{\gamma}\zeta}(s)_\alpha = r\)}
    \end{prooftree}
    and
    \begin{prooftree}\def\ScoreOverhang{1pt}
    	\AxiomC{\(\dots\)}
    	\UnaryInfC{\(\partial\resWith{\vec{\gamma}}(s')_\alpha = (\jmpc~\ell, \nu)\)}
    	\AxiomC{\(\dots\)}
    	\UnaryInfC{\(\partial\resWith{\vec{\gamma}\zeta}(\jmpc~\ell, \nu)_\alpha = r\)}
    	\BinaryInfC{\(\partial\resWith{\vec{\gamma}\zeta}(s')_\alpha = r\)}
    \end{prooftree}
    
    The backwards direction can be seen by the same argument using the fact that there exists a derivation of \(s' \hookrightarrow_\alpha^{\vec{\gamma}} r\) which is smaller than the derivation of \(s' \hookrightarrow_\alpha^{\vec{\gamma}} r\). By the above construction and the inductive hypothesis, for each derivation of \(s' \hookrightarrow_\alpha^{\vec{\gamma}} r\) there exists a strictly smaller derivation of \(s \hookrightarrow_\alpha^{\vec{\gamma}} r\).
\end{proof}
\begin{theoremrep}[\RocqLabelRef{rejectingsetaborts}] \label{rejectingsetaborts}
    A state in a resolved syntactic \coalg{\CFGKAT} \(s \in (S, \delta)\) is such that \(s \tr[\delta]{}{\alpha} \bot\) if and only if it is in some rejecting set \(N\) for $\alpha$.
    Otherwise, there exists an \(r \neq \bot\) with \(s \hookrightarrow_\alpha^\delta r\).

\end{theoremrep}
\begin{proof}
    We first show the \(\implies\) direction by observing that if \(s \tr[\delta]{}{\alpha} \bot\), then \(N = \{s\}\) is a rejecting set.

    We now show the \(\impliedby\) direction by deriving a contradiction from the derivability of \(s \hookrightarrow_\alpha^\delta r\) for some \(s \in S\) and \(r \neq \bot\). 
    Because proof sizes are natural numbers, which are well ordered, if any \(s \hookrightarrow_\alpha^\delta r\) is derivable, then there exist \(s_0 \in S\) and \(r_0 \neq \bot\) such that \(s_0 \hookrightarrow_\alpha^\delta r_0\) has the smallest proof among all \(s \in S\) and \(r \neq \bot\).
    We do a case analysis on the reason that \(s_0\) is in the rejecting set \(N\) to derive the contradiction:
    \begin{itemize}[nosep,leftmargin=*]
        \item If \(s_0 \tr[\delta]{}{\alpha} \bot\), then it directly contradicts the derivability of \(s_0 \hookrightarrow_\alpha^\delta r_0\) for \(r_0 \neq \bot\). Moreover by inspecting~\Cref{operationalsemantics} and the definition of jump resolution, we notice that no rules can derive a conclusion of the form \(s \tr[\delta]{}{\alpha} \bot\).
        Therefore, \(s_0 \tr[\delta]{}{\alpha} \bot\) if and only if \(s_0 \hookrightarrow_\alpha^\delta r_0\) is underivable.
        \item If \(s_0 \transSilent_\alpha^\delta s_0'\), then by~\Cref{smallerderivation} \(s_0' \hookrightarrow_\alpha^\delta r_0\) can be derived using a (strictly) smaller derivation, contradicting the fact that \(s_0 \hookrightarrow^\delta r_0\) has a smallest derivation among \(s \in S\) and \(r \neq \bot\).
    \end{itemize}
    Thus, we conclude there is no \(s \in S\) and \(r \neq \bot\) s.t. \(s \hookrightarrow_\alpha^\delta r\) is derivable. Therefore \(s \tr[\delta]{}{\alpha} \bot\) by the definitions of the operational semantics (\Cref{operationalsemantics}) and jump resolution (\Cref{resolutiondef}).
\end{proof} 

Additionally, silent transitions give rise to our definition of intermediate bisimulations.
Intuitively, these relax the definition of bisimulation by allowing us to defer the condition imposed for \(\alpha \in \At\) on a related pair to a pair that can be reached via a silent transition.
Although we take inspiration from similar constructs in process algebra~\cite{milner1984lectures} and interaction trees~\cite{xia_InteractionTreesRepresenting_2020}, our definition is subtly different:
(1) Intermediate bisimulations \emph{do not define equivalence}: they are proof techniques to reason about silent jumps and loop restarts.
(2) The matching of silent transitions is offered as an \emph{alternative proof goal}, allowing us to construct intermediate bisimulations from bisimulations.

\begin{definition}[Intermediate Bisimulation \RocqLabelRef{strongbisim}]\label{strongbisim}
    A relation \(R \subseteq (\Expm \times I^Y)^2\) is an intermediate bisimulation between \coalg{\CFGKAT}s \((\Expm \times I^Y, \partial)\resWith{\vec{\gamma}}\) and \((\Expm \times I^Y, \partial)\resWith{\vec{\zeta}}\) if for all \((s, u) \in R\) and \(\alpha \in \At\):
    \begin{itemize}[nosep]
        \item \(s \tr[\vec{\gamma}]{}{\alpha} r\) and \(u \tr[\vec{\zeta}]{}{\alpha} r\) for some \(r \in K\); or
        \item \(s \tr[\vec{\gamma}]{p}{\alpha} s'\) and \(u \tr[\vec{\zeta}]{p}{\alpha} u'\) such that \((s', u') \in R\); or
        \item \(s \transSilent_{\alpha}^{\vec{\gamma}} s'\) and \(u \transSilent_{\alpha}^{\vec{\zeta}} u'\) such that \((s', u') \in R\).
    \end{itemize} 
\end{definition}

\begin{theoremrep}[\RocqLabelRef{strongbisimtobisim}]\label{strongbisimtobisim}
    Any intermediate bisimulation is also a bisimulation.
\end{theoremrep}
\begin{proof}
	To show that \(R\) is a bisimulation, we will need to show that one of the following is true for all \(s, u \in \Expm \times I^Y\) and \(\alpha \in \At\):
    \begin{itemize}[nosep]
    	\item \(s \tr[\vec{\gamma}]{}{\alpha} r\) and \(u \tr[{\vec{\zeta}}]{}{\alpha} r\), for some \(r \in K\);
	\item \(s \tr[\vec{\gamma}]{p}{\alpha} s'\) and \(u \tr[{\vec{\zeta}}]{}{\alpha} u'\), for some \((s', u') \in R\).
    \end{itemize}
    The first two conditions of intermediate bisimulation match these two; therefore, we only need to show that if \(s \transSilent^{\vec{\gamma}}_\alpha s'\), \(u \transSilent^{\vec{\zeta}}_\alpha u'\) and \((s', u') \in R\), then one of the above two conditions holds.
    We do a case analysis on whether \(s\) is in some rejecting set.

    \textbf{If \(s\) is in some rejecting set}, then we take \(N\) to be the union of all rejecting sets for \(\alpha\) in \((\Expm \times I^Y)\resWith{\vec{\gamma}}\).
    Since rejecting sets are closed under union, \(N\) is the largest rejecting set for \(\alpha\) in \((\Expm \times I^Y, \partial)\resWith{\vec{\gamma}}\).
    Let \(N' \eqdef \{u \mid (s, u) \in R, s \in N\}\); we show that \(N'\) is also a rejecting set for \(\alpha\) in \((\Expm \times I^Y, \partial)\resWith{\vec{\zeta}}\).
    Take any \(u \in N'\). Then there exists \((s, u) \in R\) such that \(s \in N\); we prove the rejecting-set condition by case analysis on the witness that \(R\) is an intermediate bisimulation:
    \begin{itemize}[leftmargin=*]
        \item If \(s \tr[\vec{\gamma}]{}{\alpha} r\) and \(u \tr[{\vec{\zeta}}]{}{\alpha} r\) for some \(r \in K\), then by~\Cref{rejectingsetaborts} and \(s \in N\), \(r = \bot\).
        Therefore, \(u \tr[{\vec{\zeta}}]{}{\alpha} \bot\).
        \item If \(s \tr[\vec{\gamma}]{p}{\alpha} s'\) and \(u \tr[{\vec{\zeta}}]{p}{\alpha} u'\), by~\Cref{rejectingsetaborts} and \(s \in N\), \(s\) has to reject \(\alpha\), hence cannot transition to \(s'\), a contradiction.
        \item If \(s \transSilent^{\vec{\gamma}}_{\alpha} s'\) and \(u \transSilent^{\vec{\zeta}}_{\alpha} u'\), we will show that \(u' \in N'\), satisfying the requirement of a rejecting set.
		First, we show that \(s' \in N\), by the following chain of implications:
        \begin{align*}
            & s \tr[\vec{\gamma}]{}{\alpha} \bot
                & \text{\Cref{rejectingsetaborts}} \\
            \implies 
            & s \hookrightarrow^{\vec{\gamma}}_{\alpha} r \text{ is not derivable for any \(r\)}
                & \text{Definition} \\
            \implies
            & s' \hookrightarrow^{\vec{\gamma}}_{\alpha} r \text{ is not derivable for any \(r\)}
                & \text{\Cref{smallerderivation}} \\ 
            \implies 
            & s' \tr[\vec{\gamma}]{}{\alpha} \bot
                & \text{Definition}
        \end{align*}
        Therefore, \(s'\) is in the singleton rejecting set \(\{s'\}\), hence in the union of all rejecting sets \(N\).
        Thus, \(u \transSilent^{\vec{\zeta}}_{\alpha} u'\) and \(u, u' \in N'\), satisfying the condition required for \(N'\) to be a rejecting set.

    \end{itemize}               
    Since both \(N\) and \(N'\) are rejecting sets by~\Cref{rejectingsetaborts}, \(s \tr[{\vec{\gamma}}]{}{\alpha} \bot\) and \(u \tr[{\vec{\zeta}}]{}{\alpha} \bot\), satisfying the condition for \(R\) to be a bisimulation.

    \textbf{If \(s\) is not in any rejecting set}, then by~\Cref{rejectingsetaborts}, \(s \hookrightarrow^{\vec{\gamma}}_\alpha r\) for some \(r \neq \bot\).
    We show a general result that \(R\) satisfies the bisimulation condition for all \(s\) not in any rejecting set, by (strong) induction on the length of the shortest derivation \(s \hookrightarrow^{\vec{\gamma}}_\alpha r\):
    \begin{itemize}[leftmargin=*]
        \item If the derivation is of length 0 (i.e. \(s \hookrightarrow^{\vec{\gamma}}_\alpha r\) is reached directly): we note that there doesn't exist an \(s'\) s.t. \(s \transSilent^{\vec{\gamma}}_\alpha s'\), because \Cref{smallerderivation} implies that \(s' \hookrightarrow^{\vec{\gamma}}_\alpha r\) is reached by a strictly smaller derivation.
        Because \(R\) is an intermediate bisimulation and there is no \(s'\) such that \(s \transSilent^{\vec{\gamma}}_\alpha s'\), one of the following needs to be true:
        \begin{itemize}[nosep, leftmargin=*]
            \item \(s \tr[{\vec{\gamma}}]{}{\alpha} r\) and \(u \tr[{\vec{\zeta}}]{}{\alpha} r\), for some \(r \in K\);
            \item \(s \tr[{\vec{\gamma}}]{p}{\alpha} s'\) and \(u \tr[{\vec{\zeta}}]{p}{\alpha} u'\), for some \((s', u') \in R\);
        \end{itemize}
        satisfying the condition of a bisimulation.
        \item Otherwise, we do a case analysis on the witness of \(R\) being an intermediate bisimulation.
        If one of the following is true:
        \begin{itemize}[nosep, leftmargin=*]
            \item \(s \tr[{\vec{\gamma}}]{}{\alpha} r\) and \(u \tr[{\vec{\zeta}}]{}{\alpha} r\), for some \(r \in K\);
            \item \(s \tr[{\vec{\gamma}}]{p}{\alpha} s'\) and \(u \tr[{\vec{\zeta}}]{p}{\alpha} u'\), for some \((s', u') \in R\);
        \end{itemize}
        then the bisimulation condition is satisfied.

        The non-trivial case is when \(s \transSilent^{\vec{\gamma}}_\alpha s_1\) and \(u \transSilent^{\vec{\zeta}}_\alpha u_1\), for some \((s_1, u_1) \in R\).
        Notice that \(s_1\) cannot be in any rejecting set: if \(s_1\) is in a rejecting set \(N\) and \(s \transSilent^{\vec{\gamma}}_\alpha s_1\), then \(N \cup \{s\}\) is also a rejecting set, contradicting the premise that \(s\) is not in any rejecting set.
        Since \(s_1\) is not in any rejecting set, and \(s_1 \hookrightarrow^{\vec{\gamma}}_\alpha r\) is derivable with a smaller derivation (\Cref{smallerderivation}), the inductive hypothesis gives one of the following:
        \begin{itemize}[nosep, leftmargin=*]
            \item \(s_1 \tr[{\vec{\gamma}}]{}{\alpha} r\) and \(u_1 \tr[{\vec{\zeta}}]{}{\alpha} r\) for some \(r \in K\); by \Cref{smallerderivation}:
            \[
            s \tr[{\vec{\gamma}}]{}{\alpha} r
            \quad\text{and}\quad
            u \tr[{\vec{\zeta}}]{}{\alpha} r\ .
            \]
            \item \(s_1 \tr[{\vec{\gamma}}]{p}{\alpha} s'\) and \(u_1 \tr[{\vec{\zeta}}]{p}{\alpha} u'\) for some \((s', u') \in R\);
            by \Cref{smallerderivation},
            \[
            s \tr[{\vec{\gamma}}]{p}{\alpha} s'\ ,
            \quad
            u \tr[{\vec{\zeta}}]{p}{\alpha} u'\ ,
			\quad\text{and}\quad
			(s', u') \in R\ .
            \]
        \end{itemize}
        Both cases above satisfy the bisimulation conditions.\qed
    \end{itemize}
\end{proof}

The coinductive reasoning principles of rejecting sets and intermediate bisimulations make it easier to prove properties of (resolved versions of) the syntactic \coalg{\CFGKAT}.
They will be used extensively from this point on. 
As a first use of intermediate bisimulations, we observe that resolving an extended environment in stages is equivalent to resolving by the combined environment.

\begin{theoremrep}[\RocqLabelRef{supsyssolsubsyssynt}]\label{supsyssolsubsyssynt}
	If \(E_1 = (E, \ell=f)\), then \((e, \pi)\resWith{E_1} \sim (e, \pi)\resWith{E}\resWith{E_1}\).

\end{theoremrep}
\begin{proof}
	We show that \({\sim} \eqdef \{((e, \pi)\resWith{E_1}, (e, \pi)\resWith{E}\resWith{E_1}) \mid (e, \pi) \in \Expm \times I^Y\}\) is an intermediate bisimulation.
	
	If \((e, \pi) \tr{}{\alpha} r \notin \{\jmpc~k \mid k \in \text{dom}(\gamma)+\ell\} \times I^Y\) for \(r \in K\), then their transitions are equal by \Cref{resolutiondef}.
	
	If \((e, \pi) \tr{p}{\alpha} (e', \sigma)\) then
	\begin{align*}
		(e, \pi) &\tr[E E_1]{p}{\alpha}
 (e', \sigma) & \tCref{resolutiondef}\\
		(e, \pi) &\tr[E]{p}{\alpha}
 (e', \sigma)& \tCref{resolutiondef}
	\end{align*}
	and \((e', \sigma)\resWith{E}\resWith{E_1} \sim (e', \sigma)\resWith{E_1}\).
	
	If \((e, \pi) \tr{}{\alpha} (\jmpc~\ell, \sigma)\), since \(E_1(\ell) = f\) and \(\ell \notin \text{dom}(E)\)
	\begin{align*}
		(e, \pi) &\transSilent_\alpha^{E E_1} (f, 	\sigma) & \tCref{resolutiondef}\\
		(e, \pi) &\transSilent_\alpha^{E_1} (f, \sigma) & \tCref{resolutiondef}
	\end{align*}
	These states are related by \(\sim\).
	
	If \((e, \pi) \tr{}{\alpha} (\jmpc~k, \sigma)\) for \(k \in \text{dom}(E)\) then
	\begin{align*}
		(e, \pi) &\transSilent_\alpha^{E E_1} (E(k), \sigma)  & \tCref{resolutiondef}\\
		(e, \pi) &\transSilent_\alpha^{E_1} (E(k), \sigma)  & \tCref{resolutiondef}
	\end{align*}
 \end{proof}
\end{toappendix}
\section{Axiomatization}\label{sec:axioms}

\begin{figure*}[!tp]
	\centering
	\begingroup
	\setlength{\fboxsep}{0pt}
	\setlength{\tabcolsep}{0pt}
	\setlength{\arrayrulewidth}{\fboxrule}
	\renewcommand{\arraystretch}{0.84}%
	\newcommand{\axctxrule}[3]{\phantomsection\label{#1}\RocqLinkedLabelText{#1}{#2}\,\vcenter{\hbox{\(#3\)}}}
	\newcommand{\axctxstackinf}[2]{\displaystyle\frac{\begin{array}{@{}c@{}}#1\end{array}}{#2}}
	\newcommand{\axctxnamedstackinf}[3]{\begin{array}{@{}l@{}}\DefTirName{#1}\\[-0.35ex]\axctxstackinf{#2}{#3}\end{array}}
	\newcommand{\congcell}[1]{\begin{minipage}[t]{\hsize}\centering\vspace{-0.3em}#1\par\vspace{0.1em}\end{minipage}}%
	\newcommand{\congcellmid}[1]{\begin{minipage}[t]{\hsize}\centering\vspace{1.5em}#1\par\vspace{0.1em}\end{minipage}}%
	\newcommand{\congstacktwo}[2]{%
		\vcenter{\halign{\hfil\(\displaystyle##\)\hfil\cr
			#1\cr
			\noalign{\vskip 0.45ex}
			#2\cr}}}
		\newcommand{\congstackthree}[3]{%
			\vcenter{\halign{\hfil\(\displaystyle##\)\hfil\cr
				#1\cr
				\noalign{\vskip 0.45ex}
				#2\cr
				\noalign{\vskip 0.45ex}
				#3\cr}}}
		\newcommand{\conglabel}[2]{\phantomsection\label{#1}\RocqLinkedLabelText{#1}{\mbox{\textsc{#2}}}}%
		\newcommand{\congline}[3]{\conglabel{#1}{#2}\,#3}
		\newcommand{\axfullwidth}{\dimexpr\linewidth-2\fboxrule\relax}
	\providecommand{\dom}{\operatorname{dom}}
	\newcommand{\axname}[2]{\phantomsection\label{#1}\RocqLinkedLabelText{#1}{#2}}
	\newcommand{\axeq}[4]{\axname{#1}{#2} & \(#3\) & \(\equiv\) & \(#4\)}
	\newcommand{\axbaeq}[4]{\axname{#1}{#2} & \(#3\) & \(\eqBA\) & \(#4\)}
	\newcommand{\axinferrule}[4]{\axname{#1}{#2} & \multicolumn{3}{c}{\(\displaystyle\frac{\displaystyle #3}{\displaystyle #4}\)}}
	\newcommand{\axwhen}[1]{& \multicolumn{3}{c}{\(\text{\textit{when }}#1\)}}
	\newcommand{\axhead}[2]{\multicolumn{#1}{c}{\rule{0pt}{2.3ex}\textbf{#2}}\\[0.1ex]}
	\noindent
	\fbox{%
		\begin{minipage}[t]{\axfullwidth}
			\noindent\begin{tabular}{@{}p{0.53\linewidth}@{}|@{}p{\dimexpr0.47\linewidth-\fboxrule\relax}@{}}
				\begin{tabularx}{\linewidth}[t]{@{}X@{}r@{\hspace{0.3em}}r@{\;}c@{\;}l@{}X@{}}
					\axhead{6}{Boolean Tests}
					& \axbaeq{Axindexclusivity}{\Axindexclusivity}{(x=i)\land(x=j)}{0\,\text{\textit{when }}i{\neq}j} & \\
					& \axbaeq{Axindsum}{\Axindsum}{\bigvee_{i\in I} (x=i)}{1} & \\[0.4ex]
					\hline
					\axhead{6}{Loops}
					& \axeq{Axneutleftloop}{\Axneutleftloop}{1\loopseq e}{e} & \\
					& \axeq{Axrightloopannabort}{\Axrightloopannabort}{0\loopseq e}{0} & \\
					& \axeq{Axrightloopannret}{\Axrightloopannret}{\comReturn\loopseq e}{\comReturn} & \\
					& \axeq{Axrightloopanngoto}{\Axrightloopanngoto}{\comGoto~\ell\loopseq e}{\comGoto~\ell} & \\
					& \axeq{Axbreak}{\Axbreak}{\comBreak \loopseq e}{1} & \\
					& \axeq{Axcontinue}{\Axcontinue}{\comCont \loopseq e}{e} & \\
					& \axeq{Axloopassocact}{\Axloopassocact}{(p\seq f) \loopseq g}{p\seq (f \loopseq g)} & \\
					& \axeq{Axloopassocassign}{\Axloopassocassign}{(x\assign i\seq f)\loopseq g}{x\assign i\seq (f \loopseq g)} & \\
					& \axeq{Axloopdistributivity}{\Axloopdistributivity}{(e +_b f)\loopseq g}{e \loopseq g +_b f \loopseq g} & \\
					& \axeq{Axloopunfold}{\Axloopunfold}{e^{(b)}}{e \loopseq e^{(b)} +_b 1} & \\
					& \axinferrule{Axlooptighten}{\Axlooptighten}
						{(c \land b \seq e) \loopseq \lnot (c \land b) \equiv 0}
						{(e)^{(b)} \equiv (\lnot c \seq e)^{(b)}} & \\[0.1ex]
				\end{tabularx}
				&
				\begin{tabularx}{\linewidth}[t]{@{}X@{}r@{\hspace{0.3em}}r@{\;}c@{\;}l@{}X@{}}
					\axhead{6}{Guarded Choice}
					& \axeq{Axidempotence}{\Axidempotence}{e +_b e}{e} & \\
					& \axeq{Axskewcomm}{\Axskewcomm}{e +_b f}{f +_{\lnot b} e} & \\
					& \axeq{Axskewassoc}{\Axskewassoc}{(e +_b f) +_c g}{e +_{b\land c} (f +_c g)} & \\
					& \axeq{Axguardedness}{\Axguardedness}{e +_b f}{b \seq e +_b f} & \\
					& \axeq{Axrightdistr}{\Axrightdistr}{e \seq g +_b f \seq g}{(e +_b f) \seq g} & \\[0.4ex]
					\hline
					\axhead{6}{Sequential Composition}
					& \axeq{Axassoc}{\Axassoc}{(e \seq f)\seq g}{e \seq (f \seq g)} & \\
					& \axeq{Axabsleft}{\Axabsleft}{0 \seq e}{0} & \\
					& \axeq{Axneutleft}{\Axneutleft}{1 \seq e}{e} & \\
					& \axeq{Axneutright}{\Axneutright}{e \seq 1}{e} & \\
					& \axeq{Axtestcomb}{\Axtestcomb}{b \seq c}{b \land c} & \\
					& \axeq{Axrightseqannret}{\Axrightseqannret}{\comReturn \seq e}{\comReturn} & \\
					& \axeq{Axrightseqanncont}{\Axrightseqanncont}{\comCont \seq e}{\comCont} & \\
					& \axeq{Axrightseqannbrk}{\Axrightseqannbrk}{\comBreak \seq e}{\comBreak} & \\
					& \axeq{Axrightseqanngoto}{\Axrightseqanngoto}{\comGoto~\ell \seq e}{\comGoto~\ell} & \\
				\end{tabularx}
			\end{tabular}
			\par\nointerlineskip\hrule height\fboxrule
			\noindent\begin{tabularx}{\linewidth}{@{}X@{}r@{\hspace{0.3em}}r@{\;}c@{\;}l@{\hspace{2.2em}}r@{\hspace{0.3em}}r@{\;}c@{\;}l@{}X@{}}
				\axhead{10}{Indicators}
				& \axeq{Axindcommact}{\Axindcommact}{x\assign i\seq p}{p\seq x\assign i}
				& \axeq{Axinddistr}{\Axinddistr}{x\assign i\seq (e+_t f)}{x\assign i\seq e+_t x\assign i\seq f} & \\
				& \axeq{Axindval}{\Axindval}{x\assign i\seq x=i}{x\assign i}
				& \axeq{Axindabort}{\Axindabort}{x\assign i\seq 0}{0} & \\
				& \axeq{Axindred}{\Axindred}{x=i\seq x\assign i}{x=i}
				& \axeq{Axindindcomm}{\Axindindcomm}{x\assign i\seq x'\assign j}{x'\assign j\seq x\assign i} & \\
				& \axeq{Axindannih}{\Axindannih}{x\assign i\seq x\assign j}{x\assign j}
				& \axwhen{x \neq x'} & \\
				& \axeq{Axindret}{\Axindret}{x\assign i\seq \comReturn}{\comReturn}
				& \axeq{Axindinddistr}{\Axindinddistr}{x\assign i\seq (e+_{x'=j} f)}{x\assign i\seq e+_{x'=j} x\assign i\seq f} & \\
				& & & & & \axwhen{x \neq x'} & \\[0.15ex]
			\end{tabularx}
			\par\nointerlineskip\hrule height\fboxrule
			\begin{minipage}[t]{\linewidth}
				\vspace{0pt}
				{\centering\rule{0pt}{2.3ex}\textbf{Contextual Axioms}\par}\vspace{0.2ex}
				\mprset{fractionaboveskip=0.12ex, fractionbelowskip=0.08ex}
				\noindent\makebox[\linewidth][c]{\(
				\renewcommand{\arraystretch}{0.92}
				\begin{array}{@{}c@{\hspace{1.5em}}c@{}}
					\axctxrule{Axweaken}{\Axweaken}{
						\inferrule[Weaken]{E \vdash e \equiv f \dashv E}
							{E,\ell=e' \vdash e \equiv f \dashv E,\ell=e'}}
					&
					\axctxrule{Axreplacement}{\Axreplacement}{
						\inferrule[Replacement]{E \vdash e_1 \equiv e_2 \dashv E}
							{E,(\ell=e_1) \vdash e \equiv e \dashv E,(\ell=e_2)}}
					\\[0.45em]
					\axctxrule{Axredundancy}{\Axredundancy}{
						\inferrule[Redundancy]{E,e\text{ contain no }\comGoto\,\ell,\ \ell\in\dom(F)}
							{E,F \vdash e \equiv e \dashv E}}
					&
					\axctxrule{Axrenaming}{\Axrenaming}{
						\inferrule[Renaming]{\ell'\text{ fresh in }E\\ \ell\in\dom(E)}
							{E \vdash \comGoto\,\ell \equiv \comGoto\,\ell' \dashv E\langle\ell'/\ell\rangle}}
					\\[0.45em]
					\multicolumn{2}{@{}c@{}}{
						\axctxrule{Axsubstitution}{\Axsubstitution}{
							\inferrule[Substitution]{\comGoto\,\ell\text{ rightmost in }e}
								{E,(\ell=e_1) \vdash e[e_1/\comGoto\,\ell] \equiv e \dashv E,(\ell=e_1)}}
					}
					\\[0.45em]
					\multicolumn{2}{@{}c@{}}{
						\axctxrule{Axunprbra}{\Axunprbra}{
							\inferrule[Unproductive Branch]{E \vdash \lnot b \seq f \seq b \equiv 0 \dashv E}
								{E,(\ell=g +_b f\seq\comGoto\,\ell) \vdash e \equiv e \dashv E,(\ell=b\seq g)}}
					}
					\\[0.45em]
					\multicolumn{2}{@{}c@{}}{
						\axctxrule{Axuniqueness}{\Axuniqueness}{
							\axctxnamedstackinf{Uniqueness}{
								F \vdash \comGoto\,\ell \equiv E(\ell) \dashv F\\[-0.25ex]
								\dom(F)=\dom(E),\quad
								\forall\ell,\ell'\in\dom(E).\,
								\comGoto\,\ell\text{ guarded in }E(\ell')
							}{
								F \vdash e \equiv e \dashv E
							}
						}
					}
					\end{array}
					\)}\par\vspace{0.35em}\hrule height 0pt
				\end{minipage}%
				\par\nointerlineskip\hrule height\fboxrule
				\begin{minipage}[t]{\linewidth}
					\vspace{0pt}
					{\centering\refstepcounter{figure}\label{cong}\addtocounter{figure}{-1}{\rule{0pt}{2.3ex}}\textbf{Congruence Rules}\par}\vspace{0.35ex}
					\mprset{fractionaboveskip=0.45ex, fractionbelowskip=0.4ex}
					\noindent
					\begin{tabularx}{\linewidth}{@{}>{\hsize=0.96\hsize}Y@{\hspace{0.8em}}>{\hsize=1.1\hsize}Y@{\hspace{0.8em}}>{\hsize=0.94\hsize}Y@{}}%
						\congcell{
							\(\inferrule{E \vdash e \equiv f \dashv F\\ E \vdash g \equiv g \dashv F}{
							\congstacktwo
								{\congline{Axcongrightseq}{\Axcongrightseq}{E \vdash g\seq e \equiv g \seq f \dashv F}}
								{\congline{Axcongrightloopseq}{\Axcongrightloopseq}{E \vdash g\loopseq e \equiv g \loopseq f \dashv F}}}\)
							\par\vspace{0.6em}%
							\(\inferrule{E \vdash e \equiv e \dashv F}
							{\congline{Axcongloopbody}{\Axcongloopbody}{E \vdash e^{(b)} \equiv e^{(b)} \dashv F}}\)
						}
						&
						\congcell{
							\(\inferrule{E \vdash e \equiv f \dashv F\\ E \vdash e' \equiv f' \dashv F}
							{\congline{Axcongguardedchoice}{\Axcongguardedchoice}{E \vdash e+_b e' \equiv f +_b f' \dashv F}}\)
							\par\vspace{0.6em}%
							\(\inferrule{b \eqBA c}{
							\congstackthree
								{\congline{Axcongboolguardedchoice}{\Axcongboolguardedchoice}{e +_b f \equiv e +_c f}}
								{\congline{Axcongboolassertion}{\Axcongboolassertion}{b \equiv c}}
								{\congline{Axcongboolloop}{\Axcongboolloop}{e^{(b)} \equiv e^{(c)}}}}\)
						}
						&
						\congcellmid{
							\(\inferrule{e \equiv f}{
							\congstackthree
								{\congline{Axcongleftseqempty}{\Axcongleftseqempty}{e\seq g \equiv f \seq g}}
								{\congline{Axcongleftloopseqempty}{\Axcongleftloopseqempty}{e \loopseq g \equiv f \loopseq g}}
								{\congline{Axcongloopbodyempty}{\Axcongloopbodyempty}{e^{(b)} \equiv f^{(b)}}}}\)
						}
					\end{tabularx}\par
				\end{minipage}
			\end{minipage}%
		}
		\endgroup
		\caption{The axioms and congruence rules of \CFGKAT. Angle brackets denote componentwise substitution of
		\(\comGoto\,\ell'\) for \(\comGoto\,\ell\) in the domain of \(E\) and all terms in the range of \(E\). \RocqLabelRef{ax}}
		\label{ax}
		
	\end{figure*}

	We now axiomatize the equivalence of (reduced) \CFGKAT expressions in the context of environments. Our axioms will take the shape \({E \vdash e \equiv f \dashv F}\),
	where \(e, f \in \Expm\) and \(E, F\) are environments, that is, partial functions \(L \partialTo \Expm\).
    Often, \(E\) and \(F\) are empty; we thus abbreviate such \emph{non-contextual} axioms \(\varnothing \vdash e \equiv f \dashv \varnothing\) by writing \(e \equiv f\). 
	
	The full axiomatization is presented in \Cref{ax}. In a nutshell, we extend the \GKAT axioms with rules for loops, indicators, and environments. 
	We also carefully restrict the congruence rules, as noted in the introduction, as not all congruences are sound.
	
	A few remarks on \Cref{ax} are warranted.  
	First, note how all axioms except the contextual ones (\textsf{C}) concern equivalence in {\em empty environments}. 
	Second, the two groups of axioms labelled  (\textsf{G}) and (\textsf{S}) are similar to \GKAT, except the right-annihilation rules for non-local control flow primitives. 
	Third, the {\em loop axioms} include rules for both the loop sequencing operation \(\loopseq\) and the loop construct \((-)^{(b)}\). 
	Recall that handles the \(\comBreak\) and \(\comCont\) on its left argument; this makes \(\loopseq\) not associative in general, but only as indicated by \RefAx{\Axloopassocact}--\RefAx{\Axloopdistributivity}.
	\RefAx{\Axlooptighten} states that invariants under which the loop body is unproductive may be replaced with rejection: the premise $(c \land b \seq e) \loopseq \lnot (c \land b) \equiv 0$ asserts that there are no productive behaviors in $e$ that start from a state satisfying $c \land b$ and end in a state falsifying $c \wedge b$
    Some useful derived facts appear in~\Cref{fig:derived-facts}.
	
	\begin{figure}[t]
		\centering
		\fbox{\begingroup%
			\newcommand{\derivfact}[3]{%
				\begingroup
				\renewcommand{\theequation}{#1}%
				\refstepcounter{equation}\label{#2}%
				\begin{array}{@{}r@{\;}l@{}}
					\textup{(\RocqLinkedLabelText{#2}{\theequation})}&\displaystyle #3
				\end{array}%
				\endgroup}%
		\par\smallskip\noindent\makebox[\dimexpr\linewidth-2\fboxsep-2\fboxrule\relax][c]{\(
		\renewcommand{\arraystretch}{1}
		\begin{array}{@{}c@{\hspace{1.5em}}c@{}}%
			\derivfact{D1}{loopassoctest}{b\seq(f\loopseq g) \equiv (b\seq f) \loopseq g}
			&
			\derivfact{D2}{loopseqisseqtest}{b \loopseq e \equiv b\seq e}\\[0.15em]
			\derivfact{D3}{loopseqisseqasasign}{X \assign i \loopseq e \equiv X \assign i \seq e}
			&
			\derivfact{D4}{loopseqisseqact}{p \loopseq e \equiv p \seq e}\\[0.25em]
			\multicolumn{2}{c}{\textit{If }x=j\textit{ is not a subterm of }b\textit{ for any }j \in I\textit{, then}}\\[0.15em]
			\derivfact{D5}{inddistr}{x\assign i \seq (e +_b f) \equiv (x\assign i \seq e +_b x\assign i \seq f)}
			&
			\derivfact{D6}{indcommtest}{x\assign i \seq b \equiv b \seq x\assign i}
		\end{array}
		\)}\par\smallskip
		\endgroup}
		\caption{Derivable facts using the \CFGKAT axioms in \Cref{ax}}\label[figure]{fig:derived-facts}
	\end{figure}

\subsection{Tests and Boolean Algebra}
The standard axioms of Boolean algebra~\cite{Davey_Priestley_2002} are left implicit.
Unlike \KAT and \GKAT, our Boolean algebra is not freely generated: indicator tests on one variable are mutually exclusive, and the disjunction of all tests on one variable is valid.
The two (\textsf{B}) axioms in \Cref{ax} capture these facts.
We record soundness for the Boolean sublanguage.
\begin{theoremrep}[Soundness of Boolean Algebra \RocqLabelRef{Booleansoundness}] \label{Booleansoundness}
	\(b \eqBA c \implies \langle b \rangle = \langle c \rangle\).
\end{theoremrep}
\begin{appendixproof}
	We show this for each axiom in sequence.
	\paragraph*{\RefAx{\Axindexclusivity}}
	\begin{align*}
		\langle x = i \land x=j \rangle &= \langle x = i\rangle  \cap \langle x=j \rangle & \tCref{Booleansemantics}\\
		&= \At \times \{\pi \mid \pi(x)=i\}  \cap \At \times \{\pi \mid \pi(x)=j\} & \tCref{Booleansemantics}\\
		&= \emptyset & i\neq j\\
		&= \langle 0 \rangle & \tCref{Booleansemantics}
	\end{align*}
	
	\paragraph*{\RefAx{\Axindsum}}
	\begin{align*}
		\langle \bigvee_{i \in I} x=i \rangle &= \bigcup_{i \in I} \At \times \{\pi \mid \pi(x)=i\} & \tCref{Booleansemantics}\\
		&= \At \times I^Y & \\
		&= \langle 1 \rangle & \tCref{Booleansemantics} & \qed
	\end{align*}
\end{appendixproof}

\subsection{Contextual rules}
Contextual rules are subtle due to potential self-reference and the power of \comGoto. 
\RefAx{\Axweaken} states that equalities holding in the \emph{same} environment extend to larger environments. 
But this axiom is unsound if \(e\) and \(f\) start out with distinct environments, as the added binding may be used to link a free \(\comGoto\) to a label with distinct bindings. 

Similarly, the replacement rule \RefAx{\Axreplacement}, which states that if two terms are equivalent, then one may be replaced with the other in the environment, requires that equivalence be provable without the label \(\ell\) in the environment,
preventing self-reference from \comGoto. 
Indeed, \RefAx{\Axreplacement}  would be unsound if \(\ell\) is bound in the premise, even if both environments bind it to the same value. This would allow a replacement of \(\ell\) with \(\comGoto~\ell\). 

The redundancy axiom, \RefAx{\Axredundancy}, allows to remove labels which do not play a role in the semantics of $e$: if we can partition the environment into \(E\) and \(F\) such that no label in \(F\) may be reached from \(e\) or \(E\), then \(F\) has no impact on the semantics of \(e\).  

To describe substitution, we formally define syntactic substitution and rightmostness.
\begin{definition}[Syntactic Substitution \RocqLabelRef{synsub}] \label{synsub}
	We define \([g/\comGoto~\ell]: \Expm \to \Expm\) by:
	\[
\begin{array}{@{}r@{\;}c@{\;}l@{\qquad}l@{}}
(e\mathbin{\circ} f)[g/\comGoto\,\ell]
  &\eqdef&
  e[g/\comGoto\,\ell]\mathbin{\circ} f[g/\comGoto\,\ell]
  & \circ \in \{+_b,\seq,\loopseq\},
\\
(e^{(b)})[g/\comGoto\,\ell]
  &\eqdef&
  \bigl(e[g/\comGoto\,\ell]\bigr)^{(b)}
\\
v[g/\comGoto\,\ell]
  &\eqdef&
  \begin{cases}
    g, & v = \comGoto\,\ell,\\
    v, & v \ne \comGoto\,\ell
  \end{cases}
  & v \text{ primitive}.
\end{array}
\]%
\end{definition}
\begin{definition}[Rightmost \RocqLabelRef{rightmostdef}]\label{rightmostdef}
    For a label $\ell$, define $\Expm_r(\ell), \mathrm{RHS}(\ell) \subseteq \Expm$:
	\begin{align*}
		f, g &\in \Expm_r(\ell) \eqdef  \Sigma \mid b \in \BExp \mid \comBreak \mid \comReturn \mid \comCont \mid\\&\quad \comGoto~k \; \text{when } k\neq \ell \mid x\assign i \mid f +_{b} g \mid f\seq g \mid (f)^{(b)} \mid f\loopseq g 
		\\\
		h, j &\in \mathrm{RHS}(\ell) \eqdef  \comGoto~\ell \mid f \mid h +_{b} j \mid g\seq h \mid g\loopseq h
	\end{align*}
	We say that \(\comGoto~\ell\) is \emph{rightmost} in \(e \in \Expm\), if \(e \in \mathrm{RHS}(\ell)\).
\end{definition}
\RefAx{\Axsubstitution} allows for a rightmost substitution using the environment, inserting it into the main term.
\RefAx{\Axrenaming} allows for the renaming of labels. 
\RefAx{\Axunprbra} is the analogue of \RefAx{\Axlooptighten} for \comGoto: it allows us to remove an unproductive invariant of \comGoto. 
Note that, under bisimilarity, \(e \equiv 0\) is a strong statement: it entails that \(e\) performs no actions whatsoever.
In particular, \(p \seq 0 \not\equiv 0\), because $p \seq 0$ performs the action $p$ prior to failure.
The premise of \RefAx{\Axunprbra} thus expresses that the code at \(\ell\) silently recurses (and thus diverges) when $b$ is false.

\RefAx{\Axuniqueness} states that productive environments, treated as fixpoints, are unique. 
This is the analogue of the uniqueness axiom of \GKAT \cite{gkat} or the recursive specification principle in process algebra \cite{bergstra}. 
\RefAx{\Axuniqueness} is particularly interesting as it allows us to capture productive systems of equations in a single rule, avoiding the infinitary axiom schema found in other \GKAT variants \cite{gkat,rozowski2023probabilistic,van2025weighted}. 
For \RefAx{\Axuniqueness} to be sound we require the notion of \emph{guardedness}, which states that a \(\comGoto~\ell\) may execute only after at least one action.
Like the empty word property in the system of~Salomaa~\cite{salomaa}, guardedness lets us uniquely identify fixpoints.
In the following definition, we describe guardedness as a decidable semantic property, however, it can be characterized syntactically with~\Cref{fundthm}.
\begin{definition}[Semantic Guardedness \RocqLabelRef{guardedsemdef}] \label{guardedsemdef}
	We say that \(\comGoto~\ell\) is guarded in an expression \(e \in \Expm\) if for all \(\alpha, \sigma \in \At, \pi \in I^Y\), it holds that \((e, \pi) \hookrightarrow_{\alpha} r \neq (\jmpc~\ell, \sigma)\).
\end{definition}

The premise $F\vdash \comGoto~\ell \equiv E(\ell) \dashv F$ in \RefAx{\Axuniqueness} captures the idea that \emph{$F$ solves the equation $E$}, generalizing ``productive unrolling'' from \GKAT~\cite{gkat}: plugging in \(F\) for the label \(\ell\) on the left matches the result of plugging $F$ into the labels of $E(\ell)$ on the right.

Finally, the congruence rules are subtle. For example, in \RefAx{\Axcongrightseq}, the left term in the sequencing must be identical rather than merely equivalent.
This is necessary: consider the equivalence \((\ell = p) \vdash \comGoto~\ell \equiv p \dashv \varnothing\); 
sequencing on the right breaks equivalence because \(\comGoto\) is right annihilating whereas \(p\) is not.
Allowing mere equivalence on the left is similar to performing a non-rightmost substitution, which is indeed unsound.

\begin{toappendix}

We first need some lemmas before proving soundness.

	\begin{lemma}[\RocqLabelRef{notransitionnogoto}]\label{notransitionnogoto}
	For any \(e\in \Expm, \pi \in I^Y\), if \(\comGoto~\ell\) is not a subterm of \(e\) and \((e, \pi) \tr{p}{\alpha} (e', \sigma)\) then \(\comGoto~\ell\) is not a subterm of \(e'\) and \((e, \pi) \allowbreak\not\tr{}{\alpha} (\jmpc~\ell, \sigma)\) for any \(\sigma \in I^Y, \alpha \in \At\).

\end{lemma}
	\begin{proof}
	
	We will define \(R(e)\) as an upper bound on the set of expressions reachable from \(e\) in at least one step. That is, if some number of transitions with \((\alpha_1, \dots \alpha_n)\) starting with \((e, \pi)\) for some \(\pi\) result in \((f, \sigma)\), then \(f \in R'(e)\). Let \(R\) be defined inductively as
	\[\begin{array}{r c l}
		R'(p) &\coloneq &\{1\} \\
		R'(e+_b f)& \coloneq &R(e) \cup R(f) \\
		R'(e \loopseq f) &\coloneq &(R(e) \loopseq^\triangle \{f\}) \cup \{R(f)\}\\
		R'(e \seq f) &\coloneq &(R(e) \seq^\triangle \{f\}) \cup \{R(f)\} \\ 
		R'(e^{(b)}) &\coloneq &(R(e) \loopseq^\triangle \{e^{(b)}\})\\
		R'(e) &\coloneq & \emptyset \quad \text{else}
	\end{array}\]
	where \(\square^\triangle\) for \(\square \in \{\seq, \loopseq\}\) denotes a componentwise composition. Let \(R(e) = R'(e) \cup \{e\}\).
	
	\((f, \sigma) \in \langle e, \pi \rangle \implies f \in R(e)\). This is proven by induction on the construction of \(e\). The base cases are straightforward. Guarded choice follows from the fact that \(e+_b f\) cannot reach any states other than those reachable by \(e\) or \(f\). Both kinds of composition follow from the fact that all terms reachable from \(e \square f\) are either states in \(e\) sequenced with \(f\) or are states of \(f\). Finally the loop case follows from the fact that all states reachable from \(e^{(b)}\) are states reachable from \(e\) composed with \(e^{(b)}\) or are \(e^{(b)}\).
	
	By a straightforward inductive argument if \(e' \in R(e)\) contains \(\comGoto~k\) as a subterm, then so does \(e\). Hence by contrapositive if \(e\) does not contain \(\comGoto~k\) as a subterm, then for any \(e'\) such that \(e' \in R(e)\), \(e'\) does not contain \(\comGoto~k\) as a subterm.
	
	We now show the second part of the lemma statement by showing via induction that if there exists some \(\pi, \sigma \in I^Y, \alpha \in \At\) such that \((e, \pi) \tr{}{\alpha} (\jmpc~k, \sigma)\) then \(e\) contains \(\comGoto~k\) as a subterm. We will call the above condition ``\(e\) transitioning to \(\jmpc~k\)'' in order to omit explicitly quantifying the assignment each time.
	
	The base cases are straightforward.
	
	If \(e = f +_b g\), then either \(f\) or \(g\) transitions to \((\jmpc~k, \sigma)\) under \(\alpha\) and some \(\nu\). Hence one of them contains \(\comGoto~k\) as a subterm, and so does \(e\).
	
	If \(e = f \seq g\), then either \(f\) transitions to \((\jmpc~k, \sigma)\) under \(\alpha\) and some \(\nu\), or \(f\) transitions to \((\acc, \nu)\) for some \(\nu\), and \(g\) transitions to \((\jmpc~k, \sigma)\) under \(\alpha\) and \(\nu\). In either case one of \(f\) or \(g\) contains \(\comGoto~k\) as a subterm and so does \(e\).
	
	If \(e = f \loopseq g\), the case is similar to the one above, just accounting for \((\contc, \nu)\) in addition to \((\acc, \nu)\).
	
	If \(e = e^{(b)}\), then it must be that \(e\) transitioned to \(\jmpc~k\), and hence \(e\) contains \(\comGoto~k\) as a subterm.
	
	Hence, if \(e\) does not contain \(\comGoto~k\) as a subterm then there exists no \(\alpha \in \At, \pi, \sigma \in I^Y\) such that \((e, \pi) \tr{}{\alpha} (\jmpc~k, \sigma)\).
	\end{proof}

We will use the following notation throughout the soundness arguments.
		\begin{definition}[Lifting Relations \RocqLabelRef{def:relation-lifting}]\label{def:relation-lifting}
		For two \coalg{\CFGKAT}s \((S, \delta)\), \((U, \eta)\) and a relation \(R \subseteq S \times U\), we use the notation \(R_F\) when the following is true:
		\begin{itemize}
			\item If \(r \in K\), then \(r \sim_F r\);
			\item If \(s_1 \sim s_2\) and \(p \in \Sigma\), \((p, s_1) R_F (p, s_2)\).
		\end{itemize}
		By unfolding the definition, we know any relation \(\sim\) is a bisimulation if and only if, for all \(s_1 \sim s_2\), \(s_1 \hookrightarrow_\alpha^\delta r_1\), and \(s_2 \hookrightarrow_\alpha^\eta r_2\), we have \(r_1 \sim_F r_2\).

	\end{definition}

	The following lemma is useful for a few contextual axioms.
		\begin{lemma}[\RocqLabelRef{thm:substitute-preserve-non-jump}]\label{thm:substitute-preserve-non-jump}
		Fix a label \(\ell \in \lset\), an atom \(\alpha \in \At\), and an expression \(e\). If \(r \in K\) is not \((\jmpc~\ell, \pi)\) for any \(\pi \in I^Y\), then for all expressions \(f\),
		\begin{align*}
			(e, \sigma) \tr{}{\alpha} r 
				& \iff (e[f/\comGoto~\ell], \sigma) \tr{}{\alpha} r \\  
			(e, \sigma) \tr{p}{\alpha} (e', \sigma') 
				& \iff (e[f/\comGoto~\ell], \sigma) \tr{p}{\alpha} (e'[f/\comGoto~\ell], \sigma')
		\end{align*}
	\end{lemma}
	\begin{proof}
		We prove this by induction on the structure of \(e\), and we use ``IH'' to denote the inductive hypothesis.
					
		\textbf{Goto case:} \(e \eqdef \comGoto~\ell\), then none of the premises hold, making the equivalence trivially true. 

		\textbf{Primitive action case:} \(e \eqdef p\), then
		\[(p, \sigma) \tr{p}{\alpha} (1, \sigma) 
		\quad\text{and}\quad 
		(p[f/\comGoto~\ell], \sigma) \tr{p}{\alpha} (1[f/\comGoto~\ell], \sigma)\]
		\textbf{All other base cases} produce continuations \(r \in K\), and are preserved by the substitution, i.e. \(e = e[f/\comGoto~\ell]\).
		Therefore,
		\[(e, \sigma) \tr{}{\alpha} r 
		\quad\text{and}\quad 
		(e[f/\comGoto~\ell], \sigma) \tr{}{\alpha} r \]
		\textbf{Guarded choice case:} \(e \eqdef e_1 +_b e_2\), w.l.o.g. we assume that \(\alpha \in \langle b \rangle\), then the following implication can be derived by semantics, definition of substitution, and the inductive hypothesis:
		\begin{align*}
			(e, \sigma) \tr{}{\alpha} r &\iff (e_1, \sigma) \tr{}{\alpha} r \\
			&\iff (e_1[f/\comGoto~\ell], \sigma) \tr{}{\alpha} r\\
			&\iff (e[f/\comGoto~\ell], \sigma) \tr{}{\alpha} r \\  
			(e_1, \sigma) \tr{p}{\alpha} (e', \sigma') &\iff (e_1, \sigma) \tr{p}{\alpha} (e', \sigma')\\
			&\iff (e_1[f/\comGoto~\ell], \sigma) \tr{p}{\alpha} (e'[f/\comGoto~\ell], \sigma')\\
			&\iff (e[f/\comGoto~\ell], \sigma) \tr{p}{\alpha} (e'[f/\comGoto~\ell], \sigma')
		\end{align*}
		\textbf{Sequencing case:} \(e \eqdef e_1 \seq e_2\), then by the definition of substitution
		\[e[f/\comGoto~\ell] = e_1[f/\comGoto~\ell]\seq e_2[f/\comGoto~\ell]\ .\] 
		\begin{itemize}[nosep]
			\item If \((e_1, \sigma) \tr{}{\alpha} (\acc, \sigma')\), then for \(r \in K\) that is not \((\jmpc~\ell, \pi)\):
			\begin{align*}
				(e, \sigma) \tr{}{\alpha} r
				&\iff  (e_2, \sigma) \tr{}{\alpha} r\\
				&\iff  (e_2[f/\comGoto~\ell], \sigma) \tr{}{\alpha} r\\
				&\iff  (e[f/\comGoto~\ell], \sigma) \tr{}{\alpha} r \\
				(e, \sigma) \tr{p}{\alpha} (e', \sigma') 
				&\iff  (e_2, \sigma) \tr{p}{\alpha} (e', \sigma')\\
				&\iff  (e_2[f/\comGoto~\ell], \sigma) \tr{p}{\alpha} (e'[f/\comGoto~\ell], \sigma')\\
				&\iff  (e[f/\comGoto~\ell], \sigma) \tr{p}{\alpha} (e'[f/\comGoto~\ell], \sigma')
			\end{align*}
			\item If \((e_1, \sigma) \tr{p}{\alpha} (e_1', \sigma)\), then \(e \tr{p}{\alpha} (e_1' \seq e_2, \sigma')\), and we denote \(e' \eqdef e_1' \seq e_2\).
			\begin{align*}
				(e, \sigma) \tr{p}{\alpha} (e', \sigma')
				&\iff  (e_2, \sigma) \tr{p}{\alpha} (e', \sigma')\\
				&\iff  (e_2[f/\comGoto~\ell], \sigma) \tr{p}{\alpha} (e'[f/\comGoto~\ell], \sigma')\\
				&\iff  (e[f/\comGoto~\ell], \sigma) \tr{p}{\alpha} (e'[f/\comGoto~\ell], \sigma')
			\end{align*}
			\item Otherwise, \((e_1, \sigma) \tr{}{\alpha} r \in K\), for some continuation \(r\) that annihilates \(e_2\), like \(\brkc, \contc, \jmpc\), or \(\retc\):
			\begin{align*}
				(e_1, \sigma) \tr{}{\alpha} r
				\iff & (e_1[f/\comGoto~\ell], \sigma) \tr{}{\alpha} r
					& \text{IH}\\
				\iff & (e[f/\comGoto~\ell], \sigma) \tr{}{\alpha} r 
					& \text{\(r\) annihilates}
			\end{align*}
		\end{itemize}

		\textbf{Loop sequencing case:} \(e \eqdef e_1 \loopseq e_2\), then it follows a similar pattern to \(e_1 \seq e_2\) (above case), accounting for \(\brkc\) and \(\contc\).
		The only different case is when \((e_1, \sigma) \tr{}{\alpha} (\brkc, \sigma')\), then \((e_1 \seq e_2, \sigma) \tr{}{\alpha} (\acc, \sigma')\):
		\begin{align*}
			(e_1, \sigma) \tr{}{\alpha} (\brkc, \sigma')
			 \iff&
			(e_1[f/\comGoto~\ell], \sigma) \tr{}{\alpha} (\brkc, \sigma')
				& \text{IH} \\
			 \iff&
			(e'[f/\comGoto~\ell], \sigma) \tr{}{\alpha} (\acc, \sigma')
		\end{align*}
		\textbf{Loop case:} \(e \eqdef {e_1}^{(b)}\), we first focus on the case where \((e, \sigma) \hookrightarrow_\alpha r\) is derivable for some result \(r \neq \bot\), then we show the theorem holds by performing strong induction on the size of the derivation for \((e, \sigma) \hookrightarrow_\alpha r\): The base case is only when \((\alpha, \sigma) \notin \langle b\rangle\), then 
		\[(e, \sigma) \tr{}{\alpha} (\acc, \sigma)
		\quad\text{and}\quad 
		({e_1}^{(b)}[f/\comGoto~\ell], \sigma) \tr{}{\alpha} (\acc, \sigma)\]
		For the inductive case, we do a case analysis on the transition of \((e_1, \sigma)\):
		\begin{itemize}[leftmargin=*]
			\item We first present the non-trivial case: if \((e_1, \sigma) \tr{}{\alpha} (c, \sigma')\) where \(c \in \{\acc, \contc\}\), then by the inductive hypothesis, \((e_1[f/\comGoto~\ell], \allowbreak \sigma) \tr{}{\alpha} (c, \sigma')\).
			Thus, we have the following silent transitions
			\[
				({e_1}^{(b)}, \sigma) \transSilent (e_1^{(b)}, \sigma')
				\quad\text{and}\quad
				 ({e_1}^{(b)}[f/\comGoto~\ell], \sigma) \transSilent (e_1^{(b)}[f/\comGoto~\ell], \sigma')\ .
			\] 
			By~\Cref{smallerderivation}, \(({e_1}^{(b)}, \sigma') \hookrightarrow_\alpha r\) is derivable with a smaller derivation, giving us the following inductive hypothesis by reducing the derivation length:
			\begin{align*}
				(e, \sigma') \tr{}{\alpha} r
				 &\iff (e[f/\comGoto~\ell], \sigma') \tr{}{\alpha} r \\
				(e, \sigma') \tr{p}{\alpha} (e', \sigma_n) 
				&\iff (e[f/\comGoto~\ell], \sigma') \tr{p}{\alpha} (e'[f/\comGoto~\ell], \sigma_n)
			\end{align*}
			With~\Cref{smallerderivation} and the above induction hypothesis, we derive the desired equivalences: let \(r \in K\) and \(r \neq \bot\),
			\begin{align*}
				(e, \sigma) \tr{}{\alpha} r 
				& \iff (e, \sigma') \tr{}{\alpha} r \\
				& \iff (e[f/\comGoto~\ell], \sigma') \tr{}{\alpha} r \\
				& \iff (e[f/\comGoto~\ell], \sigma) \tr{}{\alpha} r \\
				(e, \sigma) \tr{p}{\alpha} (e', \sigma_n)
				& \iff (e, \sigma') \tr{p}{\alpha} (e', \sigma_n) \\
				& \iff (e[f/\comGoto~\ell], \sigma') \tr{p}{\alpha} (e'[f/\comGoto~\ell], \sigma_n) \\
				& \iff (e[f/\comGoto~\ell], \sigma) \tr{p}{\alpha} (e'[f/\comGoto~\ell], \sigma_n)
			\end{align*}
			\item If \((e_1, \sigma) \tr{p}{\alpha} (e_1', \sigma')\), then \((e, \sigma) \tr{p}{\alpha} (e_1' \loopseq e_2, \sigma')\), and we denote \(e' \eqdef e_1' \loopseq e_2\).
			We have the following inductive hypothesis:
			\[
				(e_1, \sigma) \tr{p}{\alpha} (e_1', \sigma') 
				\iff (e_1[f/\comGoto~\ell], \sigma) \tr{p}{\alpha} (e_1'[f/\comGoto~\ell], \sigma')
			\]
			By unfolding the definition of substitution and semantics:
			\[
				(e, \sigma) \tr{p}{\alpha} (e', \sigma') \iff (e[f/\comGoto~\ell], \sigma) \tr{p}{\alpha} (e'[f/\comGoto~\ell], \sigma')
			\]
			\item If \((e_1, \sigma) \tr{}{\alpha} r\), where \(r\) annihilates further actions, i.e. \(r \in \{\retc\} \cup \{\jmpc~\ell'\} \times I^X\) for \(\ell' \neq \ell\), then by the inductive hypothesis and unfolding the operational semantics
			\[
				({e_1}^{(b)}, \sigma) \tr{}{\alpha} r 
				\qquad\text{and}\qquad
				({e_1}^{(b)}[f/\comGoto~\ell], \sigma) \tr{}{\alpha} r  \qed
			\]
		\end{itemize}

	\end{proof}

		\begin{lemma}[\RocqLabelRef{indnotrefbool}]\label{indnotrefbool}
		If \(b\) does not reference indicator \(x\), then \((\alpha, \pi) \in \langle b \rangle \iff (\alpha, \pi[x \leftarrow i])\in \langle b \rangle\). 

	\end{lemma}
	\begin{proof}
		We prove this by structural induction on \(b\). The base cases for \(0, 1, t\) result from the fact that no indicators are ruled out by their interpretations. The base case for \(x'=j\) follows from the fact that \(\pi(x') = j \iff \pi[x \leftarrow i](x') = j\), as \(x\neq x'\). 
		
		For our first inductive case:
		\begin{align*}
			(\alpha, \pi) \in \langle b \land c\rangle &\iff (\alpha, \pi) \in \langle b \rangle \land (\alpha, \pi) \in \langle c \rangle\\
			&\iff (\alpha, \pi[x\leftarrow i]) \in \langle b \rangle \land (\alpha, \pi[x\leftarrow i]) \in \langle c \rangle& \text{ind hyp}\\
			&\iff(\alpha, \pi[x\leftarrow i]) \in \langle b \land c\rangle
		\end{align*}
		A similar argument holds for \(\lor\) and  \(\lnot b\).
	\end{proof}
		\begin{lemma}[\RocqLabelRef{renamingsemlem}]\label{renamingsemlem}
		Replacing a \(\comGoto~\ell\) with \(\comGoto~\ell'\) changes the jump continuation from \(\ell\) to \(\ell'\):
		\[(e, \pi) \tr{}{\alpha} (\jmpc~\ell, \sigma)
		\implies
		(e[\comGoto~\ell'/\comGoto~\ell], \pi) \tr{}{\alpha} (\jmpc~\ell', \sigma)\]
	\end{lemma}
	\begin{proof}
		We prove this by induction on the structure of \(e\): because of the premise \((e, \pi) \tr{}{\alpha} (\jmpc~\ell, \sigma)\), we only have a few cases to consider.
		We use ``IH'' as an abbreviation for inductive hypothesis.

		\textbf{Base case:} The only relevant case is \(e \eqdef \comGoto~\ell\); in this case \(e[\comGoto~\ell'/\comGoto~\ell] = \comGoto~\ell'\), hence
		\[(e[\comGoto~\ell'/\comGoto~\ell], \pi) \tr{}{\alpha} (\jmpc~\ell', \sigma).\]
		\textbf{Guarded choice case:} \(e \eqdef e_1 +_b e_2\). W.l.o.g. we assume \((\alpha, \pi) \in \langle b \rangle\), then by definition of substitution and the operational semantics: 
		\begin{align*}
			(e, \pi) \tr{}{\alpha} (\jmpc~\ell, \sigma)
			&\implies (e_1, \pi) \tr{}{\alpha} (\jmpc~\ell, \sigma)\\
			& \implies (e_1[\comGoto~\ell'/\comGoto~\ell], \pi) \tr{}{\alpha} (\jmpc~\ell, \sigma)\\
			& \implies (e[\comGoto~\ell'/\comGoto~\ell], \pi) \tr{}{\alpha} (\jmpc~\ell, \sigma)
		\end{align*}
		\textbf{Sequencing case:} \(e \eqdef e_1 \seq e_2\), then there are two cases:
		\begin{itemize}[leftmargin=*]
			\item First, if \((e_1, \pi) \tr{}{\alpha} (\jmpc~\ell, \sigma)\), then by the inductive hypothesis and the operational semantics:
			\begin{align*}
				(e_1, \pi) \tr{}{\alpha} (\jmpc~\ell, \sigma) & \implies (e_1[\comGoto~\ell'/\comGoto~\ell], \pi) \tr{}{\alpha} (\jmpc~\ell', \sigma) \\ 
				& \implies (e[\comGoto~\ell'/\comGoto~\ell], \pi) \tr{}{\alpha} (\jmpc~\ell', \sigma)
			\end{align*}
			\item If \((e_1, \pi) \tr{}{\alpha} (\acc, \pi')\) and \((e_2, \pi') \tr{}{\alpha} (\jmpc~\ell, \sigma)\), then by the inductive hypothesis:
			\begin{align*}
				(e_2, \pi') \tr{}{\alpha} (\jmpc~\ell, \sigma) &\implies (e_2[\comGoto~\ell'/\comGoto~\ell], \pi') \tr{}{\alpha} (\jmpc~\ell', \sigma)
			\end{align*}
			and because of~\Cref{thm:substitute-preserve-non-jump}, 
			\[(e_1[\comGoto~\ell'/\comGoto~\ell], \pi) \tr{}{\alpha} (\acc, \pi')\]
			Finally, because of the operational semantics:
			\[(e[\comGoto~\ell'/\comGoto~\ell], \pi) \tr{}{\alpha} (\jmpc, \sigma)\]
		\end{itemize}

		\textbf{Loop sequencing case:} \(e \eqdef e_1 \loopseq e_2\); this mirrors the proof of the sequencing case, except we replace \(\acc\) with either \(\acc\) or \(\contc\) in the second subcase.

		\textbf{Loop case:} \(e \eqdef {e_1}^{(b)}\); we perform a strong induction on the size of the derivation of \((e, \pi) \tr{}{\alpha} (\jmpc, \sigma)\).
		There are only two cases, both requiring \((\alpha, \pi) \in \langle b \rangle\): 
		\begin{itemize}[leftmargin=*]
			\item If \((e_1, \pi) \tr{}{\alpha} (\jmpc~\ell, \sigma)\), then by the inductive hypothesis:
			\begin{align*}
				(e_1, \pi) \tr{}{\alpha} (\jmpc~\ell, \sigma) & \implies (e_1[\comGoto~\ell'/\comGoto~\ell], \pi) \tr{}{\alpha} (\jmpc~\ell, \sigma) \\ 
				& \implies (e[\comGoto~\ell'/\comGoto~\ell], \pi) \tr{}{\alpha} (\jmpc~\ell, \sigma)
			\end{align*}
			\item If \((e_1, \pi) \tr{}{\alpha} (c, \pi')\) where \(c \in \{\acc, \contc\}\), and \((e, \pi') \tr{}{\alpha} (\jmpc~\ell, \sigma)\) then the derivation of \((e, \pi') \tr{}{\alpha} (\jmpc~\ell, \sigma)\) is a sub-derivation of \((e, \pi) \tr{}{\alpha} (\jmpc~\ell, \sigma)\), giving us the inductive hypothesis:
			\[(e[\comGoto~\ell'/\comGoto~\ell], \pi') \tr{}{\alpha} (\jmpc~\ell, \sigma).\]
			By~\Cref{thm:substitute-preserve-non-jump}, \((e_1[\comGoto~\ell'/\comGoto~\ell], \pi) \tr{}{\alpha} (c, \pi')\).
			By \(c \in \{\acc, \contc\}\), the fact above, and the operational semantics, we reach the desired conclusion:
			\[(e[\comGoto~\ell'/\comGoto~\ell], \pi) \tr{}{\alpha} (\jmpc~\ell, \sigma).\qed\]
		\end{itemize}
	\end{proof}

	\begin{remark}
		In the above lemma, we perform a nested induction in the loop case.
		Alternatively, we can also only induct on the size of the derivation of \((e, \pi) \tr{}{\alpha} (\jmpc~\ell, \sigma)\), because taking a sub-expression of \(e\) also reduces the derivation length.
		In this way, the above lemma can be proven using just a single induction, which might make the formalization cleaner. 
	\end{remark}
	
		\begin{lemma}[Common suffix lifting \RocqLabelRef{common-suffix-lifting}]\label{common-suffix-lifting}
		Let \(E,F,G\) be environments. Suppose that 
		for every \(\ell \in \lset,\nu \in I^Y,h \in \Expm\), if \(G(\ell)=h\) then
		\((h,\nu)\resWith{E}\sim (h,\nu)\resWith{F}\).  If
		\((s)\resWith{E}\sim (t)\resWith{F}\), then
		\((s)\resWith{[EG]}\sim (t)\resWith{[FG]}\), where \(s,t \in \Expm \times I^Y\).

	\end{lemma}
	\begin{proof}
		Let \(R\) relate states whose prefix resolutions are bisimilar:
		\((s, t) \in R \iff (s)\resWith{E}\sim (t)\resWith{F}\).  We show that \(R\) is
		an intermediate bisimulation after resolving both sides with \(G\).  
		
		Fix \(s, t\) and an atom \(\alpha\).  Choose a bisimulation \(S\)
		witnessing \((s)\resWith{E}\sim (t)\resWith{F}\).  Its one-step condition
		relates the \(E\)-resolved transition of \(s\) and the \(F\)-resolved
		transition of \(t\).
		
		If \(s \tr{}{\alpha} k \in K\) where \(k\) is not a jump in \(\dom(G)\), then resolution \(G\) leaves matching outputs.  
		If \(s \tr{p}{\alpha} s'\), then by the original bisimulation \(t \tr{p}{\alpha} t'\) where
		\((s')\resWith{E}\sim(t')\resWith{F}\), hence \((s',t') \in R\).
		
		It remains to consider a jump \((\jmpc~\ell,\nu)\) which is bound
		by \(G\), say \(G(\ell)=h\).  After appending \(G\), \(s \transSilent_\alpha^E (h,\nu)\).
		By the prefix bisimulation, \(t\) must make the same jump, hence \(t \transSilent_\alpha^E (h,\nu)\).
		By the side condition, \((h,\nu)\resWith{E}\sim(h,\nu)\resWith{F}\), so
		\(((h,\nu),(h,\nu)) \in R\).  Hence \(R\) is an intermediate bisimulation, and by \Cref{strongbisimtobisim} it is a bisimulation.
	\end{proof}
\end{toappendix}
Our axiomatization of \CFGKAT is {\em sound} with respect to bisimilarity.

\begin{theoremrep}[Soundness \RocqLabelRef{thm:sound}]\label{thm:sound} Let \(e,f \in \Expm\), and \(E,F\) be environments.
If \(E \vdash e \equiv f \dashv F\), then for all $\pi \in I^Y$ it holds that \((e, \pi)\resWith{E} \sim (f, \pi)\resWith{F}\).
\end{theoremrep}
The judgment \(E \vdash e \equiv f \dashv F\) denotes the least equivalence relation generated by the rules in \Cref{ax}.
In particular, observe that instantiating \RefAx{\Axcongrightseq}, \RefAx{\Axcongrightloopseq}, and \RefAx{\Axcongguardedchoice} with \(\varnothing\) as all environments yields a full congruence over expressions and operations.
\begin{appendixproof}		
	We start with the \textbf{non-contextual axioms}.
	
	For these axioms we prove this by proving the bisimilarity in the syntactic coalgebra. By \Cref{resolutiondef}, this results in their bisimilarity in the coalgebra resolved by the empty set. 
	
	We establish bisimilarity by using a distinct bisimulation for each axiom (and the following identity). 
	Unless otherwise stated we will show that the transitions of two states coincide to establish bisimulation up to equivalence closure \cite[Section 4.2]{coinductionupto}. 
	That is, to show \(e\equiv f\) is sound we show that \(\{(e, \pi), (f, \pi) \mid \pi \in I^Y\}\) is a bisimulation up to equivalence closure. 
	
	We will add one more identity for pairs, which is that \[ (e, \pi[x \leftarrow i]) \sim (x\assign i\seq e, \pi)\]
	
	\paragraph*{Equivalence axioms}
	
	These follow immediately from bisimulation up to the equivalence closure \cite[Section 4.2]{coinductionupto}.
	
	\paragraph*{Additional Identity}
	For any \(\alpha,\pi\), 
	\begin{align*}
		(x \assign i \seq e, \pi)\hookrightarrow_\alpha r \iff (e, \pi[x\leftarrow i])\hookrightarrow_\alpha r
	\end{align*}
	by \Cref{operationalsemantics}.
	
	\paragraph*{\RefAx{\Axidempotence}}
	For any \(\alpha,\pi\), suppose \((\alpha, \pi) \in \langle b \rangle\), then
	\begin{align*}
		(e, \pi) \hookrightarrow_\alpha r \iff (e+_b e, \pi) \hookrightarrow_\alpha r
	\end{align*}
	A symmetric argument holds if \((\alpha, \pi) \in \langle \lnot b \rangle\).
	\paragraph*{\RefAx{\Axskewcomm}}
	For any \(\alpha,\pi\), suppose \((\alpha, \pi) \in \langle b \rangle\), then
	\begin{align*}
		(e, \pi) \hookrightarrow_\alpha r \!&\iff\!\! (f +_{\lnot b} e, \pi) \hookrightarrow_\alpha r \\&\iff\!\! (e +_b f , \pi) \hookrightarrow_\alpha r& \tCref{operationalsemantics,Booleansemantics}
	\end{align*}
	A symmetric argument holds if \((\alpha, \pi) \in \langle \lnot b \rangle\). Hence
	\[(f +_{\lnot b} e, \pi) \hookrightarrow_\alpha r \iff (e +_b f , \pi) \hookrightarrow_\alpha r\]
	\paragraph*{\RefAx{\Axskewassoc}}
	For any \(\alpha,\pi\), suppose \((\alpha, \pi) \in \langle \lnot c \rangle\), then 
	\begin{align*}
		\MoveEqLeft[1] ((e+_b f)+_cg, \pi) \hookrightarrow_\alpha r \\
		&\iff (g, \pi) \hookrightarrow_\alpha r\\
		&\iff (f+_c g, \pi) \hookrightarrow_\alpha r\\
		&\iff (e+_{b\land c} (f+_c g), \pi) \hookrightarrow_\alpha r & \tCref{operationalsemantics,Booleansemantics}
	\end{align*}
	The cases where \((\alpha, \pi) \in \langle b \land c \rangle\), and \((\alpha, \pi) \in \langle \lnot (b) \land c \rangle\) are similar.
	\paragraph*{\RefAx{\Axguardedness}}
	For any \(\alpha,\pi\), suppose \((\alpha, \pi) \in \langle b \rangle\), \\then \((b, \pi) \tr{}{\alpha} (\acc, \pi)\) and hence
	\begin{align*}
		(e, \pi) \hookrightarrow_\alpha r &\iff (b\seq e, \pi) \hookrightarrow_\alpha r\\
		(e+_b f, \pi) \hookrightarrow_\alpha r &\iff ((b\seq e) +_b f, \pi) \hookrightarrow_\alpha r 
	\end{align*}
	Suppose \((\alpha, \pi) \in \langle \lnot b \rangle\), then 
	\begin{align*}
		(f, \pi) \hookrightarrow_\alpha r &\iff (e+_b f, \pi) \hookrightarrow_\alpha r \\
		(e+_bf, \pi) \hookrightarrow_\alpha r &\iff ((b\seq e) +_b f, \pi) \hookrightarrow_\alpha r & 
	\end{align*}
	\paragraph*{\RefAx{\Axrightdistr}}
	For any \(\alpha,\pi\), suppose \((\alpha, \pi) \in \langle b \rangle\).
	
	Further suppose that \((e, \pi) \tr{}{\alpha} k \in K  \setminus \{\acc\}\times I^Y\)
	\begin{align*}
		((e+_b f), \pi) \tr{}{\alpha} k	&\implies ((e+_b f)\seq g, \pi) \tr{}{\alpha} k\\
		(e \seq g, \pi) \tr{}{\alpha} k &\implies (e \seq g +_b f\seq g, \pi) \tr{}{\alpha} k
	\end{align*}
	Suppose \((e, \pi) \tr{}{\alpha} (\acc, \sigma)\), then
	\begin{align*}
		&((e+_b f), \pi) \tr{}{\alpha} (\acc, \sigma)\\ 
		&\implies ((e+_b f)\seq g, \pi) \hookrightarrow_\alpha r \iff (g, \sigma) \hookrightarrow_\alpha r\\
		&(e \seq g, \pi) \hookrightarrow_\alpha r \iff (g, \sigma) \hookrightarrow_\alpha r\\
		&\implies (e \seq g+_b f\seq g, \pi) \hookrightarrow_\alpha r \iff (g, \sigma) \hookrightarrow_\alpha r
	\end{align*}
	Suppose \((e, \pi) \tr{p}{\alpha} (e', \sigma)\), then
	\begin{align*}
		((e+_b f), \pi) \tr{p}{\alpha} (e', \sigma)
		&\implies ((e+_b f)\seq g, \pi) \tr{p}{\alpha} (e'\seq g, \sigma)\\
		(e \seq g, \pi) \tr{p}{\alpha} (e'\seq g, \sigma)
		&\implies (e \seq g +_b f\seq g, \pi) \tr{p}{\alpha} (e'\seq g, \sigma)
	\end{align*}
	The case where \((\alpha, \pi) \in \langle \lnot b \rangle\) is symmetric.
	\paragraph*{\RefAx{\Axassoc}}
	In this case we show the relation \(\sim \coloneq \{((e\seq(f \seq g), \pi), ((e \seq f) \seq g, \pi)) \mid \pi \in I^Y\}\) is a bisimulation up to equivalence closure \cite{generalcoinupto}.
	
	For any \(\alpha,\pi\), suppose \((e, \pi) \tr{}{\alpha} k \in K  \setminus \{\acc\} \times I^Y\). Then 
	\begin{align*}
		(e \seq (f\seq g), \pi)&\tr{}{\alpha} k\\
		((e\seq f)\seq g, \pi) &\tr{}{\alpha} k
	\end{align*}
	Suppose \((e, \pi) \tr{p}{\alpha} (e', \sigma)\). Then 
	\begin{align*}
		(e\seq f, \pi) &\tr{p}{\alpha} (e'\seq f, \sigma) \\
		((e\seq f)\seq g, \pi) &\tr{p}{\alpha} ((e'\seq f)\seq g, \sigma)
	\end{align*}
	These states are related by \(\sim\).
	
	Suppose \((e, \pi) \tr{}{\alpha} (\acc, \sigma)\) and \((f, \sigma) \tr{}{\alpha} k \in K  \setminus \{\acc\} \times I^Y\). Then 
	\begin{align*}
		(e\seq f, \pi) &\tr{}{\alpha} k \land (f \seq g, \sigma) \tr{}{\alpha} k\\
		((e\seq f)\seq g, \pi) &\tr{}{\alpha} k \land (e\seq (f\seq g), \pi) \tr{}{\alpha} k
	\end{align*}
	Suppose \((e, \pi) \tr{}{\alpha} (\acc, \sigma)\) and \((f, \sigma) \tr{p}{\alpha} (f', \nu)\). Then 
	\begin{align*}
		(e\seq f, \pi) &\tr{p}{\alpha} (f', \nu)\\
		((e\seq f)\seq g, \pi) &\tr{p}{\alpha} (f'\seq g, \nu)\\
		(f\seq g, \sigma) &\tr{p}{\alpha} (f'\seq g, \nu)\\
		(e\seq (f\seq g), \pi) &\tr{p}{\alpha} (f'\seq g, \nu)
	\end{align*}
	Finally, suppose \((e, \pi) \tr{}{\alpha} (\acc, \sigma)\) and \((f, \sigma) \tr{p}{\alpha} (\acc, \nu)\). Then
	\begin{align*}
		(e\seq f, \pi) &\tr{}{\alpha} (\acc, \nu)\\
		((e\seq f)\seq g, \pi) &\hookrightarrow_\alpha r \iff (g, \nu) \hookrightarrow_\alpha r\\
		(f \seq g, \sigma) &\hookrightarrow_\alpha r \iff (g, \nu) \hookrightarrow_\alpha r\\
		(e\seq (f\seq g), \pi) &\hookrightarrow_\alpha r \iff (g, \nu) \hookrightarrow_\alpha r
	\end{align*}
	\paragraph*{\RefAx{\Axabsleft}}
	For any \(\alpha,\pi\)
	\begin{align*}
		(0, \pi) &\tr{}{\alpha} \bot \\
		(0\seq e, \pi) &\tr{}{\alpha} \bot
	\end{align*}
	\paragraph*{\RefAx{\Axneutleft}}
	For any \(\alpha,\pi\), since \((1, \pi) \tr{}{\alpha} (\acc, \pi)\), \((e, \pi) \hookrightarrow_\alpha r \iff (1\seq e, \pi) \hookrightarrow_\alpha r\).
	\paragraph*{\RefAx{\Axneutright}}
	For this axiom we use the relation \({\sim} \coloneq \{(e, \pi), (e \seq 1, \pi) \allowbreak \mid e \in \Expm, \pi \in I^Y\}\) and show that this relation is a bisimulation.
	 
	For any \(\alpha,\pi\), suppose \((e, \pi) \tr{}{\alpha} (\acc, \sigma) \), then \((e\seq1, \pi) \tr{}{\alpha} (\acc, \sigma)\) as \((1, \sigma) \tr{}{\alpha} (\acc, \sigma)\).
	
	Suppose \((e, \pi) \tr{}{\alpha} k \in K \setminus\{\acc\}\times I^Y\), then \((e\seq1, \pi) \tr{}{\alpha} k\).
	
	Suppose \((e, \pi) \tr{p}{\alpha} (e', \sigma)\), then \((e\seq1, \pi) \tr{p}{\alpha} (e'\seq1, \sigma)\), which are related by \(\sim\).
	
	\paragraph*{\RefAx{\Axtestcomb}} 
	For any \(\alpha,\pi\), suppose
	\((\alpha, \pi) \in \langle b\land c\rangle\), then \((b, \pi) \tr{}{\alpha} (\acc, \pi)\) and \((c, \pi) \tr{}{\alpha} (\acc, \pi)\). Hence \((b\seq c, \pi) \tr{}{\alpha} (\acc, \pi)\). Similarly \((b\land c, \pi) \tr{}{\alpha} (\acc, \pi)\).
	
	Suppose \((\alpha, \pi) \in \langle \lnot (b \land c)\rangle\). If  \((\alpha, \pi) \in \langle \lnot b\rangle\) then \((b, \pi) \tr{}{\alpha} \bot\). Hence \((b\seq c, \pi) \tr{}{\alpha} \bot\). Similarly \((b\land c, \pi) \tr{}{\alpha} \bot\).
	
	Else \((\alpha, \pi) \in \langle b \land \lnot c\rangle\), then \((b, \pi) \tr{}{\alpha} (\acc, \pi)\) and \((c, \pi) \tr{}{\alpha} \bot\). Hence \((b\seq c, \pi) \tr{}{\alpha} \bot\). Similarly \((b\land c, \pi) \tr{}{\alpha} \bot\).

	\paragraph*{\RefAx{\Axrightseqannret}, \RefAx{\Axrightseqanncont}, \RefAx{\Axrightseqannbrk}, \RefAx{\Axrightseqanngoto}}
	Suppose \(e_c = \comBreak\), then for any \(\alpha,\pi\), \((e_c\seq e, \pi) \tr{}{\alpha} (\brkc, \pi)\) and \((e_c, \pi) \tr{}{\alpha} (\brkc, \pi)\). The argument is similar for the other continuations.
	
	\paragraph*{\RefAx{\Axneutleftloop}} 
	For any \(\alpha,\pi\), 
	\[
	(1\loopseq e, \pi) \hookrightarrow_\alpha r \iff (e, \pi) \hookrightarrow_\alpha r 
	\]
	
	\paragraph*{\RefAx{\Axrightloopannabort}, \RefAx{\Axrightloopannret}, \RefAx{\Axrightloopanngoto}}
	For any \(\alpha,\pi\), suppose \(e_c = 0\), then \((e_c, \pi) \tr{}{\alpha} \bot\) and \((e_c\loopseq e, \pi) \tr{}{\alpha} \bot\). The argument is similar for \comReturn and \comGoto.
	\paragraph*{\RefAx{\Axbreak}}
	For any \(\alpha,\pi\)
	\begin{align*}
		(\comBreak \loopseq e, \pi) &\tr{}{\alpha} (\acc, \pi)\\
		(1, \pi) &\tr{}{\alpha} (\acc, \pi)
	\end{align*}
	\paragraph*{\RefAx{\Axcontinue}}
	For any \(\alpha,\pi\), suppose \((e, \pi) \hookrightarrow_\alpha r\). By \Cref{operationalsemantics} \((\comCont \loopseq e, \pi) \hookrightarrow_\alpha r\).
	
	\paragraph*{\RefAx{\Axloopassocact}}
	We show that the relation \({\sim} \coloneq \{((p\seq (f\loopseq g), \pi), ((p\seq f)\loopseq g, \pi)) \mid f, g \in \Expm, \pi \in I^Y, p \in \Sigma\} \) is a bisimulation up to behavioral equivalence and equivalence closure \cite[Section 4.1]{coinductionupto}. For any \(\alpha,\pi\)
	\begin{align*}
		(p, \pi) &\tr{p}{\alpha} (1, \pi)\\
		(p\seq f, \pi) &\tr{p}{\alpha} (1\seq f, \pi)\\
		((p\seq f)\loopseq g, \pi) &\tr{p}{\alpha} ((1\seq f)\loopseq g, \pi)\\
		(p\seq (f\loopseq g), \pi) &\tr{p}{\alpha} (1\seq (f\loopseq g), \pi)
	\end{align*}
	and \(((1\seq f)\loopseq g, \pi)\), \((1\seq (f\loopseq g), \pi)\) are each behaviorally equivalent to \((f \loopseq g, \pi)\) as they all have identical transitions. More formally, \(\{((1\seq f)\loopseq g, \pi), (1\seq (f\loopseq g), \pi)\}\) is a bisimulation up to equivalence closure, hence by \Cref{coinduction} \(\final((1\seq f)\loopseq g, \pi) = \final(1\seq (f\loopseq g), \pi)\).
	
	\paragraph*{\RefAx{\Axloopassocassign}}
	For any \(\alpha,\pi\), by definition \(((x\assign i), \pi) \tr{}{\alpha} (\acc, \pi[x\leftarrow i])\).
	
	We first use this fact, then build up the rest of the term preserving the outputs as we do so by appealing to \Cref{operationalsemantics}.
	\begin{align*}
		((x\assign i\seq e), \pi) \hookrightarrow_\alpha r &\iff (e, \pi[x\leftarrow i]) \hookrightarrow_\alpha r\\
		((x\assign i\seq e)\loopseq f, \pi) \hookrightarrow_\alpha r &\iff (e\loopseq f, \pi[x\leftarrow i]) \hookrightarrow_\alpha r\\
		((x\assign i\seq e)\loopseq f, \pi) \hookrightarrow_\alpha r &\iff (x\assign i\seq (e\loopseq f), \pi) \hookrightarrow_\alpha r
	\end{align*}
	\paragraph*{\RefAx{\Axloopdistributivity}}
	For any \(\alpha,\pi\), suppose \((\alpha, \pi) \in \langle b \rangle\). We then build up the desired iff statement by starting with an identity and then using the same operations
	\begin{align*}
		(e +_b f, \pi) \hookrightarrow_\alpha r &\iff (e, \pi) \hookrightarrow_\alpha r\\
		((e +_b f)\loopseq g, \pi) \hookrightarrow_\alpha r &\iff (e\loopseq g, \pi) \hookrightarrow_\alpha r\\
		((e +_b f)\loopseq g, \pi) \hookrightarrow_\alpha r &\iff ((e\loopseq g) +_b (f\loopseq g), \pi) \hookrightarrow_\alpha r
	\end{align*}
	and by a symmetric argument, if \((\alpha, \pi) \in \langle \lnot b \rangle\), then \(
	((e +_b f) \allowbreak \loopseq g, \pi) \hookrightarrow_\alpha r \iff ((e\loopseq g) +_b (f\loopseq g), \pi) \hookrightarrow_\alpha r\). Hence
	\[((e +_b f)\loopseq g, \pi) \hookrightarrow_\alpha r \iff ((e\loopseq g) +_b (f\loopseq g), \pi) \hookrightarrow_\alpha r\]
	\paragraph*{\RefAx{\Axloopunfold}}
	For any \(\alpha,\pi\), if \((\alpha, \pi) \in \langle \lnot b \rangle\) then 
	\begin{align*}
		(e \loopseq e^{(b)} +_b 1, \pi) &\tr{}{\alpha} (\acc, \pi)\\
		(e^{(b)}, \pi) &\tr{}{\alpha} (\acc, \pi)
	\end{align*}
	For the remaining cases we assume \((\alpha, \pi) \in \langle b \rangle\). Suppose \((e, \pi) \tr{}{\alpha} (\brkc, \sigma)\). Then 
	\begin{align*}
		(e \loopseq e^{(b)}, \pi) \tr{}{\alpha} (\acc, \sigma) &\implies (e \loopseq e^{(b)} +_b 1, \pi) \tr{}{\alpha} (\acc, \sigma)\\
		(e, \pi) \tr{}{\alpha} (\brkc, \sigma) &\implies (e^{(b)}, \pi) \tr{}{\alpha} (\acc, \sigma)
	\end{align*}
	Suppose \(( e, \pi) \tr{}{\alpha} k \in \{\retc, \bot\} + \{\jmpc~\ell \mid \ell \in \lset\} \times I^Y\), then \((e^{(b)}, \pi) \tr{}{\alpha} k \in \{\retc, \bot\} + \{\jmpc~\ell \mid \ell \in \lset\} \times I^Y\). Similarly \((e\loopseq e^{(b)}, \pi) \tr{}{\alpha} k\), and by \Cref{operationalsemantics} and the fact that \((\alpha, \pi) \in \langle b \rangle\) it follows that \((e\loopseq e^{(b)}+_b 1) \tr{}{\alpha} k\).
	
	Suppose \(( e, \pi) \tr{p}{\alpha} (e', \sigma)\), then \((e^{(b)}, \pi) \tr{p}{\alpha} (e'\loopseq e^{(b)}, \sigma)\). Similarly \((e\loopseq e^{(b)}) \tr{p}{\alpha} (e'\loopseq e^{(b)}, \sigma)\), and by \Cref{operationalsemantics} and the fact that \((\alpha, \pi) \in \langle b \rangle\) it follows that \((e\loopseq e^{(b)}+_b 1) \tr{p}{\alpha} (e'\loopseq e^{(b)}, \sigma)\).
	
	Finally, suppose \(( e, \pi) \tr{}{\alpha} (c, \sigma) \in \{\acc, \contc\} \times I^Y\). Then \((e^{(b)}, \pi) \hookrightarrow_\alpha r \iff (e^{(b)}, \sigma) \hookrightarrow_\alpha r\). Furthermore by the definition of conditionals and loop composition \((e\loopseq e^{(b)}+_b 1, \pi) \hookrightarrow_\alpha r \iff (e\loopseq e^{(b)}, \pi) \hookrightarrow_\alpha r \iff (e^{(b)}, \sigma) \hookrightarrow_\alpha r\).
	\paragraph*{\RefAx{\Axlooptighten}}
	We show that \begin{align*}	
	(\sim) &\coloneq \{(((\lnot c\seq e)^{(b)}, \pi), (e^{(b)}, \pi)) \\&\qquad\mid \forall \sigma.\finalmap (c\land b \seq e \loopseq \lnot(c\land b), \sigma)=\finalmap (0, \sigma)\} \\&\quad\cup \{((e'\loopseq(\lnot c\seq e)^{(b)}, \pi), (e'\loopseq e^{(b)}, \pi)) \\&\qquad\mid e' \in \Expm, \forall \sigma.\finalmap (c\land b \seq e \loopseq \lnot(c\land b), \sigma)=\finalmap (0, \sigma)\}
	\end{align*} is an intermediate bisimulation.
	
	We start with terms that do not have the leading \(e'\).
	
	For any \(\alpha,\pi\), if \((\alpha, \pi) \in \langle \lnot b \rangle\) then \(((\lnot c\seq e)^{(b)}, \pi) \tr{}{\alpha} (\acc, \pi)\) and \(((e)^{(b)}, \pi) \tr{}{\alpha} (\acc, \pi)\). For all following cases, assume \((\alpha, \pi) \in \langle b \rangle\).
	
	Suppose \((\alpha, \pi) \in \langle c \rangle\). From the premise, \((e, \pi)\) can only transition to one of \(\acc, \bot, \contc\) under \(\alpha\), and in particular if it transitions to \((k, \sigma)\) with \(k \in \{\acc, \contc\}\), then it must be that \((\alpha, \sigma) \in \langle c \land b \rangle\) by case analysis on \Cref{operationalsemantics}. That is, \(\{(e^{(b)}, \nu) \mid (\alpha, \nu) \in \langle b \land c \rangle\}\) is a rejecting set of \(\alpha\). Hence, by \Cref{rejectingsetaborts} \((e^{(b)}, \pi) \tr{}{\alpha} \bot\), and by \Cref{operationalsemantics} \(((\lnot c \seq e)^{(b)}, \pi) \tr{}{\alpha} \bot\).
	
	Now consider the case where \((\alpha, \pi) \in \langle b \land \lnot c \rangle\). Then if \((e, \pi) \tr{}{\alpha} r\) where \(r \in \{\retc, \bot\} + (\{\brkc\} + \{\jmpc~\ell \mid \ell \in \lset\}) \times I^Y\) it holds that \((e^{(b)}, \pi) \tr{}{\alpha} r\) and \(((\lnot c\seq e)^{(b)}, \pi) \tr{}{\alpha} r\).
	
	If \((e, \pi) \tr{p}{\alpha} (e', \sigma)\), then \((e^{(b)}, \pi) \tr{p}{\alpha} (e' \loopseq e^{(b)}, \sigma)\) and \(((\lnot c\seq e)^{(b)}, \pi) \tr{p}{\alpha} (e' \loopseq (\lnot c\seq e)^{(b)}, \sigma)\) and these states are related by \(\sim\).
	
	If \(e \tr{}{\alpha} (k, \sigma)\) where \((k, \sigma) \in \{\acc, \contc\} \times I^Y\), then \((e^{(b)}, \pi) \transSilent_\alpha (e^{(b)}, \sigma)\) and  \(((\lnot c \seq e)^{(b)}, \pi) \transSilent_\alpha ((\lnot c \seq e)^{(b)}, \sigma)\), and these two states are related by \(\sim\).
	
	Finally consider the pair of terms \((e'\loopseq e^{(b)}, \pi)\) and \((e'\loopseq (\lnot c \seq e)^{(b)}, \pi)\). If \((e', \pi) \tr{}{\alpha} k \in \{\bot, \retc\} + \{\jmpc~\ell, \ell \in \lset\} \times I^Y\), then so do \((e'\loopseq e^{(b)}, \pi)\) and \((e'\loopseq (\lnot c \seq e)^{(b)}, \pi)\).
	
	If \((e', \pi) \tr{p}{\alpha} (f, \sigma)\) then \((e'\loopseq e^{(b)}, \pi) \tr{p}{\alpha} (f \loopseq e^{(b)}, \sigma)\) and \((e'\loopseq (\lnot c \seq e)^{(b)}, \pi) \tr{p}{\alpha} (f\loopseq (\lnot c\seq e)^{(b)}, \pi)\) which are related by \(\sim\).
	
	If \((e', \pi) \tr{}{\alpha} (\brkc, \sigma)\), then \((e'\loopseq e^{(b)}, \pi) \tr{}{\alpha} (\acc, \sigma)\) and \((e'\loopseq (\lnot c \seq e)^{(b)}, \pi) \tr{}{\alpha} (\acc, \sigma)\).
	
	If \((e', \pi) \tr{}{\alpha} (k, \sigma)\) where \(k \in \{\acc, \contc\}\), then \((e'\loopseq e^{(b)}, \pi) \allowbreak \transSilent_\alpha (e^{(b)}, \sigma)\) and \((e'\loopseq (\lnot c \seq e)^{(b)}, \pi) \allowbreak \transSilent_{\alpha} ((\lnot c \seq e)^{(b)}, \sigma)\). These states are related by \({\sim}\).
	
	\paragraph*{\RefAx{\Axindcommact}}
	We show the relation \({\sim}\eqdef \{(x\assign i\seq p, \pi), (p \seq x\assign i, \pi) \mid \pi \in I^Y\} \) is a bisimulation up to behavioral equivalence and equivalence closure \cite[Section 4.1/4.2]{coinductionupto}.
	
	For any \(\alpha,\pi\)
	\begin{align*}
		(x\assign i\seq p, \pi) &\tr{p}{\alpha} (1, \pi[x\leftarrow i])\\
		(p \seq x\assign i, \pi) &\tr{p}{\alpha} (1\seq x\assign i, \pi)
	\end{align*}
	These two states are behaviorally equivalent. This can be seen since the relation which relates these two particular states is a bisimulation up to equivalence closure.
	
	\paragraph*{\RefAx{\Axindval}}
	For any \(\alpha,\pi\)
	\begin{align*}
		(x \assign i, \pi[x\leftarrow i]) &\tr{}{\alpha} (\acc, \pi[x\leftarrow i])\\
		(x\assign i \seq x = i, \pi)&\tr{}{\alpha} (\acc, \pi[x\leftarrow i])
	\end{align*}
	\paragraph*{\RefAx{\Axindred}}
	For any \(\alpha,\pi\), suppose \((\alpha, \pi) \in \langle b \rangle\), then
	\begin{align*}
		(x=i, \pi) &\tr{}{\alpha} (\acc, \pi)\\
		(x=i\seq x \assign i, \pi) &\tr{}{\alpha} (\acc, \pi)
	\end{align*}
	Suppose \((\alpha, \pi) \in \langle \lnot b \rangle\), then
	\begin{align*}
		(x=i, \pi) &\tr{}{\alpha} \bot\\
		(x=i\seq x\assign i, \pi) &\tr{}{\alpha} \bot
	\end{align*}
	
	\paragraph*{\RefAx{\Axindannih}}
	For any \(\alpha,\pi\)
	\begin{align*}
		(x\assign i, \pi) &\tr{}{\alpha} (\acc, \pi[x \leftarrow i])\\
		(x\assign i \seq x\assign j, \pi) &\tr{}{\alpha} (\acc, \pi[x \leftarrow i][x \leftarrow j])
	\end{align*}
	and
	\[(\acc, \pi[x \leftarrow i][x \leftarrow j])= (\acc, \pi[x \leftarrow j]).\]%
	Moreover \((x\assign j, \pi) \tr{}{\alpha} (\acc, \pi[x \leftarrow j])\).
	\paragraph*{\RefAx{\Axindret}}
	For any \(\alpha,\pi\)
	\begin{align*}
		(x\assign i\seq\comReturn, \pi) &\tr{}{\alpha} \retc\\
		(\comReturn, \pi) &\tr{}{\alpha} \retc
	\end{align*}
	\paragraph*{\RefAx{\Axinddistr}}
	For any \(\alpha,\pi\), suppose \((\alpha, \pi) \in \langle t \rangle\), then by \Cref{indnotrefbool} \((\alpha, \pi[x\allowbreak \assign i]) \in \langle t \rangle\) and hence
	\begin{align*}
		((x\assign i\seq e) +_t (x \assign i\seq f), \pi) \hookrightarrow_\alpha r
		&\iff (x\assign i \seq e, \pi) \hookrightarrow_\alpha r\\
		&\iff (e, \pi[x \leftarrow i]) \hookrightarrow_\alpha r\\
		&\iff (e +_t f, \pi[x \leftarrow i]) \hookrightarrow_\alpha r\\
		&\iff (x\assign i\seq (e+_t f), \pi) \hookrightarrow_\alpha r
	\end{align*}
	A symmetric argument holds if \((\alpha, \pi) \in \langle b \rangle\). 
	
	\paragraph*{\RefAx{\Axindabort}}
	For any \(\alpha,\pi\)
	\begin{align*}
		(x\assign i\seq 0, \pi) &\tr{}{\alpha} \bot\\
		(0, \pi) &\tr{}{\alpha} \bot
	\end{align*}
	
	\paragraph*{\RefAx{\Axindindcomm}}
	For any \(\alpha,\pi\)
	\begin{align*}
		(x \assign j, \pi[x\leftarrow i]) &\tr{}{\alpha} (\acc, \pi[x\leftarrow i][x' \leftarrow j])\\
		\implies(x \assign i\seq x' \assign j, \pi) &\tr{}{\alpha} (\acc, \pi[x\leftarrow i][x' \leftarrow j])\\
		(x \assign i, \pi[x' \leftarrow j]) &\tr{}{\alpha} (\acc, \pi[x' \leftarrow j][x\leftarrow i])\\
		\implies(x' \assign j \seq x \assign i, \pi) &\tr{}{\alpha} (\acc, \pi[x' \leftarrow j][x\leftarrow i]) 
	\end{align*}
	and \(\pi[x\leftarrow i][x' \leftarrow j] =  \pi[x' \leftarrow j][x\leftarrow i]\) as \(x\neq x'\).
	
	\paragraph*{\RefAx{\Axindinddistr}}
	For any \(\alpha,\pi\), suppose \(\pi(x') = j\), then
	\begin{align*}
		(x\assign i \seq (e+_{x'=j} f), \pi) \hookrightarrow_\alpha r &\iff (e, \pi[x\leftarrow i]) \hookrightarrow_\alpha r\\
		&\iff (x\assign i \seq e, \pi) \hookrightarrow_\alpha r\\
		&\iff (x\assign i \seq e +_{x'=j} x\assign i\seq f, \pi) \hookrightarrow_\alpha r
	\end{align*}
	
	A symmetric argument holds if \(\pi(x') \neq j\).

	\paragraph*{Contextual soundness}\label{contextual-soundness}
	We now prove soundness of the \textbf{contextual axioms}. That is, they produce sound conclusions from sound premises.
	
	\paragraph*{\RefAx{\Axweaken}} 
	Let \(E_1 \coloneq (E, \ell=e')\).  By the premise,
	\((e,\pi)\resWith{E} \sim (f,\pi)\resWith{E}\) for every \(\pi\).  The side
	condition of \Cref{common-suffix-lifting} is immediate since the two prefixes
	are both \(E\): every suffix target \((h,\nu)\) is related to itself by
	reflexivity. Equivalently, the lifted witness uses the old bisimulation
	relation together with the identity relation for any new jumps bound by \(G\).  
	Hence, for every \(\pi\),
	\begin{align*}
		(e,\pi)\resWith{E_1}
		&\sim (e,\pi)\resWith{[E;E_1]}
		&& \tCref{supsyssolsubsyssynt}\\
		&\sim (f,\pi)\resWith{[E;E_1]}
		&& \tCref{common-suffix-lifting}\\
		&\sim (f,\pi)\resWith{E_1}
		&& \tCref{supsyssolsubsyssynt}.
	\end{align*}
	\paragraph*{\RefAx{\Axreplacement}}
	Let \(E_1 = E,(\ell = e_1)\), and \(E_2 = E,(\ell = e_2)\). We first observe that \[(e, \pi) \tr[E]{p}{\alpha} (e', \pi') \implies (e, \pi) \tr[E (E,(\ell = f))]{p}{\alpha}
 (e', \pi')\] by \Cref{resolutiondef}. Furthermore due to the bisimulation from the proof of \Cref{supsyssolsubsyssynt} being a bisimulation \[(e, \pi) \tr[E]{p}{\alpha} (e', \pi') \implies (e, \pi) \tr[E,(\ell = f)]{p}{\alpha}
 (e', \pi')\]
	We show that the following relation is an intermediate bisimulation
	\begin{align*}
		\sim_1 \eqdef \{(e, \pi)\resWith{E_1}, (f, \sigma)\resWith{E_2} \mid (e, \pi)\resWith{E} \sim_E (f, \sigma)\resWith{E}\}
	\end{align*}
	where \(\sim_E\) is the bisimulation that relates \((e_1, \pi)\resWith{E}\) and \((e_2, \pi)\resWith{E}\).
	
	The cases where \((e, \pi) \tr[E]{}{\alpha}
 r \in K\) such that \(r\) is not a jump to \(\ell\) are straightforward.
	
	If \((e, \pi) \tr[E]{p}{\alpha} (e', \pi')\) then \((f, \sigma) \tr[E]{p}{\alpha} (f', \sigma')\) such that \((e', \pi')\resWith{E} \allowbreak \sim_E (f', \sigma')\resWith{E}\) and hence
	\begin{align*}
		(e, \pi) &\tr[E_1]{p}{\alpha}
 (e', \pi')\\
		(f, \sigma) &\tr[E_2]{p}{\alpha}
 (f', \sigma')
	\end{align*}
	and  \((e', \pi')\resWith{E_1} \sim_1 (f', \sigma')\resWith{E_2}\).
	
	Finally, if \((e, \pi) \tr[E]{}{\alpha} (\jmpc~\ell, \nu)\), then \((f, \sigma) \tr[E]{}{\alpha} (\jmpc~\ell, \nu)\) since they are related by a bisimulation. Hence
	\begin{align*}
		(e, \pi) &\transSilent^{E_1} (e_1, \nu)\\
		(f, \sigma) &\transSilent^{E_2} (e_2, \nu)
	\end{align*}
	and \((e_1, \nu)\resWith{E_1} \sim_1 (e_2, \nu)\resWith{E_2}\).

	We now show that \({\sim} \eqdef {\sim_1} \cup \{(e, \pi)\resWith{E_1}, (e, \pi)\resWith{E_2}\}\) is an intermediate bisimulation. We only illustrate a pair from the second subset and defer to above for the first.

	The cases where \((e, \pi) \tr{}{\alpha} r \in K\) with \(r \neq (\jmpc~k, \sigma)\) for any \((k, \sigma) \in (\text{dom}(E) + \{\ell\}) \times I^Y\) are straightforward from \Cref{resolutiondef}.
		
	If \((e, \pi) \tr{p}{\alpha} (e', \sigma)\), then
	\begin{align*}
		(e, \pi) &\tr[E_1]{p}{\alpha}
 (e', \sigma) & \tCref{resolutiondef}\\
		(e, \pi) &\tr[E_2]{p}{\alpha}
 (e', \sigma) & \tCref{resolutiondef}
	\end{align*}	
	 and these states are related by \(\sim\).
		
	If \((e, \pi) \tr{}{\alpha} (\jmpc~\ell, \sigma)\), then
	\begin{align*}
		(e, \pi) &\transSilent_\alpha^{E_1} (e_1, \sigma) & \tCref{resolutiondef}\\
		(e, \pi) &\transSilent_\alpha^{E_2} (e_2, \sigma) & \tCref{resolutiondef}
	\end{align*}
	and these two states are related by \(\sim\).
	
	If \((e, \pi) \tr{}{\alpha} (\jmpc~k, \sigma)\) for \(k \in \text{dom}(E)\), then
	\begin{align*}
		(e, \pi) &\transSilent_\alpha^{E_1} (E(k), \sigma) & \tCref{resolutiondef}\\
		(e, \pi) &\transSilent_\alpha^{E_2} (E(k), \sigma) & \tCref{resolutiondef}
	\end{align*}
	and these two states are related by \(\sim\).

	\paragraph*{\RefAx{\Axredundancy}}
	
		Let us write \(E_1\) for the environment \((E,F)\), where \(E\) and \(F\) have disjoint domains. From the side condition no term in the range of \(E\) nor \(e\) contains \(\comGoto~\ell\) for any \(\ell \in \text{dom}(F)\) as a subterm.
	
	We establish that the relation defined by \((e, \pi)\resWith{E_1} \sim (e, \pi)\resWith{E}\), for all \((e, \pi) \in \Expm \times I^Y\) that does not contain \(\comGoto~\ell\) for any \(\ell \in \text{dom}(F)\), is an intermediate bisimulation and achieve the desired result by \Cref{coinduction}.
	
	Suppose \((e, \pi) \tr{}{\alpha} r \in K\) where \(r \neq (\jmpc~k, \sigma)\) for any \(k \in \text{dom}(E)\cup\text{dom}(F)\), \(\sigma \in I^Y\). Then
	\begin{align*}
		(e, \pi) &\tr[E_1]{}{\alpha}
 r & \tCref{resolutiondef}\\
		(e, \pi) &\tr[E]{}{\alpha}
 r & \tCref{resolutiondef}
	\end{align*} 
	
	Suppose \((e, \pi) \tr{p}{\alpha} (f, \sigma)\). Then
	\begin{align*}
		(e, \pi) &\tr[E_1]{p}{\alpha}
 (f, \sigma) & \tCref{resolutiondef}\\
		(e, \pi) &\tr[E]{p}{\alpha}
 (f, \sigma) & \tCref{resolutiondef}
	\end{align*} 
	and \((f, \sigma)\resWith{E_1} \sim (f, \sigma)\resWith{E}\). 
	Indeed, by \Cref{notransitionnogoto}, for every \(\ell \in \text{dom}(F)\) the visible successor \(f\) still contains no \(\comGoto~\ell\); hence \((f,\sigma)\) is again in the relation.
	The adjacent no-jump lemma also rules out a direct jump from \((f,\sigma)\) to any label in \(\text{dom}(F)\).
	
	Suppose \((e, \pi) \tr{}{\alpha} (\jmpc~k, \sigma)\) for some \(\sigma \in I^Y, k \in \text{dom}(E)\).
	\begin{align*}
		(e, \pi) &\transSilent_\alpha^{E} (E(k), \sigma)\\
		(e, \pi) &\transSilent_\alpha^{E_1} (E(k), \sigma)
	\end{align*}
	and these states are related by \(\sim\).

	Note that \((e,\pi)\) cannot transition to a jump in the domain of \(F\) by \Cref{notransitionnogoto}.

	\paragraph*{\RefAx{\Axrenaming}}
	Let \(E_1\) denote \(E\langle \ell'/\ell \rangle\). We claim that the relation defined by  
	\[(e, \pi)\resWith{E} \sim  (e[\comGoto~\ell'/\comGoto~\ell], \pi)\resWith{E_1}\] for all \((e, \pi) \in \Expm \times I^Y\), is an intermediate bisimulation.

	Suppose \((e, \pi) \tr{}{\alpha} r \in K\) where \(r \neq (\jmpc~k, \sigma)\) for any \(\sigma \in I^Y\) and \(k \in \text{dom}(E)\).
	\begin{align*}
		(e, \pi) &\tr[E]{}{\alpha} r & \tCref{resolutiondef}\\
		(e[\comGoto~\ell'/\comGoto~\ell], \pi) &\tr[E_1]{}{\alpha}
 r & \tCref{thm:substitute-preserve-non-jump,resolutiondef}
	\end{align*}
	Suppose \((e, \pi) \tr{p}{\alpha} (f, \sigma)\).
	\begin{align*}
		(e, \pi) &\tr[E]{p}{\alpha} (f, \sigma) & \tCref{resolutiondef}\\
		(e[\comGoto~\ell'/\comGoto~\ell], \pi) &\tr[E_1]{p}{\alpha}
 (f[\comGoto~\ell'/\comGoto~\ell], \sigma) & \eqref{thm:substitute-preserve-non-jump},\tCref{resolutiondef}
	\end{align*}
	and \((f, \sigma)\resWith{E} \sim (f[\comGoto~\ell'/\comGoto~\ell], \sigma)\resWith{E_1}\).
	
	Suppose \((e, \pi) \tr{}{\alpha} (\jmpc~k, \sigma)\) where \((k, \sigma) \in \text{dom}(E)\times I^Y\). Let \(k' = k\) if \(k \neq \ell\) or else \(\ell'\) if \(k=\ell\).
	\begin{align*}
		(e, \pi) &\transSilent_{\alpha}^E (E(k), \sigma) & \tCref{resolutiondef}\\
		(e[\comGoto~\ell'/\comGoto~\ell], \pi) &\transSilent_{\alpha}^{E_1} (E_1(k'), \sigma) & \eqref{renamingsemlem},\tCref{resolutiondef}
	\end{align*}
	but \(E_1(k') = E(k)[\comGoto~\ell'/\comGoto~\ell]\) by definition of \(\langle \ell'/\ell \rangle\). 
	Hence, 
	\[(e[\comGoto~\ell'/\comGoto~\ell], \pi) \transSilent_{\alpha}^{E_1} (E(k)[\comGoto~\ell'/\comGoto~\ell], \sigma)\]
	and \((E(k), \sigma)\resWith{E} \sim (E(k)[\comGoto~\ell'/\comGoto~\ell], \sigma)\resWith{E_1}\).

	\paragraph*{\RefAx{\Axsubstitution}}
	Let \(E_1 = E,\ell=e_1\).  Write
	\(h^\ell\) for \(h[e_1/\comGoto~\ell]\).  The side condition says that 
	\(\comGoto~\ell\) is rightmost in \(e\).
	
	Define a relation \(R\) on syntactic states by
\begin{align*}
		&((f,\nu),(g,\sigma)) \in R\\
		&\enspace\iff 
		(f = g \land \nu=\sigma)\lor ((f = g[e_1/\comGoto~\ell] \land \nu=\sigma \land \comGoto~\ell\text{ is rightmost in }g))
	\end{align*}%
	Let \(R_F\) be the usual lifting of \(R\) to outputs.  It suffices to show that whenever
	\(\comGoto~\ell\) is rightmost in \(h\) and
	\[
	(h^\ell,\nu)\hookrightarrow_\alpha^{E_1} r,
	\]
	there is an \(r'\) such that
	\[
	(h,\nu)\hookrightarrow_\alpha^{E_1} r'
	\qquad\text{and}\qquad
	(r, r') \in R_F
	\]
	We prove this by induction on the rightmost derivation for \(h\).
	If \(h=\comGoto~\ell\), then \(h^\ell=e_1\).  Since \(E_1(\ell)=e_1\),
	\begin{align*}
		(e_1,\nu)\hookrightarrow_\alpha^{E_1} r
		&\iff
		(\comGoto~\ell,\nu)\hookrightarrow_\alpha^{E_1} r
		& \ref{operationalsemantics}, \ref{resolutiondef}.
	\end{align*}
	If \(h\) contains no occurrence of \(\comGoto~\ell\), then \(h^\ell=h\).
	Thus they have identical transitions
	\[
	(h^\ell,\nu)\hookrightarrow_\alpha^{E_1} r
	\iff
	(h,\nu)\hookrightarrow_\alpha^{E_1} r,
	\]
	If \(h=f +_b g\) and \((\alpha,\nu)\in\langle b\rangle\), then
	\begin{align*}
		(h^\ell,\nu)\hookrightarrow_\alpha^{E_1} r
		&\iff
		(f^\ell+_b g^\ell,\nu)\hookrightarrow_\alpha^{E_1} r
		&& \ref{synsub} \\
		&\iff
		(f^\ell,\nu)\hookrightarrow_\alpha^{E_1} r
		&& \ref{operationalsemantics}, \ref{resolutiondef}.
	\end{align*}
	By the induction hypothesis, there is \(r'\) such that
	\[
	(f,\nu)\hookrightarrow_\alpha^{E_1} r'
	\qquad\text{and}\qquad
	(r, r') \in R_F
	\]
	Thus by \Cref{operationalsemantics} and \Cref{resolutiondef}
	\[
	(h,\nu)\hookrightarrow_\alpha^{E_1} r'
	\iff
	(f,\nu)\hookrightarrow_\alpha^{E_1} r'
	\]
	The case where \((\alpha,\nu)\in\langle\lnot b\rangle\) is identical, using \(g\).
	If \(h=f\seq g\), then \(f\) contains no occurrence of \(\comGoto~\ell\) and
	\(g\) is rightmost. Hence by \Cref{synsub} \(h^\ell=f\seq g^\ell\).
	If
	\[
	(f,\nu)\tr{}{\alpha}(\acc,\sigma),
	\]
	then
	\begin{align*}
		(f\seq g^\ell,\nu)\hookrightarrow_\alpha^{E_1} r
		&\iff
		(g^\ell,\sigma)\hookrightarrow_\alpha^{E_1} r
		&& \ref{operationalsemantics}, \ref{resolutiondef}.
	\end{align*}
	The induction hypothesis for \(g\) gives \(r'\) with
	\[
	(g,\sigma)\hookrightarrow_\alpha^{E_1} r'
	\qquad\text{and}\qquad
	(r, r') \in R_F.
	\]
	Thus by \Cref{operationalsemantics} and \Cref{resolutiondef}
	\[
	(f\seq g,\nu)\hookrightarrow_\alpha^{E_1} r'
	\]
	If the first component, \(f\), produces any non-accepting halting output, then the transition does not depend on \(g\) and the contextual transition preserves this equality, that is
	\begin{align*}
	(f\seq g^\ell,\nu)\tr[E_1]{}{\alpha} k \in K\setminus \{\acc\} \times I^Y
	&\iff (f,\nu)\tr[E_1]{}{\alpha} k \in K\setminus \{\acc\} \times I^Y && \ref{operationalsemantics},\ref{resolutiondef}\\
	&\iff (f\seq g,\nu)\tr[E_1]{}{\alpha} k \in K\setminus \{\acc\} \times I^Y && \ref{operationalsemantics},\ref{resolutiondef}
	\end{align*}
	Finally, if
	\(
	(f,\nu)\tr{p}{\alpha}(f',\sigma)
	\)
	then
	\begin{align*}
		(f\seq g^\ell,\nu)
		&\tr[E_1]{p}{\alpha}
		(f'\seq g^\ell,\sigma)
		& \ref{operationalsemantics}, \ref{resolutiondef}\\
		(f \seq g,\nu)
		&\tr[E_1]{p}{\alpha}
		(f'\seq g,\sigma)
		& \ref{operationalsemantics}, \ref{resolutiondef}
	\end{align*}
	Since \(f\) contains no \(\comGoto~\ell\), neither does the visible successor
	\(f'\) by \Cref{notransitionnogoto}; hence \(f'\seq g\) is again rightmost, and
	\[
	((f'\seq g^\ell,\sigma), (f'\seq g,\sigma)) \in R
	\]
	If \(h=f\loopseq g\), then \(f\) contains no occurrence of
	\(\comGoto~\ell\) and \(g\) is rightmost.  Hence by \Cref{synsub}
	\(h^\ell=f\loopseq g^\ell\).
	If
	\(
	(f,\nu)\tr{}{\alpha}(c,\sigma)
	\quad c\in\{\acc,\contc\},
	\)
	then by \Cref{operationalsemantics} and \Cref{resolutiondef}
	\[
	(f\loopseq g^\ell,\nu)\hookrightarrow_\alpha^{E_1} r
	\iff
	(g^\ell,\sigma)\hookrightarrow_\alpha^{E_1} r,
	\]
	The induction hypothesis for \(g\) gives \(r'\) with
	\[
	(g,\sigma)\hookrightarrow_\alpha^{E_1} r'
	\qquad\text{and}\qquad
	(r, r') \in R_F.
	\]
	Thus by \Cref{operationalsemantics} and \Cref{resolutiondef}
	\[
	(f\loopseq g,\nu)\hookrightarrow_\alpha^{E_1} r'
	\]
	If
	\(
	(f,\nu)\tr{}{\alpha}(\brkc,\sigma),
	\)
	then both sides produce \((\acc,\sigma)\).  If \(f\) produces \(r\) which is \(\retc\), \(\bot\),
	or a jump, then 
	\begin{align*}
		(f\loopseq g^\ell,\nu)\tr[E_1]{}{\alpha} r
		&\iff (f,\nu)\tr[E_1]{}{\alpha} r & \ref{operationalsemantics},\ref{resolutiondef}\\
		&\iff (f \loopseq g,\nu)\tr[E_1]{}{\alpha} r & \ref{operationalsemantics},\ref{resolutiondef}
	\end{align*}
	Finally, if
	\(
	(f,\nu)\tr{p}{\alpha}(f',\sigma),
	\)
	then
	\begin{align*}
		(f \loopseq g^\ell,\nu)
		&\tr[E_1]{p}{\alpha}
		(f'\loopseq g^\ell,\sigma)
		& \ref{operationalsemantics}, \ref{resolutiondef}\\
		(f\loopseq g,\nu)
		&\tr[E_1]{p}{\alpha}
		(f'\loopseq g,\sigma)
		& \ref{operationalsemantics}, \ref{resolutiondef}
	\end{align*}
	Again \(f'\) contains no \(\comGoto~\ell\) by \Cref{notransitionnogoto}, so
	\[
	((f'\loopseq g^\ell,\sigma), (f'\loopseq g,\sigma)) \in R
	\]
	Therefore \(R\) is a bisimulation on the syntactic automaton resolved by
	\(E_1\).  Taking \(h=e\) and \(\nu=\pi\) gives
	\[
	(e[e_1/\comGoto~\ell],\pi)\resWith{E_1}
	\sim
	(e,\pi)\resWith{E_1},
	\]
	
	\paragraph*{\RefAx{\Axunprbra}}
	Let \(E_1 = E ,(\ell = b\seq g)\), and \(E_2 = E ,(\ell = g+_b f\seq \comGoto~\ell)\).  We claim that the relation defined by 
	\begin{align*}
		{\sim} \eqdef\{(e, \pi)\resWith{E_1}, (e, \pi)\resWith{E_2} \mid (e, \pi) \in \Expm \times I^Y\}\\
		\cup \{(b\seq g, \pi)\resWith{E_1}, (g +_b f\seq \comGoto~\ell, \pi)\resWith{E_2} \mid \pi \in I^Y\}
	\end{align*} for all \((e, \pi) \in \Expm \times I^Y\), is an intermediate bisimulation and the desired result follows from \Cref{coinduction}.
	
	We now establish the intermediate bisimulation conditions on the first subset of \({\sim}\). If \((e, \pi) \tr{}{\alpha} r \in K\) where \(r \neq (\jmpc~k, \sigma)\) for any \(k \in \text{dom}(E)\cup \{\ell\}, \sigma \in I^Y\), then both terms transition to \(r\) by \Cref{resolutiondef}.
	
	If \((e, \pi) \tr{}{\alpha} (p, (e', \sigma))\), then  
	\begin{align*}
		(e, \pi) &\tr[E_1]{p}{\alpha}
 (e', \sigma) & \tCref{resolutiondef}\\
		(e, \pi) &\tr[E_2]{p}{\alpha}
 (e', \sigma)& \tCref{resolutiondef}
	\end{align*}
	and \((e', \sigma)\resWith{E_1} \sim (e', \sigma)\resWith{E_2}\).
		
	If \((e, \pi) \tr{}{\alpha} (\jmpc~\ell, \sigma)\),  then   
	\begin{align*}
		(e, \pi) &\transSilent_\alpha^{E_1} (b\seq g, \sigma)\\
		(e, \pi) &\transSilent_\alpha^{E_2} (g+_b f\seq \comGoto~\ell, \sigma)
	\end{align*} 
	and these two states are related by \(\sim\).
	
	If \((e, \pi) \tr{}{\alpha} (\jmpc~k, \sigma)\) with \(k \in \text{dom}(E)\), then 
	\begin{align*}
		(e, \pi) &\transSilent_\alpha^{E_1} (E(k), \sigma)\\
		(e, \pi) &\transSilent_\alpha^{E_2} (E(k), \sigma)
	\end{align*}
	and these states are related by \(\sim\).

	Now we check the second set in the definition of \(\sim\). 
		
	Observe first that the premise gives a tail-rejection fact. Since \(0\)
	rejects every atom, soundness of the premise and the rules for tests and
	sequential composition imply that, whenever
	\((\alpha,\pi)\in\langle\lnot b\rangle\),
	\[
	(f\seq b,\pi)\tr[E]{}{\alpha}\bot .
	\]
	Fix \(\alpha\), and define
	\[
	N_\alpha =
	\{s \mid s\tr[E]{}{\alpha}\bot\}
	\cup
	\{(g+_b f\seq\comGoto~\ell,\pi)
	\mid (\alpha,\pi)\in\langle\lnot b\rangle\}.
	\]
	We claim that \(N_\alpha\) is a rejecting set for the automaton resolved by
	\(E_2\).  The first set is rejecting by \Cref{resolutiondef,supsyssolsubsyssynt}.
	
	Now let \(s=(g+_b f\seq\comGoto~\ell,\pi)\) with
	\((\alpha,\pi)\in\langle\lnot b\rangle\).  The guarded choice takes the
	false branch, so it suffices to inspect the unresolved transition of \(f\)
	from \(\pi\).  
	
	If \(f\) rejects, then \(s\) rejects.  
	
	If \(f\) accepts with assignment \(\sigma\), the tail-rejection fact entails that
	\((\alpha,\sigma)\in\langle\lnot b\rangle\), and the following
	\(\comGoto~\ell\) is resolved by \(E_2\) to
	\((g+_b f\seq\comGoto~\ell,\sigma)\), which lies in the second component of
	\(N_\alpha\).  
	
	If \(f\) jumps to some \(k\in\dom(E)\), then \(E_2\) follows
	the same old binding to \((E(k),\sigma)\), and the tail-rejection fact plus
	\Cref{smallerderivation} puts this target in the first component.  
	
	Finally, a free jump, a productive step, or a non-rejecting result such as \retc, \brkc,
	or \contc would give the same non-rejecting behavior for
	\((f\seq b,\pi)\) under \(E\), again contradicting tail rejection.
	
	Thus \(N_\alpha\) is rejecting, and by \Cref{rejectingsetaborts},
	\[
	(g+_b f\seq\comGoto~\ell,\pi)\tr[E_2]{}{\alpha}\bot
	\qquad\text{whenever }(\alpha,\pi)\in\langle\lnot b\rangle.
	\]
	If \((\alpha, \pi) \in \langle b \rangle\), the argument is straightforward using similar cases as above due to the fact that \((b\seq g, \pi)\) and \(g +_b f \seq \comGoto~\ell\) have identical transitions under this \(\alpha\).
	
	If \((\alpha, \pi) \in \langle \lnot b \rangle\), then \((b\seq g, \pi) \tr[E_1]{}{\alpha}
 \bot\), and the rejecting-set argument above gives
	\((g +_b f\seq \comGoto~\ell, \pi) \tr[E_2]{}{\alpha}\bot\).
		
	\paragraph*{\RefAx{\Axuniqueness}}
	Write \(s\sim_F t\) for \(s\resWith{F}\sim t\resWith{F}\).  We claim that
	\[
		(s,t)\in R \quad\text{iff}\quad s\sim_F t
	\]
	is a bisimulation between resolution by \(F\) and resolution by \(E\).
	Since \((e,\pi)\sim_F(e,\pi)\), the desired conclusion follows from
	\Cref{coinduction}.
	
	We first show that for any pair of states which represent the binding of the same label in the two environments, 
	their resolved output satisfies the bisimulation condition.
	Fix \(k\in\dom(E)=\dom(F)\), and write \(E(k)=e_k\) and
	\(F(k)=f_k\).  The premise gives
	\[
		(\comGoto~k,\sigma)\sim_F(e_k,\sigma).
	\]
	Resolving the bound goto through \(F\) follows the first binding for \(k\):
	\begin{align*}
		(\comGoto~k,\sigma)
		&\transSilent_\alpha^F (f_k,\sigma)
		&& \tCref{resolutiondef}
	\end{align*}
	Thus \((f_k,\sigma)\sim_F(e_k,\sigma)\) by
	\Cref{smallerderivation}.  In particular \(((f_k,\sigma),(e_k,\sigma))\in R\).
	
	Furthermore, we show for every atom \(\alpha\) and
	resolved outputs \(r_F, r_E\),
	\[
		(f_k,\sigma)\tr[F]{}{\alpha}r_F
		\quad\text{and}\quad
		(e_k,\sigma)\tr[E]{}{\alpha}r_E
		\quad\Longrightarrow\quad
		(r_F, r_E) \in R_F
	\]
	where \(R_F\) is the standard lifting of \(R\) to outputs.
	We prove this by case analysis.
	
	If \((e_k,\sigma)\tr{}{\alpha}r\), where \(r\in K\), then guardedness rules
	out the subcase \(r=(\jmpc~j,\tau)\) with \(j\in\dom(E)\).  Hence \(r\) is
	not a bound jump in either environment, and
	\begin{align*}
		(e_k,\sigma) &\tr[F]{}{\alpha} r && \tCref{resolutiondef}\\
		(e_k,\sigma) &\tr[E]{}{\alpha} r && \tCref{resolutiondef}.
	\end{align*}
	Since \((e_k, \sigma) \sim_F (f_k, \sigma)\), it also holds that \((f_k,\sigma) \tr[F]{}{\alpha} r\).
	
	If \((e_k,\sigma)\tr{p}{\alpha}u'\), then
	\begin{align*}
		(e_k,\sigma) &\tr[F]{p}{\alpha} u' && \tCref{resolutiondef}\\
		(e_k,\sigma) &\tr[E]{p}{\alpha} u' && \tCref{resolutiondef}.
	\end{align*}
	The \(F\)-bisimilarity of \(f_k\) and \(e_k\) gives
	\((f_k,\sigma)\tr[F]{p}{\alpha}v'\) for some \(v'\) with
	\(v'\sim_F u'\), hence \((v',u')\in R\).
	
	We now show that for any single state, its resolved transitions through \(F\) and \(E\) are related by \(R_F\). That is,
	for every source state \(u\), atom \(\alpha\), and resolved outputs
	\(r_F, r_E\), we prove
	\[
		u\tr[F]{}{\alpha}r_F
		\quad\text{and}\quad
		u\tr[E]{}{\alpha}r_E
		\quad\Longrightarrow\quad
		(r_F, r_E) \in R_F
	\]
	We prove this by cases on the unresolved transition of \(u\) at \(\alpha\).
	
	If \(u\tr{}{\alpha}r\), where \(r\in K\) is not a jump bound by
	\(\dom(E)=\dom(F)\), then
	\begin{align*}
		u &\tr[F]{}{\alpha} r && \tCref{resolutiondef}\\
		u &\tr[E]{}{\alpha} r && \tCref{resolutiondef}.
	\end{align*}
	If \(u\tr{p}{\alpha}u'\), then
	\begin{align*}
		u &\tr[F]{p}{\alpha} u' && \tCref{resolutiondef}\\
		u &\tr[E]{p}{\alpha} u' && \tCref{resolutiondef}.
	\end{align*}
	The successor pair \((u',u')\) is in \(R\) by reflexivity of \(\sim_F\).
	
	If \(u\tr{}{\alpha}(\jmpc~k,\sigma)\) with \(k\in\dom(E)=\dom(F)\),
	write \(E(k)=e_k\) and \(F(k)=f_k\).  Resolution first moves to the two
	bodies:
	\begin{align*}
		u &\transSilent_\alpha^F (f_k,\sigma) && \tCref{resolutiondef}\\
		u &\transSilent_\alpha^E (e_k,\sigma) && \tCref{resolutiondef}.
	\end{align*}
	The comparison of the resolved outputs is exactly the above bound-body case.
	
	Finally we can show that a pair of states related by $R$ satisfies the bisimulation condition when resolved by \(E\) and \(F\) respectively.
	Take arbitrary \(s,t\) with \((s,t)\in R\), and fix \(\alpha\). 
	We know \(s\sim_F t\), and we want to show that the transitions of 
	\(s\) resolved by \(F\) and \(t\) resolved by \(E\)
	satisfy the bisimulation condition. Suppose
	\[
		s \hookrightarrow_\alpha^F r_s,\
		t \hookrightarrow_\alpha^F r_F,\ \text{and}\ 
		t \hookrightarrow_\alpha^E r_E
	\]
	Since \(s\sim_F t\), \((r_s, r_F) \in R_F\), and by the same-source fact \((r_F, r_E) \in R_F\).
	Hence \((r_s, r_E) \in R_F\), since \(R\) is transitive as it is defined via bisimilarity under resolution by \(F\) 
	and the standard output lifting preserves transitivity.
	
	\paragraph*{Congruence soundness}\label{congruence-soundness}
	Finally, we show the soundness of the congruence rules.
	\paragraph*{\RefAx{\Axcongrightseq}}
	Suppose \(\forall \pi.(e, \pi)\resWith{E} \sim (f, \pi) \resWith{F}\) and
	\(\forall \pi.(h, \pi)\resWith{E} \sim (h, \pi) \resWith{F}\).  For each \(\pi\), choose a
	bisimulation \(R_h\) witnessing \((h,\pi)\resWith{E}\sim
	(h,\pi)\resWith{F}\).  Define \(R\) to contain every pair
	\[
		((g\seq e,\sigma),(g\seq f,\sigma))
	\]
	such that \((g,\sigma)\,R_h\,(g,\sigma)\), together with all pairs from the
	tail premises \(\forall \nu.(e, \nu)\sim (f, \nu)\) and the premise witnesses \(R_h\).  We check a
	prefix pair by cases on the unresolved transition of \((g,\sigma)\).
	
	If
	\[
		(g,\sigma)\tr{}{\alpha} r
		\qquad
		r\in\{\bot,\retc\}+\{\brkc,\contc\}\times I^Y,
	\]
	then sequencing propagates the same output on both sides:
	\begin{align*}
		(g\seq e,\sigma) &\tr[E]{}{\alpha} r
			& \ref{operationalsemantics},\ref{resolutiondef}\\
		(g\seq f,\sigma) &\tr[F]{}{\alpha} r
			& \ref{operationalsemantics},\ref{resolutiondef}
	\end{align*}
	If \((g,\sigma)\tr{}{\alpha}(\jmpc~\ell,\nu)\), then the right component of
	a sequence is not inspected.  The resolved outputs of
	\((g\seq e,\sigma)\) and \((g\seq f,\sigma)\) are exactly the resolved
	outputs of \((g,\sigma)\) under \(E\) and \(F\), hence are related by
	\(R_h\).
	
	If \((g,\sigma)\tr{}{\alpha}(\acc,\nu)\), then the sequence proceeds to the
	right component:
	\begin{align*}
		(g\seq e,\sigma)\tr[E]{}{\alpha} r_E
			&\iff (e,\nu)\tr[E]{}{\alpha} r_E
			& \ref{operationalsemantics},\ref{resolutiondef}\\
		(g\seq f,\sigma)\tr[F]{}{\alpha} r_F
			&\iff (f,\nu)\tr[F]{}{\alpha} r_F
			& \ref{operationalsemantics},\ref{resolutiondef}.
	\end{align*}
	The lifted outputs are related by the witness to \((e, \nu)\resWith{E} \sim (f, \nu)\resWith{F}\).
	
	If \((g,\sigma)\tr{p}{\alpha}(g',\nu)\), then
	\begin{align*}
		(g\seq e,\sigma) &\tr[E]{p}{\alpha}(g'\seq e,\nu)
			& \ref{operationalsemantics},\ref{resolutiondef}\\
		(g\seq f,\sigma) &\tr[F]{p}{\alpha}(g'\seq f,\nu)
			& \ref{operationalsemantics},\ref{resolutiondef}
	\end{align*}
	The successors are in \(R\), using the successor condition from the premise
	bisimulation for \(h\).
	
	\paragraph*{\RefAx{\Axcongrightloopseq}}
	The proof for right loop sequencing is the same argument with the
	loop-sequencing transition rules.  Suppose
	\(\forall \pi.(e, \pi)\resWith{E} \sim (f, \pi) \resWith{F}\) and
	\(\forall \pi.(h, \pi)\resWith{E} \sim (h, \pi) \resWith{F}\).  Let \(R\) contain
	\[
		((g\loopseq e,\sigma),(g\loopseq f,\sigma))
	\]
	whenever \(((g,\sigma), (g, \sigma))\) are related by a witness for the \(h\) premise, 
	together with the tail-premise pairs for \(e\) and \(f\).
	
	If \((g,\sigma)\tr{}{\alpha} r\) with \(r\in\{\bot,\retc\}\), both
	loop-sequenced terms have output \(r\).  If
	\((g,\sigma)\tr{}{\alpha}(\brkc,\nu)\), both have output
	\((\acc,\nu)\).
	
	If \((g,\sigma)\tr{}{\alpha}(\jmpc~\ell,\nu)\), loop sequencing does not
	inspect the right component, so the resolved outputs are related by the
	premise bisimulation for \(h\).
	
	If \((g,\sigma)\tr{}{\alpha}(c,\nu)\) with
	\(c\in\{\acc,\contc\}\), loop sequencing continues with the right
	component:
	\begin{align*}
		(g\loopseq e,\sigma)\tr[E]{}{\alpha} r_E
			&\iff (e,\nu)\tr[E]{}{\alpha} r_E
			& \ref{operationalsemantics},\ref{resolutiondef}\\
		(g\loopseq f,\sigma)\tr[F]{}{\alpha} r_F
			&\iff (f,\nu)\tr[F]{}{\alpha} r_F
			& \ref{operationalsemantics},\ref{resolutiondef}
	\end{align*}
	The lifted outputs are related by the tail premise.
	
	If \((g,\sigma)\tr{p}{\alpha}(g',\nu)\), then
	\begin{align*}
		(g\loopseq e,\sigma) &\tr[E]{p}{\alpha}(g'\loopseq e,\nu)
			& \ref{operationalsemantics},\ref{resolutiondef}\\
		(g\loopseq f,\sigma) &\tr[F]{p}{\alpha}(g'\loopseq f,\nu)
			& \ref{operationalsemantics},\ref{resolutiondef}
	\end{align*}
	The successors are again in \(R\).

	\paragraph*{\RefAx{\Axcongloopbody}}
	Let \(R_e\) be a bisimulation witnessing
	\[
		(e,\pi)\resWith{E}
		\sim
		(e,\pi)\resWith{F}
	\]
	for each \(\pi\).  Define \(R\) to be \(R_e\), together with all loop-head
	pairs
	\[
		((e^{(b)},\rho)\resWith{E},
		(e^{(b)},\rho)\resWith{F}),
	\]
	and all prefix pairs
	\[
		((u\loopseq e^{(b)},\rho)\resWith{E},
		(v\loopseq e^{(b)},\tau)\resWith{F})
	\]
	such that
	\[
		((u,\rho)\resWith{E},(v,\tau)\resWith{F})\in R_e.
	\]
	The \(R_e\) cases hold because \(R_e\) is a bisimulation.
	
	Let \(N_\alpha\) be the greatest set of loop-head states satisfying:
	\[
	(e^{(b)},\pi)\in N_\alpha
	\quad\Longrightarrow\quad
	(\alpha,\pi)\in\langle b\rangle
	\]
	and there exist \(c\in\{\acc,\contc\}\) and \(\sigma\) such that
	\[
	(e,\pi)\tr{}{\alpha}(c,\sigma)
	\qquad\text{and}\qquad
	(e^{(b)},\sigma)\in N_\alpha .
	\]
	We extend \(N_\alpha\) by
	\begin{align*}
		N^+_\alpha \coloneq
		&N_\alpha
		\cup
		\{(h\loopseq e^{(b)},\pi)
		\\&\quad\mid
		(h,\pi)\tr{}{\alpha}(c,\sigma),\
		c\in\{\acc,\contc\},\
		(e^{(b)},\sigma)\in N_\alpha
		\}.
	\end{align*}
	For any environment \(G\), \(N^+_\alpha\) is a rejecting set for the automaton
	resolved by \(G\).  A state in \(N_\alpha\) silently advances to another state
	of \(N_\alpha\), and a state
	\((h\loopseq e^{(b)},\pi)\in N^+_\alpha\setminus N_\alpha\) silently advances
	to \((e^{(b)},\sigma)\in N_\alpha\).  Hence by
	\Cref{rejectingsetaborts}, every state in \(N^+_\alpha\) rejects under
	resolution by \(G\).
	
	For a loop-head pair, fix \(\alpha\).  If
	\((\alpha,\rho)\in\langle\lnot b\rangle\), both sides accept with assignment
	\(\rho\).  Now suppose \((\alpha,\rho)\in\langle b\rangle\).  If
	\((e^{(b)},\rho)\in N_\alpha\), then the rejecting-set argument gives
	\[
	(e^{(b)},\rho)\tr[E]{}{\alpha}\bot
	\qquad
	(e^{(b)},\rho)\tr[F]{}{\alpha}\bot .
	\]
	Otherwise \((e^{(b)},\rho)\notin N_\alpha\).  Since the set of 
	assignments is finite and \(N_\alpha\) is the greatest closed set of
	accept/continue loop iterations, there is a finite first exit from the loop
	iteration chain.  Thus, after finitely many matched accept/continue iterations,
	we reach an assignment \(\rho'\) such that either
	\((\alpha,\rho')\in\langle\lnot b\rangle\), in which case both loop heads
	output \(\acc, \rho'\), or the body has a non-iterating result.  
	
	In the latter case, the premise bisimulation for \(R_e\) gives matching
	resolved body outputs.  Reject, return, and jump outputs are propagated by loop
	sequencing, break becomes accept, and accept/continue are excluded by the choice
	of the first exit.  Finally, a visible step
	\[
	(e,\rho')\tr[E]{p}{\alpha}(u,\nu),
	\qquad
	(e,\rho')\tr[F]{p}{\alpha}(v,\mu)
	\]
	with \(((u,\nu)\resWith{E},(v,\mu)\resWith{F})\in R_e\) gives
	\[
	(e^{(b)},\rho')\tr[E]{p}{\alpha}(u\loopseq e^{(b)},\nu),
	\qquad
	(e^{(b)},\rho')\tr[F]{p}{\alpha}(v\loopseq e^{(b)},\mu),
	\]
	whose successors are in \(R\). 
	
	Now consider a prefix pair coming from
	\(((u,\rho)\resWith{E},(v,\tau)\resWith{F})\in R_e\).  If \(R_e\) gives the
	same output
	\[
	(u,\rho)\tr[E]{}{\alpha} r
	\qquad
	(v,\tau)\tr[F]{}{\alpha} r
	\]
	with \(r\in\{\bot,\retc\}+\{\jmpc~\ell\mid\ell\in\lset\}\times I^Y\), then
	both loop-sequenced programs output \(r\).  If \(r=(\brkc,\sigma)\), both
	output \((\acc,\sigma)\).
	
	If the prefix makes a jump,
	\[
	(u,\rho)\tr{}{\alpha}(\jmpc~\ell,\sigma),
	\]
	then loop sequencing ignores subsequent terms.  Hence
	the resolved transition of \((u\loopseq e^{(b)},\rho)\) is exactly the resolved
	transition of \((u,\rho)\), and similarly on the \(F\)-side.  The outputs are
	related because \(R_e\) already matches the resolved behavior of the two
	prefixes.
	
	If \(R_e\) gives the same output \((c,\sigma)\) with
	\(c\in\{\acc,\contc\}\), then there are two cases.  If
	\((e^{(b)},\sigma)\in N_\alpha\), then
	\[
	(u\loopseq e^{(b)},\rho)\in N^+_{\alpha}
	\qquad
	(v\loopseq e^{(b)},\tau)\in N^+_{\alpha},
	\]
	so both prefix states reject by \Cref{rejectingsetaborts}.  If
	\((e^{(b)},\sigma)\notin N_\alpha\), then the finite-exit argument for the
	loop-head pair at \(\sigma\) applies.  Therefore the resolved outputs of
	\((e^{(b)},\sigma)\) under \(E\) and \(F\) are related by \(R_F\), and the
	loop-sequencing rules identify these with the resolved outputs of
	\((u\loopseq e^{(b)},\rho)\) and \((v\loopseq e^{(b)},\tau)\).
	
	If \(R_e\) gives
	\[
	(u,\rho)\tr[E]{p}{\alpha}(u',\rho')
	\qquad
	(v,\tau)\tr[F]{p}{\alpha}(v',\tau')
	\]
	with \(((u',\rho')\resWith{E},(v',\tau')\resWith{F})\in R_e\), then
	\begin{align*}
		(u\loopseq e^{(b)},\rho)
		&\tr[E]{p}{\alpha}(u'\loopseq e^{(b)},\rho')
		& \ref{operationalsemantics},\ref{resolutiondef}\\
		(v\loopseq e^{(b)},\tau)
		&\tr[F]{p}{\alpha}(v'\loopseq e^{(b)},\tau')
		& \ref{operationalsemantics},\ref{resolutiondef}
	\end{align*}
	The successors are in \(R\) by definition.

	\paragraph*{\RefAx{\Axcongguardedchoice}}
	Suppose \(\forall \pi.(e, \pi)\resWith{E} \sim (f, \pi) \resWith{F}\) and
	\(\forall \pi.(e', \pi)\resWith{E} \sim (f', \pi) \resWith{F}\).  Let \(R\) contain the pairs
	\[
		((e+_b e',\pi),(f+_b f',\pi))
	\]
	for all \(\pi\), together with the bisimulation witnesses supplied by the
	two premises.  We check the first kind of pair.  If
	\((\alpha,\pi)\in\langle b\rangle\), then guarded choice selects the left
	branch on both sides:
	\begin{align*}
		(e+_b e',\pi) &\tr[E]{}{\alpha} r_E
			\iff (e,\pi)\tr[E]{}{\alpha} r_E
			& \ref{operationalsemantics},\ref{resolutiondef}\\
		(f+_b f',\pi) &\tr[F]{}{\alpha} r_F
			\iff (f,\pi)\tr[F]{}{\alpha} r_F
			& \ref{operationalsemantics},\ref{resolutiondef}.
	\end{align*}
	The lifted outputs are related by the first premise.  If
	\((\alpha,\pi)\in\langle\lnot b\rangle\), the same argument uses the right
	branches \(e'\) and \(f'\) and the second premise.  The remaining pairs of
	\(R\) are handled by their witnessing premise bisimulations.

	\paragraph*{\RefAx{\Axcongboolguardedchoice}}
	By \Cref{Booleansoundness}, \(\langle b\rangle=\langle c\rangle\).  Hence
	guarded choice selects the same branch in \(e+_b f\) and \(e+_c f\) for
	every atom and assignment.  The relation
	\[
		\{((e+_b f,\pi),(e+_c f,\pi))\mid \pi\in I^Y\}
	\]
	together with the identity relation is therefore a bisimulation, by a case
	analysis on whether \((\alpha,\pi)\in\langle b\rangle\).

	\paragraph*{\RefAx{\Axcongboolassertion}}
	Assume \(b \eqBA c\).  
	By \Cref{Booleansoundness}, for every atom and assignment, \((\alpha,\pi)\in\langle b\rangle\) iff \((\alpha,\pi)\in\langle c\rangle\).  
	Hence, by \Cref{operationalsemantics}, both assertions have exactly the same transition: they both accept with assignment \(\pi\) when \((\alpha,\pi)\) satisfies the test, and they both abort otherwise.  
	Therefore the relation \(\{((b,\pi),(c,\pi))\mid \pi\in I^Y\}\) is a bisimulation.

	\paragraph*{\RefAx{\Axcongboolloop}}
	We show that 
	\begin{align*}
		{\sim} &\coloneq \{((e' \loopseq e^{(b)}, \pi) , (e' \loopseq e^{(c)}, \pi)) \mid e' \in \Expm, \pi \in I^Y\} \\&\quad\cup \{((e^{(b)}, \pi) , (e^{(c)}, \pi)) \mid \pi \in I^Y\}	
	\end{align*} is an intermediate bisimulation and appeal to \Cref{strongbisimtobisim}. By \Cref{Booleansoundness} \(\langle b \rangle = \langle c \rangle\). Consider some \((\alpha, \pi)\).
	
	We start with the pair of terms that does not include the loop sequencing by \(e'\).
	
	Suppose \((\alpha, \pi) \in \langle \lnot b \rangle\), then \((\alpha, \pi) \in \langle \lnot c \rangle\). Hence \((e^{(b)}, \pi) \tr{}{\alpha} (\acc, \pi)\), as does \((e^{(c)}, \pi)\). 
	
	For the remaining cases suppose \((\alpha, \pi) \in \langle b \rangle\), and thus also \((\alpha, \pi) \in \langle c \rangle\).
	
	Suppose \((e, \pi) \tr{}{\alpha} k \in \{\bot, \retc\} + \{\jmpc~\ell \mid \ell \in \lset\} \times I^Y\). Then \((e^{(b)}, \pi) \tr{}{\alpha} k\) and \((e^{(c)}, \pi) \tr{}{\alpha} k\). 
	
	Suppose \((e, \pi) \tr{p}{\alpha} (e', \sigma)\). Then \((e^{(b)}, \pi) \tr{p}{\alpha} (e' \loopseq e^{(b)}, \sigma)\) and \((e^{(c)}, \pi) \tr{p}{\alpha} (e' \loopseq e^{(c)}, \sigma)\), which are related by \(\sim\).
	
	Suppose \((e, \pi) \tr{}{\alpha} (\brkc, \sigma)\). Then \((e^{(b)}, \pi) \tr{}{\alpha} (\acc, \sigma)\) and \((e^{(c)}, \pi) \tr{}{\alpha} (\acc, \sigma)\).
	
	Suppose \((e, \pi) \tr{}{\alpha} (c, \sigma)\) where \(c \in \{\acc, \contc\}\). Then \((e^{(b)}, \pi) \allowbreak \transSilent_\alpha (e^{(b)}, \sigma)\) and \((e^{(c)}, \pi) \transSilent_\alpha (e^{(c)}, \sigma)\). These states are related by \(\sim\).
	
	We now address an arbitrary pair of terms which does contain the loop sequencing by \(e'\).
	
	Suppose \((e', \pi) \tr{}{\alpha} c \in \{\bot, \retc\} + \{\jmpc~\ell \mid \ell \in \lset\} \times I^Y\), then \((e'\loopseq e^{(b)}, \pi) \tr{}{\alpha} c\) and the same for \((e'\loopseq e^{(c)}, \pi)\).
	
	Suppose \((e', \pi) \tr{}{\alpha}  (\brkc, \sigma)\), then \((e'\loopseq e^{(b)}, \pi) \tr{}{\alpha} (\acc, \sigma)\) and the same for \((e'\loopseq e^{(c)}, \pi)\).

	Suppose \((e', \pi) \tr{p}{\alpha}  (f, \sigma)\), then \((e'\loopseq e^{(b)}, \pi) \tr{p}{\alpha} (f\loopseq e^{(b)}, \sigma)\) and \((e'\loopseq e^{(c)}, \pi) \tr{p}{\alpha} (f\loopseq e^{(c)}, \sigma)\) and these states are related by \(\sim\).
	
	Suppose \((e', \pi) \tr{}{\alpha} (c, \sigma)\) where \(c \in \{\acc, \contc\}\), then \((e'\loopseq e^{(b)}, \pi) \transSilent_\alpha (e^{(b)}, \sigma)\) and \((e'\loopseq e^{(c)}, \pi) \transSilent_\alpha (e^{(c)}, \sigma)\) and these two states are related by \(\sim\).
	
	\paragraph*{\RefAx{\Axcongleftseqempty}}
	Fix an assignment \(\pi\), and let \(R_e\) be a bisimulation witnessing
	\((e,\pi)\sim(f,\pi)\) in the empty environment.  Let \(R\) contain the
	identity relation, the pairs of \(R_e\), and every pair
	\[
		((u\seq g,\nu),(v\seq g,\mu))
	\]
	such that \((u,\nu)\,R_e\,(v,\mu)\).  We check the last kind of pair.
	
	If \(R_e\) gives
	\[
		(u,\nu)\tr{}{\alpha} r
		\qquad
		(v,\mu)\tr{}{\alpha} r
		\qquad
		r\in K,\quad r\notin \{\acc\}\times I^Y,
	\]
	then sequencing propagates the same output:
	\begin{align*}
		(u\seq g,\nu) &\tr{}{\alpha} r
			&& \tCref{operationalsemantics}\\
		(v\seq g,\mu) &\tr{}{\alpha} r
			&& \tCref{operationalsemantics}.
	\end{align*}
	If \(R_e\) gives
	\[
		(u,\nu)\tr{}{\alpha}(\acc,\rho)
		\qquad
		(v,\mu)\tr{}{\alpha}(\acc,\rho),
	\]
	then both sequenced programs continue with the same right component:
	\begin{align*}
		(u\seq g,\nu)\hookrightarrow_\alpha r
			&\iff (g,\rho)\hookrightarrow_\alpha r
			&& \tCref{operationalsemantics}\\
			&\iff (v\seq g,\mu)\hookrightarrow_\alpha r
			&& \tCref{operationalsemantics}.
	\end{align*}
	The identity pairs in \(R\) handle any successor of \(g\).
	
	If \(R_e\) gives
	\[
		(u,\nu)\tr{p}{\alpha}(u',\rho)
		\qquad
		(v,\mu)\tr{p}{\alpha}(v',\tau)
		\qquad
		(u',\rho)\,R_e\,(v',\tau),
	\]
	then
	\begin{align*}
		(u\seq g,\nu) &\tr{p}{\alpha}(u'\seq g,\rho)
			&& \tCref{operationalsemantics}\\
		(v\seq g,\mu) &\tr{p}{\alpha}(v'\seq g,\tau)
			&& \tCref{operationalsemantics}.
	\end{align*}
	The successors are in \(R\) by definition.

	\paragraph*{\RefAx{\Axcongleftloopseqempty}}
	Fix \(\pi\), and let \(R_e\) witness \((e,\pi)\sim(f,\pi)\).  Let \(R\)
	contain the identity relation, the pairs of \(R_e\), and every pair
	\[
		((u\loopseq g,\nu),(v\loopseq g,\mu))
	\]
	such that \((u,\nu)\,R_e\,(v,\mu)\).  Again we check only the loop-sequenced
	pairs.
	
	If \(R_e\) gives the same output
	\[
		(u,\nu)\tr{}{\alpha} r
		\qquad
		(v,\mu)\tr{}{\alpha} r
	\]
	with \(r\in\{\bot,\retc\}+\{\jmpc~\ell\mid\ell\in\lset\}\times I^Y\), then
	loop sequencing propagates \(r\) on both sides.  If instead
	\(r=(\brkc,\rho)\), both loop-sequenced programs output \((\acc,\rho)\).
	
	If \(R_e\) gives
	\[
		(u,\nu)\tr{}{\alpha}(c,\rho)
		\qquad
		(v,\mu)\tr{}{\alpha}(c,\rho)
		\qquad
		c\in\{\acc,\contc\},
	\]
	then both programs continue with the same loop body:
	\begin{align*}
		(u\loopseq g,\nu)\hookrightarrow_\alpha r
			&\iff (g,\rho)\hookrightarrow_\alpha r
			&& \tCref{operationalsemantics}\\
			&\iff (v\loopseq g,\mu)\hookrightarrow_\alpha r
			&& \tCref{operationalsemantics}.
	\end{align*}
	The identity pairs in \(R\) handle the resulting behavior of \(g\).
	
	If \(R_e\) gives
	\[
		(u,\nu)\tr{p}{\alpha}(u',\rho)
		\qquad
		(v,\mu)\tr{p}{\alpha}(v',\tau)
		\qquad
		(u',\rho)\,R_e\,(v',\tau),
	\]
	then
	\begin{align*}
		(u\loopseq g,\nu) &\tr{p}{\alpha}(u'\loopseq g,\rho)
			&& \tCref{operationalsemantics}\\
		(v\loopseq g,\mu) &\tr{p}{\alpha}(v'\loopseq g,\tau)
			&& \tCref{operationalsemantics}.
	\end{align*}
	The successors are in \(R\) by definition.

	\paragraph*{\RefAx{\Axcongloopbodyempty}}
	It is sufficient to show that
	\begin{align*}
		&{\sim} \eqdef \{(e^{(b)}, \pi), (f^{(b)}, \pi) \mid \pi \in I^Y\}\\
		&\quad\cup \{(e'\loopseq e^{(b)}, \sigma), (f'\loopseq f^{(b)}, \sigma) \mid (e', \sigma) {\sim'} (f', \sigma)\}
	\end{align*}
	where \({\sim'}\) is the bisimulation which relates \((e, \pi)\) and \((f, \pi)\), is an intermediate bisimulation.

	We start with the first subset of the proposed bisimulation. The case where \((\alpha, \pi) \in \langle \lnot b \rangle\) is straightforward. We assume \((\alpha, \pi) \in \langle \lnot b \rangle\).
	
	The cases where \((e, \pi) \tr{}{\alpha} r \in \{\bot, \retc\} + (\{\jmpc~\ell \mid \ell \in \lset\} + \{\brkc\}) \times I^Y\) are straightforward.
	
	If \((e, \pi) \tr{p}{\alpha} (e', \sigma)\), then \((f, \pi) \tr{p}{\alpha} (f', \nu)\) such that \((e', \sigma)\) and \((f', \nu)\) are related by some bisimulation.
	\begin{align*}
		(e^{(b)}, \pi) &\tr{p}{\alpha} (e' \loopseq e^{(b)}, \sigma)\\
		(f^{(b)}, \pi) &\tr{p}{\alpha} (f' \loopseq e^{(b)}, \nu)
	\end{align*}
	and these states are related by \(\sim\).
	
	If \((e, \pi) \tr{}{\alpha} (c, \sigma)\) where \(c \in \{\acc, \contc\}\), then \((f, \pi) \tr{}{\alpha} (c, \sigma)\) 
	\begin{align*}
		(e^{(b)}, \pi) &\transSilent_\alpha (e^{(b)}, \sigma)\\
		(f^{(b)}, \pi)&\transSilent_\alpha (f^{(b)}, \sigma)
	\end{align*}
	These states are related by \(\sim\).
	
	We continue with the second subset of the proposed bisimulation. The cases where \((e', \sigma) \tr{}{\alpha} r \in K\) where \(r\) is not \((\acc, \sigma')\) or \((\contc, \sigma')\) are straightforward.
	
	If \((e', \sigma) \tr{p}{\alpha} (e'', \sigma')\), then \((f', \nu) \tr{p}{\alpha} (f'', \nu')\) where \((e'', \sigma')\) and \((f'', \nu')\) are related by some bisimulation. Hence,
	\begin{align*}
		(e'\loopseq e^{(b)}, \sigma) &\tr{p}{\alpha} (e''\loopseq e^{(b)}, \sigma')\\
		(f'\loopseq f^{(b)}, \nu) &\tr{p}{\alpha} (f''\loopseq f^{(b)}, \nu')
	\end{align*}
	and these states are related by \(\sim\).
	
	If \((e', \sigma) \tr{}{\alpha} (c, \sigma')\) where \(c \in \{\acc, \contc\}\), then \((f, \pi) \tr{}{\alpha} (c, \sigma')\) 
	\begin{align*}
		(e'\loopseq e^{(b)}, \sigma) &\transSilent_{\alpha} (e^{(b)}, \sigma')\\
		(f'\loopseq f^{(b)}, \nu) &\transSilent_{\alpha} (f^{(b)}, \sigma'')
	\end{align*}
	and these states are related by \(\sim\).
\end{appendixproof}

	\begin{toappendix}
		\begin{lemmarep}\label{derivedfacts}
		All the facts in~\Cref{fig:derived-facts} are derivable using the \CFGKAT axioms in \Cref{ax}.
	\end{lemmarep}
	\begin{proof}
		We prove each in sequence.
		\paragraph{\ref{loopassoctest}} By direct proof
		\begin{align*}
			b\seq(f\loopseq g) &\equiv b\seq(f\loopseq g) +_b b\seq(f\loopseq g) & \RefAx{\Axidempotence}\\
			&\equiv (f\loopseq g) +_b \lnot b \seq (b\seq(f\loopseq g)) & \RefAx{\Axguardedness}, \RefAx{\Axskewcomm}\\
			&\equiv (f\loopseq g) +_b (\lnot b \seq b) \seq(f\loopseq g) & \RefAx{\Axassoc}\\
			&\equiv f\loopseq g +_b 0 & \RefAx{\Axtestcomb}, \BA, \RefAx{\Axabsleft}\\
			&\equiv f\loopseq g +_b 0\loopseq g & \RefAx{\Axrightloopannabort}\\
			&\equiv (f +_b 0)\loopseq g & \RefAx{\Axloopdistributivity}\\
			&\equiv (f +_b 0\seq f)\loopseq g & \RefAx{\Axabsleft}\\
			&\equiv (b\seq f +_b 0\seq f)\loopseq g & \RefAx{\Axguardedness}\\
			&\equiv (b\seq f +_b \lnot b \seq (b \seq f)) \loopseq g & \BA, \RefAx{\Axtestcomb}, \RefAx{\Axassoc}\\
			&\equiv (b\seq f +_b b \seq f)\loopseq g & \RefAx{\Axskewcomm}, \RefAx{\Axguardedness}\\
			&\equiv (b\seq f)\loopseq g & \RefAx{\Axidempotence}
		\end{align*}
		
		\paragraph{\ref{loopseqisseqtest}} By direct proof
		\begin{align*}
			b\loopseq e &\equiv (b\seq 1) \loopseq e & \RefAx{\Axneutleft}\\
			&\equiv b \seq (1 \loopseq e) & \eqref{loopassoctest}\\
			&\equiv b \seq e & \RefAx{\Axneutleftloop}
		\end{align*}
		\ref{loopseqisseqasasign} and \ref{loopseqisseqact} follow by similar arguments using \Cref{loopassocassign+} and \RefAx{\Axloopassocact}.
		
		\paragraph{\ref{inddistr}} We argue by induction on the construction of \(b\). If \(b\) is \(t\) (a generator) or \(y=j\), then the desired fact follows immediately from \RefAx{\Axinddistr} and \RefAx{\Axindinddistr}. If \(b = \lnot b'\), then the result follows from \RefAx{\Axskewcomm} and the inductive hypothesis. The cases for \(1\) and \(0\) follow from \Cref{neutralright} and \RefAx{\Axskewcomm}. If \(b = c \land d\), then 
		\begin{align*}
			&x \assign i \seq (e +_{c \land d} f) \\
			&\equiv x \assign i \seq (e +_{c \land d} (f +_d f)) & \RefAx{\Axidempotence}\\ 
			&\equiv x \assign i \seq ((e +_{c} f) +_d f) & \RefAx{\Axskewassoc}\\ 
			&\equiv ((x \assign i \seq e +_{c} x \assign i \seq f) +_d x \assign i \seq f) & \text{ind hyp}\\
			&\equiv (x \assign i \seq  e +_{c \land d} (x \assign i \seq f +_d x \assign i \seq f)) & \RefAx{\Axskewassoc}\\
			&\equiv (x \assign i \seq e +_{c\land d} x \assign i\seq f) & \RefAx{\Axidempotence}
		\end{align*}
		If \(b = c \lor d\), then 
		\begin{align*}
			&x \assign i \seq (e +_{c \lor d} f) \\
			&\equiv x \assign i \seq ((e+_c e) +_{c \lor d} f) & \RefAx{\Axidempotence}\\ 
			&\equiv x \assign i \seq (e+_c (e +_{d} f)) & \eqref{antiskewcom}\\ 
			&\equiv (x \assign i \seq e+_c (x \assign i \seq e +_{d} x \assign i \seq f)) & \text{ind hyp}\\
			&\equiv ((x \assign i \seq e+_c x \assign i \seq e) +_{c \lor d} x \assign i \seq f) & \eqref{antiskewcom}\\ 
			&\equiv (x \assign i \seq e +_{c \lor d} x \assign i\seq f) & \RefAx{\Axidempotence}
		\end{align*}
		
		\paragraph{\ref{indcommtest}} By direct proof
		\begin{align*}
			x\assign i \seq b &\equiv x\assign i \seq (b\seq 1) & \RefAx{\Axneutright}\\
			&\equiv x\assign i \seq (1+_b 0) & \eqref{neutralright}\\
			&\equiv (x\assign i \seq 1+_b x\assign i \seq 0) & \eqref{inddistr}\\
			&\equiv (x\assign i +_b 0) & \RefAx{\Axneutright}, \RefAx{\Axindabort}\\
			&\equiv b\seq x\assign i & \eqref{neutralright}
			&\qed
		\end{align*}
	\end{proof}
	\end{toappendix}

\subsection{Example Transformations} \label{sec:sctransform-derivation}

We end this section by demonstrating the use of our axioms with two modest but non-trivial worked examples; more complex case studies appear in \Cref{sec:eh-loop}.

\begin{example}
First, we derive correctness of the short-circuit transformation (\Cref{sctransform}).
Here, we assume that \(\ell \neq k\), and that both labels are fresh in \(\hat e\) and \(\hat f\), and that \(\hat e\) and \(\hat f\) are both well-formed programs.
Let \(e\) and \(f\) be the reduced forms of \(\hat e\) and \(\hat f\), respectively.
After applying label extraction and reduction, we need to prove the following equality:
\[\sigma(\hat e),\sigma(\hat f) \vdash e +_{ (a \land b) \lor c} f \equiv (1\seq e+_b 1\seq (\comGoto~\ell +_c f)) +_a \comGoto~k \dashv \sigma(\hat e),\sigma(\hat f), E\]
where \(E(\ell) = 1\seq e\) and \(E(k) =  1\seq (\comGoto~\ell +_{c} f)\). 
The first part of the proof follows from the guarded choice axioms, lifted to the right environment by weakening:
\begin{align*}
	\sigma(\hat e),\sigma(\hat f) &\vdash e +_{ (a \land b) \lor c} f\RocqQedRight{sctransform-derivation}
	\\&\equiv(e+_b (e +_{c} f)) +_{a} (e +_{c} f) \dashv \sigma(\hat e),\sigma(\hat f) \tag*{\RefAx{\Axidempotence}/\RefAx{\Axskewcomm}/\RefAx{\Axskewassoc}, \RefAx{\Axweaken}}
	\\&\equiv (e+_b (e +_{c} f)) +_{a} (e +_{c} f) \dashv \sigma(\hat e),\sigma(\hat f), E \tag*{\RefAx{\Axredundancy}}
	\\&\equiv (1\seq e+_b 1\seq(1\seq e +_{c} f)) +_{a} (1\seq e +_{c} f) \dashv \sigma(\hat e),\sigma(\hat f), E \tag*{\RefAx{\Axneutleft}}
	\\&\equiv (1\seq e+_b 1\seq(\comGoto~\ell +_{c} f)) +_{a} (\comGoto~\ell +_{c} f) \dashv \sigma(\hat e),\sigma(\hat f), E \tag*{\RefAx{\Axsubstitution}}
	\\&\equiv (1\seq e+_b 1\seq(\comGoto~\ell +_{c} f)) +_{a} \comGoto~k \dashv \sigma(\hat e),\sigma(\hat f), E \tag*{\RefAx{\Axsubstitution}, \RefAx{\Axneutleft}}
\end{align*}
\end{example}

\begin{example}
Next, we demonstrate \RefAx{\Axuniqueness} with a simple example that converts between loop and \comGoto. 
Consider \(\varnothing \vdash p^{(1)} \equiv \comGoto~\ell \dashv \ell = p\seq \comGoto~\ell\), 
which states that an infinite loop that repeatedly executes \(p\) is equivalent to a program that \(p\) and jumps back to its beginning.
To prove this, we let $E(\ell) = p\seq \comGoto~\ell$ and $F(\ell) = p^{(1)}$, and derive:
\begin{align*}
	F \vdash \comGoto~\ell &\equiv p^{(1)} \dashv F \tag*{\RefAx{\Axsubstitution}}\\
	&\equiv p\loopseq p^{(1)} \dashv F \tag*{\RefAx{\Axloopunfold}, \RefAx{\Axneutleft}, \cite[U6]{gkat}}\\
	&\equiv p\seq p^{(1)} \dashv F \tag*{\ref{loopseqisseqact}}\\
	&\equiv p\seq \comGoto~\ell \dashv F \tag*{\RefAx{\Axsubstitution}}
\intertext{%
	Thus, since $F \vdash \comGoto~\ell \equiv E(\ell) \dashv F$ and the other side-conditions of \RefAx{\Axuniqueness} hold, we find:
}
	E \vdash \comGoto~\ell &\equiv \comGoto~\ell \dashv F \tag*{\RefAx{\Axuniqueness}}\\
	&\equiv p^{(1)} \dashv \varnothing \tag*{\RefAx{\Axsubstitution}, \RefAx{\Axredundancy}}
\end{align*}
\end{example}

The latter example demonstrates that \RefAx{\Axuniqueness} is crucial to completeness, because under the right circumstances it allows us to swap one environment for another.

\section{Completeness}\label{sec:completeness}

Completeness shows the effectiveness of axiomatic systems: it states that \emph{all} semantically sound equalities are derivable using the given axioms. 
We now present a completeness result for \CFGKAT. 
Because this result is highly technical, we present our construction in three steps.
First, in~\Cref{sec:fundamental-theorem}, we develop a \emph{fundamental theorem} that associates each one-step transition of \((\Expm \times I^{Y}, \partial)\) with a non-contextual equation.
Second, in~\Cref{sec:contextual-normal-form}, we use this equation to prove that an expression $e$ in environment $E$ is equivalent to a \comGoto statement in a \emph{canonical environment} derived from said data.
Lastly, we use \RefAx{\Axuniqueness} to relate the canonical environments of semantically equivalent expressions and environments.
In total, this shows that semantically equivalent expressions with environments can be proved equivalent by way of their canonical environments.

Although the proof strategy seems similar to other works in \GKAT~\cite{van2025weighted,rozowski2023probabilistic,gkatcoequations}, the complexity of our system poses significant challenges.
Indeed, reasoning about indicator variables and goto statements is highly non-trivial.
Unlike previous systems, a \CFGKAT program can perform an arbitrary number of indicator assignments and jumps before executing an action.
Even worse, programs can toggle indicator assignments and jump or loop indefinitely, leading to unproductive recursive behavior, \(\bot\).
To reason about programs like these, we prove several interesting invariant theorems with~\RefAx{\Axlooptighten} and~\RefAx{\Axunprbra}.

\subsection{Fundamental Theorem}\label{sec:fundamental-theorem}
We begin the completeness argument by constructing a non-contextual normal form
which will match precisely the semantics of the given reduced expression in the empty environment. 
To do that, we define complete assignments and tests similar to those found in NetKAT~\cite{netkat}.
These constructs allow us to assign to and test all variables at once.
\begin{definition} [Complete Tests and Assignments \RocqLabelRef{complassign/test}]\label{complassign/test}
	Let \(X = \{x_1, \dots, x_n\} \subseteq Y\) and let \(\pi \in I^Y\). 
	The \emph{assignment} and \emph{test} of \(\pi\) on \(X\) are
	\[
		X \leftarrow \pi \eqdef x_1 \assign \pi(x_1) \seq \cdots \seq x_n \assign \pi(x_n)
		\qquad
		X = \pi \eqdef x_1 = \pi(x_1) \land \cdots \land x_n = \pi(x_n),
	\]
	with \(\emptyset \leftarrow \pi \eqdef 1\) and \((\emptyset = \pi) \eqdef 1\).
	These terms are well-defined up to \(\equiv\) due to \RefAx{\Axindindcomm} and \RefAx{\Axassoc}.
	For \(\pi \in I^Y\), we call \(Y \leftarrow \pi\) and \(Y = \pi\) the \emph{complete} assignment and test.
\end{definition}
Complete tests and assignments are the syntactic counterparts to indicator states: just like each state can only have one indicator state, the conjunction of two complete tests is false.
Additionally, every state has to be at \emph{some} indicator state, reflected by the fact that all complete tests sum to \(1\).
Besides these unique properties, complete tests and assignments inherit desirable properties of their single counterparts, like commutativity with actions and non-indicator tests, and analogous properties to \RefAx{\Axindval}, \RefAx{\Axindred}, \RefAx{\Axindannih}, and \RefAx{\Axloopassocassign}.
With this gadget, we prove the completeness of our Boolean algebra by inductively unfolding each expression into a disjunction of complete tests with atoms, respectively accounting for the semantics of indicator tests and primitive tests.

\begin{toappendix}
	\begin{lemma}[\RocqLabelRef{atomsum}]\label{atomsum}
		\(\bigvee_{\alpha \in \At} \alpha \eqBA 1\)

	\end{lemma}
	\begin{lemma}[\RocqLabelRef{atomsdisagree}]\label{atomsdisagree}
		For any \(\alpha, \beta \in \At\) if \(\alpha \neq \beta\), then \(\alpha \land \beta \eqBA 0\).

	\end{lemma}
	\begin{lemma}[\RocqLabelRef{indboollem}] \label{indboollem}
		For any \(X \subseteq Y\), \(\bigwedge_{x \in X} \bigvee_{i \in I} x=i \eqBA \bigvee_{\pi \in I^X} X=\pi \eqBA 1\).

	\end{lemma}
	\begin{proof}
		We prove the first equality by induction on the size of \(X\). If \(X = \emptyset\), the result follows immediately from unrolling the definitions. If \(X = X' \cup \{x\}\), then 
		\begin{align*}
			\bigwedge_{x' \in X} \bigvee_{i \in I} x'=i &\eqBA \left(\bigvee_{i \in I} x=i\right) \land \bigwedge_{x' \in X'} \bigvee_{i \in I} x'=i &\\
			&\eqBA \left(\bigvee_{i \in I} x=i\right) \land \bigvee_{\pi \in I^{X'}} X'= \pi & \text{Ind Hyp}\\
			&\eqBA \bigvee_{\pi \in I^{X'}}  \left(\bigvee_{i \in I} x=i\right) \land X'= \pi & \text{Distr}\\
			&\eqBA \bigvee_{\pi \in I^{X'}}  \bigvee_{i \in I} x=i \land X'= \pi & \text{Distr}\\
			&\eqBA \bigvee_{(\pi, i) \in I^{X'} \times I} x=i \land X'= \pi & \text{Assoc, Comm}\\
			&\eqBA \bigvee_{\pi \in I^X} X= \pi & \text{Reindex}
		\end{align*}

		For the second equality, each $\bigvee_{i \in I} x=i \eqBA 1$ by $\RefAx{\Axindsum}$, so their conjunction is $1$.
	\end{proof}
	\begin{lemma}[\RocqLabelRef{inddisagree}] \label{inddisagree}
		For any \(\pi, \sigma \in I^Y\) such that \(\pi \neq \sigma\) it holds that \(Y=\sigma \land Y=\pi \eqBA 0\).

	\end{lemma}
	\begin{proof}
		There exists some \(x \in Y\) such that \(\pi(x) \neq \sigma(x)\). By idempotence, commutativity and associativity \(Y=\sigma \land Y=\pi \eqBA (x=\pi(x) \land x=\sigma(x)) \land (Y = \pi) \land (Y = \sigma) \eqBA 0\) by \RefAx{\Axindexclusivity} and boundary.
	\end{proof}
	\begin{lemma}[\RocqLabelRef{boolnormleqlem}] \label{boolnormleqlem}
		If \((\alpha, \pi) \in \langle b \rangle \implies \alpha \land Y=\pi \leqBA b\) and \((\alpha, \pi) \in \langle \lnot b \rangle \implies b\land \alpha \land Y=\pi \eqBA 0\) then \(b \eqBA \bigvee_{(\alpha, \pi) \in \langle b \rangle} \alpha \land Y=\pi\).

	\end{lemma}
	\begin{proof}
		For the first conjunct suppose \((\alpha, \pi) \in \langle b \rangle\)
		\begin{align*}
			b &\eqBA b \land 1 \land 1 & \text{ident}\\
			&\eqBA  b \land  (\bigvee_{\alpha \in \At}\alpha) \land (\bigvee_{\pi \in \At} Y=\pi)  & \tCref{indboollem,atomsum}\\
			&\eqBA  b \land  (\bigvee_{\alpha \in \At}\bigvee_{\pi \in \At} \alpha \land Y=\pi)   & \text{distr}\\
			&\eqBA  b \land  (\bigvee_{(\alpha, \pi) \in \At\times I^Y} \alpha \land Y=\pi)   & \text{comm, assoc}\\
			&\eqBA  \bigvee_{(\alpha, \pi) \in \At\times I^Y} b \land  (\alpha \land Y=\pi)   & \text{distr}\\
			&\eqBA \bigvee_{(\alpha, \pi) \in \langle b \rangle} \alpha \land Y=\pi &
		\end{align*}
		where the last step follows because for each \((\alpha, \pi)\), either \((\alpha, \pi) \in \langle b \rangle\) or \(\langle \lnot b \rangle\). In the first case we use the premise to see \(b \land (\alpha \land Y=\pi) \eqBA \alpha \land Y=\pi\) by \BA. Otherwise \(\alpha \land Y= \pi \land b \eqBA 0\). The disjunction then collapses to \(\langle b \rangle\) by identity.
	\end{proof}
	\begin{lemma}[\RocqLabelRef{Booleannormalform}]\label{Booleannormalform}
		\(b \eqBA \bigvee_{(\alpha, \pi) \in \langle b \rangle} \alpha \land Y=\pi\).

	\end{lemma}
	\begin{proof}
	We prove by structural induction on \(b \in \BExp\) that \((\alpha, \pi) \in \langle b \rangle \implies \alpha \land Y=\pi \leqBA b\) and \((\alpha, \pi) \notin \langle b \rangle \implies b\land \alpha \land Y=\pi \eqBA 0\), then we apply \Cref{boolnormleqlem}. The case \(b=0\) follows immediately from the definition of \(\bigvee\).
	\paragraph*{case \(1\)}
	\[1 \lor (\alpha \land Y=\pi) \eqBA 1\]
	by boundary, and all \((\alpha, \pi) \in \langle 1 \rangle\).
	
	\paragraph*{case \(x=i\)}
	For the first fact \((\alpha, \pi) \in \langle x=i \rangle\) iff \(\pi(x) = i\) and hence
	\begin{align*}
			x=i \lor (\alpha \land Y=\pi) &\eqBA x=i \lor( x=i \land (\alpha \land Y=\pi))  & \text{idemp, comm, distr}\\
			&\eqBA x=i  & \text{absorption}
		\end{align*}
	For the second fact \((\alpha, \pi) \notin \langle x=i \rangle\) iff \(\pi(x) \neq i\) and hence
	\begin{align*}
			x=i \land \bigvee_{(\alpha, \pi) \in \langle b \rangle} \alpha \land Y=\pi 
			&\eqBA \bigvee_{(\alpha, \pi) \in \langle b \rangle} x=i \land \alpha \land Y=\pi  & \text{distr}\\
			&\eqBA 0  & \text{comm}, \RefAx{\Axindexclusivity}, \text{ident}
		\end{align*}

	\paragraph*{case \(t\)}
	Follows by a similar argument to the previous case.
	
	\paragraph*{case \(\lor\)}
	We start with the first fact. Given some \((\alpha, \pi) \in \langle b \lor c \rangle\) then w.l.o.g. suppose \((\alpha, \pi) \in \langle b \rangle\)
	\begin{align*}
		(b \lor c) \lor (\alpha \land Y= \pi)
		&\eqBA c \lor (b \lor (\alpha \land Y= \pi)) & \text{assoc, comm}\\
		&\eqBA c \lor b & \text{ind hyp}
	\end{align*}
	For the second fact suppose \((\alpha, \pi) \notin \langle b \lor c \rangle\).  By \Cref{Booleansemantics}  \((\alpha, \pi)\) is in neither the semantics of \(b\) nor \(c\).
	\begin{align*}
		(b \lor c) \land (\alpha \land Y= \pi)
		&\eqBA (b \land (\alpha \land Y= \pi)) \lor (c \land (\alpha \land Y= \pi)) & \text{distr}\\
		&\eqBA0 \lor 0 & \text{ind hyp}\\
		&\eqBA0& \text{idemp}
	\end{align*}
	
	\paragraph*{case \(\land\)}
	We start with the first fact. Given some \((\alpha, \pi) \in \langle b \land c \rangle\) then
	\begin{align*}
		(b \land c) \lor (\alpha \land Y= \pi)
		&\eqBA (b \lor (\alpha \land Y= \pi)) \land (c \lor (\alpha \land Y= \pi)) & \text{distr}\\
		&\eqBA b \land c & \text{ind hyp}
	\end{align*}
	Now suppose \((\alpha, \pi) \notin \langle b \land c \rangle\). W.l.o.g. suppose \((\alpha, \pi) \notin \langle c \rangle\), then
	\begin{align*}
		(b \land c) \land (\alpha \land Y= \pi)
		&\eqBA b \land (c \land (\alpha \land Y= \pi)) & \text{comm, assoc}\\
		&\eqBA c \land 0 & \text{ind hyp}\\
		&\eqBA0 & \text{bnd}
	\end{align*}
	\paragraph{case \(\lnot\)}
	
	For the first fact suppose \((\alpha, \pi) \in \langle \lnot b\rangle\) then by \Cref{Booleansemantics} \((\alpha, \pi) \notin \langle b\rangle\), by the inductive hypothesis \(\alpha \land Y=\pi \land b \eqBA 0\), and by \BA \(\alpha \land Y=\pi \leqBA \lnot b\).

	For the second fact suppose \((\alpha, \pi) \notin \langle \lnot b\rangle\) then by \Cref{Booleansemantics} \((\alpha, \pi) \in \langle b\rangle\), by the inductive hypothesis \(\alpha \land Y= \pi \leqBA b\) hence by \BA \(\alpha \land Y= \pi \land \lnot b \eqBA 0\).
 	\end{proof}
	\begin{corollaryrep}[\RocqLabelRef{atomscorr}] \label{atomscorr}
		The atoms of the Boolean algebra generated by \(t \in T, x = i \) for \(x \in X, i \in I\) are \(\{\alpha \land Y=\pi \mid \alpha \in \At, \pi \in I^Y\}\).

	\end{corollaryrep}
	\begin{proof}
		Consider a term of the form \(\alpha \land Y=\pi\). Now suppose some \(b \leqBA \alpha \land Y=\pi\). By \Cref{Booleannormalform} for each \((\beta, \sigma) \in \langle b \rangle\), \(\beta \land Y= \sigma \leqBA \alpha \land Y=\pi\). Hence \((\beta \land \alpha) \neqBA 0\) and \(Y=\sigma \land Y=\pi \neqBA 0\) hence \((\alpha, \pi) = (\beta, \sigma)\). So \(\langle b \rangle\) may only contain \((\alpha, \pi)\). Hence \(\langle b \rangle\) is either  \(\{(\alpha, \pi)\}\) or the empty set. By \Cref{Booleannormalform} \(b \eqBA \alpha \land Y=\pi\) or \(b \eqBA 0\).
		
		Consider an atom \(b\). By \Cref{Booleannormalform} \(\alpha \land Y=\pi \leqBA b\) for all \((\alpha, \pi) \in \langle b \rangle\). Since \(\alpha \land Y=\pi \neqBA 0\) by \Cref{Booleansoundness} it holds that \(b \eqBA \alpha \land Y=\pi\).
	\end{proof}
\end{toappendix}

\begin{theoremrep}[Completeness of the Boolean Algebra \RocqLabelRef{Booleancompleteness}] \label{Booleancompleteness}
\(\langle b \rangle = \langle c \rangle \implies b \eqBA c\)

\end{theoremrep}
\begin{appendixproof}
\begin{align*}
	b &\eqBA \bigvee_{(\alpha, \pi) \in \langle b \rangle} \alpha \land Y=\pi & \tCref{Booleannormalform}\\
	&\eqBA \bigvee_{(\alpha, \pi) \in \langle c \rangle} \alpha \land Y=\pi & \text{premise}\\
	&\eqBA c & \tCref{Booleannormalform} &\qed
\end{align*}
\end{appendixproof}

\begin{figure}[t]
  \centering
	\begin{tikzpicture}[
		baseline=(current bounding box.center),
		node distance=.2cm and .75cm,
		every node/.style={
			align=center,
			draw,
			rounded corners,
			inner sep=3pt,
			outer sep=0pt,
			font=\scriptsize
		},
	]
	\node (x-tt-start) {$e, \color{teal}{x \mapsto 1}$};
	\node (p-trans) [right=of x-tt-start, yshift=.25cm]
		{$1 \loopseq e, {\color{purple}x \mapsto 1}$};

	\node (acc-x-tt) [right=of x-tt-start, yshift=-.4cm, draw=none]
		{$\acc, {\color{purple}~x \mapsto 1}$};

	\draw[transVia] (x-tt-start) -- node[above, draw=none] {${\color{teal}t}/{\color{MidnightBlue}p}$} (p-trans);
	\draw[transOut] (x-tt-start) -- node[below, draw=none] {$\color{teal}{\overline{t}}$} (acc-x-tt);

	\node (x-tt-big-box) [
		fit=(x-tt-start)(p-trans)(acc-x-tt),
		thick, 
		draw opacity=0.5,
		draw=gray,
		dashed,
		rounded corners,
		label={[anchor=south west]north west:{\tiny Transitions of \(e\) with valuation \(x \mapsto 1\)}}
	] {};

	\node (x-ff-start) [right=3cm of x-tt-start] {$e, {\color{teal}x \mapsto 0}$};

	\node (bot-x-ff) [right=of x-ff-start, yshift=.25cm, draw=none]
		{$\bot$};

	\node (acc-x-ff) [right=of x-ff-start, yshift=-.4cm, draw=none]
		{$\acc, {\color{purple}~x \mapsto 0}$};
	
	\draw[transOut] (x-ff-start) -- node[above, draw=none] {${\color{teal}t}$} (bot-x-ff);
	\draw[transOut] (x-ff-start) -- node[below, draw=none] {${\color{teal}\overline{t}}$} (acc-x-ff);

	\node[
		fit=(x-ff-start)(bot-x-ff)(acc-x-ff),
		thick, 
		draw opacity=0.5,
		draw=gray,
		dashed,
		rounded corners,
		label={[anchor=south west]north west:{\tiny Transitions of \(e\) with indicator state \(x \mapsto 0\)}}
	] {};
	\end{tikzpicture}%
	\hspace{2em}%
	{\scriptsize\(
	\begin{aligned}
	e \equiv{} 
	& {\color{purple}x \assign 1} \seq 
		{\color{MidnightBlue}p} \seq 
		(1 \loopseq e) 
		+_{{\color{teal}t} \land {\color{teal}x=1}} {}\\
	& {\color{purple}x \assign 1} 
		+_{{\color{teal}\overline{t}} \land {\color{teal}x=1}} {}\\
	& 0 
		+_{{\color{teal}t} \land {\color{teal}x=0}} {} \\
	& {\color{purple}x \assign 0}
		+_{{\color{teal}\overline{t}} \land {\color{teal}x=0}} 0\\[-10pt]
	\end{aligned}\)}

  \caption{
    The transitions (left) and fundamental theorem (right) of expression \(e \eqdef (p +_{x=1} x \assign 0)^{(t)}\).
    Starting indicator states and atoms are marked in {\color{teal}teal}, the actions in {\color{MidnightBlue}blue}, and ending indicator states in {\color{purple}purple}.
    When the starting indicator assignment maps \(x\) to \(0\) and the test \(t\) is satisfied, $e$ will reassign \(0\) to \(x\) (leaving $t$ intact), causing an infinite loop.
    In general, loops like this can go through an arbitrary number of reassignments, making them hard to reason about.
    Yet, as we will show in~\Cref{L11lemmaFTloop}, we can provably reduce all such loops to \(0\) using our calculus.
  }
  \label{fig:fundamental-theorem-example}
\end{figure}

\begin{toappendix}
\begin{definition}[Generalized guarded sum~\cite{gkat} \RocqLabelRef{ggdedsum}]\label{ggdedsum}
	Given \(\Phi \subseteq \At \times I^Y\) and an indexed family \(\{e_{\alpha, \pi}\}_{(\alpha, \pi) \in \Phi} \subseteq \Expm\), we define a \emph{generalized guarded sum} as
\[
		\bigplus_{(\alpha, \pi) \in \Phi} e_{(\alpha, \pi)} \eqdef
		\begin{dcases}
			0 & \text{if } \Phi = \emptyset\\
			e_{(\beta, \sigma)} +_{\beta \land Y=\sigma} \bigplus_{(\alpha, \pi) \in \Phi \setminus (\beta, \sigma)} e_{(\alpha, \pi)} & \text{if } (\beta, \sigma) \in \Phi
		\end{dcases}
	\]
	We define generalized guarded sums over \(\Phi \subseteq \At\) or \(\Phi \subseteq I^Y\) similarly.

\end{definition}
\begin{lemmarep}[{\cite[Rem.~3.6]{gkat}} \RocqLabelRef{gensumwd}]\label{gensumwd}
	Generalized guarded sums are well-defined up to \(\equiv\).
\end{lemmarep}
\end{toappendix}
\begin{appendixproof}
	This follows from the same proof as in \GKAT for all three kinds of index sets.
\end{appendixproof}

\begin{toappendix}
	\begin{remark}
		For disjoint \(I\) and \(J\) we will write \(\bigplus_{i \in I} e_i + \bigplus_{j \in J} e_j\) to denote \(\bigplus_{i \in I \cup J} g_i\) where \(g_i = e_i\) if \(i \in I\) and \(g_j = e_j\) else, so that we may reason about the \(I\) and \(J\) branches separately.
	\end{remark}
	\begin{lemma}[GKAT B.2 \RocqLabelRef{biplusboollem}]\label{biplusboollem}
		\begin{align*}
			b\seq \bigplus_{(\alpha, \pi) \in \Phi} e_{(\alpha, \pi)} \equiv \bigplus_{(\alpha, \pi) \in \Phi \cap \langle b \rangle} e_{(\alpha, \pi)}
		\end{align*}
	\end{lemma}
	\begin{proof}
		This follows from the same proof as in \GKAT, using the fact that the atoms of this Boolean algebra are pairs of atoms generated by \(T\) and complete tests (\Cref{atomscorr}).
	\end{proof}
	\begin{lemma}[\cite{gkat}]
		\begin{align}
			b\seq (e+_b f) &\equiv b \seq e & \tag{\RocqLinkedLabelText{branchsel}{U8}} \label{branchsel} \\
			b \seq (e +_c f) &\equiv b \seq e +_c b \seq f & \tag{\RocqLinkedLabelText{testdistr}{U5'}} \label{testdistr}\\
			e +_b (f +_c g) &\equiv (e+_b f) +_{b+c} g & \tag{\RocqLinkedLabelText{antiskewcom}{U3'}} \label{antiskewcom}\\
			e+_b 0 &\equiv b \seq e & \tag{\RocqLinkedLabelText{neutralright}{U6}} \label{neutralright}
		\end{align}
	\end{lemma}
\end{toappendix}

\begin{toappendix}
	\begin{lemma}[Derived Facts for Generalized Terms]
		\begin{align}
			\bigplus_{(\alpha, \pi) \in \At \times I^Y} e &\equiv e&\label{idem+}\RocqLabelRefMath{idem+}\\
			X\leftarrow \pi &\equiv  X\leftarrow \pi \seq X=\pi&\label{indval+}\RocqLabelRefMath{indval+}\\
			X= \pi &\equiv  X= \pi \seq X\leftarrow\pi&\label{indred+}\RocqLabelRefMath{indred+}\\
			X \leftarrow \pi \seq X \leftarrow \sigma &\equiv X \leftarrow \sigma&\label{indannih+}\RocqLabelRefMath{indannih+}\\
			\bigplus_{(\alpha, \pi) \in \Phi} e_{(\alpha, \pi)} &\equiv 
			\bigplus_{(\alpha, \pi) \in \Phi} (\alpha \land Y=\pi) \seq e_{(\alpha, \pi)}&\label{guardedness+}\RocqLabelRefMath{guardedness+}\\
			(X \leftarrow \pi \seq e) \loopseq f &\equiv X\leftarrow \pi \seq (e\loopseq f)&\label{loopassocassign+}\RocqLabelRefMath{loopassocassign+}\\
			b \seq \bigplus_{(\alpha, \pi) \in \Phi} e_{\alpha, \pi} &\equiv \bigplus_{(\alpha, \pi) \in \Phi} b \seq e_{\alpha, \pi} & \label{testdistr+}\RocqLabelRefMath{testdistr+}
		\end{align}
	\end{lemma}
	\begin{proof}
		We prove each individually.
	\paragraph{\Cref{idem+}}
		This is a special case of \cite[B.3]{gkat}.
	\paragraph{\Cref{indval+}}
		We prove this by induction on the size of \(X\). The base case for \(|X| = 0\) is trivial as the assignment/test are empty. Suppose it is true for all \(|X| = k\). Then for \(X' = X \cup \{x'\}\)
		\begin{align*}
			X' \leftarrow \pi &\equiv X\leftarrow \pi \seq x' \leftarrow \pi(x') & \tCref{complassign/test}\\
			&\equiv X\leftarrow \pi \seq x' \leftarrow \pi(x') \seq x' = \pi(x') & \RefAx{\Axindval}\\
			&\equiv X\leftarrow \pi \seq X = \pi \seq x' \leftarrow \pi(x') \seq x' = \pi(x')& \text{ind hyp}\\
			&\equiv X\leftarrow \pi \seq x' \leftarrow \pi(x')\seq X = \pi \seq x' = \pi(x') & \eqref{indcommtest}\\
			&\equiv X'\leftarrow \pi \seq (X' = \pi) & \RefAx{\Axtestcomb}
		\end{align*}  
		\paragraph{\Cref{indred+}}
		We prove this by induction on the size of \(X\). The base case for \(|X| = 0\) is trivial as the assignment/test are empty. Suppose it is true for all \(|X| = k\). Then for \(X' = X \cup \{x'\}\)
		\begin{align*}
			X' = \pi &\equiv X= \pi \seq x' = \pi(x') & \tCref{complassign/test}\\
			&\equiv X= \pi \seq x' = \pi(x') \seq x' \leftarrow \pi(x') & \RefAx{\Axindred}\\
			&\equiv X= \pi \seq X \leftarrow \pi \seq x' = \pi(x') \seq x' \leftarrow \pi(x') & \text{ind hyp}\\
			&\equiv X= \pi \seq x' = \pi(x')\seq X \leftarrow \pi \seq x' \leftarrow \pi(x') & \eqref{indcommtest}\\
			&\equiv X' = \pi \seq X' \leftarrow \pi & \RefAx{\Axtestcomb}
		\end{align*}  
	\paragraph{\Cref{indannih+}}
		We prove this by induction on the size of \(X\). The base case for \(|X| = 0\) is trivial as the assignments are empty. Suppose it is true for all \(|X| = k\). Then for \(X' = X \cup \{x'\}\)
		\begin{align*}
			X' \leftarrow \pi \seq X' \leftarrow \sigma &\equiv x' \assign \pi(x') \seq X \leftarrow \pi \seq X \leftarrow \sigma \seq x' \assign \sigma(x') & \RefAx{\Axindindcomm}, \RefAx{\Axassoc}\\
			&\equiv x' \assign \pi(x') \seq X \leftarrow \sigma \seq x' \assign \sigma(x') & \text{ind hyp}, \RefAx{\Axassoc}\\
			&\equiv x' \assign \pi(x') \seq x' \assign \sigma(x') \seq X \leftarrow \sigma & \RefAx{\Axindindcomm}, \RefAx{\Axassoc}\\
			&\equiv x' \assign \sigma(x') \seq X \leftarrow \sigma & \RefAx{\Axindannih}\\
			&\equiv X' \leftarrow \sigma & \eqref{complassign/test}
		\end{align*}  
	\paragraph{\Cref{guardedness+}}
		As in \cite[B.4]{gkat}.
	\paragraph{\Cref{loopassocassign+}}
		By induction on the size of \(X\). If it is \(0\), then this follows immediately from \Cref{complassign/test}. If \(X' = X \cup \{x'\}\) then
		\begin{align*}
			(X' \leftarrow \pi \seq e) \loopseq f &\equiv ((x' \assign \pi(x')\seq X \leftarrow \pi) \seq e) \loopseq f  & \tCref{complassign/test}\\
			&\equiv (x' \assign \pi(x')\seq (X \leftarrow \pi \seq e)) \loopseq f  & \RefAx{\Axassoc}\\
			&\equiv x' \assign \pi(x')\seq ((X \leftarrow \pi \seq e) \loopseq f)  & \RefAx{\Axloopassocassign}\\
			&\equiv x' \assign \pi(x')\seq (X \leftarrow \pi \seq (e \loopseq f))  & \text{ind hyp}\\
			&\equiv (x' \assign \pi(x')\seq X \leftarrow \pi) \seq (e \loopseq f)  & \RefAx{\Axassoc}\\
			&\equiv (X' \leftarrow \pi) \seq (e \loopseq f)  & \tCref{complassign/test}
		\end{align*}
		
	\paragraph{\ref{testdistr+}}
		This follows from a straightforward inductive argument using \Cref{testdistr} and \RefAx{\Axtestcomb} for the \(0\) branch.
	\end{proof}
	
	\begin{lemma}[Branch Selection+ \RocqLabelRef{branchsel+}]\label{branchsel+}
		If \((\alpha, \pi) \in \Phi\), then
		\begin{align*}
			(\alpha \land Y=\pi) \seq \bigplus_{(\beta, \sigma) \in \Phi} e_{\beta, \sigma} \equiv (\alpha \land Y=\pi) \seq e_{\alpha, \pi}
		\end{align*}
	\end{lemma}
	\begin{proof}
\begin{align*}
			\MoveEqLeft[1] (\alpha \land Y=\pi) \seq \bigplus_{(\beta, \sigma) \in \Phi} e_{\beta, \sigma} \\
			&\equiv (\alpha \land Y=\pi) \seq \bigl(e_{(\alpha, \pi)} +_{\alpha \land Y=\pi} \bigplus_{(\beta, \sigma) \in \Phi\setminus \{(\alpha, \pi)\}} e_{\beta, \sigma} \bigr) & {\tCref{ggdedsum}}\\
			&\equiv (\alpha \land Y=\pi) \seq e_{\alpha, \pi}  & \tCref{branchsel} &\qed
		\end{align*}%
	\end{proof}
\end{toappendix}

We need one more ingredient before stating the {\em fundamental theorem} which, intuitively, states that an expression can be provably decomposed into the sum of its one-step transitions in the syntactic coalgebra \((\Expm \times I^{Y}, \partial)\); see~\Cref{fig:fundamental-theorem-example} for an example.
\begin{definition}[Casting \RocqLabelRef{sys}]\label{sys}
	Define \(\sys\colon K + \Sigma \times (\Expm \times I^Y) \to \Expm\) as follows:
	\[
	\begin{array}{@{}r@{\;}c@{\;}l@{\qquad}r@{\;}c@{\;}l@{}}
	\sys(\bot)        &\eqdef& 0
	& \sys(\retc)         &\eqdef& \comReturn
	\\[0.25em]
	\sys(\acc,\pi)    &\eqdef& Y \leftarrow \pi
	& \sys(\brkc,\pi)     &\eqdef& Y \leftarrow \pi \seq \comBreak
	\\[0.25em]
	\sys(\contc,\pi)  &\eqdef& Y \leftarrow \pi \seq \comCont
	& \sys(\jmpc\,\ell,\pi)
							&\eqdef& Y \leftarrow \pi \seq \comGoto\,\ell
	\\[0.25em]
	\sys(p,(f,\pi))   &\eqdef& Y \leftarrow \pi \seq p \seq f
	\end{array}
	\]
\end{definition}
\begin{toappendix}
	\begin{lemma}[\RocqLabelRef{sysannih}] \label{sysannih}
		With \sys as defined in \Cref{sys}, for any \(Y, \pi, r\)
		\[Y \leftarrow \pi \seq \sys(r) \equiv \sys(r)\]
	\end{lemma}
	\begin{proof}
		By case analysis on \sys and \RefAx{\Axindret}, \RefAx{\Axindabort}, \Cref{indannih+}.
	\end{proof}
\end{toappendix}

\begin{theoremrep}[Fundamental Theorem \RocqLabelRef{fundthm}] \label{fundthm}
	The following holds for all \(e \in \Expm\), 
	\[e \equiv \bigplus_{(\alpha, \pi) \in \At \times I^Y} \sys(\partial(e, \pi)_\alpha)\]
\end{theoremrep}
\begin{appendixproof}
	We prove this by structural induction on \(e\). We will use the \CongRulesRef implicitly since there are no non-empty environments in this argument.
	
 	We prove by induction that \(\alpha \land Y= \pi \seq e \equiv \alpha \land Y=\pi \seq \sys(\partial(e, \pi)_\alpha)\). Then observe that the desired result follows from 
 	\begin{align*}
 		e &\equiv \bigplus_{(\alpha, \pi) \in \At \times I^Y}\alpha \land Y=\pi \seq e & \eqref{idem+}, \eqref{guardedness+}\\
 		&\equiv \bigplus_{(\alpha, \pi) \in \At \times I^Y}\alpha \land Y=\pi \seq \sys(\partial(e, \pi)_\alpha) &\\
 		&\equiv \bigplus_{(\alpha, \pi) \in \At \times I^Y}\sys(\partial(e, \pi)_\alpha) & \eqref{guardedness+}
 	\end{align*}
	\paragraph*{Base Cases}
		The base cases are all similar; we illustrate a few.
		
	\textbf{Actions}
	\begin{align*}		
		\alpha \land Y=\pi \seq p &\equiv \alpha \land Y=\pi \seq p\seq 1 & \RefAx{\Axneutright}\\
		&\equiv \alpha \land Y=\pi \seq Y\assign \pi \seq p\seq 1 & \RefAx{\Axtestcomb},\eqref{indred+}\\
		&\equiv \alpha \land Y=\pi \seq \sys(\partial(p, \pi)_\alpha)& \eqref{sys}
	\end{align*}
	\textbf{Goto}
	\begin{align*}		
		\alpha \land Y=\pi \seq \comGoto~\ell 
		&\equiv \alpha \land Y=\pi \seq Y\assign \pi \seq \comGoto~\ell & \RefAx{\Axtestcomb},\eqref{indred+}\\
		&\equiv \alpha \land Y=\pi \seq \sys(\partial(\comGoto~\ell, \pi)_\alpha)& \eqref{sys}
	\end{align*}
	\textbf{Return}
	\begin{align*}		
		&\alpha \land Y=\pi \seq \comReturn \equiv \alpha \land Y=\pi \seq\sys(\partial(\comReturn, \pi)_\alpha) 
	\end{align*}
	\textbf{Tests}
	\begin{align*}		
		\alpha \land Y=\pi \seq b &\equiv \alpha \land Y=\pi \land b & \RefAx{\Axtestcomb}\\
		&\equiv \begin{cases}\alpha \land Y=\pi & (\alpha, \pi) \in \langle b \rangle\\ 0 & (\alpha, \pi) \in \langle \lnot b \rangle \end{cases} & \tCref{boolnormleqlem}\\
		&\equiv \begin{cases}\alpha \land Y=\pi \seq Y \assign \pi & (\alpha, \pi) \in \langle b \rangle\\ 0 & (\alpha, \pi) \in \langle \lnot b \rangle \end{cases} & \RefAx{\Axtestcomb}, \eqref{indred+}\\
		&\equiv\begin{cases}\alpha \land Y=\pi \seq  Y \assign \pi & (\alpha, \pi) \in \langle b \rangle\\ \alpha \land Y=\pi \seq 0 & (\alpha, \pi) \in \langle \lnot b \rangle \end{cases} & \BA, \RefAx{\Axtestcomb}\\
		&\equiv\alpha \land Y=\pi \seq  \sys(\partial(b, \pi)_\alpha)& \eqref{sys}
	\end{align*}
	
	\paragraph*{Guarded Choice}
	W.l.o.g. suppose \((\alpha, \pi) \in \langle b \rangle\)
	\begin{align*}		
		\alpha \land Y=\pi \seq (e +_b f) &\equiv \alpha \land Y=\pi \seq e & \eqref{branchsel}\\
		&\equiv \alpha \land Y=\pi \seq \sys(\partial(e, \pi)_\alpha) & \text{ind hyp}\\
		&\equiv \alpha \land Y=\pi \seq \sys(\partial(e+_b f, \pi)_\alpha) & \tCref{operationalsemantics}
	\end{align*}	
	
	\paragraph*{Sequencing}
	
	We start with
	\begin{align*}		
		&\alpha \land Y= \pi \seq e \seq  f \equiv \alpha \land Y= \pi \seq \sys(\partial(e, \pi)_\alpha) \seq  f & \text{ind hyp}
	\end{align*}
	There are now three cases 
	\begin{align*}
		\partial(e, \pi)_\alpha &= (\acc, \sigma)\\
		\partial(e, \pi)_\alpha &= (p, (e', \sigma))\\
		\partial(e, \pi)_\alpha &\in K \setminus \{\acc\} \times I^Y
	\end{align*}
	In the first case 
\begin{align*}		
		\MoveEqLeft[1] \alpha \land Y= \pi \seq \sys(\partial(e, \pi)_\alpha) \seq  f \\
		& \equiv \alpha \land Y= \pi \seq Y\leftarrow \sigma \seq  f\\			
		& \equiv Y= \pi \seq \alpha \seq Y\leftarrow \sigma \seq  f & \RefAx{\Axtestcomb}\\			
		& \equiv Y= \pi \seq Y\leftarrow \sigma \seq \alpha \seq f & \eqref{indcommtest}\\
		& \equiv Y= \pi \seq Y\leftarrow \sigma \seq Y =\sigma \land \alpha \seq f & \eqref{indval+}, \RefAx{\Axtestcomb}\\
		& \equiv Y= \pi \seq Y\leftarrow \sigma \seq Y =\sigma \land \alpha \seq \sys(\partial(f, \sigma)_\alpha) & \text{ind hyp}\\
		& \equiv \alpha \land Y= \pi \seq Y\leftarrow \sigma \seq Y =\sigma \seq \sys(\partial(f, \sigma)_\alpha) & \RefAx{\Axtestcomb}, \eqref{indcommtest}\\
		& \equiv \alpha \land Y= \pi \seq Y\leftarrow \sigma \seq \sys(\partial(f, \sigma)_\alpha) & \eqref{indval+}\\
		& \equiv \alpha \land Y= \pi \seq Y\leftarrow \sigma \seq \sys(\partial(e\seq f, \pi)_\alpha) & \tCref{operationalsemantics}\\
		& \equiv \alpha \land Y= \pi \seq \sys(\partial(e\seq f, \pi)_\alpha) & \eqref{sysannih}
	\end{align*}%
	In the second case
	\begin{align*}		
		\alpha \land Y= \pi \seq \sys(\partial(e, \pi)_\alpha) \seq  f		
		&\equiv\alpha \land Y= \pi \seq Y\leftarrow \sigma \seq p \seq e' \seq f & \text{ind hyp}\\		
		&\equiv \alpha \land Y= \pi \seq \sys(\partial(e\seq f, \pi)_\alpha) & \tCref{operationalsemantics}, \eqref{sys}
	\end{align*}
	In the third
	\begin{align*}		
		\alpha \land Y= \pi \seq \sys(\partial(e, \pi)_\alpha) \seq  f
		&\equiv \alpha \land Y= \pi \seq \sys(\partial(e, \pi)_\alpha) & \RefAx{\Axrightseqannret}/ \RefAx{\Axrightseqanncont}/\RefAx{\Axrightseqannbrk}/\RefAx{\Axrightseqanngoto}, \RefAx{\Axabsleft}, \eqref{sys}\\		
		&\equiv \alpha \land Y= \pi \seq \sys(\partial(e\seq f, \pi)_\alpha) & \tCref{operationalsemantics}
	\end{align*}
	
	\paragraph*{Loop sequencing}
	
	We start with
	\begin{align*}		
		\alpha \land Y= \pi \seq (e \loopseq f) &\equiv (\alpha \land Y= \pi \seq e) \loopseq f & \eqref{loopassoctest}\\
		&\equiv \alpha \land Y= \pi \seq (\sys(\partial(e, \pi)_\alpha) \loopseq  f) & \text{ind hyp}, \eqref{loopassoctest}
	\end{align*}
	There are four cases 
	\begin{align*}
		\partial(e, \pi)_\alpha &\in \{\bot, \retc\}+\{\jmpc\}\times I^Y\\
		\partial(e, \pi)_\alpha &= (\brkc, \sigma)\\
		\partial(e, \pi)_\alpha &= (c, \sigma)\quad c \in \{\acc, \contc\}\\
		\partial(e, \pi)_\alpha &= (p, (e', \sigma))
	\end{align*}
	In the first case
	\begin{align*}		
		\alpha \land Y= \pi \seq (\sys(\partial(e, \pi)_\alpha) \loopseq  f )
		&\equiv \alpha \land Y= \pi \seq \sys(\partial(e, \pi)_\alpha)& \RefAx{\Axrightloopannabort}/\RefAx{\Axrightloopannret}/\RefAx{\Axrightloopanngoto}, \eqref{sys}\\		
		&\equiv \alpha \land Y= \pi \seq \sys(\partial(e\loopseq f, \pi)_\alpha) & \tCref{operationalsemantics}
	\end{align*}
	In the second case
	\begin{align*}		
		\alpha \land Y= \pi \seq \sys(\partial(e, \pi)_\alpha) \loopseq  f	
		&\equiv \alpha \land Y= \pi \seq (Y\leftarrow \pi\seq \comBreak) \loopseq f& \eqref{sys}\\		
		&\equiv \alpha \land Y= \pi \seq Y\leftarrow \pi\seq (\comBreak \loopseq f)& \eqref{loopassocassign+}\\		
		&\equiv \alpha \land Y= \pi \seq Y\leftarrow \pi\seq 1 & \RefAx{\Axbreak}\\		
		&\equiv \alpha \land Y= \pi \seq Y\leftarrow \pi & \RefAx{\Axneutright}\\		
		&\equiv \alpha \land Y= \pi \seq \sys(\partial(e\loopseq f)_\alpha) &\tCref{operationalsemantics}, \eqref{sys}
	\end{align*}
	In the third case
	\begin{align*}		
		\alpha \land Y= \pi &\seq \sys(\partial(e, \pi)_\alpha) \loopseq  f\\
		&\equiv\alpha \land Y= \pi \seq \begin{cases}Y \leftarrow \sigma\loopseq f &c = \acc \\ (Y \leftarrow \sigma \seq \comCont )\loopseq f&c = \contc \end{cases}& \tCref{operationalsemantics}, \eqref{sys}
	\end{align*}
	In the first case
	\begin{align*}
		Y \leftarrow \sigma\loopseq f & \equiv Y \leftarrow \sigma \seq f & \eqref{loopseqisseqasasign}
	\end{align*}
	In the second \((Y \leftarrow \sigma \seq \comCont )\loopseq f \equiv Y \leftarrow \sigma \seq f\) by \ref{loopassocassign+} and \RefAx{\Axcontinue}. We continue
	\begin{align*}
		&\alpha \land Y= \pi \seq \begin{cases}Y \leftarrow \sigma\loopseq f &c = \acc \\ (Y \leftarrow \sigma \seq \comCont )\loopseq f&c = \contc \end{cases}\\	
		&\equiv \alpha \land Y= \pi \seq Y \leftarrow \sigma\seq f\\
		&\equiv \alpha \land Y= \pi \seq Y \leftarrow \sigma\seq Y=\sigma \seq f& \eqref{indval+}\\
		&\equiv Y= \pi \seq Y \leftarrow \sigma\seq \alpha \land Y=\sigma \seq f& \RefAx{\Axtestcomb}, \eqref{indcommtest}\\
		&\equiv \alpha \land Y= \pi \seq Y \leftarrow \sigma \seq \alpha \land Y=\sigma\seq \sys(\partial(f, \sigma)_\alpha)& \text{ind hyp}\\
		&\equiv \alpha \land Y= \pi \seq Y \leftarrow \sigma \seq Y=\sigma\seq \sys(\partial(f, \sigma)_\alpha)& \RefAx{\Axtestcomb}, \eqref{indcommtest}\\
		&\equiv \alpha \land Y= \pi \seq\sys(\partial(f, \sigma)_\alpha)& \eqref{indval+},\eqref{sysannih}\\
		&\equiv \alpha \land Y= \pi \seq\sys(\partial(e\loopseq f, \pi)_\alpha)& \tCref{operationalsemantics}
	\end{align*}
	Finally, in the fourth case
	\begin{align*}		
		\alpha \land Y= \pi \seq \sys(\partial(e, \pi)_\alpha) \loopseq  f
		&\equiv \alpha \land Y= \pi \seq (Y\leftarrow \sigma\seq (p\seq e')) \loopseq f & \eqref{sys}\\		
		&\equiv\alpha \land Y= \pi \seq Y\leftarrow \sigma\seq ((p\seq e') \loopseq f) & \eqref{loopassoctest}, \eqref{loopassocassign+}\\	
		&\equiv \alpha \land Y= \pi \seq Y\leftarrow \sigma\seq (p\seq (e' \loopseq f))   & \RefAx{\Axloopassocact}\\
		&\equiv \alpha \land Y= \pi \seq \sys(\partial(e\loopseq f)_\alpha)	& \tCref{operationalsemantics}, \eqref{sys}	
	\end{align*}
	
	\paragraph{Loops} 
	First we apply \Cref{L11lemmaFTloop} using the inductive hypothesis
	\begin{align*}
		\alpha \land Y=\pi \seq e^{(b)}
		&\equiv \alpha \land Y=\pi \seq (\lnot C^+\seq e)^{(b)} & \eqref{L11lemmaFTloop}\\
		&\equiv \alpha \land Y=\pi \seq((\lnot C^+\seq e) \loopseq (\lnot C^+\seq e)^{(b)}+_b 1) & \RefAx{\Axloopunfold}
	\end{align*}
	We now do a case analysis on \(\alpha, \pi\).  
	
	If \(\alpha \land Y=\pi \leqBA \lnot b\) then
	\begin{align*}
		\alpha \land Y=\pi \seq((\lnot C^+\seq e) \loopseq (\lnot C^+\seq e)^{(b)}+_b 1)
		&\equiv (\alpha\land Y=\pi)\seq 1& \eqref{branchsel}\\
		&\equiv \alpha\land Y=\pi& \RefAx{\Axneutright}\\
		&\equiv \alpha\land Y=\pi \seq Y \assign \pi& \RefAx{\Axtestcomb}, \eqref{indred+}\\
		&\equiv \sys(\partial(e^{(b)}, \pi)_\alpha) & \tCref{operationalsemantics}
	\end{align*}
	If \(\alpha \land Y=\pi \leqBA b\) then we simplify before further case analysis
\begin{align*}
		\MoveEqLeft[1] \alpha \land Y=\pi \seq((\lnot C^+\seq e) \loopseq (\lnot C^+\seq e)^{(b)}+_b 1) \\
		&\equiv \alpha\land Y=\pi\seq (\lnot C^+\seq e \loopseq (\lnot C^+\seq e)^{(b)})& \eqref{branchsel}\\
		&\equiv (\alpha \land Y=\pi \land \lnot C^+\seq e) \loopseq (\lnot C^+\seq e)^{(b)}& \eqref{loopassoctest}, \RefAx{\Axtestcomb}
	\end{align*}%
	If also \(\alpha \land Y=\pi \leqBA C^+\), then we continue with
	\begin{align*}
		(\alpha \land Y=\pi \land \lnot C^+\seq e) \loopseq (\lnot C^+\seq e)^{(b)}
		&\equiv 0 \loopseq (\lnot C^+\seq e)^{(b)} &\BA, \RefAx{\Axabsleft}\\
		&\equiv  0 & \RefAx{\Axrightloopannabort}\\
		&\equiv  \alpha \land Y=\pi \seq 0 & \BA, \RefAx{\Axtestcomb}\\
		&\equiv  \alpha \land Y=\pi \seq \sys(\partial(e^{(b)}, \pi)_\alpha) & \tCref{operationalsemantics}
	\end{align*}
	where we know that \(\partial(e^{(b)}, \pi)_\alpha = \bot\) because the property of being an invariant of \(e\) prevents a derivation of a transition for the loop \(e^{(b)}\).
	
	Finally, if  \(\alpha \land Y=\pi \leqBA \lnot C^+\) we simplify first, and then do a further case analysis on the transitions of \(e\)
\begin{align*}
		\MoveEqLeft[1] (\alpha \land Y=\pi \land \lnot C^+\seq e) \loopseq (\lnot C^+\seq e)^{(b)} \\
		&\equiv(\alpha \land Y=\pi \seq e) \loopseq (\lnot C^+\seq e)^{(b)}& \BA\\
		&\equiv(\alpha \land Y=\pi \seq \sys(\partial(e, \pi)_\alpha)) \loopseq (\lnot C^+\seq e)^{(b)}& \text{ind hyp}\\
		&\equiv\alpha \land Y=\pi \seq (\sys(\partial(e, \pi)_\alpha) \loopseq (\lnot C^+\seq e)^{(b)})& \eqref{loopassoctest}
	\end{align*}%
	Suppose \(\partial(e, \pi)_\alpha \in \{\bot, \retc\}+\{\jmpc\}\times I^Y\), then
\begin{align*}
		\MoveEqLeft[1] \alpha \land Y=\pi \seq (\sys(\partial(e, \pi)_\alpha) \loopseq (\lnot C^+\seq e)^{(b)}) \\
		&\equiv \alpha \land Y=\pi \seq (\sys(\partial(e, \pi)_\alpha)) & \RefAx{\Axrightloopannabort}/\RefAx{\Axrightloopanngoto}/\RefAx{\Axrightloopannret}, \eqref{sys}\\
		&\equiv \sys(\partial(e^{(b)}, \pi)_\alpha) & \tCref{operationalsemantics}, \eqref{sys}
	\end{align*}%
	Suppose \(\partial(e, \pi)_\alpha = (\brkc, \sigma)\), then
	\begin{align*}
		&\alpha \land Y=\pi \seq (\sys(\partial(e, \pi)_\alpha) \loopseq (\lnot C^+\seq e)^{(b)})\\
		&\equiv \alpha \land Y=\pi \seq ((Y\leftarrow \sigma \seq \comBreak) \loopseq (\lnot C^+\seq e)^{(b)}) & \eqref{sys}\\
		&\equiv \alpha \land Y=\pi \seq (Y\leftarrow \sigma \seq (\comBreak \loopseq (\lnot C^+\seq e)^{(b)})) & \eqref{loopassocassign+}\\
		&\equiv \alpha \land Y=\pi \seq (Y\leftarrow \sigma \seq 1) & \RefAx{\Axbreak}\\
		&\equiv \alpha \land Y=\pi \seq Y\leftarrow \sigma & \RefAx{\Axneutright}\\
		&\equiv \alpha \land Y=\pi \seq \sys(\partial(e^{(b)}, \pi)_\alpha) &\tCref{operationalsemantics}, \eqref{sys}
	\end{align*}
	Suppose \(\partial(e, \pi)_\alpha = (p, (e', \sigma))\), then
	\begin{align*}
		&\alpha \land Y=\pi \seq (\sys(\partial(e, \pi)_\alpha) \loopseq (\lnot C^+\seq e)^{(b)})\\
		&\equiv \alpha \land Y=\pi \seq ((Y\leftarrow \sigma \seq (p \seq e')) \loopseq (\lnot C^+\seq e)^{(b)}) & \eqref{sys}\\
		&\equiv \alpha \land Y=\pi \seq (Y\leftarrow \sigma \seq (p \seq (e' \loopseq (\lnot C^+\seq e)^{(b)}))) & \eqref{loopassocassign+}, \RefAx{\Axloopassocact}\\
		&\equiv \alpha \land Y=\pi \seq (Y\leftarrow \sigma \seq (p \seq (e' \loopseq e^{(b)}))) & \eqref{L11lemmaFTloop}\\
		&\equiv \sys(\partial(e^{(b)}, \pi)_\alpha) & \tCref{operationalsemantics}, \eqref{sys}
	\end{align*}
	Finally, suppose \(\partial(e, \pi)_\alpha \in \{\acc, \contc\}\times I^Y\), then
	\begin{align*}
		&\alpha \land Y=\pi \seq (\sys(\partial(e, \pi)_\alpha) \loopseq (\lnot C^+\seq e)^{(b)})\\
		&\equiv \alpha \land Y=\pi\seq \begin{cases} Y \leftarrow \sigma \loopseq (\lnot C^+\seq e)^{(b)} & c = \acc\\ (Y \leftarrow \sigma \seq \comCont) \loopseq (\lnot C^+\seq e)^{(b)} & c = \contc\end{cases} & \eqref{sys}
	\end{align*}
	These two cases follow from the fact that \(C^+\) is an invariant. We simplify this to \[\alpha \land Y=\pi \seq Y \leftarrow \sigma \seq (\lnot C^+\seq e)^{(b)}\] by the same argument as previously using \RefAx{\Axcontinue}/\ref{loopassocassign+} for the \contc case, and \eqref{loopseqisseqasasign} for the \acc case.
	
	We continue
	\begin{align*}
		&\alpha \land Y=\pi \seq Y \leftarrow \sigma \seq (\lnot C^+\seq e)^{(b)}\\
		&\equiv \alpha \land Y=\pi \seq Y \leftarrow \sigma \seq Y=\sigma \seq (\lnot C^+\seq e)^{(b)} & \eqref{indred+}\\
		&\equiv Y=\pi \seq Y \leftarrow \sigma \seq (\alpha \land Y=\sigma \seq (\lnot C^+\seq e)^{(b)}) & \RefAx{\Axtestcomb}, \eqref{indcommtest}
	\end{align*}

	Observe that this inner term is the same as the initial term, with \(\sigma\) taking the place of \(\pi\). If \(\partial(e, \sigma)_\alpha \notin \{\acc, \contc\} \times I^Y\) or \(\alpha \land Y=\sigma \leqBA \lnot b\), then by the previous cases we would see that 
	\begin{align*}
		&Y=\pi \seq Y \leftarrow \sigma \seq (\alpha \land Y=\sigma \seq (\lnot C^+\seq e)^{(b)})\\
		&\equiv Y=\pi \seq Y \leftarrow \sigma \seq (\alpha \land Y=\sigma \seq \sys(\partial(e^{(b)}, \sigma))_\alpha) & \text{supposition}\\
		&\equiv \alpha \land Y=\pi \seq Y \leftarrow \sigma \seq Y=\sigma \seq \sys(\partial(e^{(b)}, \sigma))_\alpha & \RefAx{\Axtestcomb}, \eqref{indcommtest}\\
		&\equiv \alpha \land Y=\pi \seq Y \leftarrow \sigma \seq \sys(\partial(e^{(b)}, \sigma))_\alpha & \eqref{indval+}\\
		&\equiv \alpha \land Y=\pi \seq \sys(\partial(e^{(b)}, \sigma))_\alpha & \eqref{sysannih}\\
		&\equiv \alpha \land Y=\pi \seq \sys(\partial(e^{(b)}, \pi))_\alpha & \tCref{operationalsemantics}
	\end{align*}
	Otherwise it is \((c, \sigma_1)\) for \(c \in \{\acc, \contc\}\) and some \(\sigma_1\). We make the same argument for \((c, \sigma_1) \in \{\acc, \contc\} \times I^Y\), to see
	\begin{align*}
		&Y=\pi \seq Y \leftarrow \sigma \seq \alpha \land Y=\sigma \seq (\lnot C^+\seq e)^{(b)}\\
		&\equiv Y=\pi \seq Y \leftarrow \sigma \seq Y=\sigma \seq Y \leftarrow \sigma_1 \seq \alpha \land Y=\sigma_1 \seq (\lnot C^+\seq e)^{(b)}\\
		&\equiv \alpha \land Y=\pi \seq Y \leftarrow \sigma \seq Y=\sigma \seq Y \leftarrow \sigma_1 \seq Y=\sigma_1 \seq (\lnot C^+\seq e)^{(b)} & \RefAx{\Axtestcomb}, \BA, \eqref{indcommtest}\\
		&\equiv \alpha \land Y=\pi \seq Y \leftarrow \sigma \seq Y \leftarrow \sigma_1 \seq (\lnot C^+\seq e)^{(b)} & \eqref{indval+}\\
		&\equiv \alpha \land Y=\pi \seq Y \leftarrow \sigma_1 \seq (\lnot C^+\seq e)^{(b)} & \eqref{indannih+}\\
		&\equiv Y=\pi \seq Y \leftarrow \sigma_1 \seq \alpha \land Y=\sigma_1 \seq (\lnot C^+\seq e)^{(b)} & \eqref{indval+}, \RefAx{\Axtestcomb}, \eqref{indcommtest}
	\end{align*}
	Observe that this is precisely the term we started with, except \(\sigma_1\) replaces \(\sigma\).
	 
	We repeat this argument for arbitrarily many \(\sigma_i\). If there were a \(\sigma_n\) such that \(\partial(e, \sigma_n)_\alpha = (c, \sigma_{j}) \in \{\acc, \contc\} \times I^Y\) for \(j<n\), then \(C \eqdef (\alpha \land Y=\pi) \lor (\alpha \land Y=\sigma) \lor  \bigvee_{1 \leq i \leq n}\alpha \land Y=\sigma_i\) is an invariant of \(e\) such that \(\alpha \land Y=\pi \leqBA C \leqBA C^+\). But we assumed that \(\alpha \land Y=\pi \leqBA \lnot C^+\). Therefore each \(\sigma_i\) must be distinct. Since \(I^Y\) is finite, eventually there is some \(\sigma_n\) such that \(\partial(e, \sigma_n)_\alpha \notin \{\acc, \contc\} \times I^Y\) or \(\alpha \land Y=\sigma_n \leqBA \lnot b\) and hence
	\begin{align*}
		&Y=\pi \seq Y \leftarrow \sigma \seq \alpha \land Y=\sigma \seq (\lnot C^+\seq e)^{(b)}\\
		&\vdots\\ 
		&\equiv Y=\pi \seq Y \leftarrow \sigma_n \seq \alpha \land Y=\sigma_n \seq (\lnot C^+\seq e)^{(b)}\\ 
		&\equiv \alpha \land Y=\pi \seq \sys(\partial(e^{(b)}, \sigma_n)_\alpha)&\\
		&\equiv \alpha \land Y=\pi \seq \sys(\partial(e^{(b)}, \pi)_\alpha)& \tCref{operationalsemantics}
	\end{align*}
	Having examined each case, we obtain
	\[\alpha \land Y=\pi \seq e^{(b)} \equiv \alpha \land Y=\pi \seq \sys(\partial(e^{(b)}, \pi)_\alpha) 
	\qed\]
\end{appendixproof}

Generalized sums $\bigplus$ are formally defined in the appendix. The proof proceeds by structural induction on \(e\) and, perhaps unsurprisingly, the most interesting inductive case is the loop case.
As noted before, loops may contain unproductive cycles with an arbitrary number of reassignments; a simple example can be found in the \(t \land x = 0\) branch in~\Cref{fig:fundamental-theorem-example}.
To deal with this, the proof of \Cref{fundthm} uses \Cref{L11lemmaFTloop}, which allows to remove unproductive cycles from a loop, and thereby match the behavior of the transition function when a loop diverges.
Its statement relies on the notion of \emph{invariants}.%

\begin{definition}[Invariant \RocqLabelRef{acc-cont-loop-invariant}]\label{acc-cont-loop-invariant}
	We call \(C \in \BExp\) an \emph{invariant} of \(e\) if for all \((\alpha, \pi) \in \langle C \rangle\), \(\partial(e, \pi)_\alpha = (c, \sigma)\) where \(c \in \{\acc, \contc\}\) and \((\alpha, \sigma) \in \langle C \rangle\).

\end{definition}

\begin{lemmarep}[\RocqLabelRef{L11lemmaFTloop}]\label{L11lemmaFTloop}
	Let \(e\) be such that for all $\alpha \in \At$ and $\pi \in I^Y$, we have \(\alpha \land (Y=\pi) \seq e \equiv \alpha \land (Y=\pi) \seq \sys(\partial(e, \pi)_\alpha)\).
    Then \( e^{(b)} \equiv (\lnot C^+ \seq e)^{(b)} \) where \(C^+\) is the largest invariant of \(e\) (w.r.t. Boolean implication, $\leqBA$) such that \(C^+ \leqBA b\).
\end{lemmarep}
\begin{proofsketch}
	We first show that the maximum invariant \(C^{+}\) exists and is well-defined up to equivalence.
	Our test fragment has a join operation defined as the disjunction \(\lor\), and invariants are closed under join.
	There can be only finitely many invariants because the free Boolean algebra over primitive and indicator tests is finite, and our fragment is a refinement of that.
    Thus, \(C^{+}\) can be defined as the disjunction of all invariants.

	We then remove the maximum invariant \(C^+ \leqBA b\) from the loop \(e^{(b)}\).
	We use \(\RefAx{\Axlooptighten}\) to reduce the goal to \((C^+\land b \seq e) \loopseq (\lnot C^+\land b) \equiv 0\).
	Because \(C^+ \land b \eqBA C^+\), the goal is \((C^+ \seq e) \loopseq \lnot C^+ \equiv 0\).
	By \RefAx{\Axidempotence} and induction, this term is equivalent to a guarded sum of copies of itself, one for each atom and complete test; furthermore each copy may be prefixed with its own guard:
	\[(C^+ \seq e) \loopseq \lnot C^+ 
	\equiv 
	\bigplus_{(\alpha, \pi) \in \At \times I^Y} \alpha \land (Y=\pi) \seq ((C^+ \seq e) \loopseq \lnot C^+)\]
	We show that each term in the guarded sum on the right is equivalent to \(0\). 
	Since \(\alpha \land (Y=\pi)\) is an atom, we can do a case analysis on whether \(\alpha \land (Y=\pi) \leqBA C^+\) or \(\alpha \land (Y=\pi) \leqBA \lnot C^+\).
    In the second case \(\alpha \land (Y=\pi) \seq C^+\) immediately produces \(0\).
	In the first case, since \(C^+\) is an invariant, \(\partial(e, \pi)_\alpha = (c, \sigma)\) where \(c \in \{\acc, \contc\}\) and \(\alpha \land (Y= \sigma) \leqBA C^+\).
	By definition, \(\sys(\partial(e, \pi)_\alpha)\) can only be \(Y\leftarrow \sigma\) or \(Y\leftarrow \sigma \seq \comCont\). 
	By the premise, we can decompose \(\alpha \land (Y=\pi) \seq e\) into \(\alpha \land (Y=\pi) \seq ((C^+ \seq \sys(\partial(e, \pi)_\alpha)) \loopseq \lnot C^+)\), which simplifies to \(\alpha \land (Y=\pi) \seq C^+ \seq Y \leftarrow \sigma \seq \lnot C^+\).
	From here we commute \(\alpha\) and use \RefAx{\Axindval} to modify the rightmost test to \((Y=\sigma)\land \alpha \land \lnot C^+\), which is \(0\) because \((Y=\sigma)\land \alpha \leqBA C^+\). 
	This annihilates the rest of the term.
\end{proofsketch}
\begin{appendixproof}
	Let \(C^+\) be the disjunction of all invariants \(C\) of \(e\) such that \(C \leqBA b\), as there are finitely many of them up to equivalence. \(C^+\) is also an invariant because for each \((\alpha, \pi) \in \langle C^+ \rangle\), \((\alpha, \pi) \in \langle C_k \rangle\) for some \(C_k\) in the disjunction. By the definition of invariant \(\partial(e, \alpha)_\pi = (c, \sigma)\) for \(c \in \{\acc, \contc\}\) with \((\alpha, \pi) \in \langle C_k \rangle \subseteq \langle C^+ \rangle\).
	
	We use \RefAx{\Axlooptighten} to remove this invariant. In order to do so, we now show that this largest invariant, \(C^+\), satisfies the premise of \RefAx{\Axlooptighten}.
	
	It is sufficient to show that  for all \((\alpha, \pi)\)
	\[\alpha \land Y=\sigma \seq ((C^+ \land b) \seq e \loopseq \lnot(C^+ \land b)) \equiv 0\]
	because from this we can derive
	\begin{align*}
		&(C^+ \land b) \seq e \loopseq \lnot(C^+ \land b)\\
		&\equiv \bigplus_{(\alpha, \pi) \in \At \times I^Y} (C^+ \land b) \seq e \loopseq \lnot(C^+ \land b) & \eqref{idem+}\\
		&\equiv \bigplus_{(\alpha, \pi) \in \At \times I^Y} \alpha \land Y=\pi \seq (C^+ \land b) \seq e \loopseq \lnot(C^+ \land b) & \eqref{guardedness+}\\
		&\equiv \bigplus_{(\alpha, \pi) \in \At \times I^Y} 0 & \text{supposition}\\
		&\equiv 0 & \eqref{idem+}
	\end{align*}
	Furthermore, since \(C^+ \land b \equiv C^+\), it is sufficient to show
	\[\alpha \land Y=\sigma \seq (C^+ \seq e \loopseq \lnot C^+) \equiv 0\]
	We proceed by a case analysis on \((\alpha, \pi)\) to show the required fact.
	If \(\alpha \land Y=\pi \leqBA \lnot C^+\), then
	\begin{align*}
		\alpha \land Y=\sigma \seq (C^+ \seq e \loopseq \lnot C^+)
		&\equiv (\alpha \land Y=\sigma \land C^+) \seq (e \loopseq \lnot C^+) & \RefAx{\Axassoc}, \RefAx{\Axtestcomb}\\
		&\equiv 0 \seq (e \loopseq \lnot C^+) & \BA\\
		&\equiv 0 & \RefAx{\Axabsleft}
	\end{align*}
	If \(\alpha \land Y=\pi \leqBA C^+\), then
	\begin{align*}
		&\alpha \land Y=\sigma \seq (C^+ \seq e \loopseq \lnot C^+)\\
		&\equiv (\alpha \land Y=\pi \seq C^+) \seq (e \loopseq \lnot C^+) & \RefAx{\Axtestcomb}\\ 
		&\equiv \alpha \land Y=\pi \seq (e \loopseq \lnot C^+) & \BA\\ 
		&\equiv (\alpha \land Y=\pi \seq e) \loopseq \lnot C^+ & \eqref{loopassoctest}\\ 
		&\equiv (\alpha \land Y=\pi \seq \sys(\partial(e, \pi)_\alpha)) \loopseq \lnot C^+ & \text{premise}\\
		&\equiv \alpha \land Y=\pi \seq (\sys(\partial(e, \pi)_\alpha) \loopseq \lnot C^+) & \eqref{loopassoctest}\\ 
		&\equiv \alpha \land Y=\pi \seq \begin{cases}Y \assign \sigma\loopseq \lnot C^+  & \partial(e, \pi)_\alpha = (\acc, \sigma)\\Y \assign \sigma \seq \comCont \loopseq \lnot C^+  & \partial(e, \pi)_\alpha = (\contc, \sigma) \end{cases} &\eqref{sys}&
	\end{align*}
	where the last step follows from the fact that \((\alpha, \pi) \in \langle C^+ \rangle\) and \(C^+\) is an invariant of \(e\).
	
	In the first case
	\begin{align*}
		\alpha \land Y=\pi \seq (Y \assign \sigma \loopseq \lnot C^+)
		&\equiv \alpha \land Y=\pi \seq Y \assign \sigma \seq \lnot C^+& \eqref{loopseqisseqasasign}\\
		&\equiv \alpha \land Y=\pi \seq Y \assign \sigma \seq Y = \sigma \seq \lnot C^+& \eqref{indred+}\\
		&\equiv Y=\pi \seq Y \assign \sigma \seq \alpha \land Y = \sigma \seq \lnot C^+& \RefAx{\Axtestcomb}, \eqref{indcommtest}\\
		&\equiv Y=\pi \seq Y \assign \sigma \seq 0& \RefAx{\Axtestcomb}, \BA\\
		&\equiv 0& \RefAx{\Axindabort}, \RefAx{\Axtestcomb}, \BA
	\end{align*}
	In the second case
	\begin{align*}
		&\alpha \land Y=\pi \seq ((Y \assign \sigma \seq \comCont) \loopseq \lnot C^+) \\
		&\equiv \alpha \land Y=\pi \seq (Y \assign \sigma \seq (\comCont \loopseq \lnot C^+))& \eqref{loopassocassign+} \\
		&\equiv \alpha \land Y=\pi \seq Y \assign \sigma \seq \lnot C^+ & \eqref{loopassoctest}, \RefAx{\Axcontinue}\\
		&\equiv 0 & \text{as case 1}
	\end{align*}
	Hence, we apply \RefAx{\Axlooptighten} to obtain \(e^{(b)} \equiv (\lnot C^+\seq e)^{(b)}\).
\end{appendixproof}

\subsection{A Contextual Normal Form}\label{sec:contextual-normal-form}
	We now construct a contextual normal form, which is the core piece of the completeness result, and it enables us to show that all behaviors have a canonical environment which captures their behavior.
\begin{toappendix}
	\begin{lemma}[\RocqLabelRef{ggdedsumprodinj}] \label{ggdedsumprodinj}
		\begin{align*}
			\bigplus_{\alpha \in \Phi\subseteq \At} e_\alpha &\equiv \bigplus_{(\alpha, \pi) \in \Phi \times I^Y} e_\alpha\\
			\bigplus_{\pi \in \Phi\subseteq I^Y} e_\pi &\equiv \bigplus_{(\alpha, \pi) \in \At \times \Phi} e_\pi
		\end{align*}
	\end{lemma}
	\begin{proof}
		We prove the first fact, and note that the second follows by a symmetric argument. We prove this by induction on the size of \(\Phi\).
		
		Suppose \(\Phi = \emptyset\), then the desired result follows from unrolling \Cref{ggdedsum}.
		
		Suppose \(\Phi = \Phi' \cup \{\alpha\}\), then 
		\begin{align*}
			&\bigplus_{\alpha \in \Phi} e_\alpha \equiv e_\alpha +_\alpha \bigplus_{\beta \in \Phi'} e_\beta &\eqref{ggdedsum}\\
			&\equiv e_\alpha +_\alpha \bigplus_{(\beta, \sigma) \in \Phi' \times I^Y} e_\beta & \text{ind hyp}\\
			&\equiv \bigplus_{(\beta, \pi) \in \{\alpha\} \times I^Y} e_\beta +_\alpha \bigplus_{(\beta, \sigma) \in \Phi' \times I^Y} e_\beta & \eqref{idem+}, \RefAx{\Axguardedness}, \eqref{biplusboollem}\\
			&\equiv \alpha \seq \bigplus_{(\beta, \pi) \in \At \times I^Y} e_\beta +_\alpha \lnot \alpha \seq \bigplus_{(\beta, \sigma) \in \Phi \times I^Y} e_\beta & \eqref{biplusboollem}\\
			&\equiv \bigplus_{(\beta, \pi) \in \Phi \times I^Y} e_\beta +_\alpha \bigplus_{(\beta, \sigma) \in \Phi \times I^Y} e_\beta & \RefAx{\Axguardedness}, \RefAx{\Axskewcomm}\\
			&\equiv \bigplus_{(\beta, \pi) \in \Phi \times I^Y} e_\alpha & \RefAx{\Axidempotence}
		\end{align*} 
		
		The second fact follows by an analogous argument.
	\end{proof}
	
	This next lemma looks almost immediate from \Cref{resolutiondef}; however, it is subtly different. In the case that \(r \neq \bot\) it is immediate from \Cref{resolutiondef}. However, if \(r = \bot\) then what we are saying is that stepping to a resolved term which aborts means that had we resolved the current term it would abort as well. That is, silently transitioning to a state which has no derivation means that the starting state also has no derivation.
	\begin{lemma}[\RocqLabelRef{iteratedloweringlem}]\label{iteratedloweringlem}
		If \(\delta(s)_\alpha = (\jmpc~\ell, \pi)\) and \(\delta\resWith{\gamma}(\gamma(\ell, \pi))_\alpha = r\), then \(\delta\resWith{\gamma}(s)_\alpha = r\).

	\end{lemma}
	\begin{proof}
		There are two cases. If \(\delta\resWith{\gamma}(\gamma(\ell, \pi))_\alpha = r\) was derived from finitely many applications of the cases in \Cref{resolutiondef}, then by the same derivation \(\delta\resWith{\gamma}(s)_\alpha = r\).
		
		Otherwise, there was an infinite sequence of labels and assignments \(\ell_i, \sigma_i\) where \(\ell_1 = \ell, \sigma_1 = \pi\) such that \(\delta\gamma(\ell_i, \sigma_i) = (\jmpc~\ell_{i+1}, \sigma_{i+1})\). Since \(\delta(s)_\alpha = (\jmpc~\ell, \pi)\), it also has the same infinite sequence. Hence \(\delta\resWith{\gamma}(s)_\alpha = \bot = \delta\resWith{\gamma}(\gamma(\ell, \pi))_\alpha\).
	\end{proof}
	\begin{lemma}[\RocqLabelRef{exprsubcofinite}] \label{exprsubcofinite}
		For any \(e \in \Expm\), let \(D(e)\) be the least set containing \(e\) such that
		\(g \in D(e)\) whenever \(f \in D(e)\) and \(\partial(f,\pi)_\alpha = (p,(g,\sigma))\)
		for some \(\pi, \alpha, p, \sigma\). Then \(D(e)\) is finite.

	\end{lemma}
	\begin{proof}
		By structural induction on \(e\). For a primitive action \(p\), \(D(p)=\{p,1\}\);
		for every other base expression, \(D(e)=\{e\}\).
		For guarded choice, we have
		\[
		D(f+_b g) \subseteq \{f+_b g\} \cup D(f) \cup D(g).
		\]
		For either composition operator \(\square \in \{\seq,\loopseq\}\), we have
		\[
		D(f\square g) \subseteq \{f\square g\} \cup D(g) \cup \{h\square g \mid h\in D(f)\}.
		\]
		Finally, for loops we have
		\[
		D(f^{(b)}) \subseteq \{f^{(b)}\} \cup \{h\loopseq f^{(b)} \mid h\in D(f)\}.
		\]
		The transition rules show that each bounding set contains its initial expression and is
		closed under action successors for arbitrary valuations, using induction on a finite
		transition derivation in the loop case. Each bound is finite by the inductive hypothesis,
		so the least such set \(D(e)\) is finite.
	\end{proof}
	
	\begin{lemma}[\RocqLabelRef{gtnflem1}]\label{gtnflem1}
		For any \(E,e\)
		\begin{align*}
			&E \vdash e \equiv e \dashv E'
		\end{align*}
		where 
		\begin{align*}
			E'(k) = \bigplus_{(\alpha, \pi) \in \At \times I^Y} \sys(\partial\resWith{E}(E(k), \pi)_\alpha)
		\end{align*}
	\end{lemma}
\begin{proof}
	We will use \RefAx{\Axcongguardedchoice} and \RefAx{\Axcongrightseq} implicitly throughout this argument. Furthermore by \RefAx{\Axcongleftseqempty} and \RefAx{\Axweaken} we can see that \(\varnothing \vdash e \equiv f \dashv \varnothing\) implies \(E \vdash e \seq g \equiv f \seq g \dashv E\), which is a fact we will use to apply non-contextual equivalences on the left of a sequencing operation. 
	
	First we apply \Cref{fundthm} and \Cref{guardedness+} to all terms in the environment to get the environment \(E_n\) such that
	\[
	E_n(k) =  \bigplus_{(\alpha, \pi) \in \At \times I^Y} \alpha \land Y=\pi \seq \sys(\partial(E(k), \pi)_\alpha)
	\]
	By \RefAx{\Axweaken},\RefAx{\Axreplacement} \(\forall e. E \vdash e \equiv e \dashv E_n\).
	
	We use \(\restrict{f}{s}\) to denote the restriction of the function \(f\) to the subset \(s\) of its domain. We first show that \(\restrict{E_n}{\{\ell\}} \vdash \alpha \land Y=\pi \seq Y \leftarrow \sigma \seq \comGoto~\ell \equiv \sys(\partial(E(\ell), \sigma)_\alpha) \dashv \restrict{E_n}{\{\ell\}}\) if \(k\neq \ell\).
	\begin{align*}
		&\restrict{E_n}{\{\ell\}} 
		\vdash \alpha \land Y=\pi \seq Y \leftarrow \sigma \seq \comGoto~\ell\\
		& \equiv Y=\pi \seq \alpha \seq Y \leftarrow \sigma \seq Y=\sigma \seq \comGoto~\ell \dashv \restrict{E_n}{\{\ell\}} & \RefAx{\Axtestcomb}, \RefAx{\Axassoc}, \eqref{indred+}\\
		& \equiv Y=\pi \seq Y \leftarrow \sigma \seq Y=\sigma\land\alpha \seq \comGoto~\ell \dashv \restrict{E_n}{\{\ell\}} & \RefAx{\Axassoc}, \eqref{indcommtest}, \RefAx{\Axtestcomb}\\
		& \equiv Y=\pi \seq Y \leftarrow \sigma \seq \alpha \land Y = \sigma \seq \sys(\partial(E(\ell), \sigma)_\alpha) \dashv \restrict{E_n}{\{\ell\}} & \RefAx{\Axsubstitution}, \eqref{branchsel+}\\
		& \equiv \alpha \land Y=\pi \seq Y \leftarrow \sigma \seq Y=\sigma \seq \sys(\partial(E(\ell), \sigma)_\alpha) \dashv \restrict{E_n}{\{\ell\}} & \BA, \RefAx{\Axtestcomb}, \eqref{indcommtest}\\
		& \equiv \alpha \land Y=\pi \seq \sys(\partial(E(\ell), \sigma)_\alpha) \dashv \restrict{E_n}{\{\ell\}} & \eqref{indred+}, \eqref{sysannih}
	\end{align*}
	We will apply this to branches of a generalized sum by invoking \eqref{guardedness+} to create the leading test and folding it back into the guard of the sum.

	By
	\[
	(E(k_1),\pi)\transSilent^+_{\alpha}(E(k_2),\sigma)
	\]
	we mean that there exist \(n\ge 1\), labels
	\[
	\ell_1,\ldots,\ell_n\in\dom(E),
	\]
	and assignments
	\[
	\pi_1,\ldots,\pi_n\in I^Y
	\]
	such that
	\[
	(\ell_n,\pi_n)=(k_2,\sigma),
	\]
	\[
	(E(k_1),\pi)\tr{}{\alpha}(\jmpc~\ell_1,\pi_1),
	\]
	and, for every \(1\le i<n\),
	\[
	(E(\ell_i),\pi_i)\tr{}{\alpha}(\jmpc~\ell_{i+1},\pi_{i+1}).
	\]
	We additionally require the reached label/valuation pairs to be simple:
	\[
	(\ell_i,\pi_i)\neq(\ell_j,\pi_j)
	\qquad
	\text{for all }1\le i<j\le n.
	\]
	By
	\[
	(E(k_1),\pi)\prescript{}{k}\transSilent^+_{\alpha}(E(k_2),\sigma)
	\]
	we mean such a \(\transSilent^+_{\alpha}\)-chain whose proper intermediate
	reached labels are not \(k\):
	\[
	\ell_i\neq k
	\qquad
	\text{for all }1\le i<n.
	\]
	Thus the endpoint is allowed to have label \(k\).
	
	By
	\[
	(E(k_1),\pi)\prescript{}{k^+}\transSilent^+_{\alpha}(E(k_2),\sigma)
	\]
	we mean such a \(\transSilent^+_{\alpha}\)-chain whose reached labels never
	include \(k\):
	\[
	\ell_i\neq k
	\qquad
	\text{for all }1\le i\le n.
	\]
	Annotations are read conjunctively.  For instance,
	\[
	(E(k_1),\pi)\prescript{}{k,\ell^+}\transSilent^+_{\alpha}(E(k_2),\sigma)
	\]
	means both
	\[
	(E(k_1),\pi)\prescript{}{k}\transSilent^+_{\alpha}(E(k_2),\sigma)
	\]
	and
	\[
	(E(k_1),\pi)\prescript{}{\ell^+}\transSilent^+_{\alpha}(E(k_2),\sigma).
	\]
	Equivalently,
	\[
	\ell_i\neq k \text{ for }1\le i<n
	\qquad\text{and}\qquad
	\ell_i\neq \ell \text{ for }1\le i\le n.
	\]
	Fix \(k\in\dom(E)\).  For each atom \(\alpha\), define the partial
	first-return map
	\[
	\rho_{\alpha,k}: I^Y \rightharpoonup I^Y
	\]
	by setting \(\rho_{\alpha,k}(\pi)=\sigma\) iff
	\[
	(E(k),\pi)\prescript{}{k}\transSilent^+_\alpha(E(k),\sigma).
	\]
	That is, starting from \((E(k),\pi)\), the silent bound-jump chain returns
	to label \(k\) at valuation \(\sigma\), and no intermediate state in the
	chain has label \(k\).
	
	Let \(d_k = \bigvee\{\alpha \land Y= \pi \mid \rho_{\alpha, k}(\pi)\ \text{defined}\}\).
	
	A test \(C\) is a \(k\)-goto invariant of \(E\) if
	\(
	C\leqBA d_k,
	\)
	and \(C\) is closed under first returns to \(k\) if
	\[
	(\alpha,\pi)\in\langle C\rangle
	\quad\text{and}\quad
	\rho_{\alpha,k}(\pi)=\sigma
	\quad\Longrightarrow\quad
	(\alpha,\sigma)\in\langle C\rangle .
	\]
	Let \[C^+_k \coloneq \bigvee\{C \mid C\ \text{is a}\ k\text{-}\comGoto\ \text{invariant of}\ E\}\]
	We start by inductively applying the syntactic branch step to only the branches \((\alpha, \pi) \in C^+_k\) in the environment with \(k\) omitted and push the new form into the environment with \RefAx{\Axweaken} and \RefAx{\Axreplacement}.
	\[
		E_{nk}(k) =
		\bigplus_{(\alpha, \pi) \in \At \times I^Y} \begin{cases} 
			Y \assign \sigma \seq \comGoto~k & (\alpha, \pi) \in C^+_k, \rho_{\alpha, k}(\pi) = \sigma\\
			\sys(\partial(E(k), \pi)_\alpha) & (\alpha, \pi) \notin C^+_k \end{cases}
	\] 
	And \(E \vdash e \equiv e \dashv E_{nk}\) for all \(e\) by \RefAx{\Axweaken} and \RefAx{\Axreplacement}.
	
	We now want to apply \RefAx{\Axunprbra}.
	
	First observe, by \RefAx{\Axidempotence}, \RefAx{\Axguardedness}, \Cref{biplusboollem}, that:
	\begin{align*}
		&\bigplus_{(\alpha, \pi) \in \At \times I^Y} \begin{cases} 
				Y \assign \sigma \seq \comGoto~k & (\alpha, \pi) \in C^+_k, \rho_{\alpha, k}(\pi) = \sigma\\
				\sys(\partial(E(k), \pi)_\alpha) & (\alpha, \pi) \notin C^+_k \end{cases}
		\\&\equiv \bigplus_{(\alpha, \pi) \in \langle \lnot C^+_k \rangle} \sys(\partial(E(k), \pi)_\alpha)
		+_{\lnot C^+_k} \bigplus_{(\alpha, \pi) \in \langle C^+_k \rangle} Y \assign \sigma_{\alpha, \pi} \seq \comGoto~k
		\end{align*}
	Thus we can create the environment \(E_{nk'}\) such that \(E_{nk'}(\ell) = E_{n\ell}(\ell)\) for \(\ell \neq k\) and 
	\[
		E_{nk'}(k) = \bigplus_{(\alpha, \pi) \in \langle \lnot C^+_k \rangle} \sys(\partial(E(k), \pi)_\alpha)
		+_{\lnot C^+_k} \bigplus_{(\alpha, \pi) \in \langle C^+_k \rangle} 
		Y \assign \sigma_{\alpha, \pi} \seq \comGoto~k 
	\]
	and by \(\RefAx{\Axreplacement}\), \(\RefAx{\Axweaken}\) along with transitivity \(\forall e. E\vdash e \equiv e \dashv E_{nk'}\). 
	
	We now show the premise of \RefAx{\Axunprbra} holds for \(C^+_k\) and apply \RefAx{\Axunprbra} to \(E_{nk'}\).
	\begin{align*}
		&E_{nk'} \vdash C^+_k \seq \left(\bigplus_{(\alpha, \pi) \in \langle C^+_k \rangle} Y \assign \sigma_{\alpha, \pi}\right) \seq \lnot C^+_k\\
		&\equiv  \left(\bigplus_{(\alpha, \pi) \in \langle C^+_k \rangle} Y \assign \sigma_{\alpha, \pi}\right) \seq \lnot C^+_k \dashv E_{nk'} & \eqref{biplusboollem}\\ 
		&\equiv  \left(\bigplus_{(\alpha, \pi) \in \langle C^+_k \rangle} Y \assign \sigma_{\alpha, \pi}  \seq \lnot C^+_k\right)  \dashv E_{nk'} & \RefAx{\Axrightdistr}\\ 
		&\equiv  \left(\bigplus_{(\alpha, \pi) \in \langle C^+_k \rangle} \alpha \seq Y \assign \sigma_{\alpha, \pi}  \seq \lnot C^+_k\right) \dashv E_{nk'} & \eqref{guardedness+}, \BA, \RefAx{\Axtestcomb}\\ 
		&\equiv  \left(\bigplus_{(\alpha, \pi) \in \langle C^+_k \rangle}Y \assign \sigma_{\alpha, \pi}  \seq \alpha \land Y=\sigma_{\alpha, \pi} \land \lnot C^+_k\right) \dashv E_{nk'} & \eqref{indcommtest}, \eqref{indred+}, \RefAx{\Axtestcomb}\\ 
		&\equiv  \bigplus_{(\alpha, \pi) \in \langle C^+_k \rangle}Y \assign \sigma_{\alpha, \pi}  \seq 0  \dashv E_{nk'}& \BA\\ 
		&\equiv  C^+_k \seq \bigplus_{(\alpha, \pi) \in \At \times I^Y}Y \assign \sigma_{\alpha, \pi}  \seq 0  \dashv E_{nk'} & \eqref{biplusboollem}\\ 
		&\equiv  0  \dashv E_{nk'}& \RefAx{\Axindabort}, \eqref{idem+}, \RefAx{\Axtestcomb}, \BA
	\end{align*}
	Hence by \RefAx{\Axunprbra}, for all \(e\) \(E \vdash e \equiv e \dashv E_{nkp}\) where  \(E_{nkp}(\ell) = E_{nk'}(\ell)\) for \(\ell \neq k\) and
	\[
		E_{nkp}(k)= \lnot C^+_k \seq \bigplus_{(\alpha, \pi) \in \langle \lnot C^+_k \rangle} \sys(\partial(E(k), \pi)_\alpha)
	\]
	We then apply \Cref{biplusboollem} and \RefAx{\Axreplacement} to obtain \(E_{nkp'}(k)\), again such that it agrees with \(E_{nk'}\) on all \(e\), and its only difference is at \(k\).
	\[
		E_{nkp'}(k)=\bigplus_{(\alpha, \pi) \in \langle \lnot C^+_k \rangle} \sys(\partial(E(k), \pi)_\alpha)
	\]
	We now want to repeat this argument for all other labels in the environment. Let \(E_P\) be the environment with self cycles at all \(k \in P \subseteq \dom(E)\) trimmed. However, the inductive \comGoto resolution step may terminate early with \(0\) if it encounters a previously trimmed cycle at another label, hence we get
	
\begin{align*}
		E_{P}(\ell) &= 
		\bigplus_{(\alpha, \pi) \in \langle \lnot C^+_\ell \rangle} \sys(\partial(E(\ell), \pi)_\alpha) \\
		&+_{\lnot C^+_\ell} \bigplus_{(\alpha, \pi) \in \langle C^+_\ell \rangle} 
		\begin{cases}
			0 & \begin{array}[t]{@{}l@{}}\exists k \sigma.(E(\ell), \pi) \prescript{}{k, \ell^+}\transSilent^+_\alpha (E(k), \sigma),\\ (\alpha, \sigma) \in \langle C^+_k \rangle\end{array}\\
			Y \assign \sigma_{\alpha, \pi} \seq \comGoto~\ell & \rho_{\alpha, \ell}(\pi) = \sigma
		\end{cases} 
	\end{align*}%
	In the first case, before encountering \(\comGoto~\ell\), the resolution encounters some \((k, \sigma)\) which was already trimmed as a self cycle. The argument for the premise of \RefAx{\Axunprbra} in \(E_P\) is unchanged, as this branch is already \(0\).
	
	By iterating this construction over all labels, we obtain
	\[
		E_{P'}(k)=\bigplus_{(\alpha, \pi) \in \langle \lnot C^+_k \rangle} \sys(\partial(E(k), \pi)_\alpha)
	\]
	for all \(k\), and by transitivity \(E \vdash e \equiv e \dashv E_{P'}\) for all \(e\).
	
	We now want to apply \RefAx{\Axuniqueness}. Consider the environment \(E'\) defined for \(k \in \dom(E)\)
	\[
		E'(k) = \bigplus_{(\alpha, \pi) \in \langle \lnot C^+_k \rangle} \begin{cases}
			Y \assign \sigma \seq p \seq f & \partial\resWith{E}(E(k), \pi)_\alpha = (p, (f, \sigma))\\
			\sys(\partial\resWith{E}(E(k), \pi)_\alpha) & \text{else} \end{cases}
	\]
	We now show the key premise of uniqueness; pick some arbitrary \(k \in \dom(E)\)
\begin{align*}
		\MoveEqLeft[1] E_{P'} \vdash \comGoto~k \\
		&\equiv \bigplus_{(\alpha, \pi) \in \langle \lnot C^+_k \rangle} 
		\begin{cases}
			Y \assign \sigma \seq p \seq f & \partial(E(k), \pi)_\alpha = (p, (f, \sigma))\\
			Y \assign \sigma \seq \comGoto~\ell & \partial(E(k), \pi)_\alpha = (\jmpc~\ell, \sigma)\\
			\sys(\partial(E(k), \pi)_\alpha) & \text{else} 
		\end{cases}  \dashv E_{P'} & \RefAx{\Axsubstitution}
	\end{align*}%
	We inductively resolve each \(\comGoto~\ell \in \dom(E)\) syntactically as previously. 

	Since all \(C^+_k\)-branches have been trimmed, no remaining nonzero
	bound-jump branch can follow bound jumps forever. To see why: Fix \(\alpha\).   
	If a remaining branch followed bound jumps forever, 
	then its label/assignment orbit would eventually repeat
	some pair \((k,\sigma)\) as the domain of $E$ and $I^Y$ are finite. 
	The segment between two consecutive occurrences
	of label \(k\) gives a nonempty first return to \(k\). 
	That is, it generates a nonempty \(k\)-\comGoto\ invariant.  
	Hence that branch lies in
	\(C^+_k\). But all \(C^+_k\)-branches have already been rewritten to \(0\),
	contradiction.

	Hence, each remaining nonzero bound-jump branch has a finite
	resolution derivation, and therefore finitely many substitutions and branch
	selections reach either an action transition or a non-bound output.
	
	\begin{align*}
		\dots &\equiv \bigplus_{(\alpha, \pi) \in \langle \lnot C^+_k \rangle} \begin{cases}
			Y \assign \sigma \seq p \seq f & \partial\resWith{E}(E(k), \pi)_\alpha = (p, (f, \sigma))\\
			\sys(\partial\resWith{E}(E(k), \pi)_\alpha) & \text{else} 
		\end{cases} \dashv E_{P'}
	\end{align*}
	We now want to bring back the invariant branches as reject branches. First observe that \((\alpha, \pi) \in C^+_k \implies \partial \resWith{E}(E(k), \pi)_\alpha = \bot\). 
\begin{align*}
		\dots &\equiv \lnot C^+_k \seq \bigplus_{(\alpha, \pi) \in \At \times I^Y} \begin{cases}
			Y \assign \sigma \seq p \seq f & \partial\resWith{E}(E(k), \pi)_\alpha = (p, (f, \sigma))\\
			\sys(\partial\resWith{E}(E(k), \pi)_\alpha) & \text{else} 
		\end{cases} \dashv E_{P'} \ehreason{\ref{biplusboollem}}\\
		&\equiv \bigplus_{(\alpha, \pi) \in \At \times I^Y} (\alpha \land Y=\pi \land \lnot C^+_k) \\&\quad \seq \begin{cases}
			Y \assign \sigma \seq p \seq f & \partial\resWith{E}(E(k), \pi)_\alpha = (p, (f, \sigma))\\
			\sys(\partial\resWith{E}(E(k), \pi)_\alpha) & \text{else} 
		\end{cases} \dashv E_{P'} \ehreason{\ref{testdistr+}, \ref{guardedness+}, \RefAx{\Axtestcomb}}
	\end{align*}%
	On the branches \(\alpha, \pi \in \langle \lnot C^+_k \rangle\), \(\alpha \land Y=\pi \land \lnot C^+_k \eqBA \alpha \land Y=\pi\); on the other branches \(\alpha \land Y=\pi \land \lnot C^+_k \eqBA 0\) which by \RefAx{\Axtestcomb} and \BA leaves \(\sys(\partial\resWith{E}(E(k), \pi)_\alpha)\) untouched (it is already \(0\) by \Cref{resolutiondef} and \Cref{sys}). Hence, we continue
	
\begin{align*}
		&\equiv \bigplus_{(\alpha, \pi) \in \At \times I^Y} \alpha \land Y=\pi \begin{cases}
			Y \assign \sigma \seq p \seq f & \partial\resWith{E}(E(k), \pi)_\alpha = (p, (f, \sigma))\\
			\sys(\partial\resWith{E}(E(k), \pi)_\alpha) & \text{else} 
		\end{cases} \dashv E_{P'}\\
		&\equiv \bigplus_{(\alpha, \pi) \in \At \times I^Y} \begin{cases}
			Y \assign \sigma \seq p \seq f & \partial\resWith{E}(E(k), \pi)_\alpha = (p, (f, \sigma))\\
			\sys(\partial\resWith{E}(E(k), \pi)_\alpha) & \text{else} 
		\end{cases} \dashv E_{P'} \ehreason{\ref{guardedness+}, \BA, \ref{guardedness+}}
	\end{align*}%
	But recall
	\[
		E'(k) = \bigplus_{(\alpha, \pi) \in \At \times I^Y} \begin{cases}
		Y \assign \sigma \seq p \seq f & \partial\resWith{E}(E(k), \pi)_\alpha = (p, (f, \sigma))\\
		\sys(\partial\resWith{E}(E(k), \pi)_\alpha) & \text{else} 
	\end{cases}
	\]
	Hence, having shown \(E_{P'} \vdash \comGoto~k \equiv E'(k) \dashv E_{P'}\), and as all \(\comGoto~k \in \dom(E)\) branches in any image in \(E'\) are guarded since they are resolved transitions, and the two environments have the same domain, we apply \RefAx{\Axuniqueness} and obtain \(\forall e.E' \vdash e \equiv e \dashv E_{P'}\). So given the original \(e\)
	\begin{align*}
		E \vdash e & \equiv e \dashv E_{P'}\\
		& \equiv e \dashv E'
	\end{align*}
\end{proof}	
\end{toappendix}
 Analogously to what we had to do in the previous section to state the fundamental theorem, we start by defining a function to turn the semantics of an expression into a canonical syntax.

\begin{definition}[\RocqLabelRef{syscYdef}]\label{syscYdef}
	Let \(\sysc_Y: K + \Sigma \times (\Expm \times I^Y) \to \Expm\) be
	\begin{mathpar}
		\sysc_Y(p, (f, \pi)) \coloneq p\seq \comGoto~\ell_{f, \pi} \and \sysc_Y(k) \eqdef \sys(k) \quad k \in K
	\end{mathpar}
\end{definition}
\begin{toappendix}
	\begin{definition}[\RocqLabelRef{syscdef}]\label{syscdef}
		Let \(\sysc: K + \Sigma \times (\Expm \times I^Y) \to \Expm\) be
		\begin{mathpar} 
			\sysc(p, (f, \pi)) \coloneq Y \assign \pi \seq p \seq \comGoto~\ell_f\and \sysc(k) \eqdef \sys(k) \quad k \in K
		\end{mathpar}
	\end{definition}
	\begin{lemma}[\RocqLabelRef{gtnflem2}] \label{gtnflem2}
		For any \(E,e\), let \(D_{E,e} \eqdef D(e) \cup \bigcup_{k \in \dom(E)} D(E(k))\), with \(D\) as in \Cref{exprsubcofinite}.
		Then
		\begin{align*}
			&E \vdash e \equiv e \dashv E'
		\end{align*}
		where, for all \(f \in D_{E,e}\),
		\begin{align*}
			&E^\prime(\ell_f) = \bigplus_{(\alpha, \pi) \in \At \times I^Y} \sysc(\partial\resWith{E}(f, \pi)_\alpha)
		\end{align*}
		and for \(k\) in \(\text{dom}(E)\)
		\begin{align*}
			&E^\prime(k) = \bigplus_{(\alpha, \pi) \in \At \times I^Y} \sysc(\partial\resWith{E}(E(k), \pi)_\alpha)
		\end{align*}
		Moreover \(E' \vdash e \equiv \comGoto~\ell_e \dashv E'\).

	\end{lemma}
	\begin{proof}
		For any \(E, e\) we create the environment from \Cref{gtnflem1} and call it \(E_1\). W.l.o.g. we assume that no \(\ell_f\) for any \(f \in \Expm\) is an element of \(\text{dom}(E)\), nor does any \comGoto~in \(e\) or \(E\) use these labels.
		
		\(D_{E,e}\) is finite (\Cref{exprsubcofinite}), so we can create a label for each of its
		expressions and each existing label, giving a finite domain. Moreover, \(D_{E,e}\) is closed
		under resolved steps from any valuation: if \(f \in D_{E,e}\) and
		\(\partial\resWith{E}(f,\pi)_\alpha = (p,(g,\sigma))\), then \(g \in D_{E,e}\). Indeed, by
		\Cref{resolutiondef} such a step is an unresolved step of \(f\), or of some \(E(k)\) reached
		through jumps, and \(D(f), D(E(k)) \subseteq D_{E,e}\).
		
		We define the new environment \(E_{f}\) which agrees with \(E_1\) on its domain, but further includes for each \(f \in D_{E,e}\)
		\begin{align*}
			E_{f}(\ell_f) &= \bigplus_{(\alpha, \pi) \in \At \times I^Y} \sys(\partial(f, \pi)_\alpha)
		\end{align*}
		Note that by \Cref{fundthm} \(\varnothing \vdash E_{f}(\ell_f) \equiv f \dashv \varnothing\). 
		
		By using \Cref{guardedness+}, \RefAx{\Axsubstitution}, and \Cref{branchsel+} to substitute in values from \(E_1(k)\) for the \(\jmpc\) branches, we derive
		\begin{align*}
			&\restrict{E_f}{\{k \mid k \in \text{dom}(E_1)\}} \vdash \bigplus_{(\alpha, \pi) \in \At \times I^Y} \sys(\partial(f, \pi)_\alpha)\\
			&\equiv \bigplus_{(\alpha, \pi) \in \At \times I^Y} \sys(\partial\resWith{E}(f, \pi)_\alpha) \dashv \restrict{E_f}{\{k \mid k \in \text{dom}(E_1)\}}
		\end{align*}
		since the jump resolution agrees with any immediate productive behavior.
		
		We push this term into every \(E_f(\ell_f)\) with \RefAx{\Axreplacement} and \RefAx{\Axweaken} to create the environment \(E_{f'}\) such that \(E_f \vdash e \equiv e \dashv E_{f'}\) for all \(e\).
		
		Furthermore, we can derive from the above that
		\begin{align*}
			E_f \vdash f &\equiv \bigplus_{(\alpha, \pi) \in \At \times I^Y} \sys(\partial(f, \pi)_\alpha) \dashv E_f & \tCref{fundthm}, \RefAx{\Axweaken}\\
			&\equiv \bigplus_{(\alpha, \pi) \in \At \times I^Y} \sys(\partial\resWith{E}(f, \pi)_\alpha) \dashv E_f & \text{see above}, \RefAx{\Axweaken}\\
			&\equiv \bigplus_{(\alpha, \pi) \in \At \times I^Y} \sys(\partial\resWith{E}(f, \pi)_\alpha) \dashv E_{f'} & \text{see above}\\
			&\equiv E_{f'}(\ell_f) \dashv E_{f'} 
		\end{align*}
		And \(E_f \vdash g \equiv g \dashv E_{f'}\).
		
		We now want to use \RefAx{\Axuniqueness} to achieve the desired environment. Let \(E'\) be the system from the lemma statement. All \comGoto~in \(E'\) whose labels lie in \(\dom(E')\) are guarded, and \(\text{dom}(E') = \text{dom}(E_{f})\). So we now show \(E_{f'} \vdash \comGoto~\ell \equiv E'(\ell) \dashv E_{f'}\). 
		We start with an arbitrary \(\ell_f\) and use \RefAx{\Axsubstitution}, the fact that \(E_{f'} \vdash g \equiv E_{f'}(\ell_g) \dashv E_{f'}\) (here \(\ell_g\) exists because \(D_{E,e}\) is closed under resolved steps from any valuation), and the fact that \(\sysc(r) = \sys(r)\)
		\begin{align*}
			E_{f'} \vdash E'(\ell_f) &\equiv
			\bigplus_{(\alpha, \pi) \in \At \times I^Y}
			\begin{cases}
				 Y \assign \sigma \seq p \seq E_{f'}(\ell_g) & \partial\resWith{E}(f, \pi)_\alpha = (p, (g, \sigma))\\
				\sys(r) & \partial\resWith{E}(f, \pi)_\alpha = r \in K
			\end{cases} \dashv E_{f'}\\
			&\equiv \bigplus_{(\alpha, \pi) \in \At \times I^Y}
			\begin{cases}
				Y \assign \sigma \seq p\seq g & \partial\resWith{E}(f, \pi)_\alpha = (p, (g, \sigma))\\
				\sys(r) & \partial\resWith{E}(f, \pi)_\alpha = r \in K
			\end{cases} \dashv E_{f'}\\
			&\equiv \bigplus_{(\alpha, \pi) \in \At \times I^Y}
			\begin{cases}
				\sys(p, (g, \sigma)) & \partial\resWith{E}(f, \pi)_\alpha = (p, (g, \sigma))\\
				\sys(r) & \partial\resWith{E}(f, \pi)_\alpha = r \in K
			\end{cases} \dashv E_{f'}\\
			&\equiv E_{f'}(\ell_f) \dashv E_{f'}
		\end{align*}
		Finally, applying \RefAx{\Axsubstitution} results in \(\comGoto~\ell_f \dashv E_{f'}\). The case for \(k \in \text{dom}(E_1)\) follows from a similar argument.
		By \RefAx{\Axuniqueness} \(\forall e. E_{f'} \vdash e \equiv e \dashv E'\).
		Now since the original \(e, E_1\) do not reference any labels of the form \(\ell_f\)
		\begin{align*}
			E \vdash e &\equiv e \dashv E_1 & \tCref{gtnflem1}\\
			&\equiv e \dashv E_{f} & \RefAx{\Axredundancy}\\
			&\equiv e \dashv E_{f'}& \text{above}\\
			&\equiv e \dashv E'& \text{above}
		\end{align*}
		For the final fact, we observe that
		\begin{align*}
			E' \vdash e &\equiv e \dashv E_{f'} & \text{above}\\ 
			&\equiv e \dashv E_{f}& \text{above}\\ 
			&\equiv \comGoto~\ell_e \dashv E_{f} & \tCref{fundthm}, \RefAx{\Axsubstitution}\\ 
			&\equiv \comGoto~\ell_e \dashv E_{f'} & \text{above}\\ 
			&\equiv \comGoto~\ell_e \dashv E' & \text{above}
			&\qed
		\end{align*}
\end{proof}	
\end{toappendix}
We can now state the contextual normal form theorem for expressions.
\begin{theoremrep}[Goto Normal Form \RocqLabelRef{gtnfthm}] \label{gtnfthm}
	For any \(E,e\):
	\begin{align*}
		&\text{let}&
			E^\bot_Y(\ell_{f, \pi}) & = \bigplus\nolimits_{\alpha \in \At} \sysc_Y(\partial\resWith{E}(f, \pi)_\alpha)
			& \text{for all \((f, \pi) \in \langle e, \sigma \rangle\resWith{E}\) for some \(\sigma\),} \\
		&\text{then}&
			E \vdash e & \equiv \bigplus\nolimits_{\pi \in I^Y} \comGoto~\ell_{e, \pi} \dashv E^\bot_Y
	\end{align*}
    where \(\langle e, \sigma \rangle\resWith{E}\) denotes the subcoalgebra of \((\Expm \times I^Y, \partial)\resWith{E}\) generated by \((e,\sigma)\).
\end{theoremrep}
\begin{appendixproof}	
	This proof will push indicator values into labels, creating a system which does not explicitly assign or test indicator values anywhere. First we apply \Cref{gtnflem2} to get the system \(E'\) from that lemma statement.
	
	Let us introduce the environment \(E'_Y\) which agrees with \(\text{dom}(E')\) on all labels in its domain, but introduces \(u_\pi\) for all \(\pi \in I^Y\) and \(u \in \text{dom}(E')\). Let \(E'_Y(u_\sigma) = Y \leftarrow \sigma \seq E'(u)\). W.l.o.g.  assume all \(u_\pi\) are fresh in \(E'\). Otherwise use \RefAx{\Axrenaming} to rename any conflicting labels. By \RefAx{\Axredundancy} these two systems agree on all terms which do not reference a label annotated with an assignment. That is \(\forall e. E' \vdash e \equiv e \dashv E'_Y\) as long as \(e\) does not reference any \(u_\pi\).
	
	Consider another environment \(E_Y\) with labels \(\ell_f\) and \(\ell_{f,\pi}\) for
	\(f\in D_{E,e}\), and labels \(k\) and \(k_\pi\) for \(k\in\dom(E)\), in each case
	for all \(\pi\in I^Y\). Define
	\begin{align*}
		E_Y(\ell_{f, \pi}) &= \bigplus_{\alpha \in \At} \sysc_Y(\partial\resWith{E}(f, \pi)_\alpha)\\
		E_Y(\ell_f) &=\bigplus_{(\alpha, \pi) \in \At \times I^Y} \sysc_Y(\partial\resWith{E}(f, \pi)_\alpha)
	\end{align*} 
	with analogous images for \(k, k_\pi\). Since \(D_{E,e}\) is closed under resolved steps from every
	valuation, every successor label \(\ell_{g,\sigma}\) occurring in these bodies exists.
	
	We now want to apply \RefAx{\Axuniqueness}. 
	Clearly \(E'_Y\) and \(E_Y\) have the same domain. 
	Furthermore all \comGoto~to labels in \(\dom(E_Y)\) are guarded by the above construction.
	Pick some \(\ell_{f}\); we now show that
	\[E'_Y \vdash \comGoto~u \equiv E_Y(u) \dashv E'_Y\]
	We start with \(u = \ell_f\). By the definitions of \(E'\), \(E'_Y\) and \RefAx{\Axsubstitution}, \RefAx{\Axindcommact}:
	\begin{align*}
		E'_Y \vdash E_Y(\ell_f)
		&\equiv \bigplus_{(\alpha, \pi) \in \At \times I^Y} 
		\begin{cases}
			Y \assign \sigma \seq p\seq E'(\ell_g) & \partial\resWith{E}(f, \pi)_\alpha = (p, (g, \sigma))\\
			\sysc_Y(r) & \partial\resWith{E}(f, \pi)_\alpha = r \in K
		\end{cases}\dashv E'_Y
	\end{align*} 
	We now address the first case
	\begin{align*}
		E'_Y \vdash Y \assign \sigma \seq p\seq E'(\ell_g) &\equiv Y \assign \sigma \seq p\seq \comGoto~\ell_g \dashv E'_Y& \RefAx{\Axsubstitution}
	\end{align*}
	Hence since \(\sysc_Y(r) = \sysc(r)\),
	\begin{align*}
		E'_Y \vdash E_Y(\ell_f)
		&\equiv \bigplus_{(\alpha, \pi) \in \At \times I^Y} 
		\begin{cases}
			\sysc(p, (g, \sigma)) & \partial\resWith{E}(f, \pi)_\alpha = (p, (g, \sigma))\\
			\sysc(r) & \partial\resWith{E}(f, \pi)_\alpha = r \in K
		\end{cases}\dashv E'_Y
	\end{align*} 
	This is precisely \(E'_Y(\ell_f)\), so by \RefAx{\Axsubstitution} we see this is equivalent to \(\comGoto~\ell_f \dashv E'_Y\).
	
	We continue with \(u = \ell_{f, \pi}\). By the definitions of \(E'_Y\) and \(E_Y\) and \RefAx{\Axsubstitution}
	\begin{align*}
		E'_Y \vdash E_Y(\ell_{f,\pi})
		&\equiv \bigplus_{\alpha \in \At} 
		\begin{cases}
			p\seq E'_Y(\ell_{g,\sigma}) & \partial\resWith{E}(f, \pi)_\alpha = (p, (g, \sigma))\\
			\sysc_Y(r) & \partial\resWith{E}(f, \pi)_\alpha = r \in K
		\end{cases}\dashv E'_Y
	\end{align*}
	We now examine the first case
	\begin{align*}
		E'_Y \vdash p \seq E'_Y(\ell_{g,\sigma})
		&\equiv p\seq Y \assign \sigma \seq E'(\ell_{g}) \dashv E'_Y & \text{def of }E'_Y\\
		&\equiv p\seq Y\assign \pi \seq Y \assign \sigma \seq E'(\ell_{g}) \dashv E'_Y & \eqref{indannih+}\\
		&\equiv Y\assign \pi \seq Y \assign \sigma \seq  p\seq E'(\ell_{g}) \dashv E'_Y & \RefAx{\Axindcommact}
	\end{align*}
	In the second case by \Cref{sysannih} and the fact that \(\sys(r) = \sysc(r) = \sysc_Y(r)\) 
	\[\sysc_Y(r) \equiv Y\assign \pi \seq \sysc_Y(r)\] Hence we simplify the overall term to
	\begin{align*}
		E'_Y \vdash E_Y(\ell_{f,\pi})\equiv \bigplus_{\alpha \in \At} Y \assign \pi \seq
		\begin{cases}
			Y \assign \sigma \seq p\seq E'_Y(\ell_{g}) & \partial\resWith{E}(f, \pi)_\alpha = (p, (g, \sigma))\\
			\sysc(r) & \partial\resWith{E}(f, \pi)_\alpha = r \in K
		\end{cases}\dashv E'_Y
	\end{align*}
	and by \eqref{inddistr}, \RefAx{\Axsubstitution} and the definition of \sysc/\(E_Y'\)
	\begin{align*}
		\dots &\equiv Y \assign \pi \seq \bigplus_{\alpha \in \At} 
		\begin{cases}
			Y \assign \sigma \seq p\seq \comGoto~\ell_{g} & \partial\resWith{E}(f, \pi)_\alpha = (p, (g, \sigma))\\
			\sysc(r) & \partial\resWith{E}(f, \pi)_\alpha = r \in K
		\end{cases}\dashv E'_Y\\
		&\equiv Y \assign \pi \seq \bigplus_{\alpha \in \At} \sysc(\partial\resWith{E}(f, \pi)_\alpha) \dashv  E'_Y
	\end{align*}
	We now adjust the indexing
	\begin{align*}
		&E'_Y\vdash Y \assign \pi \seq \bigplus_{\alpha \in \At} \sysc(\partial\resWith{E}(f, \pi)_\alpha)\\
		&\equiv Y \assign \pi \seq \bigplus_{(\alpha, \sigma) \in \At \times I^Y} \sysc(\partial\resWith{E}(f, \pi)_\alpha) \dashv E'_Y & \eqref{ggdedsumprodinj}\\
		&\equiv Y \assign \pi \seq \bigplus_{(\alpha, \sigma) \in \At \times \{\pi\}} \sysc(\partial\resWith{E}(f, \pi)_\alpha) \dashv E'_Y & \eqref{indred+}, \eqref{biplusboollem}\\
		&\equiv Y \assign \pi \seq \bigplus_{(\alpha, \sigma) \in \At \times I^Y} \sysc(\partial\resWith{E}(f, \sigma)_\alpha) \dashv E'_Y &\eqref{biplusboollem}, \eqref{indred+}\\
		&\equiv Y \assign \pi \seq E'(\ell_f) \dashv E'_Y &\text{def of }E'\\
		&\equiv E'_Y(\ell_{f, \pi}) \dashv E'_Y &\text{def of }E'_Y\\
		&\equiv \comGoto~\ell_{f,\pi} \dashv E'_Y &\RefAx{\Axsubstitution}
	\end{align*}
	where we use \Cref{indred+} and \Cref{biplusboollem} twice to make the indexing match. A similar argument holds for \(k\) and \(k_\pi\). By \RefAx{\Axuniqueness}, we obtain \(E_Y \vdash e \equiv e \dashv E'_Y\) for all \(e\). Hence with the original expression \(e\), 
	\begin{align*}
		&E\vdash e \equiv e \dashv E' & \tCref{gtnflem2}\\
		&\equiv \comGoto~\ell_e \dashv E' & \tCref{gtnflem2}\\
		&\equiv \comGoto~\ell_e \dashv E'_Y &\RefAx{\Axredundancy}\\
		&\equiv \comGoto~\ell_e \dashv E_Y & \text{above fact}\\
		&\equiv \bigplus_{(\alpha, \sigma)\in \At\times I^Y} \comGoto~\ell_{e} \dashv E_Y & \eqref{idem+}\\
		&\equiv \bigplus_{(\alpha, \sigma)\in \At\times I^Y} Y=\sigma\seq \comGoto~\ell_{e} \dashv E_Y & \eqref{guardedness+}, \BA, \RefAx{\Axtestcomb}\\
		&\equiv \bigplus_{(\alpha, \sigma)\in \At\times I^Y} Y=\sigma\seq \bigplus_{(\beta, \pi) \in \At \times I^Y} \sysc_Y(\partial\resWith{E}(e, \sigma)_\beta) \dashv E_Y & \RefAx{\Axsubstitution},\eqref{biplusboollem},\eqref{biplusboollem}\\
		&\equiv \bigplus_{(\alpha, \sigma)\in \At\times I^Y} \bigplus_{\{(\beta, \pi) \in \At \times I^Y\}} \sysc_Y(\partial\resWith{E}(e, \sigma)_\beta) \dashv E_Y & \eqref{guardedness+}, \BA, \RefAx{\Axtestcomb}\\
		&\equiv \bigplus_{(\alpha, \sigma)\in \At\times I^Y} \bigplus_{\{\beta \in \At \}} \sysc_Y(\partial\resWith{E}(e, \sigma)_\beta) \dashv E_Y & \eqref{ggdedsumprodinj}\\
		&\equiv \bigplus_{(\alpha, \sigma)\in \At\times I^Y} \comGoto~\ell_{e, \sigma} \dashv E_Y & \RefAx{\Axsubstitution}\\
		&\equiv \bigplus_{\sigma \in I^Y} \comGoto~\ell_{e, \sigma} \dashv E_Y & \eqref{ggdedsumprodinj}\\
		&\equiv \bigplus_{\sigma \in I^Y} \comGoto~\ell_{e, \sigma} \dashv E^\bot_Y & \RefAx{\Axredundancy}
	\end{align*}
	For the last step, let
	\[
	S\eqdef\{\ell_{f,\pi}\mid\exists\sigma\in I^Y.\ (f,\pi)\in\langle e,\sigma\rangle\resWith{E}\}.
	\]
	Then \(E^\bot_Y=\restrict{E_Y}{S}\). Every initial label \(\ell_{e,\sigma}\) belongs to \(S\),
	and every action successor of a retained state has its label in \(S\). Any remaining jump
	output targets an original free label outside \(\dom(E)\), which freshness keeps distinct
	from all introduced labels. Thus neither the final expression nor any retained body
	contains a \comGoto~to a removed binding, as required by \RefAx{\Axredundancy}.
	\end{appendixproof}
    \begin{toappendix}
	\begin{corollary}[Goto-Indicator Normal Form \RocqLabelRef{gtnfcor}] \label{gtnfcor}
		For any \(E,e, \pi\), we have that
		\begin{align*}
			&E \vdash Y\leftarrow \pi \seq e \equiv \comGoto~\ell_{e, \pi} \dashv E^\bot_Y
	\end{align*}
\end{corollary}
\begin{proof}
	First, observe that \sysc annihilates leading assignments for the same reason as \sys.
	\begin{align*}
		E \vdash Y\leftarrow \pi \seq e &\equiv Y \leftarrow \pi \seq \bigplus_{\sigma \in I^Y} \comGoto~\ell_{e, \sigma} \dashv E^\bot_Y & \eqref{gtnfthm}, \RefAx{\Axcongrightseq} \\
		&\equiv Y \leftarrow \pi \seq \comGoto~\ell_{e, \pi} \dashv E^\bot_Y & \eqref{indval+}, \eqref{branchsel}, \eqref{indval+}
	\end{align*}
	As shown in the proof of \Cref{gtnfthm}, the uniqueness argument gives
	\(E^\bot_Y \vdash h \equiv h \dashv E'_Y\) for every term \(h\) whose
	gotos use only assignment-indexed labels. Hence it is enough to
	remove the leading assignment in \(E'_Y\).  There,
	\(E'_Y(\ell_{e,\pi}) = Y\assign\pi \seq E'(\ell_e)\), so
	\begin{align*}
		E_Y^\bot \vdash Y \assign \pi \seq \comGoto~\ell_{e, \pi} &\equiv 
		Y\assign\pi \seq \comGoto~\ell_{e,\pi} \dashv E'_Y\\
		&\equiv
		Y\assign\pi \seq Y\assign\pi \seq E'(\ell_e) \dashv E'_Y
		& \RefAx{\Axsubstitution}\\
		&\equiv
		Y\assign\pi \seq E'(\ell_e) \dashv E'_Y
		& \eqref{indannih+}\\
		&\equiv
		\comGoto~\ell_{e,\pi} \dashv E'_Y
		& \RefAx{\Axsubstitution}\\
		&\equiv
		\comGoto~\ell_{e,\pi} \dashv E_Y^\bot
	\end{align*}
	\end{proof}
    \end{toappendix}

    \begin{proofsketch}  The construction is obtained in three steps. 
	The first syntactically resolves jumps analogously to \Cref{resolutiondef}. It uses the fundamental theorem to normalize the environment before doing any contextual reasoning, which 
	 is critical to inductively resolving jumps via {\em syntactic} reasoning. It also uses \RefAx{\Axunprbra} to remove unproductive sequences of \comGoto, and \RefAx{\Axuniqueness} to resolve self-reference. 
	The second step pulls apart each term in the environment such that each term either terminates or executes a single action, \(p\), followed by a \comGoto. 
	The third step removes indicator assignments by replacing them with control flow via expanding the environment with new labels.
\end{proofsketch}

Note how in the normal form the indicator assignments only appear at the beginning (in a generalized guarded choice) and at the end when an output is given. 
Intuitively, this means we have internalized the control flow from the flags into our system, that is, we have rewritten the expression and environment such that they behave the same way {\em but do not use the assignment}. Instead, the structure of the program replicates the branching behavior of the assignment. 
Observe also that this normal form corresponds precisely to the semantics, \((e, \pi)\resWith{E}\). 
In fact, for any \coalg{\CFGKAT}, we may construct an environment where each state is represented as a label with its transitions encoded in the corresponding expression.
Furthermore, this \textit{canonical environment} is the same as $E_Y^\bot$ when defined for a (resolved) syntactic coalgebra.
\begin{theoremrep}[\RocqLabelRef{canonicalenv}] \label{canonicalenv}
	Given a finite \coalg{\CFGKAT} \((X, \chi)\), for any \(\pi \in I^Y\) and \(x \in X\), we have that \((\comGoto~\ell_x, \pi)\resWith{E} \sim x\),
	where
	\begin{align*}
		E(\ell_{x}) &= \bigplus_{\alpha \in \At} \begin{cases} p \seq \comGoto~\ell_{x'} & \chi(x)_\alpha = (p, x')\\ \textit{sys}(c)&  \chi(x)_\alpha = c \in K \end{cases}.
	\end{align*}
\end{theoremrep}
\begin{appendixproof}
	We show the relation \((\comGoto~\ell_x, \pi)\resWith{E} \sim x\) for all \(x \in X\) is a bisimulation.
	
	We choose labels \(\ell_{x}\) that are fresh for all \(x\), that is, no state transitions to a \(\jmpc~\ell_{x}\) for any assignment.
	\[
	(\comGoto~\ell_x, \pi) \hookrightarrow_{\alpha}^E r \iff (E(\ell_x), \pi) \hookrightarrow_\alpha^E r \quad \tCref{resolutiondef}
	\]
	If \(\chi(x)_\alpha = c \in K\), then \((E(\ell_x), \pi) \tr[E]{}{\alpha}
 \sys(c)\). No jump can be resolved since \(c\) cannot be a \(\comGoto~\ell_x \in \text{dom}(E)\).
	
	Suppose \(\chi(x)_\alpha = (p, x')\), then \((E(\ell_x), \pi) \tr[E]{p}{\alpha} (\comGoto~\ell_{x'}, \pi)\).
\end{appendixproof}

Our goal is to turn a bisimulation between two expressions into an equational proof of their equivalence.
To this end, we view a bisimulation \(R\) as a \coalg{\CFGKAT} whose states are the related pairs, with projections into the two original coalgebras~\cite{universalcoalg}.
These projections are homomorphisms, meaning that they preserve transitions.
\begin{definition}[Homomorphism \RocqLabelRef{hom}]\label{cfgkat-hom}
    Let \((X,\chi)\) and \((Z,\zeta)\) be \coalg{\CFGKAT}s.
	A function \(h: X \to Z\) is a \emph{homomorphism} when for every \(x \in X\) and \(\alpha \in \At\),
    \[
        \zeta(h(x))_\alpha
        =
        \begin{cases}
            k & \text{if } \chi(x)_\alpha = k \in K,\\
            (p,h(x')) & \text{if } \chi(x)_\alpha = (p,x') \in \Sigma \times X.
        \end{cases}
    \]
\end{definition}

Since the contextual normal form corresponds to the semantics of an expression, we can establish that our proof system suffices to show that the normal form of any \coalg{\CFGKAT} and its image under any homomorphism are provably equivalent.

\begin{lemmarep}[Homomorphisms preserve solutions \RocqLabelRef{hompressol}] \label{hompressol}
	If \((X, \chi)\) and \((Z, \zeta)\) are finite \coalg{\CFGKAT}s and \(h: X \to Z\) is a homomorphism, then \(E_X \vdash \comGoto~\ell_x \equiv \comGoto~\ell_{h(x)} \dashv E_Z\),
	where \(E_X\) and \(E_Z\) are the canonical environments as in \Cref{canonicalenv}.
\end{lemmarep}
\begin{appendixproof}
    We construct two new environments, \(E_X'\) and \(E_Z'\):
    \begin{mathpar}
        E_X'(\ell_x) = E_X(\ell_x)
        \and
        E_X'(\ell_z) = E_Z(\ell_z)
        \\
        E_Z'(\ell_x) = E_Z(\ell_{h(x)})
        \and
        E_Z'(\ell_z) = E_Z(\ell_z)
    \end{mathpar}
    We now claim that for all \(y \in X \cup Z\), it holds that
    \begin{equation}\label{eq:hom-lifted}
        E_Z' \vdash \comGoto~\ell_y \equiv E_X'(\ell_y) \dashv E_Z'
    \end{equation}
    When \(y \in Z\), the proof is easy:
    \begin{align*}
        E_Z' \vdash \comGoto~\ell_y
            &\equiv \comGoto~\ell_y \dashv E_Z \tag{\RefAx{\Axredundancy}} \\
            &\equiv E_Z(\ell_y) \dashv E_Z \tag{\RefAx{\Axsubstitution}} \\
            &\equiv E_X'(\ell_y) \dashv E_Z' \tag{\RefAx{\Axredundancy}, def. \(E_X'\)}
    \end{align*}
    When \(y \in X\), our claim can be reduced via \RefAx{\Axsubstitution} and the definitions of \(E_X\), \(E_Z\), \(E_X'\) and \(E_Z'\) to the following equivalence:
    \begin{align*}
        E_Z' &\vdash \bigplus_{\alpha \in \At}
            \begin{cases}
            p; \comGoto~\ell_{x'} & \chi(x)_\alpha = (p, x') \\
            \sys(\chi(x)_\alpha) & \chi(x)_\alpha \in K
            \end{cases} \\
            &\equiv
            \bigplus_{\alpha \in \At}
            \begin{cases}
            p; \comGoto~\ell_{z'} & \zeta(h(x))_\alpha = (p, z') \\
            \sys(\zeta(h(x))_\alpha) & \zeta(h(x))_\alpha \in K
            \end{cases}
            \dashv E_Z'
    \end{align*}
    We show equivalence for each component of the sum, which suffices for the above equivalence, by \RefAx{\Axcongguardedchoice}.
    First, if \(\chi(x)_\alpha \in K\), then \(\chi(x)_\alpha = \zeta(h(x))_\alpha\), because \(h\) is a homomorphism; thus \(\sys(\chi(x)_\alpha) = \sys(\zeta(h(x))_\alpha)\).
    Otherwise, if \(\chi(x)_\alpha = (p, x')\), then \(\zeta(h(x))_\alpha = (p, h(x'))\), again because \(h\) is a homomorphism; thus:
    \begin{align*}
    E_Z' \vdash p; \comGoto~\ell_{h(x')}
        &\equiv p; E_Z'(\ell_{h(x')}) \dashv E_Z'   \tag{\RefAx{\Axsubstitution}} \\
        &\equiv p; E_Z'(\ell_{x'}) \dashv E_Z       \tag{def. \(E_Z'\)} \\
        &\equiv p; \comGoto~\ell_{x'} \dashv E_Z'   \tag{\RefAx{\Axsubstitution}}
    \end{align*}
    Having established \eqref{eq:hom-lifted}, we can apply \RefAx{\Axuniqueness} to derive that \(E_X' \vdash \comGoto~\ell_x \equiv \comGoto~\ell_x \dashv E_Z'\).
    With this in hand, we can then derive:
    \begin{align*}
    E_X \vdash \comGoto~\ell_x
        &\equiv \comGoto~\ell_x \dashv E_X'         \tag{\RefAx{\Axredundancy}} \\
        &\equiv \comGoto~\ell_x \dashv E_Z'         \tag{the above} \\
        &\equiv E_Z'(\ell_x) \dashv E_Z'            \tag{\RefAx{\Axsubstitution}} \\
        &\equiv E_Z'(\ell_{h(x)}) \dashv E_Z'       \tag{def. \(E_Z'\)} \\
        &\equiv \comGoto~\ell_{h(x)} \dashv E_Z'    \tag{\RefAx{\Axsubstitution}} \\
        &\equiv \comGoto~\ell_{h(x)} \dashv E_Z     \tag{\RefAx{\Axredundancy}}
    \end{align*}
\end{appendixproof}
\begin{toappendix}
	\begin{corollaryrep}[\RocqLabelRef{ebotsemenv}]\label{ebotsemenv}
		For any \(e, \psi\), \(E\), and \((f, \sigma) \in \langle e, \psi \rangle \resWith{E}\)
		\begin{align*}
			E_\bot^Y(\ell_{f, \sigma}) = E_{\langle e, \psi \rangle\resWith{E}}(\ell_{f, \sigma})
		\end{align*}
		where the annotation means the canonical environment (\ref{canonicalenv}) of the given coalgebra.

	\end{corollaryrep}
	\begin{proof}
		Immediate from the definition of \(E_\bot^Y\) and \(E_{\langle e, \psi \rangle\resWith{E}}\).
	\end{proof}	
\end{toappendix}
For completeness, suppose two expressions are bisimilar, and let \(R\) be a bisimulation relating them.
Applying \Cref{hompressol} to the two projections out of \(R\) relates the canonical environment of \(R\) to those of the two related coalgebras.
Finally, we relate the contextual normal form of an expression to the canonical environment of its coalgebra.

We now have all we need to prove the main technical result of our paper: the axiomatization we presented of \CFGKAT is complete with respect to bisimilarity.

\begin{theoremrep}[Completeness \RocqLabelRef{completeness}]\label{completeness} Let \(e,f \in \Expm\), and \(E,F\) be environments. 
    If for all $\pi \in I^Y$ we have that $(e, \pi)\resWith{E} \sim (f, \pi)\resWith{F}$, then $E \vdash e \equiv f \dashv F$.
\end{theoremrep}
\begin{proofsketch}
Using \Cref{gtnfthm}, we first argue that for any $\pi \in I^Y$, we have that $E \vdash Y \leftarrow \pi; e \equiv \comGoto~\ell_{e, \pi} \dashv E_{\langle e, \pi\rangle\resWith{E}}$, where $E_{\langle e, \pi \rangle\resWith{E}}$ is the canonical environment of the coalgebra $\langle e, \pi \rangle \resWith{E}$.
Using \Cref{hompressol}, we can furthermore show that for any $\pi$, we have $E_{\langle e, \pi\rangle\resWith{E}} \vdash \comGoto~\ell_{e, \pi} \equiv \comGoto~\ell_{(e, \pi), (f, \pi)} \dashv E_R$, where $E_R$ is the canonical environment of the bisimulation between $\langle e, \pi \rangle\resWith{E}$ and $\langle f, \pi \rangle\resWith{F}$, which is itself a \coalg{\CFGKAT}.
This establishes for any $\pi$ that $E \vdash Y \leftarrow \pi; e \equiv \comGoto~\ell_{(e, \pi), (f, \pi)} \dashv E_R$, and analogously we find that $F \vdash Y \leftarrow \pi; f \equiv \comGoto~\ell_{(e, \pi), (f, \pi)} \dashv E_R$ --- hence $E \vdash Y \leftarrow \pi; e \equiv Y \leftarrow \pi; f \dashv F$ for any $\pi$ by transitivity.
The claim then follows by standard properties of the generalized sum, quantifying over all $\pi$.
\end{proofsketch}
\begin{appendixproof}
	We start by showing \(E \vdash Y \leftarrow \pi\seq e \equiv Y \leftarrow \pi \seq f \dashv F\) for all \(\pi\).
	\begin{align*}
		E \vdash Y\leftarrow \pi\seq e &\equiv \comGoto~\ell_{e, \pi} \dashv E_\bot^Y & \tCref{gtnfcor}\\
		&\equiv \comGoto~\ell_{e, \pi} \dashv \restrict{E_\bot^Y}{\text{dom}(E_{\langle e, \pi \rangle\resWith{E}})} & \RefAx{\Axredundancy}\\
		&\equiv \comGoto~\ell_{e, \pi} \dashv E_{\langle e, \pi \rangle\resWith{E}} & \tCref{ebotsemenv}
	\end{align*}
	By the premise, there exists a bisimulation \((R, \rho), \pi_1: R \to X, \pi_2: R \to Z\) which relates \((e,\pi)\resWith{E}\) and \((f,\pi)\resWith{F}\) where \(X\) denotes the subcoalgebra of \(\Expm \times I^Y\) generated by \((e,\pi)\resWith{E}\) and \(Z\) \((f,\pi)\resWith{F}\) and \(R \subseteq X \times Z\). Then 
	\begin{align*}
		E_R \vdash \comGoto~\ell_{(e, \pi), (f, \pi)}
		&\equiv\comGoto~\ell_{\pi_1((e, \pi), (f, \pi))} \dashv E_{\langle e, \pi \rangle \resWith{E}} & \tCref{hompressol}\\
		&\equiv \comGoto~\ell_{e, \pi} \dashv E_{\langle e, \pi \rangle \resWith{E}}\\
		& \equiv Y\leftarrow \pi \seq e \dashv E & \text{transitivity}
	\end{align*}
	but also
	\begin{align*}
		E_R \vdash \comGoto~\ell_{(e, \pi), (f, \pi)}
		&\equiv \comGoto~\ell_{\pi_2((e, \pi), (f, \pi))} \dashv E_{\langle f, \pi \rangle \resWith{F}} & \tCref{hompressol}\\
		&\equiv \comGoto~\ell_{f, \pi} \dashv E_{\langle f, \pi \rangle \resWith{F}}\\
		& \equiv Y\leftarrow \pi \seq f \dashv F & \text{transitivity}
	\end{align*}
	Hence, for all \(\pi\)
	\[E \vdash Y\leftarrow\pi \seq e \equiv Y\leftarrow\pi\seq f \dashv F\]
	Finally 
	\begin{align*}
		e &\equiv \bigplus_{(\alpha, \pi) \in \At \times I^Y} e & \eqref{idem+}\\
		&\equiv \bigplus_{(\alpha, \pi) \in \At \times I^Y} \alpha \land Y=\pi \seq e & \eqref{guardedness+}\\
		&\equiv \bigplus_{(\alpha, \pi) \in \At \times I^Y} Y=\pi \seq e  & \BA, \RefAx{\Axtestcomb}, \eqref{guardedness+}\\
		&\equiv \bigplus_{(\alpha, \pi) \in \At \times I^Y} (Y=\pi\seq Y\leftarrow \pi) \seq e  & \eqref{indred+}\\
		&\equiv \bigplus_{(\alpha, \pi) \in \At \times I^Y} \alpha \land Y=\pi\seq Y\leftarrow \pi \seq e & \eqref{guardedness+}, \RefAx{\Axtestcomb}, \BA\\
		&\equiv \bigplus_{(\alpha, \pi) \in \At \times I^Y} Y\leftarrow \pi \seq e & \RefAx{\Axassoc}, \eqref{guardedness+}\\
		&\equiv \bigplus_{Y=\pi \mid \pi \in I^Y} Y\leftarrow \pi \seq e & \eqref{ggdedsumprodinj}
	\end{align*}
	Hence by \RefAx{\Axweaken}, and the last two facts
	\begin{align*}
		E \vdash e &\equiv \bigplus_{\pi \in I^Y} Y\leftarrow \pi \seq e \dashv E\\
		&\equiv \bigplus_{\pi \in I^Y} Y\leftarrow \pi \seq f \dashv F & \RefAx{\Axcongguardedchoice}\\
		&\equiv f \dashv F
	\end{align*}
	where the last step follows from an analogous derivation for \(f\).
\end{appendixproof}

\section{Case Studies}\label{sec:examples}
The individual axioms we have chosen for CF-GKAT are fairly intuitive, but powerful enough to prove any fact that is semantically true.
In this section, we demonstrate the use of our framework, ultimately deriving validity of the Erosa--Hendren loop transformation (\Cref{fig:eh-loop-transform}), which involves both local and non-local control flow.

\subsection{Useful Facts}\label{sec:useful-facts}
Before we get started on proving the correctness of the Erosa--Hendren loop transformation, however, it pays off to derive a number of helpful properties, which are also applicable in derivations beyond the objective for this section; these are laid out in \Cref{fig:eh-useful-facts}, and can be explained as follows:
\begin{itemize}[leftmargin=*]
    \item
    \Cref{looptighten_invariant} says that a non-productive part of a loop (which gets stuck, and implicitly rejects) can be ignored.
    In particular, it holds when $C$ is the \emph{largest} invariant of $e$ below $b$.
    \item
    \Cref{loopuniqueafterlargest} is, in essence, a special case of \RefAx{\Axuniqueness}, which says that if $g$ and $h$ can be written using similarly structured (guarded) recursion, then they must be equivalent.
    \item
    \Cref{loopseqdnr} and \Cref{assignmentduplication} concern programs that assign a variable only once, at the beginning; they say that this assignment can be repeated later on without changing its meaning.
    \item
    \Cref{partialselection}, \Cref{assignmentbranchselectiontrue} and \Cref{assignmentbranchselectionfalse} are properties that can be proven in \GKAT alone, relating to the interaction between Boolean assertions and branching.
    \item
    \Cref{assignmentprefixenvironment} says that, if the value of $x$ when jumping out of $h$ is $i$, and the environment never changes $x$, then the programs in the environment can start by assigning $i$ to $x$.
    \item
    \Cref{indremoval}, also known as \emph{ghost variable elimination}, allows us to eliminate an indicator variable $x$, if it never influences the control flow and is not observable as an output.
    \item
    Finally, \Cref{iteratedenvironmentreplacement} allows us to modify the environment using laws that do not themselves depend on it; e.g., we can apply laws such as \RefAx{\Axidempotence} in the environment with impunity.
\end{itemize}

To prove these facts we can reason algebraically; for instance, \Cref{loopuniqueafterlargest} can be proved by pure algebraic derivation, roughly as follows.
First, we recursively define the sequence of expressions $R_n$ given by $R_0 = \comGoto~\ell$ (for a fresh $\ell$) and $R_{n+1} = (\neg C^+; e) \loopseq R_n +_b r$.
By repeatedly applying \RefAx{\Axsubstitution}, \RefAx{\Axweaken}, and the premise that $g \equiv (\neg C^+; e) \loopseq g +_b r$ one can show that $\ell = g \vdash \comGoto~\ell \equiv R_n \dashv \ell = g$, for all $n$.
Because $C^+$ is the largest invariant of $e$, there must be a sufficiently large $n$ such that $\comGoto~\ell$ is guarded in $R_n$.
If we then choose $E(\ell) = R_n$, we can use $\ell = g \vdash \comGoto~\ell \equiv R_n \dashv \ell = g$ to derive:
\[
	g 
	\hspace{1em}\stackrel{\RefAx{\Axredundancy}}{\equiv}\hspace{1em}
	g \dashv \ell = g 
	\hspace{1em}\stackrel{\RefAx{\Axsubstitution}}{\equiv}\hspace{1em}
	\comGoto~\ell \dashv \ell = g 
	\hspace{1em}\stackrel{\RefAx{\Axuniqueness}}{\equiv}\hspace{1em} 
	\comGoto~\ell \dashv E 
\]
Analogously, $\varnothing \vdash h \equiv \comGoto~\ell \dashv E$; thus $\varnothing \vdash g \equiv h \dashv \varnothing$ ($g \equiv h$) follows by transitivity.

When a claim is predicated on some kind of syntactic condition, for instance pertaining to whether or not a variable is assigned, the better approach is often to reason semantically, by constructing a bisimulation between the relevant terms.
For example, to prove \Cref{indremoval}, it suffices to find a bisimulation witnessing the following for all terms $f$ and environments $E$ which (1)~are well nested, (2)~end with \comReturn, (3)~do not test $x$, and (4)~have no free \comGoto:
\[
    (f,\pi)\resWith{E}
    \sim
    (f[1/x\assign j \mid j \in I],\pi[x\leftarrow i])\resWith{E[1/x\assign j \mid j \in I]}.
\]
While verifying this claim involves a bit of bookkeeping, it is quite doable with the help of a proof assistant.
We can then instantiate it with $f = e$, and set $i = \pi(x)$ to find
\[
    (e,\pi)\resWith{E}
    \sim
    (e[1/x\assign j \mid j \in I],\pi)\resWith{E[1/x\assign j \mid j \in I]},
\]
which by completeness (\Cref{completeness}) allows us to conclude the claim made by \Cref{indremoval}.

A semantic proof can be more compact when a purely algebraic argument would require a careful application of \RefAx{\Axuniqueness} to deal with loops (e.g., in \Cref{loopseqdnr}).
More commonly, however, bisimulation-based proofs help when a side condition does not lend itself to inductive argumentation.
\Cref{indremoval} is a prime example: the first condition may be true for $e$ itself, but not all of its subterms.
Nevertheless, all side conditions are maintained by derivatives, and therefore \Cref{indremoval} lends itself well to a coinductive (bisimilarity-based) argument.
This highlights the added value of completeness: \emph{it lets us package difficult coinductive arguments as compact equations that can later be used syntactically}.

\begin{figure*}[t]
\begingroup
\newcommand{\ehtag}[2]{%
	\begingroup
	\renewcommand{\theehfact}{#1}%
	\refstepcounter{ehfact}\label{#2}%
	\textup{(\RocqLinkedLabelText{#2}{#1})}%
	\endgroup}
\newcommand{\eheq}[2]{\begin{array}{@{}r@{\;}c@{\;}l@{}}#1&\equiv&#2\end{array}}
\newcommand{\ehctxeq}[4]{\begin{array}{@{}r@{\;}c@{\;}l@{}}#1\vdash #2&\equiv&#3\dashv #4\end{array}}
\newcommand{\ehprem}[1]{\begin{array}{@{}c@{}}#1\end{array}}
\newcommand{\ehrule}[4]{%
	\displaystyle\inferrule{#1}{#2}\;\ehtag{#4}{#3}}
\newcommand{\ehheading}[1]{{\centering\textbf{#1}\par}\vspace{0.3ex}}
\newcommand{\ehrow}[2]{\noindent\hspace*{\fill}#1\hspace{\fill}#2\hspace*{\fill}\par}
\newcommand{\ehgap}{\par\vspace{0.5ex}}
\renewcommand{\arraystretch}{1.05}
\mprset{fractionaboveskip=0.35ex,fractionbelowskip=0.35ex}
\ehheading{Loop/fixpoint}
\ehrow{\(\ehrule{
	\ehprem{C\text{ is an invariant of }e\\\text{with }C\leqBA b}%
}{
\eheq{e^b}{(\lnot C\seq e)^b}
}{looptighten_invariant}{EH1}\)}{
\(\ehrule{
	\ehprem{
	C^+\text{ is the greatest invariant of }e\text{ with }C^+\leqBA b\\
	((\lnot C^+\seq e)\loopseq g)+_b r\equiv g\\
	((\lnot C^+\seq e)\loopseq h)+_b r\equiv h}
}{
\eheq{g}{h}
}{loopuniqueafterlargest}{EH2}\)}
\vspace{0.5ex}
\ehheading{Indicators}
\ehrow{\(\ehrule{
	e\text{ does not assign }x
}{
\eheq{(x\assign i\seq e)\loopseq f}
	{(x\assign i\seq e)\loopseq(x\assign i\seq f)}
}{loopseqdnr}{EH3}\)}{
\(\ehrule{
	e\text{ does not assign }x
}{
\eheq{x\assign i\seq e}{x\assign i\seq(e\seq x\assign i)}
}{assignmentduplication}{EH4}\)}
\vspace{0.5ex}
\noindent
\begin{minipage}[t]{0.375\linewidth}%
\ehheading{Branch selection}
\centering
\(\ehrule{
	d\land c\eqBA d\land c'
}{
\eheq{d\seq(e+_c f)}{d\seq(e+_{c'} f)}
}{partialselection}{EH5}\)
\ehgap
\(\ehrule{
	x=i\land c\eqBA x=i
}{
\eheq{x\assign i\seq(e+_c f)}{x\assign i\seq e}
}{assignmentbranchselectiontrue}{EH6}\)
\ehgap
\(\ehrule{
	x=i\land c\eqBA0
}{
\eheq{x\assign i\seq(e+_c f)}{x\assign i\seq f}
}{assignmentbranchselectionfalse}{EH7}\)
\end{minipage}\hfill
\begin{minipage}[t]{0.615\linewidth}%
\ehheading{Environments}
\centering
\(\ehrule{
	\ehprem{
  \forall k.\ E(k) \text{ assigns only } i \text{ to } x\hspace{1.2em}%
  h \text{ does not assign } x%
}
}{
\eheq{E\vdash x\assign i \seq h}{x \assign i \seq h\dashv(x\assign i\seq E)}
}{assignmentprefixenvironment}{EH8}\)
\ehgap
\(\ehrule{
	\ehprem{
	e,E\text{ are well nested and end with }\comReturn\\
	e,E\text{ do not test }x\text{ and have no free }\comGoto
	}
}{
\ehctxeq{E}{e}{e[1/x\assign j \mid j \in I]}{E[1/x\assign j \mid j \in I]}
}{indremoval}{EH9}\)
\ehgap
\(\ehrule{
	\forall k.\ E(k)\equiv F(k)
}{
\ehctxeq{E}{h}{h}{F}
}{iteratedenvironmentreplacement}{EH10}\)
\end{minipage}
\endgroup
\caption{Useful derived facts for the Erosa--Hendren loop case.
We interpret operations on environments pointwise, e.g., \(e\seq E\) denotes the environment \(k \mapsto e\seq E(k)\).
Conditions on environments are also pointwise; e.g., \(E\) ends with return when \(E=F\seq\comReturn\) for some environment \(F\).
A term is \emph{well nested} if \comBreak and \comGoto appear only in loops, or inside the left operand of $\loopseq$ \RocqLabelRef{wellnesteddef}.
\(E(k)\equiv F(k)\) means that both sides are undefined, or both are defined and equivalent.}
\label{fig:eh-useful-facts}
\end{figure*}

\begin{toappendix}
\begin{lemma}[\RocqLabelRef{semanticlemmaindrem}]\label{semanticlemmaindrem}
	Suppose \(e\) does not contain a test referencing \(x\). Let \(e_x \assign e[1/x\assign j \mid j \in I]\), that is all assignments \(x\assign j\) are replaced by \(1\) for all \(j\), and
	\begin{align*}
		(e, \pi) \tr{p}{\alpha} (f, \sigma) &\implies \forall i. (e_x, \pi[x\leftarrow i]) \tr{p}{\alpha} (f_x, \sigma[x \leftarrow i])\\
		(e, \pi) \tr{}{\alpha} c \in \{\bot, \retc\} &\implies \forall i.(e_x, \pi[x\leftarrow i]) \tr{}{\alpha} c\\
		(e, \pi) \tr{}{\alpha} (c, \sigma) &\implies \forall i.(e_x, \pi[x\leftarrow i]) \tr{}{\alpha} (c, \sigma[x \leftarrow i])
	\end{align*}
\end{lemma}
\begin{proof}
	We prove this by structural induction on \(e\). 
	
	\paragraph{Base Cases}
	For all base terms other than \(x \assign i\) the proof is trivial as \(e_x = e\), and no such term affects the value of \(x\).
	For \(x \assign j\):
	\begin{align*}
		(x\assign j, \pi) & \tr{}{\alpha} (\acc, \pi[x\leftarrow j])\\
		(1, \pi[x\leftarrow i]) & \tr{}{\alpha} (\acc, \pi[x\leftarrow i])
	\end{align*}
	because \(\pi[x \leftarrow j][x \leftarrow i] = \pi[x \leftarrow i]\), so it satisfies the required implications.
	
	\paragraph{Guarded Choice}
	Suppose \((\alpha, \pi) \in \langle b \rangle\), then \(b\) does not reference \(x\). Hence
	\begin{align*}
		(e+_b f, \pi) \tr{}{\alpha} r &\implies (e, \pi) \tr{}{\alpha} r\\
		&\implies (e_x, \pi[x\leftarrow i]) \tr{}{\alpha} r[x\leftarrow i] & \text{ind hyp}\\
		&\implies (e_x+_b f_x, \pi[x\leftarrow i]) \tr{}{\alpha} r[x\leftarrow i]
	\end{align*}
	The argument for \((\alpha, \pi) \in \langle \lnot b \rangle\) is symmetric.
	
	\paragraph{Sequential Composition}
	Suppose \((e, \pi) \tr{}{\alpha} r \neq (\acc, \sigma)\), then
	\begin{align*}
		(e\seq f, \pi) \tr{}{\alpha} r &\iff (e, \pi) \tr{}{\alpha} r\\
		&\iff (e_x, \pi) \tr{}{\alpha} r[x\leftarrow i] & \text{ind hyp}\\
		&\iff (e_x \seq f_x, \pi) \tr{}{\alpha} r[x\leftarrow i]
	\end{align*}
	where \(r[x\leftarrow i]\) is \((p, (e', \pi'[x \leftarrow i]))\) if \(r \eqdef (p, (e', \pi'))\); \(r[x\leftarrow i]\) is \((c, \pi'[x \leftarrow i])\) if \(r \eqdef (c, \pi')\); and \(r[x \leftarrow i]\) is \(r\) otherwise.
	
	Suppose \((e, \pi) \tr{}{\alpha} (\acc, \sigma)\), then by the inductive hypothesis \((e_x, \pi[x\leftarrow i]) \tr{}{\alpha} (\acc, \sigma[x\leftarrow i])\)
	\begin{align*}
		(e\seq f, \pi) \hookrightarrow_\alpha r &\iff (f, \sigma) \hookrightarrow_\alpha r\\
		&\iff (f_x, \sigma[x\leftarrow i]) \hookrightarrow_\alpha r[x\leftarrow i] & \text{ind hyp}\\
		&\iff (e_x \seq f_x, \pi[x\leftarrow i]) \hookrightarrow_\alpha r
	\end{align*}
	Suppose \((e, \pi) \tr{p}{\alpha} (g, \sigma)\), then by the inductive hypothesis \((e_x, \pi[x\leftarrow i]) \tr{}{\alpha} (g_x, \sigma[x\leftarrow i])\). In this case
	\begin{align*}
		(e\seq f, \pi) \tr{p}{\alpha} (g\seq f, \sigma) &\iff 
		(e_x \seq f_x, \pi[x\leftarrow i]) \tr{}{\alpha} (g_x \seq f_x, \sigma[x\leftarrow i]) & \tCref{operationalsemantics}
	\end{align*}
	
	\paragraph{Loop Composition}
	Suppose \((e, \pi) \tr{}{\alpha} (c, \sigma),\) s.t. \(c \in \{\bot, \retc\}\), then
	\begin{align*}
		(e\loopseq f, \pi) \tr{}{\alpha} r
		\iff & (e, \pi) \tr{}{\alpha} r\\
		\iff & (e_x, \pi[x \leftarrow i]) \tr{}{\alpha} r[x\leftarrow i] & \text{ind hyp}\\
		\iff & (e_x \loopseq f_x, \pi[x \leftarrow i]) \tr{}{\alpha} r[x\leftarrow i]
	\end{align*}
	Suppose \((e, \pi) \tr{}{\alpha} (c, \sigma)\) where \(c \in \{\acc, \contc\}\), then by the inductive hypothesis \((e_x, \pi[x\leftarrow i]) \tr{}{\alpha} (c, \sigma[x\leftarrow i])\)
	\begin{align*}
		(e\loopseq f, \pi) \hookrightarrow_\alpha r
		\iff& (f, \sigma) \hookrightarrow_\alpha r\\
		\iff& (f_x, \sigma[x\leftarrow i]) \hookrightarrow_\alpha r[x\leftarrow i] & \text{ind hyp}\\
		\iff& (e_x \loopseq f_x, \pi[x\leftarrow i]) \hookrightarrow_\alpha r[x \leftarrow i]
	\end{align*}
	Suppose \((e, \pi) \tr{p}{\alpha} (g, \sigma)\), then by the inductive hypothesis and operational semantics:
	\begin{align*}
		\MoveEqLeft[1] (e_x, \pi[x\leftarrow i]) \tr{p}{\alpha} (g_x, \sigma[x\leftarrow i]) \\
		&\iff (e_x \loopseq f_x, \pi[x\leftarrow i]) \tr{}{\alpha} (g_x \loopseq f_x, \sigma[x\leftarrow i]) 
	\end{align*}%
	Suppose \((e, \pi) \tr{}{\alpha} (\brkc, \sigma)\), then by the inductive hypothesis \((e_x, \pi[x\leftarrow i]) \tr{}{\alpha} (\brkc, \sigma[x\leftarrow i])\), and by operational semantics, the only possible result satisfies the required implication:
	\begin{gather*}
		(e \loopseq f, \pi[x\leftarrow i]) \tr{}{\alpha} (\acc, \sigma[x\leftarrow i])
		\\ \text{and}\quad
		(e_x \loopseq f_x, \pi[x\leftarrow i]) \tr{}{\alpha} (\acc, \sigma[x\leftarrow i])
	\end{gather*}%
	
	\paragraph{Loop}
	Suppose \((\alpha, \pi) \in \langle \lnot b \rangle\), then
	\[(e^{(b)}, \pi) \tr{}{\alpha} (\acc, \pi) \quad\text{and by \Cref{indnotrefbool}}\quad (e^{(b)}_x, \pi[x\leftarrow i]) \tr{}{\alpha} (\acc, \pi[x\leftarrow i]).\]%
	For the remaining cases we suppose \((\alpha, \pi) \in \langle \lnot b \rangle\).

	Suppose \((e, \pi) \tr{}{\alpha} r\) for \(r \in \{\retc, \bot\}\), then
	\begin{align*}
		(e^{(b)}, \pi) \tr{}{\alpha} r &\iff (e, \pi) \tr{}{\alpha} r\\
		&\iff (e_x, \pi) \tr{}{\alpha} r[x\leftarrow i] & \text{ind hyp}\\
		&\iff (e_x^{(b)}, \pi) \tr{}{\alpha} r[x\leftarrow i]
	\end{align*}
	Suppose \((e, \pi) \tr{p}{\alpha} (g, \sigma)\), then by the inductive hypothesis \((e_x, \pi[x\leftarrow i]) \tr{}{\alpha} (g_x, \sigma[x\leftarrow i])\). In this case the only possible result satisfies the required implication:
	\begin{align*}
		(e^{(b)}, \pi) & \tr{p}{\alpha} (g\loopseq e^{(b)}, \sigma) \text{ and} \\
		(e_x^{(b)}, \pi[x\leftarrow i]) & \tr{}{\alpha} (g_x \loopseq e_x^{(b)}, \sigma[x\leftarrow i])
	\end{align*}
	Suppose \((e, \pi) \tr{}{\alpha} (\brkc, \sigma)\), then by the inductive hypothesis \((e_x, \pi[x\leftarrow i]) \tr{}{\alpha} (\brkc, \sigma[x\leftarrow i])\), and the only possible result satisfies the required implication:  
	\begin{align*}
		(e^{(b)}, \pi) & \tr{}{\alpha} (\acc, \sigma) \text{ and} \\
		 (e_x^{(b)}, \pi[x\leftarrow i]) & \tr{}{\alpha}(\acc, \sigma[x\leftarrow i])
	\end{align*}
	Suppose \((e, \pi) \tr{}{\alpha} (c, \sigma)\) where \(c \in \{\acc, \contc\}\), then by the inductive hypothesis \((e_x, \pi[x\leftarrow i]) \tr{}{\alpha} (\contc, \sigma[x\leftarrow i])\). So \((e^{(b)}, \pi) \hookrightarrow_\alpha r \iff (e^{(b)}, \sigma) \hookrightarrow_\alpha r\), and \((e_x^{(b)}, \pi[x\leftarrow i]) \hookrightarrow_\alpha r \iff (e_x^{(b)}, \sigma[x\leftarrow i]) \hookrightarrow_\alpha r\). By a straightforward induction either these sequences terminate in one of the above cases in which case by transitivity we have the desired result, or they do not, in which case both transition to \(\bot\).

\end{proof}
\end{toappendix}
\begin{toappendix}
\begin{lemma}[\RocqLabelRef{nestinglem}] \label{nestinglem}
	Well-nested terms paired with assignments form a subcoalgebra of \((\Expm \times I^Y)\) such that for any well-nested term \(e\) \(\forall (\alpha, \pi) \in \At \times I^Y\; (e, \pi) \hookrightarrow_\alpha r \not\in \{\contc, \brkc\} \times I^Y\).

\end{lemma}
\begin{proof}
	We prove this by induction on the grammar. All terms in \(\Expm\) which do not contain \(\comCont\) and \(\comBreak\) satisfy the desired property by a similar argument to \Cref{notransitionnogoto}.
	
	Consider \(f \loopseq w\). If \((f, \pi) \hookrightarrow_\alpha r\) such that \(r \not\in \{\acc, \contc, \brkc\} \times I^Y\), then so does \(f \loopseq w\). If \((f, \pi) \tr{}{\alpha} (\acc, \sigma)\), then by the inductive hypothesis \((w, \sigma) \hookrightarrow r \not\in \{\contc, \brkc\} \times I^Y\). Hence by the operational semantics \((f \loopseq w, \pi) \hookrightarrow r\). A similar argument holds if \((f, \pi) \tr{}{\alpha} (\contc, \sigma)\). If \((f, \pi) \tr{}{\alpha} (\brkc, \sigma)\), then  \((f \loopseq w, \pi) \tr{}{\alpha} (\acc, \sigma)\).
	
	If \((f\loopseq w, \pi) \tr{p}{\alpha} (e', \sigma)\) then either \(e' = f' \loopseq w\) in which case the term is still well nested because \(w\) is, or \((f, \pi) \tr{}{\alpha} (\acc, \nu)\) and \((w, \nu) \tr{p}{\alpha} (w^\prime, \sigma)\), in which case by the inductive hypothesis \(w^\prime\) is well nested.
	
	Consider \(f^{(b)}\). If \((\alpha, \pi) \in \langle \lnot b\rangle\), then \((f^{(b)}, \pi) \tr{}{\alpha} (\acc, \pi)\). For the following cases we suppose \((\alpha, \pi) \in \langle b\rangle\). If \((f, \pi) \tr{}{\alpha} r \not\in \{\acc, \contc, \brkc\} \times I^Y\), then so does \((f^{(b)}, \pi)\). If \((f, \pi) \tr{}{\alpha} (\brkc, \sigma)\), then  \((f^{(b)}, \pi) \tr{}{\alpha} (\acc, \sigma)\). If \((f, \pi) \tr{p}{\alpha} (f', \sigma)\), then  \((f^{(b)}, \pi) \tr{p}{\alpha} (f' \loopseq f^{(b)}, \sigma)\) which is well nested. Finally, if \((f, \pi) \tr{}{\alpha} (c, \sigma)\) with \(c \in \{\acc, \contc\}\), then there is either an infinite sequence of \(\pi_i\) with \(\pi_1= \sigma\) and \((f, \pi_i) \tr{}{\alpha} (c_{i+1}, \pi_{i+1})\) where \(c_{i+1} \in \{\acc, \contc\}\), or there exists some \(n\) such that \((f, \pi_n) \hookrightarrow r \notin \{\acc, \contc\} \times I^Y\). If it is infinite, then \((f^{(b)}, \pi) \tr{}{\alpha} \bot\). If the sequence is finite, then \((f^{(b)}, \pi) \tr{}{\alpha} r\) for \(r\) such that the sequence ends with \(\pi_n\) and \((f^{(b)}, \pi_n) \hookrightarrow r \notin \{\acc, \contc\} \times I^Y\) which satisfies the lemma statement by the previous cases of loop.
	
	 For \(e = w\seq v\), if \((e, \pi) \tr{p}{\alpha} (e', \sigma)\), then either \((w, \pi) \tr{}{\alpha} (w', \sigma)\) and \(e' = w' \seq v\) or \((w, \pi) \tr{}{\alpha} (\acc, \nu)\), \((v, \nu) \tr{p}{\alpha} (v', \sigma)\), and \(e' = v'\). By the inductive hypothesis \(v'\) is well nested. Similarly \(w' \seq v\) is well nested by the inductive hypothesis and the fact that two sequenced well-nested terms are also well nested. We note that by the inductive hypothesis \((w, \pi) \tr{}{\alpha} c \implies c \notin \{\contc, \brkc\} \times I^Y\), \(\forall \beta, \sigma. (v, \sigma) \tr{}{\beta} c \implies c \notin \{\contc, \brkc\} \times I^Y\). Hence by the operational semantics \((e, \pi) \tr{}{\alpha} c \implies c \notin \{\contc, \brkc\} \times I^Y\).
	 
	 For \(e = w +_b v\), w.l.o.g.  suppose \((\alpha, \pi) \in \langle b \rangle\), then \((e, \pi) \hookrightarrow_\alpha r \iff (w, \pi) \hookrightarrow_\alpha r\) such that \(r\) satisfies the desired property by the inductive hypothesis.
\end{proof}
\end{toappendix}
\begin{toappendix}
\begin{proof}[of \Cref{indremoval}]
		Let $E_x \coloneq E[1/x\assign j \mid j \in I]$.

		We prove this by arguing that the relation \((f, \pi)\resWith{E} \sim (f_x, \pi[x\leftarrow i])\resWith{E_x}\) for all \(i\) and \((f,\pi)\) in \(\langle e, \sigma\rangle\resWith{E}\) for some \(\sigma\), is an intermediate bisimulation. We then invoke \Cref{completeness}.
		
		By \Cref{nestinglem} and \Cref{notransitionnogoto} for any \((f, \sigma)\) in \(\langle e, \pi \rangle\resWith{E}\) for any \(\pi\), \((f, \sigma) \not\tr{}{\alpha} (c, \nu)\)  where \(c \in \{\brkc, \contc\} + \{\jmpc~\ell \mid \ell \in \lset\}\) for any \(\nu\). 
		
		Hence \((f, \sigma) \hookrightarrow_\alpha \bot/\retc/(\jmpc~\ell\in\text{dom}(E), \nu)/(p, (f', \nu))\), since the term ends in return. 
		
		If \((f, \pi) \tr{}{\alpha} c \in \{\bot, \retc\}\) then by \Cref{resolutiondef} \((f, \pi) \tr[E]{}{\alpha}
 c\). Moreover by \Cref{semanticlemmaindrem}, \((f_x, \pi[x\leftarrow i]) \tr[E_x]{}{\alpha}
 c\). 
		
		If \((f, \pi) \tr{p}{\alpha} (f', \sigma)\) then by \Cref{resolutiondef} \((f, \pi) \tr[E]{p}{\alpha} (f', \sigma)\). By \Cref{semanticlemmaindrem} \((f_x, \pi[x\leftarrow i]) \tr[E_x]{p}{\alpha}
 (f_x', \sigma[x\leftarrow i])\), and these two terms are related by \(\sim\).
		
			If \((f,\pi) \tr{}{\alpha} (\jmpc~\ell, \sigma)\), then \(\ell \in \text{dom}(E)\) and by \Cref{resolutiondef} \((f,\pi) \transSilent_\alpha^E (E(\ell), \sigma)\). By \Cref{semanticlemmaindrem}, \((f_x, \pi) \transSilent_\alpha^{E_x} (E(\ell)_x, \sigma[x\allowbreak\leftarrow i])\) and these two states are related by \({\sim}\).
\end{proof}
\end{toappendix}

\subsection{The Erosa--Hendren Loop Case}\label{sec:eh-loop}
\label{sec:mechanized-eh-loop}
\newcommand{\ehreason}[1]{\tag*{\(\displaystyle #1\)}}

We now turn our attention to our main case study: the transformation described in \Cref{fig:eh-loop-transform}.
This is one of several steps in an algorithm due to Erosa and Hendren~\cite{taming}, aimed at eliminating all occurrences of \comGoto from a program by inserting auxiliary variables, which is achieved by pushing each \comGoto through the structure of the program until it can be eliminated.
To demonstrate that \CFGKAT is powerful enough to verify these transformations, we verify the correctness of this particular step.
More precisely, we will prove the following equivalence in \CFGKAT:
\begin{align*}
&(\comGoto~\ell+_c1)\seq \hat e\seq (\hat f\seq\comLabel~\ell\seq \hat g)^b\seq\comReturn \\
&\progeq (x\assign1+_c x\assign0)\seq(\hat e +_{x=0}1) \\
&\hphantom{{}\progeq{}}\seq \bigl((\comGoto~\ell+_{x=1}1)\seq \hat f\seq \comLabel~\ell\seq x\assign0\seq \hat g\bigr)^{(b\lor x=1)} \seq\comReturn .
\end{align*}%
We will write the reduced forms of \(\hat e\), \(\hat f\), and \(\hat g\) as \(e\), \(f\), and \(g\) respectively.
In the sequel we will call the source program \(P_s\), and the target \(P_t\).

Before we proceed, we must make some minor assumptions.
First, observe that we have appended \comReturn to both programs.
This is necessary to make the equivalence ignore the value of $x$ upon termination.
Second, we assume that \(x\) is fresh in \(c,b,e,f,g\), that is, that none of the constituent programs reference \(x\) in any way.
Finally, we assume that neither program has any free \comGoto or duplicate labels, and that both are well nested.

With these assumptions in place, we can refine our goal by stating it in our calculus, as follows: \(\sigma(P_s) \vdash (P_s)_{-} \equiv (P_t)_{-} \dashv \sigma(P_t)\).
Before delving into our algebraic proof, we pause for a moment to consider the alternatives.
For instance, one might ask why we do not invoke the decision procedure from Zhang et al.~\cite{cfgkat}.
The problem is that this decision procedure checks fully instantiated programs, whereas our goal quantifies over all program fragments \(e,f,g\) and tests \(b,c\).
What we want to prove is therefore a metatheorem about the transformation, not a check of one particular instance.

Complicating matters further, and in contrast with regular expressions, such metatheorem cannot be reduced to a claim about concrete programs 
(e.g., with actions $p$, $q$, and $r$ in place of $e$, $f$ and $g$).
The reason is that \CFGKAT equivalence is not closed under arbitrary substitution: as discussed in \Cref{sec:axioms}, substitution and congruence must respect rightmostness and environment side conditions.
A symbolic decision procedure for such parametric terms might be possible, but neeeds to account for these substitution and context restrictions, and would not expose the structure of the transformation.

Alternatively, one might consider a semantic proof, as was done for \Cref{indremoval}.
However, such an approach would produce a large bisimulation relation whose cases obscure the refinement steps.
This method becomes increasingly difficult as the transformation grows, and the resulting proof is not modular in the same way as an algebraic derivation.

Our strategy is instead hybrid, in that we use semantically derived facts such as \Cref{indremoval,assignmentduplication} as lemmas inside an algebraic proof.
The resulting derivation exposes the structure of the transformation, and can be divided into three steps:
\begin{enumerate*}[label={(\arabic*)}]
    \item
    we first show that both loops agree on all runs that start with $x$ equal to $0$ (\Cref{ehzeroentryloopequiv});
    \item
    we refactor the reduced target and source expressions to a common form (\Cref{ehmeetinthemiddle});
    \item
    we align the two environments, completing the proof (\Cref{erosa-hendren-loop-case}).
\end{enumerate*}

We write $B$ and $L$ for the (reduced) bodies and loops of the source and target:
\[
B_t=(\comGoto~\ell+_{x=1}1)\seq f\seq x\assign0\seq g,
\qquad
L_t=B_t^{(b\lor x=1)},
\qquad
L_s=(f\seq g)^b
\]
With these, we write the reduced source program $(P_s)_-$ as \((\comGoto~\ell+_c1)\seq e\seq L_s\seq\comReturn\), and the reduced target program $(P_t)_-$ as \((x\assign1+_c x\assign0)\seq(e+_{x=0}1)\seq
  L_t\seq \comReturn\). %

We begin with the first step, which is to argue that the loops behave the same when they are started from a configuration where $x$ is $0$.
Formally, this can be shown as follows.
\begin{lemma}[Zero-entry loop equivalence \RocqLabelRef{ehzeroentryloopequiv}]\label{ehzeroentryloopequiv}
The two loops agree when started with $x$ equal to $0$ --- that is to say, \(x\assign0\seq L_s \equiv x\assign0\seq L_t\).
\end{lemma}
\begin{proof}
We will prove this via an application of \Cref{loopuniqueafterlargest} to the source and target loops.
Specifically, we will show that both $x \assign 0 \seq L_s$ and $x \assign 0 \seq L_t$ are solutions to $y$ in 
\begin{equation}\label{eq:source-target-fixpoint}
y \equiv \bigl((x\assign0\seq \lnot C_s^+\seq f\seq g) \loopseq y\bigr)+_b x\assign0
\end{equation}
where \(C_s^+\) is the largest invariant for \(f\seq g\) below \(b\).
We first expose the source loop:
\begin{align}
x\assign0\seq L_s
&\equiv
x\assign0\seq(\lnot C_s^+\seq f\seq g)^b
\ehreason{\tCref{looptighten_invariant}}\\
&\equiv
x\assign0\seq
\bigl((\lnot C_s^+\seq f\seq g)\loopseq
(\lnot C_s^+\seq f\seq g)^b+_b1\bigr)
\ehreason{\RefAx{\Axloopunfold}}\\
&\equiv
x\assign0\seq
\bigl((\lnot C_s^+\seq f\seq g)\loopseq
(\lnot C_s^+\seq f\seq g)^b\bigr)
+_b x\assign0
\ehreason{\eqref{inddistr},\RefAx{\Axneutright}}
\intertext{%
    Focusing on the left branch of the conditional, we can derive:
}
&x\assign0\seq
\bigl((\lnot C_s^+\seq f\seq g)\loopseq
(\lnot C_s^+\seq f\seq g)^b\bigr)\tag*{} \\
&\equiv
(x\assign0\seq \lnot C_s^+\seq f\seq g)
\loopseq(\lnot C_s^+\seq f\seq g)^b
\ehreason{\RefAx{\Axloopassocassign}}\\
&\equiv
(x\assign0\seq \lnot C_s^+\seq f\seq g)
\loopseq (x\assign0\seq(\lnot C_s^+\seq f\seq g)^b)
\ehreason{\tCref{loopseqdnr}}\\
&\equiv
(x\assign0\seq \lnot C_s^+\seq f\seq g)
\loopseq (x\assign0\seq L_s)
\ehreason{\tCref{looptighten_invariant}}
\intertext{%
    Putting this together, we can conclude that $x \assign 0 \seq L_s$ solves for $y$ in \eqref{eq:source-target-fixpoint}, i.e.:
}\label{eq:source-zero-fixed}
x\assign0\seq L_s
&\equiv
\bigl((x\assign0\seq \lnot C_s^+\seq f\seq g)
\loopseq (x\assign0\seq L_s)\bigr)+_b x\assign0
\end{align}
Let \(C_t^+\) be the largest invariant for \(B_t\) below \(b\lor x=1\).
To achieve a similar property for $x \assign 0 \seq L_t$, it helps to prove that $C^+_s$ and $C^+_t$ are equivalent, given that $x$ is $0$.
To this end, first note that one execution of \(B_t\) after $x \assign 0$ is just one execution of \(f\seq g\), that is:
\begin{align}\label{eq:zero-entry-body}
	x\assign0\seq B_t &\equiv x\assign0\seq f\seq g
\intertext{%
    We can derive this as follows:
}
x\assign0\seq B_t
&\equiv
x\assign0\seq
(1+_{\lnot(x=1)}\comGoto~\ell)\seq f\seq x\assign0\seq g
\ehreason{\RefAx{\Axskewcomm}}\\
&\equiv
x\assign0\seq f\seq x\assign0\seq g
\ehreason{\tCref{assignmentbranchselectiontrue},\BA, \RefAx{\Axneutleft}}\\
&\equiv
x\assign0\seq f\seq g
\ehreason{\tCref{assignmentduplication}}
\end{align}
With this in hand, we argue that \((x=0)\land C_t^+\eqBA(x=0)\land C_s^+\), by maximality.

Since \(C_t^+\leqBA b\lor x=1\) and \(1\neq0\), it follows that \((x=0)\land C_t^+ \leqBA b\).
Moreover \((x=0)\land C_t^+\) is an invariant for the source body \(f\seq g\): from a state satisfying \((x=0)\land C_t^+\), $B_t$ takes an accepting or continuing step that returns to \(C_t^+\), which \eqref{eq:zero-entry-body} identifies with a step of \(f\seq g\).
Also, because \(f\) and \(g\) do not reference \(x\), the resulting assignment still satisfies \(x=0\), per \Cref{assignmentduplication}.
Hence \((x=0)\land C_t^+\) is an invariant of the source loop below \(b\), and by maximality \((x=0)\land C_t^+\leqBA C_s^+\). 
Because \((x=0)\land C_t^+ \leqBA (x=0)\), it then follows that \((x=0)\land C_t^+ \leqBA (x=0) \land C_s^+\).

Conversely, since \((x=0)\land C_s^+ \leqBA b\), we have \((x=0)\land C_s^+ \leqBA b\lor x=1\).
By definition of $C^+_s$, from a state satisfying $(x = 0) \land C_s^+$, \(f\seq g\) takes an accepting or continuing step to \(C_s^+\); \eqref{eq:zero-entry-body} implies that the target body makes the same step, and the no-reference assumptions again preserve \(x=0\).
Thus \((x=0)\land C_s^+\) is an invariant of \(B_t\) below \(b\lor x=1\), so maximality gives \((x=0)\land C_s^+ \leqBA C_t^+\), and hence \((x=0)\land C_s^+ \leqBA (x=0) \land C_t^+\).
Put together, the two inequalities yield
\begin{equation}\label{eq:zero-invariants-agree}
x=0\land C_t^+\eqBA x=0\land C_s^+
\end{equation}
We now expose the target loop as a fixed point, bringing out the body and invariant:
\begin{align*}
x\assign0\seq L_t
&\equiv
x\assign0\seq(\lnot C_t^+\seq B_t)^{(b\lor x=1)}
\ehreason{\tCref{looptighten_invariant}}\\
&\equiv
x\assign0\seq
\bigl((\lnot C_t^+\seq B_t)\loopseq
(\lnot C_t^+\seq B_t)^{(b\lor x=1)}+_{b\lor x=1}1\bigr)
\ehreason{\RefAx{\Axloopunfold}}\\
&\equiv
x\assign0\seq
\bigl((\lnot C_t^+\seq B_t)\loopseq L_t +_{b\lor x=1}1\bigr)
\ehreason{\tCref{looptighten_invariant}}
\intertext{%
    Next, use \Cref{partialselection} to remove the \(x=1\) part of the loop guard:
}
&\equiv
x\assign0\seq x=0\seq
\bigl((\lnot C_t^+\seq B_t)\loopseq
L_t +_{b\lor x=1}1\bigr)
\ehreason{\RefAx{\Axindval}}\\
&\equiv
x\assign0\seq x=0\seq
\bigl((\lnot C_t^+\seq B_t)\loopseq
L_t+_b1\bigr)
\ehreason{\tCref{partialselection},\BA.}
\intertext{%
    The leading assignment now distributes over the guarded choice:
}
&\equiv
x\assign0\seq
\bigl((\lnot C_t^+\seq B_t)\loopseq
L_t +_b1\bigr)
\ehreason{\RefAx{\Axindval}}\\
&\equiv
x\assign0\seq
\bigl((\lnot C_t^+\seq B_t)\loopseq
L_t \bigr)
+_b x\assign0
\ehreason{\eqref{inddistr},\RefAx{\Axneutright}.}
\end{align*}
We now focus on the then-branch of the guarded choice above, pushing the assignment into the left operand of \(\loopseq\), and copying it into the right operand.
\begin{align*}
x\assign0\seq
\bigl((\lnot C_t^+\seq B_t)\loopseq L_t \bigr)
&\equiv
(x\assign0\seq \lnot C_t^+\seq B_t) \loopseq L_t
\ehreason{\RefAx{\Axloopassocassign}}\\
&\equiv
(x\assign0\seq \lnot C_t^+\seq B_t)
\loopseq (x\assign0\seq L_t)
\ehreason{\tCref{loopseqdnr}}
\end{align*}
We then transform the left operand of $\loopseq$ to use the source loop body and invariant:
\begin{align*}
x\assign0\seq \lnot C_t^+\seq B_t
&\equiv
x\assign0\seq x=0\seq \lnot C_t^+\seq B_t
\ehreason{\RefAx{\Axindval}}\\
&\equiv
x\assign0\seq x=0\seq\lnot C_s^+\seq B_t
\ehreason{\eqref{eq:zero-invariants-agree},\BA}\\
&\equiv
x\assign0\seq x=0\seq\lnot C_s^+\seq
(1+_{\lnot(x=1)}\comGoto~\ell)\seq f \seq x\assign0\seq g
\ehreason{\RefAx{\Axskewcomm}}\\
&\equiv
x\assign0\seq x=0\seq\lnot C_s^+\seq f\seq x\assign0\seq g
\ehreason{\RefAx{\Axtestcomb},\tCref{partialselection},\BA, \RefAx{\Axneutleft}} \\
&\equiv
x\assign0\seq\lnot C_s^+\seq f\seq x\assign0\seq g
\ehreason{\RefAx{\Axindval}}\\
&\equiv
x\assign0\seq \lnot C_s^+\seq f\seq g
\ehreason{\tCref{assignmentduplication}}
\end{align*}
Combining the equivalences derived above then gives the equivalence:
\begin{equation}\label{eq:target-zero-fixed}
x\assign0\seq L_t
\equiv
\bigl((x\assign0\seq \lnot C_s^+\seq f\seq g)
\loopseq (x\assign0\seq L_t)\bigr)+_b x\assign0.
\end{equation}
Together, \eqref{eq:source-zero-fixed} and \eqref{eq:target-zero-fixed} tell us that $x \assign 0 \seq L_s$ and $x \assign 0 \seq L_t$ are solutions to $y$ in \eqref{eq:source-target-fixpoint}.
The claim then follows by \Cref{loopuniqueafterlargest}.
\end{proof}

This completes the first step of the plan sketched above; next, we rewrite the expressions for the two programs into a common form, \emph{within their respective environments}.

\begin{lemma}[\RocqLabelRef{ehmeetinthemiddle}]\label{ehmeetinthemiddle}
We can massage $(P_s)_{-}$ and $(P_t)_{-}$ to a common expression:
\begin{align*}
\sigma(P_s) \vdash (P_s)_{-} &\equiv
x \assign 0 \seq (g\loopseq L_s +_c e\seq L_s)\seq\comReturn \dashv \sigma(P_s)
\\
\sigma(P_t) \vdash (P_t)_{-} &\equiv
x \assign 0 \seq (g\loopseq L_s +_c e\seq L_s)\seq\comReturn \dashv \sigma(P_t).
\end{align*}
\end{lemma}
\begin{proof}
The first equivalence can be proved easily enough:
\begin{align*}
\sigma(P_s) \vdash (P_s)_{-}
&\equiv
(\comGoto~\ell+_c1)\seq e\seq L_s\seq\comReturn \dashv \sigma(P_s)
\tag{def.\ reduction} \\
&\equiv \comGoto~\ell\seq e\seq L_s\seq\comReturn +_c e\seq L_s\seq\comReturn \dashv \sigma(P_s)
\ehreason{\RefAx{\Axrightdistr},\RefAx{\Axneutleft}}\\
&\equiv \comGoto~\ell+_c e\seq L_s\seq\comReturn \dashv \sigma(P_s)
\ehreason{\RefAx{\Axrightseqanngoto}} \\
&\equiv (g\loopseq L_s)\seq\comReturn +_c e\seq L_s\seq\comReturn \dashv \sigma(P_s)
\ehreason{\RefAx{\Axsubstitution}} \\
&\equiv x \assign 0 \seq ((g\loopseq L_s)\seq\comReturn +_c e\seq L_s\seq\comReturn) \dashv \sigma(P_s)
\tag*{\Cref{indremoval}, \RefAx{\Axneutleft}} \\
&\equiv x \assign 0 \seq (g\loopseq L_s +_c e\seq L_s)\seq\comReturn \dashv \sigma(P_s)
\tag*{\RefAx{\Axrightdistr}}
\end{align*}

For the second equivalence, we first split the initial \(c\)-choice.
\begin{align*}
(P_t)_{-}
&\equiv (x\assign1+_c x\assign0)\seq(e+_{x=0}1)\seq L_t\seq\comReturn\\
&\equiv
x\assign1\seq(e+_{x=0}1)\seq L_t\seq\comReturn+_c
x\assign0\seq(e+_{x=0}1)\seq L_t\seq\comReturn
\ehreason{\RefAx{\Axrightdistr}}
\end{align*}
If \(x=1\), the entry choice skips \(e\).
We unroll the loop and select the goto $\ell$ branch:
\begin{align*}
\MoveEqLeft[1] x\assign1\seq(e+_{x=0}1)\seq L_t\seq\comReturn \\
&\equiv
x\assign1\seq L_t\seq\comReturn
\ehreason{\tCref{assignmentbranchselectionfalse}, \RefAx{\Axneutleft}} \\
&\equiv
x\assign1\seq(B_t\loopseq L_t)\seq\comReturn
\ehreason{\RefAx{\Axloopunfold},\tCref{assignmentbranchselectiontrue}}\\
&\equiv
x\assign1\seq
((\comGoto~\ell\seq f\seq x\assign0\seq g) \loopseq L_t)
\seq\comReturn
\ehreason{\RefAx{\Axloopassocassign},\tCref{assignmentbranchselectiontrue}} \\
&\equiv
x\assign1\seq\comGoto~\ell
\ehreason{\RefAx{\Axrightseqanngoto},\RefAx{\Axrightloopanngoto}}
\intertext{%
If \(x=0\), the entry choice selects \(e\).
    We copy the assignment and apply \Cref{ehzeroentryloopequiv}:
}
\MoveEqLeft[1] x\assign0\seq(e+_{x=0}1)\seq L_t\seq\comReturn \\
&\equiv
x\assign0\seq e\seq L_t\seq\comReturn
\ehreason{\tCref{assignmentbranchselectiontrue}} \\
&\equiv
x\assign0\seq e\seq x\assign0\seq L_t\seq\comReturn
\ehreason{\tCref{assignmentduplication}}\\
&\equiv
x\assign0\seq e\seq x\assign0\seq L_s\seq\comReturn
\ehreason{\tCref{ehzeroentryloopequiv}}\\
&\equiv
x\assign0\seq e\seq L_s\seq\comReturn
\ehreason{\tCref{assignmentduplication}}
\end{align*}%
Putting this together, we fill in the program \(\bigl((x\assign0\seq g)\loopseq L_t\bigr)\seq\comReturn\) at $\ell$ in $\sigma(P_t)$:
\begin{align*}
& \sigma(P_t) \vdash (P_t)_{-}\\
&\equiv
x\assign1\seq\comGoto~\ell
+_c x\assign0\seq e\seq L_s\seq\comReturn\\
&\equiv
x\assign1\seq
\bigl((x\assign0\seq g)\loopseq L_t\bigr)\seq\comReturn +_c x\assign0\seq e\seq L_s\seq\comReturn
\dashv \sigma(P_t)
\ehreason{\RefAx{\Axsubstitution}}\\
&\equiv
(x\assign1\seq
\bigl((x\assign0\seq g)\loopseq L_t\bigr) +_c x\assign0\seq e\seq L_s) \seq\comReturn
\dashv \sigma(P_t)
\ehreason{\RefAx{\Axrightdistr}}
\intertext{%
    The two leading assignments collapse:
}
&\equiv
((x\assign0 \seq g) \loopseq L_t +_c x\assign0\seq e \seq L_s)\seq\comReturn
\dashv \sigma(P_t)
\ehreason{\RefAx{\Axloopassocassign},\RefAx{\Axindannih}}
\intertext{%
    We copy the zero assignment across \(\loopseq\), then use the zero-entry loop equivalence again:
}
&\equiv
(\bigl((x\assign0\seq g)\loopseq (x\assign0\seq L_t)\bigr) 
+_c x\assign0 \seq e\seq L_s) \seq\comReturn \dashv \sigma(P_t)
\ehreason{\tCref{loopseqdnr}}\\
&\equiv
(\bigl((x\assign0\seq g)\loopseq (x\assign0\seq L_s)\bigr) 
+_c x\assign0\seq e\seq L_s) \seq\comReturn \dashv \sigma(P_t)
\ehreason{\tCref{ehzeroentryloopequiv}}\\
&\equiv
x\assign0\seq (g\loopseq L_s +_c e\seq L_s)\seq\comReturn \dashv \sigma(P_t)
\ehreason{\tCref{loopseqdnr}, \RefAx{\Axloopassocassign}, \text{\eqref{inddistr}}}
\end{align*}%
This completes the proof of the lemma.
\end{proof}

The final step of the proof is to show that we can swap the environments generated by both programs; we then arrive at our desired theorem.

\begin{theorem}[Erosa--Hendren loop case \RocqLabelRef{erosa-hendren-loop-case}]\label{erosa-hendren-loop-case}
For \(P_s\) and \(P_t\) as displayed above, suppose that $x$ is fresh in $P_s$ and $P_t$, that both programs have no free \comGoto or duplicate labels, and that each program is well nested.
Then \(\sigma(P_s)\vdash (P_s)_{-} \equiv (P_t)_{-} \dashv\sigma(P_t)\).
\end{theorem}
\begin{proof}
As suggested, \Cref{ehmeetinthemiddle} gets us most of the way there.
The only obstacle is that the environments on the right-hand side differ.
To overcome this, we note that by \Cref{assignmentprefixenvironment},
\begin{align*}
\sigma(P_s) \vdash (P_s)_{-} &\equiv
x \assign 0 \seq (g\loopseq L_s +_c e\seq L_s) \seq\comReturn \dashv (x \assign 0 \seq \sigma(P_s))
\\
\sigma(P_t) \vdash (P_t)_{-} &\equiv
x \assign 0 \seq (g\loopseq L_s +_c e\seq L_s) \seq\comReturn \dashv (x \assign 0 \seq \sigma(P_t))
\end{align*}
It suffices to show that we can replace the environment $x \assign 0 \seq \sigma(P_s)$ with $x \assign 0 \seq \sigma(P_t)$, that is, that we can satisfy the premise of \Cref{iteratedenvironmentreplacement}.
There are two cases: labels extracted from \(e\), \(f\), or \(g\), and the distinguished label \(\ell\).
For a label \(\ell'\) extracted from \(e\),
\begin{align*}
x\assign0\seq \sigma(P_t)_{\ell'} &\equiv
x\assign0\seq e'\seq L_t\seq\comReturn
\ehreason{\tCref{labelextraction}}\\
&\equiv
x\assign0\seq e'\seq x\assign0\seq L_t\seq\comReturn
\ehreason{\tCref{assignmentduplication}}\\
&\equiv
x\assign0\seq e'\seq x\assign0\seq L_s\seq\comReturn
\ehreason{\tCref{ehzeroentryloopequiv}}\\
&\equiv
x\assign0\seq \sigma(P_s)_{\ell'}
\ehreason{\tCref{assignmentduplication}}
\end{align*}
The derivations for labels extracted from \(f\) and \(g\) are analogous.
For the distinguished label \(\ell\), both calculations begin with the
post-label body \(x\assign0\seq g\).
\begin{align*}
x\assign0\seq \sigma(P_t)_\ell &\equiv
x\assign0\seq
\bigl((x\assign0\seq g)\loopseq L_t\bigr) \seq\comReturn
\ehreason{\tCref{labelextraction}}\\
&\equiv
\bigl((x\assign0\seq g)\loopseq L_t\bigr)\seq\comReturn
\ehreason{\RefAx{\Axloopassocassign},\RefAx{\Axindannih}}\\
&\equiv
\bigl((x\assign0\seq g)\loopseq (x\assign0\seq L_t)\bigr)\seq\comReturn
\ehreason{\tCref{loopseqdnr}}\\
&\equiv
\bigl((x\assign0\seq g)\loopseq (x\assign0\seq L_s)\bigr)\seq\comReturn
\ehreason{\tCref{ehzeroentryloopequiv}}\\
&\equiv
\bigl((x\assign0\seq g)\loopseq L_s\bigr)\seq\comReturn
\ehreason{\tCref{loopseqdnr}}\\
&\equiv
x\assign0\seq ((g\loopseq L_s)\seq\comReturn)
\ehreason{\RefAx{\Axloopassocassign}}\\
&\equiv
x\assign0\seq \sigma(P_s)_\ell
\ehreason{\tCref{labelextraction}}
\end{align*}
This completes the proof.
\end{proof}

Thus, the Erosa--Hendren loop transformation is correct for any program which satisfies the side conditions of \Cref{erosa-hendren-loop-case}.
This result shows that \CFGKAT is sufficiently powerful to verify real-world control-flow transformations in a generic way.

\section{Related Work}\label{sec:related}

Structured programming has long been a topic of interest, and its relation to the \comGoto~statement has been studied extensively \cite{kosaraju,bohmjacopini,petersonCapabilitiesRepeatExit1973}. 
In particular, a key motivation is to eliminate the \comGoto~statements, which are often ``considered harmful''~\cite{gotoharmful}. 
Despite this, \comGoto is still central to many low-level languages, like intermediate representations and assembly, making reasoning about it and other control-flow structures essential to verify and validate transformations performed by (de)compilers~\cite{taming,jumpthreading,transformationspaper,sailr}. 

\noindent
\textit{\bf Control-Flow Validation.} 
Validation techniques check the equivalence of \emph{concrete programs} in a (semi-)automated manner.
For example, translation validation~\cite{pnueli_TranslationValidation_1998,kasampalisTranslationValidationCompilation2021} proves the equivalence of programs by leveraging SMT to establish equivalence of program states at several points in the code.
Such automation can check the correctness of individual instances of (de)compilation~\cite{berger_TranslationValidationLLVMs_2025,dasguptaScalableValidationBinary2020}, fuzzing~\cite{liu_HowFarWe_2020,gorzynski_FuzzFleshRandomisedTesting_2025}, and testing~\cite{leCompilerValidationEquivalence2014}.
A recent effort based on \CFGKAT found a bug in the Ghidra decompiler~\cite{symb-cfgkat}.
However, validation tools cannot establish the correctness of transformation algorithms in general.

\noindent
\textit{\bf Control-Flow Verification.} 
Verification techniques, on the other hand, prove \emph{generic correctness results} of transformation algorithms.
Although these results are general enough to subsume validations, their proofs are known to be challenging. 
Current verification techniques for both compilers~\cite{cheng_DenotationbasedCompositionalCompiler_2026,10.1145/1538788.1538814} and decompilers~\cite{verbeekFormallyVerifiedLifting2022,naus_PosterFormallyVerified_2024,verbeekVerifiablyCorrectLifting2024} often require a deep understanding of the semantics, and often do not support compositional reasoning in the presence of \(\comGoto\).
Indeed, despite the algorithm of~Erosa and Hendren~\cite{taming} being known for more than three decades, we are not aware of any other proof of correctness.
Our paper presents a syntactic, compositional, and complete axiomatization to complement the existing semantic methods, making verification proofs more approachable.

\noindent
\textit{\bf Kleene Algebra With Tests. }
Our work builds upon work on \GKAT~\cite{gkat,gkatcoequations} and \KAT~\cite{kozenkat97}. 
The closest work to ours is the study of non-local control flow by~Kozen~\cite{kozennonlocal}, which compiles \comGoto-programs into \KAT expressions and leverages the semantics and axioms of \KAT to prove equivalences. In contrast, our semantics and equational theory are defined directly on \comGoto-programs.
Besides this, differences with~\cite{kozennonlocal} also arise from fundamental differences between \KAT and \GKAT:
\begin{enumerate*}[label={(\arabic*)}]
	\item Unlike \KAT and \CFGKAT, \GKAT cannot express all deterministic behaviors~\cite{kozenbohmjacopini,tencate_AlgebrasDeterministicComputation_2025}, preventing the reduction from \CFGKAT to \GKAT.
	\item Besides trace equivalence, \GKAT also admits bisimilarity equivalence~\cite{gkatcoequations}, which is a stronger equivalence that matches actions even in non-terminating branches.
\end{enumerate*}
Thus, our proofs and axioms necessarily deviate from those of~\cite{kozennonlocal}.

Another notable feature of \CFGKAT is indicator variables, a feature essential in control-flow transformations~\cite{nomoregoto} and goto removal~\cite{taming}.
\KATb~\cite{katb!} extends \KAT with ``mutable tests'', which are a special case of indicator variables with only two possible values, and \NetKAT~\cite{netkat} allows for an arbitrary set of indicator values. However, unlike \KATb and \CFGKAT, the indicator assignments in \NetKAT do not commute with all actions.
Thus, our work is the first complete axiomatization that supports both arbitrary indicator values and commutativity with primitive tests and actions.

\section{Conclusion and Future Work}\label{sec:conclusion}
We extended \CFGKAT with \comCont and multiple mutable variables, provided a derivative-based operational semantics, and a definition of bisimulation between suitable types of automata. 
We then proposed a set of axioms to reason about equivalence, and on the way to proving their completeness we also proved a fundamental theorem for \CFGKAT expressions and a contextual normal form theorem.
We used completeness to derive within our calculus the correctness of ghost-variable elimination and the loop case of the Erosa--Hendren \comGoto\ elimination algorithm. 
We provide a mechanization of all results in Rocq, which was LLM-assisted from human-written proofs and a Rocq skeleton, and enables a platform for further syntactic verification.
Our results answer questions left open by Zhang et al.~\cite{cfgkat} and Kozen~\cite{kozennonlocal}.

Our axiomatization conserves the local equational reasoning of \GKAT, even in the presence of non-locality.
Due to the weakening axiom (\RefAx{\Axweaken}), non-contextual axioms and congruences may be applied in matching environments, as if reasoning in \GKAT.
This restricts the subtlety of non-local reasoning to the application of the contextual rules.

\begin{credits}
\subsubsection{\ackname}
We are especially grateful to Wojciech R\'{o}\.{z}owski for many helpful discussions about this work.
\end{credits}

\nocite{generalcoinupto,coinductionupto,gkat,universalcoalg,jacobs2017introduction}
\bibliography{bibliography}

\crefalias{section}{appendix}

\end{document}